\documentstyle[12pt,german,catmac,twoside,fleqn]{report}

\catcode`@=11
\ifcase \@ptsize 

\font\tenmsx=msam10
\font\sevenmsx=msam7
\font\fivemsx=msam5

\font\tenmsy=msbm10
\font\sevenmsy=msbm7
\font\fivemsy=msbm5

\font\teneufm=eufm10
\font\seveneufm=eufm7
\font\fiveufm=eufm5

\newfam\msxfam
\newfam\msyfam
\newfam\eufmfam

\textfont\msxfam=\tenmsx  \scriptfont\msxfam=\sevenmsx
\scriptscriptfont\msxfam=\fivemsx

\textfont\msyfam=\tenmsy  \scriptfont\msyfam=\sevenmsy
\scriptscriptfont\msyfam=\fivemsy

\textfont\eufmfam=\teneufm \scriptfont\eufmfam=\seveneufm
\scriptscriptfont\eufmfam=\fiveufm

\or 

\font\elvmsx=msam10 scaled 1100
\font\egtmsx=msam7 scaled 1200
\font\fivemsx=msam5

\font\elvmsy=msbm10 scaled 1100
\font\egtmsy=msbm7 scaled 1200
\font\fivemsy=msbm5

\font\elveufm=eufm10 scaled 1100
\font\egteufm=eufm7 scaled 1200
\font\fiveufm=eufm5

\newfam\msxfam
\newfam\msyfam
\newfam\eufmfam

\textfont\msxfam=\elvmsx  \scriptfont\msxfam=\egtmsx
\scriptscriptfont\msxfam=\fivemsx

\textfont\msyfam=\elvmsy  \scriptfont\msyfam=\egtmsy
\scriptscriptfont\msyfam=\fivemsy

\textfont\eufmfam=\elveufm \scriptfont\eufmfam=\egteufm
\scriptscriptfont\eufmfam=\fiveufm

\or 

\font\twlmsx=msam10 scaled 1200
\font\egtmsx=msam7 scaled 1200
\font\fivemsx=msam5

\font\twlmsy=msbm10 scaled 1200
\font\egtmsy=msbm7 scaled 1200
\font\fivemsy=msbm5

\font\twleufm=eufm10 scaled 1200
\font\egteufm=eufm7 scaled 1200
\font\fiveufm=eufm5

\newfam\msxfam
\newfam\msyfam
\newfam\eufmfam

\textfont\msxfam=\twlmsx  \scriptfont\msxfam=\egtmsx
\scriptscriptfont\msxfam=\fivemsx

\textfont\msyfam=\twlmsy  \scriptfont\msyfam=\egtmsy
\scriptscriptfont\msyfam=\fivemsy

\textfont\eufmfam=\twleufm \scriptfont\eufmfam=\egteufm
\scriptscriptfont\eufmfam=\fiveufm
\fi

\def\hexnumber@#1{\ifcase#1 0\or1\or2\or3\or4\or5\or6\or7\or8\or9\or
        A\or B\or C\or D\or E\or F\fi }
\edef\msx@{\hexnumber@\msxfam}
\edef\msy@{\hexnumber@\msyfam}
\edef\eufm@{\hexnumber@\eufmfam}
\def\Bbb{\ifmmode\let\next\Bbb@\else
 \def\next{\errmessage{Use \string\Bbb\space only in math mode}}\fi\next}
\def\Bbb@#1{{\Bbb@@{#1}}}
\def\Bbb@@#1{\fam\msyfam#1}

\mathchardef\AA="0\msy@41
\mathchardef\BB="0\msy@42
\mathchardef\CC="0\msy@43
\mathchardef\DD="0\msy@44
\mathchardef\EE="0\msy@45
\mathchardef\FF="0\msy@46
\mathchardef\GG="0\msy@47
\mathchardef\HH="0\msy@48
\mathchardef\II="0\msy@49
\mathchardef\JJ="0\msy@4A
\mathchardef\KK="0\msy@4B
\mathchardef\LL="0\msy@4C
\mathchardef\MM="0\msy@4D
\mathchardef\NN="0\msy@4E
\mathchardef\OO="0\msy@4F
\mathchardef\PP="0\msy@50
\mathchardef\QQ="0\msy@51
\mathchardef\RR="0\msy@52
\mathchardef\SS="0\msy@53
\mathchardef\TT="0\msy@54
\mathchardef\UU="0\msy@55
\mathchardef\VV="0\msy@56
\mathchardef\WW="0\msy@57
\mathchardef\XX="0\msy@58
\mathchardef\YY="0\msy@59
\mathchardef\ZZ="0\msy@5A

\mathchardef\frA="0\eufm@41
\mathchardef\frB="0\eufm@42
\mathchardef\frC="0\eufm@43
\mathchardef\frD="0\eufm@44
\mathchardef\frE="0\eufm@45
\mathchardef\frF="0\eufm@46
\mathchardef\frG="0\eufm@47
\mathchardef\frH="0\eufm@48
\mathchardef\frI="0\eufm@49
\mathchardef\frJ="0\eufm@4A
\mathchardef\frK="0\eufm@4B
\mathchardef\frL="0\eufm@4C
\mathchardef\frM="0\eufm@4D
\mathchardef\frN="0\eufm@4E
\mathchardef\frO="0\eufm@4F
\mathchardef\frP="0\eufm@50
\mathchardef\frQ="0\eufm@51
\mathchardef\frR="0\eufm@52
\mathchardef\frS="0\eufm@53
\mathchardef\frT="0\eufm@54
\mathchardef\frU="0\eufm@55
\mathchardef\frV="0\eufm@56
\mathchardef\frW="0\eufm@57
\mathchardef\frX="0\eufm@58
\mathchardef\frY="0\eufm@59
\mathchardef\frZ="0\eufm@5A

\mathchardef\fra="0\eufm@61
\mathchardef\frb="0\eufm@62
\mathchardef\frc="0\eufm@63
\mathchardef\frd="0\eufm@64
\mathchardef\fre="0\eufm@65
\mathchardef\frf="0\eufm@66
\mathchardef\frg="0\eufm@67
\mathchardef\frh="0\eufm@68
\mathchardef\fri="0\eufm@69
\mathchardef\frj="0\eufm@6A
\mathchardef\frk="0\eufm@6B
\mathchardef\frl="0\eufm@6C
\mathchardef\frm="0\eufm@6D
\mathchardef\frn="0\eufm@6E
\mathchardef\fro="0\eufm@6F
\mathchardef\frp="0\eufm@70
\mathchardef\frq="0\eufm@71
\mathchardef\frr="0\eufm@72
\mathchardef\frs="0\eufm@73
\mathchardef\frt="0\eufm@74
\mathchardef\fru="0\eufm@75
\mathchardef\frv="0\eufm@76
\mathchardef\frw="0\eufm@77
\mathchardef\frx="0\eufm@78
\mathchardef\fry="0\eufm@79
\mathchardef\frz="0\eufm@7A
\catcode`@=11

\if@twoside
\def\ps@headings{\let\@mkboth\markboth
 \def\@oddfoot{}\def\@evenfoot{}
 \def\@evenhead{\protect\rule[-1.5mm]{\textwidth}{0.1mm}\hspace{-\textwidth
}\rm \thepage\hfil  \leftmark}
 \def\@oddhead{{\protect\rule[-1.5mm]{\textwidth}{0.1mm}\hspace{-\textwidth
} \rightmark}\hfil \rm\thepage}
 \def\chaptermark##1{\markboth {{\ifnum \c@secnumdepth >\m@ne
\thechapter. \ \fi ##1}}{}}%
 \def\sectionmark##1{\markright {{\ifnum \c@secnumdepth >\z@
   \thesection. \ \fi ##1}}}}
\else
\def\ps@headings{\let\@mkboth\markboth
\def\@oddfoot{}\def\@evenfoot{}
\def\@oddhead{{\protect\rule[-1.5mm]{\textwidth}{0.1mm}\hspace{-\textwidth}
 \rightmark}\hfil \rm\thepage}
\def\chaptermark##1{\markright {{\ifnum \c@secnumdepth >\m@ne
\thechapter. \ \fi ##1}}}}
\fi

\def\ps@plain{\let\@mkboth\@gobbletwo
 \def\@oddfoot{}\def\@evenfoot{}
 \def\@evenhead{\protect\rule[-1.5mm]{\textwidth}{0.1mm}\hspace{-\textwidth
}\rm \thepage\hfil }
 \def\@oddhead{{\protect\rule[-1.5mm]{\textwidth}{0.1mm}\hspace{-\textwidth
} }\hfil \rm\thepage}}

\def\chapter{\clearpage   \thispagestyle{empty}%
\global\@topnum\z@   \@afterindentfalse   \secdef\@chapter\@schapter}
\def\tableofcontents{\@restonecolfalse
  \if@twocolumn\@restonecoltrue\onecolumn\fi
  \chapter*{\contentsname
        \@mkboth{\contentsname}{\contentsname}}%
 \@starttoc{toc}\if@restonecol\twocolumn\fi}
\if@twoside
\newcommand{\mymark}[1]{\markboth{#1}{}}
\else
\newcommand{\mymark}[1]{\markboth{#1}{#1}}
\fi
\def\thebibliography#1{\chapter*{\bibname\@mkboth
  {\bibname}{}}\list
  {\@biblabel{\arabic{enumiv}}}{\settowidth\labelwidth{\@biblabel{#1}}%
    \leftmargin\labelwidth
    \advance\leftmargin\labelsep
    \usecounter{enumiv}%
    \let\p@enumiv\@empty
    \def\theenumiv{\arabic{enumiv}}}%
    \def\newblock{\hskip .11em plus.33em minus.07em}%
    \sloppy\clubpenalty4000\widowpenalty4000
    \sfcode`\.=\@m}
\newtheorem{Def}{Definition}[section]
\newtheorem{satz}[Def]{Satz}

\newtheorem{kor}[Def]{Korollar}
\newtheorem{lem}[Def]{Lemma}

\newtheorem{bem}[Def]{Bemerkung}

\newtheorem{beis}[Def]{Beispiel}
\newtheorem{ver}[Def]{Vermutung}
\newcommand{\ben}{\begin{equation}}
\newcommand{\een}{\end{equation}}
\newcommand{\benn}{\begin{equation*}}
\newcommand{\eenn}{\end{equation*}}
\newcommand{\bea}{\begin{eqnarray}}
\newcommand{\eea}{\end{eqnarray}}
\def\lra{\longrightarrow}

\def\lmt{\longmapsto}
\def\wt{\widetilde}
\def\bgamma{\mbox{\boldmath$\gamma$}}
\newcommand{\Vir}{\mathop{\rm Vir}\nolimits}
\newcommand{\pr}{\mathop{\rm pr}\nolimits}
\newcommand{\Lin}{\mathop{\rm Lin}\nolimits}
\newcommand{\ord}{\mathop{\rm ord}\nolimits}
\newcommand{\End}{\mathop{\rm End}\nolimits}
\newcommand{\Id}{\mathop{\rm Id}\nolimits}
\newcommand{\Hom}{\mathop{\rm Hom}\nolimits}
\newcommand{\Rad}{\mathop{\rm Rad}\nolimits}
\newcommand{\plus}{\mathop{\oplus}\limits}
\newcommand{\mod}{\mathop{\rm mod}\nolimits}
\newcommand{\im}{\mathop{\rm im}\nolimits}
\renewcommand{\Re}{\mathop{\rm Re}\nolimits}

\newcommand{\coker}{\mathop{\rm koker}\nolimits}
\newcommand{\dst}{\displaystyle}
\newcommand{\restrict}[1]{\raisebox{-0.7ex}{$\Big|_{\textstyle{#1}}$}}

\newcommand{\strich}{\relbar\joinrel\relbar\joinrel}

\newcommand{\quot}[2]%
{\begingroup\setbox254\vbox{$#1$}     \setbox253\vbox{$#2$}
 \dimen255\ht254 \advance\dimen255 by \ht253
 \dimen254\dimen255 \divide\dimen254 by 2 \dimen253 1ex
 \advance\dimen253 by \dimen254 \advance\dimen253 by -\ht254
 \dimen252\dimen253 \advance\dimen252 by -\ht253
 \dimen251\dimen253 \advance\dimen251 by \dimen252 \divide\dimen251 by 2
 \mbox{${\raisebox{\dimen253}{$#1$}}\mkern-5mu%
\raisebox{\dimen251}{$\left/\rule{0pt}{0.75\dimen255}\right.$}%
\mkern-5mu{\raisebox{\dimen252}{$#2$}}$}\endgroup}

\newcommand{\qquot}[2]{\,{}^\ifinner{#1}\!\!\big/\!_{#2}\else%
{\displaystyle #1}\!\!\Big/\!_{\displaystyle #2}\fi}

\begin{document}
\thispagestyle{empty}
\vspace*{20mm}
\begin{center}
{\LARGE \bf Vertex--Operatoren,\\
Darstellungen der Virasoro--Algebra \\[3mm]
und konforme Quantenfeldtheorie}
\\[17mm]
Dissertation\\zur Erlangung des Doktorgrades\\der Naturwissenschaften
\\[22mm]
vorgelegt beim Fachbereich Mathematik\\der Johann Wolfgang
Goethe--Universit"at\\in Frankfurt am Main\\[27mm]
von\\ Wolfram Boenkost\\aus Frankfurt
\vfill
Frankfurt 1994\\
{\scriptsize (D F 1)}
\end{center}
\newpage\thispagestyle{empty}
\vspace*{13cm}
\noindent vom Fachbereich Mathematik der\\
Johann Wolfgang Goethe--Universit"at als Dissertation angenommen.
\vfill
\noindent Dekan: Prof. W. Schwarz\\[5mm]
Gutachter: Prof. F. Constantinescu und Prof. H. F. de Groote\\[5mm]
Datum der Disputation: 11.7.1994
\newpage\thispagestyle{empty}
\vfill
\noindent{\bf Zusammenfassung}\\[5mm]
In dieser Arbeit werden die mathematischen Grundlagen zur Konstruktion der
prim"aren Felder der minimalen Modelle der konformen Quantenfeldtheorie
beschrieben.
Wir untersuchen Verma-- und Fock--Moduln der Virasoro--Algebra und
klassifizieren diese Moduln bez"uglich der Struktur der (ko--) singul"aren
Vektoren. Wir definieren die Vertex--Operatoren zwischen gewissen
Fock--Moduln (die eine kanonische Hilbert\-raumstruktur besitzen) und
beweisen verschiedene Eigenschaften dieser Operatoren: Unter bestimmten
Voraussetzungen sind Vertex--Operatoren  dicht definierte, nicht
abschlie\3bare  Operatoren zwischen den Fock--Moduln. Radialgeordnete
Produkte von Vertex--Operatoren existieren auf einem dichten Teilraum. Wir
beweisen Kommutatorrelationen zwischen Vertex--Operatoren und den
Generatoren der Virasoro--Algebra. Dann definieren wir die integrierten
Vertex--Operatoren und zeigen, da\3 diese Operatoren im wesentlichen wieder
die Eigenschaften der nichtintegrierten Vertex--Operatoren haben. Gewisse
integrierte Vertex--Operatoren k"onnen mit konformen Felder identifiziert
werden. Ein unter den Vertex--Operatoren invarianter Unterraum der
Fock--Moduln kann mit dem physikalischen Zustandsraum identifiziert
werden.
\vfill
\newpage\thispagestyle{empty}
\phantom{.}
\newpage

\pagenumbering{roman}
\tableofcontents
\newpage
\pagenumbering{arabic}

\newpage
\section*{Einleitung}
\addcontentsline{toc}{chapter}{Einleitung}
\mymark{Einleitung}
\subsection*{Physikalischer Hintergrund}
Systeme der klassischen statistischen Mechanik am kritischen Punkt
besitzen keine nat"urliche L"angenskala, sie sind skaleninvariant. Nach der
Wilson'schen Idee der Re\-nor\-mierungs\-gruppe wird das Verhalten des
Systems am kritischen Punkt durch einen Fixpunkt der Renormierungsgruppe
 beschrieben.

Man kennt verschiedene Beispiele, in denen Modelle der  klassischen
statistischen Mechanik
im Kontinuumlimes  euklidischen Quantenfeldtheorien
entsprechen.  Falls das System an einem kritischen Punkt ist, so
entspricht der Renormierungsgruppenfixpunkt nach einer Vermutung von
Polyakov  einer nicht nur skaleninvarianten, sondern sogar konform
invarianten euklidischen Quantenfeldtheorie. Von dieser konformen
Quantenfeldtheorie
 kann man Osterwalder--Schrader--Positivit"at annehmen, falls das
statistische System eine lokale Wechselwirkung hat (z.~B. n"achste
Nachbarn) und eine selbstadjungierte Transfermatrix besitzt.

Eine weitere Hypothese in der Theorie der kritischen Ph"anomene besagt, da\3
die ``kritischen Exponenten'', die das Verhalten der sogenannten
Ordnungsparameter in der N"ahe des kritischen Punktes beschreiben, nicht
mehr von speziellen Eigenschaften des Systems wie z.~B. der genauen Form
der Wechselwirkung abh"angen, vielmehr haben ganze Klassen von Modellen die
gleichen kritischen Exponenten und damit das gleiche Verhalten am
kritischen Punkt. Dies wird mit der Universalit"at der kritischen
Exponenten bezeichnet.

Der springende Punkt ist nun, da\3 der Formalismus der konformen
Quantenfeldtheorie in zwei Dimensionen es erlaubt, die kritischen
Exponenten zu berechnen, man kann auf diese Weise bestimmte konforme
Theorien mit Modellen der statistischen Mechanik, deren kritische
Exponenten bekannt sind, identifizieren. Hier seien dabei das
Isingmodell, das trikritische Isingmodell und das 3---state Potts--Modell
genannt \cite{car1,ZZ}. Aus diesem Grund gab (und gibt) es ein gro\3es
Interesse an der Klassifikation aller konformen Theorien, auf diesem Wege
lie\3en sich dann auch alle  kritischen Punkte klassifizieren.  Diese
Klassifikation aller
 konformen Quantenfeldtheorien ist bis heute nicht vollendet, es wurden
aber eine ganze Reihe von Serien konformer Theorien gefunden, siehe z.~B.
 \cite{BG}.
Die erste Serie von konformen Modellen, die von A.~Belavin, A.~Polyakov
 und A.~B.~Zamolodchikov \cite{BPZ} entdeckte Serie der minimalen
Modelle, ist der Gegenstand dieser Arbeit.

Einen weiteren Grund f"ur das  Interesse an konformer Quantenfeldtheorie
in zwei Dimensionen ist die Stringtheorie. Man erhofft sich, durch die
Untersuchung von konformen Quantenfeldtheorien einen "Uberblick "uber
m"ogliche Stringmodelle zu verschaffen. Jede konforme Quantenfeldtheorie
liefert einen m"oglichen Grundzustand f"ur ein Stringmodell.
Bemerkenswerterweise spielen auch hier die zweidimensionalen Theorien eine
entscheidende Rolle; die Weltfl"ache eines Strings ist eine
zweidimensionale Mannigfaltigkeit mit einer komplexen Struktur, d.~h.
eine Riemannsche Fl"ache. Auch die Feynman--Graphen werden in der
String--Theorie durch Riemannsche Fl"achen ersetzt. Deshalb interessiert
man sich in diesem Rahmen insbesondere f"ur konforme Theorien auf
Riemannschen Fl"achen von beliebigem Geschlecht. Wir werden aber nur
Theorien in der komplexen Ebene bzw. auf $\PP^1$ betrachten.

Wir wollen nun etwas genauer auf die konforme Invarianz einer Theorie
eingehen, vgl. hierzu \cite{st.a,car1,gin}. Ein Diffeomorphismus  einer
(Pseudo--) Riemannschen Mannigfaltigkeit $M$ hei\3t konform, wenn er die
Metrik nur um einen $C^{\infty}(M)$--Faktor "andert. Deshalb enthalten die
konformen Transformationen auch die Poincar\`e--Transformationen, da diese
die Metrik invariant lassen. F"ur $M=\RR^d$, $d\ge 2$ mit kanonischer
Metrik $g$ der Signatur $(p,q)$ hat die Gruppe der (globalen) konformen
Transformationen die Dimension $\frac{(d+1)(d+2)}{2}$, ist also
insbesondere endlichdimensional. $M=\RR^2$ mit der euklidischen Metrik ist
von besonderer Bedeutung, die konformen Transformationen entsprechenden
dann den holomorphen und antiholomorphen Funktionen, die Symmetriegruppe
ist also unendlichdimensional.
Nur die Moebius--Transformationen (und ihre antiholomorphen Partner)
erzeugen globale konforme Transformationen, jede (anti--)holomorphe
Funktion erzeugt aber eine lokale konforme Transformation.

Um den Unterschied zwischen lokalen und globalen Transformationen zu
eliminieren, geht man zu den infinitesimalen Generatoren konformer
Transformationen "uber, die durch $l_n=-z^{n+1}\frac{\partial}{\partial z}$
f"ur $n\in \ZZ$ gegeben sind. $l_n$ definieren meromorphe Vektorfelder auf
$\PP^1$ und erf"ullen die Witt--Algebra
\ben\label{g01}
[l_n,l_m]=(n-m) l_{n+m}.
\een
$l_{-1},l_0,l_1$ sind die infinitesimalen Generatoren der
Moebius--Transformationen, sie bilden eine $sl(2,\CC)$--Unteralgebra.
Analog erzeugen die antiholomorphen Funktionen eine Witt-Algebra, deren
Generatoren wir mit $\bar{l}_n$ bezeichnen.

Auf dem Quantenlevel erh"alt man als Symmetriealgebra die
Virasoro--Algebra, die die eindimensionale zentrale Erweiterung der
Witt-Algebra ist. Die zentrale Erweiterung kann man als Schwinger--Term
oder Anomalie interpretieren, bei der Quantisierung wird die klassische
Symmetrie zerst"ort. Die Relationen der Virasoro--Algebra sind
\ben\label{g02}
[e_n,e_m]=(n-m)e_{n+m}+\frac{n(n^2-1)}{12} z \delta_{n,-m}.
\een
Die komplexe Lie--Algebra, erzeugt durch $e_n,\;(n\in\ZZ)$ und $z$
bezeichnen wir mit Vir.
Bemerkenswerterweise bilden $e_{-1},e_0,e_1$ wieder eine
$sl(2,\CC)$--Unteralgebra, die globale konforme Symmetrie wird also durch
den Schwinger--Term nicht beeinflu\3t. Die volle Symmetrie--Algebra ist
(bei den von uns untersuchten Modellen) die direkte Summe $\Vir\oplus
\overline{\Vir}$.
\subsection*{Konforme Quantenfeldtheorie $\cap$ Mathematik}
\noindent Der Zustandsraum $\HH$ der konformen Quantenfeldtheorie ist  ein
Virasoro--Modul.\\
Aus physikalischen Gr"unden untersucht man graduierte H"ochstgewichtsmoduln
von Vir, denn es soll im Zustandsraum einen zyklischen Vektor (den
Vakuumvektor) minimaler Energie geben. Die Graduierung wird von
 $e_0$--Eigenr"aumen erzeugt.

Die geeignete Klasse von Darstellungen von Vir liefern die Verma--Moduln
$V(h,c)=\plus_{n\ge 0} V(h,c)_n$ f"ur $(h,c)\in\CC^2$, wobei $z$ durch
$c\Id$ dargestellt wird  und f"ur $x\in V(h,c)_n$ $e_0x=(h+n)x$ gilt.
Moduln mit diesen Eigenschaften hei\3en Moduln vom Typ $(h,c)$. $V(h,c)_n$
haben die Dimension $p(n)$, wobei $p(n)$ die Partitionsfunktion ist.
Die Bausteine des Zustandraumes sind die irreduziblen Quotienten von
 Verma--Moduln, diese Moduln bezeichnen wir mit $L(h,c)$. $c$ ist durch
das Modell festgelegt, der Zustandsraum hat also die Form
\ben\label{G03}
\HH=\plus_{i\in I} L(h_i,c)\otimes L(\bar{h}_i,c).
\een
F"ur Modelle ohne Spin (wie wir sie hier ausschlie\3lich betrachten) gilt
$h_i=\bar{h}_i$. Eine der Eigenschaften der konformen Quantenfeldtheorie
ist es, da\3 zu jedem Summand in (\ref{G03}) ein spezielles ``prim"ares''
Feld assoziert ist. Ein prim"ares Feld ist dabei "uber eine spezielle Form
des Kommutators mit den Elementen von Vir charakterisiert. Die prim"aren
Felder kann man analog zu (\ref{G03}) zerlegen, die Faktoren, die man
dabei erh"alt, nennt man konforme Felder. A.~Belavin, A.~Polyakov und
A.~B.~Zamolodchikov l"osten in der bahnbrechenden Arbeit \cite{BPZ} das
Problem, f"ur bestimmte Werte von $c$ endliche Indexmengen $I$ zu finden,
so da\3 die von den prim"aren Feldern erzeugte Operatoralgebra schlie\3t.
Diese so erhaltenen Modelle sind genau die minimalen Modelle.

Aus der vorhergegangenen Diskussion ergeben sich zwei Aufgaben f"ur die
mathematische Behandlung der konformen Quantenfeldtheorie:\\
Man mu\3 die Darstellungstheorie der Symmetrie--Algebra, in diesem Fall der
Virasoro--Algebra untersuchen, und danach mu\3 man die konformen Felder
zwischen bestimmten Moduln  der Virasoro--Algebra konstruieren.

\newpage
\noindent Zun"achst zur Darstellungstheorie von Vir:

Auf den Verma--Moduln gibt es eine kanonische kontravariante Form, die
Shapovalov--Form. Diese Form zerf"allt in eine direkte Summe von Formen
auf $V(h,c)_n$.  Aussagen "uber die Irreduzibilit"at der Verma--Moduln
lassen sich auf Aussagen "uber die Shapovalov--Form zur"uckf"uhren. Diese
Form induziert eine Form auf $L(h,c)$ und somit auch auf $\HH$. Eine
Frage, die in \cite{BPZ} nicht beantwortet wurde war die, ob diese Form
positiv definit ist und man damit $\HH$ zu einem Hilbertraum machen kann.
In diesem Fall w"are die konforme Quantenfeldtheorie auch positiv im Sinne
von Osterwalder--Schrader.

Gerade diese Bedingung der Unitarit"at hat die Entwicklung der  konformen
Quantenfeldtheorie im Sinne der axiomatischen Feldtheorie verz"ogert
\cite{Mack}, es ist ein hochgradig nichttriviales Problem, unit"are
Darstellungen der Virasoro--Algebra zu finden. Dieses Problem wurde erst
von \cite{GKO,FQS} gel"ost. Es zeigte sich, da\3 die unit"aren Modelle eine
Teilmenge der minimalen Modelle sind. Die erw"ahnten Modelle der
statistischen Mechanik lassen sich mit bestimmten unit"aren Modellen
identifizieren.

V. Kac stellte in \cite{Kac1} eine Vermutung "uber  die Determinanten der
Shapovalov--Form als Funktion von $(h,c)$ auf. Aufgrund der nichtlinearen
Abh"angigkeit von $n$ im Kommutator (\ref{g02}) treten in dieser
Determinante kompliziertere Terme auf als in vergleichbaren
Determinantenformeln f"ur Kac--Moody--Algebren \cite{Kac2}.

B. Feigin und D. Fuks gelang in \cite{FF} ein Beweis dieser
Determinantenformel. Inzwischen gibt es eine ganze Reihe von Beweisen f"ur
diese Formel \cite{KR,TK}. Wir geben einen von \cite{TK,RC1,Thorn}
motivierten Beweis unter Verwendung von Ergebnissen aus Kapitel 4. (Die
Ergebnisse dieses Kapitels sind selbstverst"andlich unabh"angig von der
Kac--Formel.)

\noindent Zur Konstruktion der konformen Felder:

V.~Dotsenko und V.~Fateev gelang es in \cite{DF1}, die konformen Felder
zu realisieren. Dabei gingen sie vom freien bosonischen Feld $\Phi$ aus
und fanden, da\3 die Korrelationen von
$:e^{\Phi}:$ denen der konformen Theorie  mit $c=1$ entsprechen. (: :
bezeichnet die Wick--Ordnung, ohne die das Exponential nicht existiert.)
$:e^{\Phi}:$ bezeichnet man als (freien) Vertex--Operator. Durch geeignete
 Abschirmungen der Vertex--Operatoren konnten sie Realisierungen der
Korrelationen konformer Felder f"ur alle minimalen Modelle finden. In
\cite{DF2} verwendeten sie diese Realisierung, um Integraldarstellungen
f"ur die Vierpunktfunktionen der minimalen Modelle anzugeben.

G. Felder ging in \cite{Fe1} weiter und realisierte die
Vertex--Operatoren im bosonischen Fock--Raum, wie es "ublicherweise in der
Stringtheorie gemacht wird. Auch die Virasoro--Algebra l"a\3t sich im
Fock--Raum "uber Erzeuger-- und Vernichter--Operatoren realisieren, man
kann auf diese Weise Fock--Moduln vom Typ $(h,c)$ definieren.
Es liegt nun nahe, die Fock--Moduln als Ersatz von $L(h,c)$ zu verwenden.

Der Vorteil dieser Moduln ist, da\3 man eine kanonische
Hilbertraumstruktur geschenkt bekommt und deshalb die Frage der
Unitarit"at von Darstellungen zun"achst keine Rolle spielt. Andererseits
sind die Fock--Moduln komplizierter als die Verma--Moduln, sie sind i.~A.
auch keine H"ochstgewichtsmoduln.

Bestimmte abgeschirmte Vertexoperatoren erf"ullen dieselben Kommutatoren
mit den Elementen von Vir wie die konformen Felder. Der entscheidende
Fortschritt von G.~Felder war nun, da\3 er die Beziehung zwischen den
Verma--Moduln (oder den Moduln $L(h,c)$) und den Fock--Moduln der
minimalen Modelle auf elegante Weise kl"aren konnte: Er konnte zeigen, da\3
ein Ko--Rand--Operator $Q$ zwischen bestimmten Fock--Moduln existiert, so
da\3 $L(h,c)$ eine Kohomologiegruppe einer Sequenz von Fock--Moduln ist.
$Q$ wird auch als BRST--Operator bezeichnet (BRST steht f"ur Becchi, Rouet,
Stora und Tyupin), in der Stringtheorie werden BRST--Operatoren verwendet
um den Zustandsraum frei von ``Ghost--Zust"anden'' zu machen \cite{kaku}.
Formal hat $Q$ auch hier genau diese Funktion, aber nicht die
physikalische Interpretation aus der Stringtheorie.

Weiter konnte G.~Felder zeigen, da\3 die abgeschirmten Vertex--Operatoren
Ketten--Abbildungen sind und sie damit Abbildungen auf $L(h,c)$
induzieren, die mit den konformen Feldern identifiziert werden k"onnen.

Zur mathematischen Genauigkeit der Arbeiten von V.~Dotsenko, V.~Fateev und
G.~Felder ist folgendes zu sagen: Die Arbeiten \cite{DF1,DF2} sind
physikalischer Natur, die erw"ahnte Identifizierung verwendet zur
Berechung von Korrelationen von Vertex--Operatoren das Funktionalintegral.
In der Arbeit \cite{Fe1} wird auf das Funktionalintegral zugunsten der
Realisierung der Vertex--Operatoren auf Fock--Moduln verzichtet. Die
Vertex--Operatoren selber werden aber nur auf formale Weise behandelt.

Dies war der Ansatzpunkt und die Motivation dazu, die Vertex--Operatoren
als Operatoren im Hilbertraum zu untersuchen und allgemeine Eigenschaften
dieser Operatoren zu beweisen, die es unter anderem erlauben, die von
G.~Felder angegebene Konstruktion auf mathematisch korrekte Weise
durchzuf"uhren.

\noindent Wir beschreiben nun kurz den Inhalt der einzelnen Kapitel.

In Kapitel 1 f"uhren wir die Verma-- und Fock--Moduln ein. Wir definieren
die Shapovalov--Form und geben Determinantenformeln an und geben einen
Beweis, der Ergebnisse aus Kapitel 4 verwendet. Au\3erdem behandeln wir
kurz die Ergebnisse "uber die Unitarit"at von Vir--Moduln.

In Kapitel 2 verwenden wir die Kac--Determinantenformel, um die Verma--
und Fock--Moduln genauer zu analysieren. Es gibt in diesen Moduln eine
ausgezeichnete Menge von Vektoren, die (ko--)singul"aren Vektoren. Im
Falle der Verma--Moduln, die nur singul"are Vektoren enthalten, erzeugt
jeder singul"are Vektor einen Untermodul, der wieder isomorph zu einem
Verma--Modul ist. Mit Hilfe der Jantzen--Filtration klassifizieren wir die
Verma--Moduln nach der Struktur der singul"aren Vektoren, gleichzeitig
erh"alt man die Charakterisierung aller Untermoduln von und Homomorphismen
zwischen Verma--Moduln.\\
Danach untersuchen wir die Fock--Moduln. Hier ist die Situation
komplizierter, da die Fock--Moduln  auch ko--singul"are Vektoren enthalten
k"onnen und i.~allg. auch keinen zyklischen Vektor besitzen. Auch die
Fock--Moduln  k"onnen wir nach der Struktur der (ko--)singul"aren Vektoren
klassifizieren.\\
Die Ergebnisse von Kapitel 2 sind bis auf einen Fehler in der
Klassifikation der Fock--Moduln in \cite{FF} enthalten. Der Beweis der
Klassifikation der Verma--Moduln verwendet aber andere (elementare)
Methoden, die von \cite{RCW} motiviert sind. Bei der Klassifikation der
Fock--Moduln verwenden wir die Ideen aus \cite{FF}. Diesen Teil kann man
als Ausarbeitung von \cite{FF} sehen.

In Kapitel 3 definieren wir die Vertex--Operatoren als unbeschr"ankte
Operatoren im Fock--Raum. Wir zeigen, da\3 diese Operatoren unter gewissen
Voraussetzungen dicht definiert sind und Produkte dieser Operatoren auf
dichten Teilmengen existieren. Die Beziehung zur Virasoro--Algebra
entsteht, wenn wir Kommutatoren zwischen (Produkten von)
Vertex--Operatoren und Elementen der Virasoro--Algebra beweisen.
Dann zeigen wir eine n"utzliche Faktorisierung der Vertex--Operatoren in
einen Hilbert--Schmidt--Operator und einen diagonalen, selbstadjungierten
Operator. Damit k"onnen wir die Vertex--Operatoren auf einen einfach zu
charakterisierenden Definitionsbereich fortsetzen. Wir beenden dieses
Kapitel mit einem Beweis, da\3 die Vertex--Operatoren nicht abschlie\3bar
sind.\\
Die Ergebnisse von Kapitel 3 sind teilweise ver"offentlicht in
\cite{BC,Boe}.

In Kapitel 4 f"uhren wir die integrierten Vertex--Operatoren ein, die auch
als abgeschirmt bezeichnet werden. Zun"achst konstruieren wir so einen
nichttrivialen Inter\-twiner zwischen gewissen Fock--Moduln; die Existenz
dieses Intertwiners erlaubt den Beweis der Kac--Determinantenformel. Dann
f"uhren wir allgemeinere integrierte Vertex--Operatoren ein und zeigen,
da\3 das Produkt von abgeschirmten Vertex--Operatoren unter bestimmten
Voraussetzungen existiert.

In Kapitel 5 kommen wir zur Anwendung unserer Ergebnisse auf die minimalen
Modelle der konformen Quantenfeldtheorie. Nach einer kurzen Einleitung
"uber die konforme Quantenfeldtheorie und die minimalen Modelle definieren
wir die konformen Felder zwischen den Fock--Moduln. Dann definieren wir
den Ko--Rand--Operator $Q$, der im wesentlichen der Intertwiner aus
Kapitel 4 ist. Wir zeigen, da\3 eine Sequenz von Fock--Moduln existiert,
deren Kohomologiegruppe der entsprechende irreduzible Vir--Modul ist. Wir
definieren eine Hilbertraumstruktur auf der Kohomologiegruppe und zeigen,
da\3 die konformen Felder mit $Q$ kommutieren und dicht definierte
Operatoren in den Kohomologiegruppen induzieren. Diese k"onnen wir mit den
physikalischen konformen Feldern identifizieren, wir haben somit die
prim"aren Felder konstruiert.
\newpage\thispagestyle{plain}\hbox{}\newpage
\chapter{Die Virasoro--Algebra}
\section{Definitionen}
Die Virasoro--Algebra wurde von Physikern in der String--Theorie als die
Moden--Algebra des Energie--Impuls Tensors entdeckt, vgl. S. Mandelstam,
\cite{Mandel}. Genau
diese Rolle spielt sie auch in der konformen Quantenfeldtheorie. Konkret
ist die
Virasoro--Algebra die (eindimensionale) zentrale Erweiterung der
Witt--Algebra, der Lie--Algebra der polynominalen Vektorfelder auf $S^1$.
Die Virasoro--Algebra ist die komplexe
Lie--Algebra gegeben durch
\ben\label{virrel}
\Vir = \plus_{n\in\ZZ} \CC\, e_n\plus \CC\, z, \qquad
[e_n,e_m]=(n-m)e_{n+m}+\frac{n^3-n}{12} z \delta_{n,-m}.
\een
Mit $\Vir_{\pm}=\plus_{n{>\atop <} 0} \CC \,e_n$ und $H=\CC \,z \plus
\CC\, e_0$ haben wir eine Cartan--Zerlegung $\Vir=\Vir_-\plus H \plus
\Vir_+$. Vir wird eine ($\ZZ$--)graduierte Lie--Algebra, wenn wir $\deg
e_n=-n$ und $\deg z=0$ setzen.

Wir wiederholen nun einige Begriffe der Darstellungstheorie.\\
Sei $M$ ein Vir--Modul. $M$ hei\3t graduiert, wenn $M=\plus_{k\in\ZZ} M_k$
mit $e_n(M_k)\subset M_{k-n}$ gilt. Ein graduierter Modul $M$ hei\3t von
endlichem Typ, falls $\dim M_k <\infty$ f"ur alle $k$ gilt. Zu jedem Modul
$M$ von endlichem Typ definieren wir den dualen Modul $M'$ durch $e'_n =
(e_n : M_k \lra M_{k-n})'$ und den kontragredienten Modul $\overline{M}$
 durch $\bar{e}_n : (e_{-n} : M_k \lra M_{k+n})'$.

Sei $\frU(\Vir)$ die universelle einh"ullende Algebra von Vir. $M$ hei\3t
ein H"ochstgewichtsmodul, wenn ein Vektor $v\in M_0$  mit $\frU(\Vir)v =  M$
und $\Vir_+v=0$ existiert.  $v$ wird als
H"ochstgewichtsvektor oder Vakuumvektor bezeichnet.

Seien $h,c \in \CC$. Ein graduierter Vir--Modul $M$ hei\3t vom Typ $(h,c)$,
wenn $zm=cm$ f"ur alle $m\in M$ und  $e_0m=(h+j)m$ f"ur $m\in M_j$ gilt.
Jeder irreduzible graduierte Vir--Modul von endlichem Typ ist ein Modul
vom Typ $(h,c)$ f"ur geeignete $h,c\in \CC$.
Ein von Null verschiedener Vektor $w\in M$ hei\3t singul"arer Vektor (vom
Typ $(h,c)$), falls $\Vir_+w=0$, $e_0 w=hw$ und $zw=cw$ gilt. Dual dazu
hei\3t
ein Vektor $w\in M$ vom Grad $n$ kosingul"arer Vektor, falls $w\notin
\Vir_-(\plus_{k<n} M_k)$ gilt.
\section{Die Verma--Moduln und die Kac--Formel}
Wir definieren nun Verma--Moduln: F"ur $h,c\in \CC$ ist $V(h,c)$ der Modul
 vom Typ $(h,c)$, der als $\Vir_-$--Modul ein freier Modul mit einem
erzeugendem Element $v_{h,c}\in V(h,c)_0$ ist. Die Vektoren $e_{-j_1}
\cdots e_{-j_s}v_{h,c}$, $j_1 \ge\cdots \ge j_s >0$, $s\ge 0$ bilden dann
eine
Basis von $V(h,c)$. Es gilt $\dim V(h,c)_n=p(n)$, wobei $p(n)$ die
Partitionsfunktion ist. Die Verma--Moduln haben die bemerkenswerte
Eigenschaft der Ko--Universalit"at, d.h zu jedem Modul $M$ mit einem
singul"aren Vektor $v$ vom Typ $(h,c)$ gibt es genau einen
Vir--Homomorphismus $V(h,c)\lra M$, der $v_{h,c}$ auf $v$ abbildet. Dual
dazu sind die kontragredienten Verma--Moduln universell, d.h. zu jedem
Modul $M$ mit einem singul"aren
Vektor $v$ vom Typ $(h,c)$ gibt es genau einen Homomorphismus $M\lra
\overline{V}(h,c)$, der $v$ auf $\bar{v}_{h,c}$ abbildet.

Es gibt genau einen irreduziblen H"ochstgewichtsmodul vom Typ $(h,c)$,
diesen Modul bezeichnen wir mit $L(h,c)$. Mit Hilfe der Verma--Moduln
k"onnen wir diesen Modul konstruieren:
Die Verma--Moduln sind, da sie von einem zyklischen Vektor erzeugt
werden,
unzerlegbar. Man "uberlegt sich leicht, da\3 $V(h,c)$
 einen gr"o\3ten echten Untermodul $M(h,c)$ enth"alt, der durch
\[
M(h,c)=\left\{x\in V(h,c):\frU(\Vir) x \subset \plus_{n>0}
V(h,c)_n\right\}
\]
gegeben ist. Es gilt dann $L(h,c)=V(h,c)/M(h,c)$.

Das wichtigste Hilfsmittel zur Untersuchung der Verma--Moduln ist die
Shapovalov--Form bzw. die Shapovalov--Abbildung. Letztere ist einfach die
kanonische Abbildung $S(h,c):V(h,c)\lra \overline{V}(h,c)$. Setzen wir
diese  Abbildung in die (bilineare) duale Paarung $\overline{V}(h,c)
\times V(h,c) \lra
\CC$ ein, erhalten wir eine symmetrische Bilinearform $V(h,c)\times
V(h,c)\lra \CC$, die Shapovalov--Form, die wir ebenfalls mit $S(h,c)$
bezeichnen. Die fundamentale Eigenschaft der Shapovalov--Form ist
\ben\label{rad}
\Rad S(h,c)=M(h,c).
\een
$S(h,c)$ zerf"allt in eine direkte Summe von Bilinearformen auf $V(h,c)_n$,
die wir mit $S_n(h,c)$ bezeichnen.
Wegen (\ref{rad}) ist es "au\3erst interessant, genaue Information "uber
$S(h,c)$ zu bekommen.
Kac gelang es in \cite{Kac1}, die Determinante der Shapovalov--Formen
$S_n(h,c)$  als Funktion von $h$ und $c$ anzugeben, der erste Beweis der
Formel stammt von \cite{FF}. Im Gegensatz zu dem ``fermionischen'' Beweis
von \cite{FF} werden wir einen ``bosonischen'' Beweis der Kac--Formel
f"uhren, der von
\cite{TK}, \cite{RC2} und \cite{Thorn} motiviert ist. Es gibt zwei
Varianten dieses Theorems.
\begin{satz}\label{kacdet1}Es gilt mit  Konstanten $K_n\ne 0$
\ben\label{kacformel}
 \det S_n(h,c)^2= K_n^2 \prod_{r,s>0,\, rs\le n} \Phi_{r,s}(h,c)^{p(n-rs)}
\een
mit
\bea\label{phirs}
\Phi_{r,s}(h,c)&=&\left(h
+\frac{1}{24}(r^2-1)(c-13)+\frac{1}{2}(rs-1)\right)\times\nonumber\\
&&\quad
\left(h+\frac{1}{24}(s^2-1)(c-13)+\frac{1}{2} (rs-1)\right)
+\frac{1}{16}(r^2-s^2)^2.
\eea
\end{satz}
Die Variante ist:
\begin{satz}\label{kacdet2} Es gilt mit $K_n$ aus Satz \ref{kacdet1}
\ben\label{kacformel2}
\det S_n(h,c)=K_n \prod_{r,s>0,\,rs\le n} (h-h_{r,s})^{p(n-rs)}
\een
mit
{\mathindent0mm\bea\label{hrs}
h_{r,s}&=&\frac{1}{4a}\left( (a+1)^2-(ar+s)^2\right)\nonumber\\
&=&\frac{1}{48}\left(
(13-c)(r^2+s^2)+\sqrt{(c-1)(c-25)}(r^2-s^2)-24rs+2c-2\right)
\eea}
wobei $a=1/12 (c-13)-1/12 \sqrt{(c-1)(c-25)}$ gesetzt wurde.
\end{satz}
Die gemeinsame Idee der Beweise zu Satz \ref{kacdet1} bzw. \ref{kacdet2}
ist die Konstruktion von Intertwinern zwischen gewissen Vir--Moduln. Der
Beweis, den wir angeben werden, f"allt mehr oder weniger aus den
Resultaten aus Kapitel 3 und 4 ab. Insbesondere die Frage nach der
Nichttrivialit"at der
konstruierten Intertwiner ist im Vergleich zu \cite{TK} einfacher
beantwortet. Wir werden eine weitere Klasse von Vir--Moduln ben"otigen,
 die wir nun einf"uhren.
\section{Fock--Darstellungen der Virasoro--Algebra}
Sei
\[
\frA=\plus_{n\in\ZZ} \CC\, a_n \plus \CC\, b \mbox{ mit } [a_n,a_m]=n
\delta_{n,-m}\mbox{ und } [a_n,b]=0
\]
die Heisenbergalgebra. $\frA$ hat irreduzible, unit"are Darstellungen auf
Hilbertr"aumen $\left({\cal H}(\alpha,\beta), \langle \cdot , \cdot \rangle
\right)$ mit folgenden Eigenschaften  (\cite{Put}):
\begin{itemize}
\item[(i)] Alle $a_n$ ($n\ne0$) werden dargestellt durch unbeschr"ankte,
abgeschlossene, dicht definierte Operatoren, die wir wieder mit $a_n$
bezeichnen. Es gibt einen dichten Teilraum ${\cal F}(\alpha,\beta) \subset
\cap_{n\in\ZZ} D(a_n)$, der invariant unter allen Elementen von $\frA$
ist.
\item[(ii)] Es existiert ein zyklischer Vektor $v_{\alpha,\beta}\in {\cal
F}(\alpha,\beta)$ mit $a_0 v_{\alpha,\beta}=\alpha v_{\alpha,\beta}$ und
$bv_{\alpha,\beta}=\beta v_{\alpha,\beta}$.
\item[(iii)] Die Vektoren
\[
\Phi_{\eta} := (\eta!I^{\eta})^{-1/2} a_{-k}^{\eta_k} \cdots
a_{-1}^{\eta_1}
v_{\alpha,\beta}
\]
f"ur die Multiindizes $\eta=(\eta_1,\eta_2,\ldots )$ mit $\eta_i \ge 0$
und $\|\eta \| := \sum_{i=1}^{\infty} i\eta_i < \infty$ bilden eine
Orthonormalbasis von ${\cal H}(\alpha,\beta)$. Dabei haben wir $\eta!=
\prod \eta_i!$ und $I^{\eta}=\prod i^{\eta_i}$ gesetzt.
\item[(iv)] Es gilt f"ur $k>0$ und $e_k=(0,\ldots,0,1,0,\ldots)$
\[
a_k \Phi_{\eta}=\sqrt{k \eta_k} \Phi_{\eta-e_k} \mbox{ und }
a_{-k}\Phi_{\eta}=\sqrt{k(\eta_k+1)}\Phi_{\eta+e_k}
\]
\item[(v)] \ben\label{dak}
D(a_{\pm k})=\{ \Phi=\sum_{\eta} c_{\eta} \Phi_{\eta}\; : \;
\sum_{\eta} |c_{\eta}|^2 (\eta_k+1) < \infty \}.\een
\end{itemize}
Als Hilbertr"aume unterscheiden sich die ${\cal H}(\alpha,\beta)$ nicht,
der einzige Unterschied liegt in der Darstellung von $a_0$ und $b$ auf
${\cal H}(\alpha,\beta)$, die durch Multiplikation mit $\alpha$ bzw.
$\beta$ gegeben ist. Das ist auch der Grund, weshalb das Skalarprodukt in
${\cal H}(\alpha,\beta)$ nicht mit $\alpha$ und $\beta$ indiziert ist.

Wir setzten ${\cal F}(\alpha,\beta)=\Lin\{\Phi_{\eta}: \|\eta\| <
\infty\}$. $\frA$ besitzt eine kanonische Graduierung und durch $\deg
\Phi_{\eta}:= \|\eta\|$ wird ${\cal H}(\alpha,\beta)$ zu einem graduierten
$\frA$--Modul. Es gilt ${\cal H}(\alpha,\beta)_n={\cal F}(\alpha,\beta)_n
=\Lin \{ \Phi_{\eta}: \|\eta\| = n\}$ und $a_k:{\cal F}(\alpha,\beta)_n
\lra {\cal F}(\alpha,\beta)_{n-k}$. ${\cal H}(\alpha,\beta)$ wird als
Fockraum bezeichnet. Wir definieren nun auf jedem ${\cal H}(\alpha,\beta)$
eine Darstellung der Virasoro--Algebra (damit wird die Bedeutung der bis
jetzt unwichtigen Parameter $\alpha$ und $\beta$ klar). Wir setzen $D
(L_n(\alpha, \beta))= {\cal F} (\alpha, \beta)$ und
\bea\label{fockrep}
L_n(\alpha,\beta)&:=& \frac{1}{2} \sum_{k\in\ZZ} a_{-k} a_{n+k} + n \beta
a_n \qquad (n\neq 0) \nonumber\\
L_0(\alpha,\beta)&:=&
\frac{1}{2}(\alpha^2-\beta^2)\Id+\sum_{k=1}^{\infty}
a_{-k}a_k, \qquad z(\beta):=(1-12\beta^2)\Id.
\eea
${\cal F}(\alpha ,\beta)$ ist invariant unter allen Operatoren
$L_n(\alpha ,\beta)$.
Die Abbildung $e_n\mapsto L_n(\alpha,\beta)$, $z\mapsto z(\beta)$ liefert
eine graduierte Darstellung von Vir vom Typ
\ben\label{typ}
(h,c)=(\frac{1}{2}(\alpha^2-\beta^2), 1-12\beta^2),
\een
was zum Beispiel in \cite{KR} nachgerechnet wird. Der nichttriviale
Anteil von $L_0(\alpha,\beta)$, $N=\sum_{k>0} a_{-k}a_k$, ist nichts
anderes als der Teilchenzahloperator in ${\cal H}(\alpha,\beta)$, es gilt
$N\Phi=n \Phi$ f"ur $\Phi \in {\cal H}(\alpha,\beta)_n$ und $\bar{N}$ ist
selbstadjungiert \cite{Put}, wobei $\bar{N}$ den Abschlu\3 von $N$ als
Operator in ${\cal H}(\alpha,\beta)$ bezeichnet. Wir werden im folgenden
kurz $L_n$ f"ur $L_n(\alpha,\beta)$ schreiben, wenn klar ist, in welchem
der R"aume ${\cal H}(\alpha,\beta)$ sie operieren.

Bei den folgenden darstellungstheoretischen Untersuchungen der
Vir--Moduln ${\cal H}(\alpha,\beta)$   werden wir uns meist auf den
dichten Teilraum ${\cal F}(\alpha,\beta)$ einschr"anken. Es gilt wie bei
den Verma--Moduln $\dim {\cal F}(\alpha,\beta)_n=p(n)$, trotzdem besteht
ein wesentlicher Unterschied zu den Verma--Moduln: Es ist nicht klar (und
im allgemeinen auch falsch), da\3 $v_{\alpha,\beta}$ ein
H"ochstgewichtsvektor f"ur Vir ist, d.h da\3 $\frU(\Vir)
v_{\alpha,\beta}={\cal F}(\alpha,\beta)$ gilt. Wir
werden diese Fragen noch im Detail untersuchen. Obwohl diese Moduln also
i.~allg. keine H"ochstgewichtsmoduln f"ur Vir sind, haben wir, falls
$(h,c)$
 und $(\alpha,\beta)$ durch (\ref{typ}) verkn"upft sind, auf Grund der
Ko--Universalit"at der Verma--Moduln kanonische Homomorphismen
$S'(\alpha,\beta): V(h,c)\lra {\cal F}(\alpha,\beta)$, die durch $X
v_{h,c}\mapsto X v_{\alpha,\beta}$ f"ur $X\in \frU(\Vir)$ gegeben sind.
Aus der
Uni\-ver\-sali\-t"at von $\overline{V}(h,c)$ erhalten wir Homomorphismen
$S''(\alpha,\beta):{\cal F}(\alpha,\beta)\lra \overline{V}(h,c)$. Die
Komposition $S''(\alpha,\beta)\circ S'(\alpha,\beta):V(h,c)\lra
\overline{V}(h,c)$ ist ein Homomorphismus mit der Eigenschaft $v_{h,c}
\mapsto
\bar{v}_{h,c}$, es  gilt folglich die Faktorisierung der
Shapovalov--Abbildung
\ben\label{shapofaktor}
S(h,c)=S''(\alpha,\beta)\circ S'(\alpha,\beta) \mbox{ mit }
h=\frac{1}{2}(\alpha^2-\beta^2) \mbox{ und } c=1-12\beta^2.
\een
Wenn wir im folgenden von $\det S'_n$ sprechen, meinen wir dabei die
Determinante der Matrix, der $S'_n$ bez"uglich der Basen $\{L_{-n_k}\cdots
L_{-1} v_{h,c}\}$ von $V(h,c)$ und $\{ \Phi_{\eta}\}$ von ${\cal
F}(\alpha,\beta)$ entspricht. Es gelten die folgenden
Determinantenformeln.
\begin{satz}\label{fockdet} Es gilt mit Konstanten $K_n',K_n''\ne 0$
\begin{eqnarray}
\label{fock1}
\det S'_n(\alpha, \beta) &=& K_n' \prod_{r,s \geq 0,rs \leq n}
\Phi'_{r,s}
(\alpha,\beta)^{p (n-rs)} \\
\label{fock2}
\det S''_n (\alpha , \beta) &=& K_n'' \prod_{r,s \ge 0, rs\le n}
\Phi''_{r,s} (\alpha,\beta)^{p(n-rs)}\\
\label{fock3}
\mbox{ mit } \Phi'_{r,s} (\alpha,\beta) &=& \alpha + \frac{r}{2} \gamma_+
+ \frac{s}{2} \gamma_- \\
\label{fock4}
\mbox{ und } \Phi''_{r,s} (\alpha,\beta) &=& \alpha - \frac{r}{2} \gamma_+
- \frac{s}{2} \gamma_-
\end{eqnarray}
wobei $\gamma_{\pm} $ die L"osungen der Gleichung $\beta=
\displaystyle{\frac{1}{\gamma} - \frac{\gamma}{1}} $ sind.
\end{satz}
Es ist klar, da\3 wegen (\ref{shapofaktor}) die Determinantenformeln eng
zusammenh"angen, denn es gilt nat"urlich auch $\det S_n(h,c)=\det
S_n'(\alpha,\beta) \det S''(\alpha,\beta)$, falls $(h,c)$ und
$(\alpha,\beta)$ wie in (\ref{shapofaktor}) zusammenh"angen. Satz
\ref{fockdet} wurde zuerst von \cite{TK} in
dieser Form bewiesen, implizit ist er schon in \cite{FF} enthalten.

\section{Unit"are Darstellungen}
Aus physikalischen Gr"unden ist man  an unit"aren Darstellungen der
Virasoro--Algebra interessiert. In diesem Fall ist  der
Energie--Impuls--Tensor $T(z)= \sum_{n= -\infty}^{\infty} z^{-n+2} L_n$
f"ur $|z|=1$ hermitesch und f"ur beliebige $z \in \CC$ gilt
$T(z)^*=T(\frac{1}{\bar{z}})$. (Diese Gleichung stimmt nat"urlich nur im
Sinne der graduierten Moduln, im Rahmen der Hilbertraumtheorie gilt
 i.~allg.
nur ``$\subset$''. Es ist auch a priori nicht klar, ob $T(z)$ f"ur $|z|
\geq 1$ "uberhaupt einen dicht definierten Operator in ${\cal
H}(\alpha,\beta)$ definiert.)\\
Wir definieren eine anti--lineare Anti--Involution $\omega$ auf $\Vir$
durch $\omega (L_n) = L_{-n},\: \omega (z) = z$ und
$\omega([x,y])=[\omega(x), \omega(y)]$.
\begin{Def}
Eine sesquilineare Form $(\cdot , \cdot)$ auf einem $\Vir$--Moduln $M$
hei\3t kontravariant (bzgl. $\omega$), wenn
\ben
(X.m_1, m_2)= (m_1, \omega(X).m_2) \qquad (m_1,m_2 \in M, X \in \frU
(\Vir))
\een
gilt. Falls zus"atzlich $(m,m)>0$ f"ur alle $m\not= 0$ gilt, hei\3t $M$
unit"arer $\Vir$--Modul.
\end{Def}
Die Unitarit"at einer Darstellung kann aus zwei Gr"unden verletzt sein:
\begin{itemize}
\item[(i)] Es existiert ein $m\in M$ mit $(m,m)<0$.
\item[(ii)] Es existiert ein $m \in M,\:m \not= 0$ mit $(m,m)=0$.
\end{itemize}
Existieren in $M$ nur Vektoren des zweiten Typs, so ist $\quot{M}{\Rad
(\cdot,\cdot)}$
ein unit"arer Modul. $M$ hei\3t dann unitarisierbar. Wir k"onnen auf triviale
Weise die bilineare Shapovalov--Form zu einer Sesquilinearform machen,
indem wir der kanonischen Abbildung $S(h,c):V(h,c)\lra \overline{V}(h,c)
$  die von $\lambda L_n \longmapsto \bar{\lambda} L_n$ erzeugte
antilineare
Involution vorschalten. Die so erzeugte Form auf $V(h,c)$ bezeichnen wir
mit $(\cdot,\cdot)$. Die Frage, ob die Verma--Moduln unit"ar bzw.
unitarisierbar sind, ist nicht ganz einfach zu beantworten. Elementar
k"onnen wir sehen, da\3 f"ur $c \geq 1$ und $h \geq 0 \; ( \cdot, \cdot)$
positiv--semi--definit ist:\\
Es gilt, wenn wir in ${\cal F} (\alpha,\beta)$ als direkte Summe endlich
dimensionaler Vektorr"aume adjungieren,
\ben
L_n (\alpha,\beta)^* =L_{-n}(\bar{\alpha}, -\bar{\beta}),
\een
d.h. falls $\alpha \in \RR, \; \beta \in i\RR$ ist, gilt
$L_n(\alpha,\beta)^* = L_{-n} (\alpha,\beta)$. In diesem Fall ist $\langle
\cdot, \cdot \rangle$ eine kontravariante Form und ${\cal F}
(\alpha,\beta)$ ein unit"arer $\Vir$--Modul.\\
Das folgende Diagramm ist kommutativ:
\ben\label{K14}\begin{array}{c}
\xext=1700\yext=500
\begin{picture}(\xext,\yext)\label{konkomm}
\setsqparms[1`1`1`1;1000`500]
\putsquare(700,0)[V(h,c)\times V(h,c)`\CC`{\cal F}(\alpha,\beta)\times
{\cal F}(\alpha,\beta)`\CC;`\scriptstyle S'(\alpha,\beta)\times
S'(\alpha,\beta)``]
\put(0,500){\makebox(0,0){$(\cdot,\cdot):$}}
\put(0,0){\makebox(0,0){$\langle \cdot,\cdot\rangle :$}}
\end{picture}
\end{array}
\een
$\langle S'(\alpha,\beta)\cdot, S'(\alpha,\beta)\cdot \rangle$ definiert
eine weitere kontravariante Form auf $V(h,c)$, da diese aber (bis auf
Normierung) eindeutig ist und $\langle
S'(\alpha,\beta)v_{h,c},S'(\alpha,\beta)v_{h,c}\rangle=\langle
v_{\alpha,\beta},v_{\alpha,\beta}\rangle=1=(v_{h,c},v_{h,c})$ gilt, m"ussen
beide Formen "ubereinstimmen. Da $\langle \cdot ,\cdot \rangle > 0$ gilt,
folgt $( \cdot, \cdot)\geq 0$ f"ur $h=\frac{1}{2} (\alpha^2-\beta^2)$, $
c=1-12 \beta^2$, d.h. f"ur $h \geq 0$ und $c \geq 1$.\\
Dies sind aber nicht alle unitarisierbaren Verma--Moduln:\\
Sei f"ur $r,s,m \in \NN,\;m \geq 2$,
\ben\label{4.3}
c(m) =1- \frac{\dst 6}{\dst m(m+1)} \mbox{ und } h_{r,s}=\frac{\dst
((m+1)r-ms)^2-1}{\dst 4m(m+1)}.
\een
Es gilt der
\begin{satz}\label{unirep}
$V(h,c)$ ist genau dann unit"ar bzw. unitarisierbar, wenn entweder $h \geq
0,\; c\geq 1$ oder $(h,c)$ von der Form (\ref{4.3}) ist.
\end{satz}
Es ist offensichtlich $h \geq 0$ und $c\geq 0$ eine notwendige
Bedingung f"ur Unitarisierbarkeit, denn $(e_{-n} v_{h,c},e_{-n}
v_{h,c})=2nh+c \dst\frac{n^3-n}{12}$ ist nur unter dieser Bedingung
positiv f"ur alle $n \geq 0$. Da\3 alle Verma--Moduln vom Typ $h \geq 0,\;
c\geq 1$ unitarisierbar sind, haben wir bereits gesehen.

Im Falle $c<1$ folgt aus einer detaillierten Untersuchung  der
Kac--De\-ter\-mi\-nan\-ten\-formel in einer Arbeit von D.~Friedan, Z.~Qiu
und Z. Shenker \cite{FQS}, da\3 alle Moduln, deren Typen nicht in
(\ref{4.3}) vorkommen, Vektoren negativer L"ange enthalten .

Die Unitarisierbarkeit der Moduln vom Typ (\ref{4.3}) folgt aus der
Goddard--Kent--Olive--Konstruktion \cite{GKO},\cite{KR}. Dort werden mit
Hilfe der Sugawara Konstruktion aus unit"aren Darstellungen von affinen
Algebren unit"are Darstellungen von $\Vir$ vom Typ (\ref{4.3})
konstruiert.
Diese Konstruktion spielt aber f"ur unsere  weiteren Untersuchungen keine
Rolle, wir werden sie deswegen hier nicht genauer beschreiben.

Wenden wir uns nun den Fock--Moduln zu, die (\ref{4.3}) entsprechen, d.h.
insbesondere $\alpha,\beta \in \RR$. In diesem Fall ist $\langle \cdot,
\cdot \rangle$ keine kontravariante Form f"ur $\Vir$. Wir k"onnen aber eine
kontravariante Form auf ${\cal H}(\alpha,\beta)$ folgenderma\3en
einf"uhren:\\
Sei $J$ der selbstadjungierte idempotente Operator definiert durch
$J\Phi_{\mu}= (-1)^{|\mu|} \Phi_{\mu}$ (wobei wie "ublich
$|\mu| =\sum \mu_i$ ist). Wir definieren eine neue Sesquilinearform
auf ${\cal H}(\alpha,\beta)$ durch
\ben\label{4.4}
\langle \cdot,\cdot \rangle_{J}:= \langle \cdot,J\cdot\rangle.
\een
$\langle \cdot,\cdot \rangle_{J}$ ist zwar nicht positiv definit, aber
immerhin nicht ausgeartet,  d.h. falls $\langle x,y\rangle=0$ f"ur alle
$y\Longrightarrow x=0$. $\left({\cal H}(\alpha,\beta), \langle \cdot,
\cdot \rangle_{J}\right)$ ist ein Krein--Raum (\cite{Bo}), denn es gilt
${\cal H}(\alpha,\beta)={\cal H}_+ \plus {\cal H}_-$, wobei ${\cal
H}_+=\overline{\Lin \{ \Phi_{\mu}\,:\, |\mu| \mbox{ gerade}\}}$ und
${\cal H}_-=\overline{\Lin \{ \Phi_{\mu}\,:\, |\mu| \mbox{ ungerade}\}}$
ist. \\
Wir k"onnen leicht den $J$--adjungierten Operator von $a_n$, den wir mit
$a_n^{\dagger}$ bezeichnen,  berechnen. Es gilt (\cite{Bo}, Lemma VI 2.1)
$a_n^{\dagger}= Ja_n^* J=-a_{-n}$ f"ur $n \not= 0$ und $a_0^{\dagger}=a_0$.
Damit folgt $L_n(\alpha,\beta)^{\dagger} \supset
L_{-n}(\bar{\alpha},\bar{\beta})$, d.h. f"ur reelle $\alpha,\beta$ ist
$\langle \cdot, \cdot \rangle_J$ kontravariant.\\
Wenn wir mit ${\cal H} (\alpha,\beta)_0$ die H"ochstgewichtskomponente von
${\cal H}(\alpha,\beta)$, d.h. ${\cal H}(\alpha,\beta)_0=\overline{\frU
(\Vir).v_{\alpha,\beta}}$ bezeichnen, gilt f"ur alle $\alpha,\beta$, die
(\ref{4.3}) entsprechen,
\[
\langle \cdot, \cdot\rangle_J \restrict{{\cal H}(\alpha,\beta)_0 \times
{\cal H}(\alpha,\beta)_0} \geq 0,
\]
da wir wieder ein kommutatives Diagramm wie in (\ref{K14})
erhalten. Diese Eigenschaft ist allerdings schwer zu verwenden, da wir
keine handliche Beschreibung von ${\cal H}(\alpha,\beta)_0$ durch die
Heisenbergalgebra besitzen. Diese Beschreibung werden wir erst in Kapitel
5 erhalten, wenn wir die Konstruktion von Felder vorstellen.

An dieser Stelle sei noch angemerkt, da\3 wir, wie das folgende Beispiel
zeigt, f"ur nicht unitarisierbare Verma--Moduln i.~allg. keine Krein--Raum
Struktur bez"uglich der Shapovalov--Form erwarten k"onnen.
\begin{beis}\label{notkrein} Sei $c=\frac{22}{5}$ und $h=-\frac{1}{5}$.
Dann gilt
\[
(L_{-2} v_{h,c},L_{-2} v_{h,c})=\frac{7}{5},\quad (L_{-1} L_{-1}
v_{h,c},L_{-1}L_{-1} v_{h,c})=-\frac{12}{5}
\]
und
\[
(L_{-1}L_{-1} v_{h,c},L_{-2} v_{h,c})=-\frac{6}{5}.
\]
$V(-\frac{1}{5},\frac{22}{5})$ ist folglich nicht zerlegbar in die
orthogonale Summe der Unterr"aume positiver bzw. negativer Vektoren.
\end{beis}
\section{Beweise der Determinantenformeln}
Wir werden die Formeln (\ref{kacdet1}), (\ref{kacdet2}), (\ref{fock1})
und
(\ref{fock2}) mit dem folgenden Satz beweisen, der ein Ergebnis aus
Kapitel 4 ist, und die wesentliche Schwierigkeit im Beweis der
Determinantenformel darstellt.\\
Sei $\gamma\in \CC$, $r,s \in \NN$. Es gibt Operatoren zwischen den
Fockr"aumen
\[
Q( \gamma ; r,s): {\cal H} (\alpha -r \gamma, \beta) \lra {\cal H}
(\alpha
,\beta),\quad D(Q(\gamma;r,s))={\cal F}(\alpha-r\gamma,\beta)
\]
 vom Grad $rs$, es gilt also genauer
\[
Q (\gamma ; r,s) : {\cal F} (\alpha -r \gamma , \beta )_n \lra {\cal F}
(\alpha , \beta)_{n+rs}.
\]
Wir werden in Kapitel 4 eine explizite Konstruktion dieser Operatoren
angeben und den folgenden Satz beweisen.
\begin{satz}\label{Inter}
Sei $\gamma \in \CC$ mit $\gamma^2 \notin \QQ$. F"ur $r,s \in \NN$ setze
$\beta= {\dst\frac{1}{\gamma}- \frac{\gamma}{2}}$,
$\alpha={\dst\frac{r}{2} \gamma - \frac{s}{\gamma}}$. Dann ist
\[
Q( \gamma ; r,s) : {\cal F} (\alpha- r \gamma ,\beta) \lra {\cal F}(\alpha
,\beta)
\]
ein nichttrivialer Intertwiner vom Grad $rs$, und
\[
Q(\gamma ; r,-s):{\cal F}(-\alpha,\beta) \lra {\cal F}(-\alpha +r \gamma,
\beta)
\]
ein nichttrivialer Intertwiner vom Grad $-rs$.
\end{satz}
Aus Satz \ref{Inter} folgt, da\3 unter den genannten Voraussetzungen ${\cal
F}(\alpha,\beta)$ und ${\cal F}(-\alpha,\beta)$ einen singul"aren Vektor
vom Grad $rs$ enthalten.\\
Den konkreten Nutzen von Satz \ref{Inter} f"ur die Verma--Moduln zeigt:
\begin{lem}\label{Aequi}
Seien $(h,c)$ und $(\alpha,\beta)$ durch (\ref{typ}) verkn"upft. Die
folgenden Aussagen sind "aquivalent.
\begin{itemize}
\item[(i)]
$V(h,c)$ ist irreduzibel.
\item[(ii)]
$V(h,c)$ enth"alt keinen singul"aren Vektor von positivem Grad.
\item[(iii)]
Die Shapovalov--Abbildung $S(h,c):V(h,c) \lra \overline{V} (h,c)$ ist ein
Isomorphismus.
\item[(iv)]
$\overline{V}(h,c)$ enth"alt keinen kosingul"aren Vektor von positivem Grad.
\item[(v)]
Es gilt $\det S_n (h,c)\not= 0$ f"ur alle $n \geq 0$.
\item[(vi)]
Die Abbildungen $S'(\alpha,\beta):V(h,c)\lra {\cal F}(\alpha,\beta)$ und
$S''(\alpha,\beta):{\cal F}(\alpha,\beta) \lra \overline{V} (h,c)$ sind
Isomorphismen.
\item[(vii)]
Es gilt $\det S'_n(\alpha,\beta) \not= 0$ und $\det S''_n (\alpha,\beta)
\not=0$ f"ur alle $n \geq 0$.
\item[(viii)]
Es gibt keine singul"aren oder kosingul"aren Vektoren in ${\cal
F}(\alpha,\beta)$.
\item[(ix)]
${\cal F}(\alpha,\beta)$ ist irreduzibel.
\end{itemize}
\end{lem}
Der Beweis von Lemma \ref{Aequi} ist eine einfache Folgerung aus
$\dim V(h,c)_n =\dim \overline{V}(h,c)_n=\dim {\cal F}(\alpha,\beta)_n $ und
(\ref{rad}).\\
Nun ist klar, da\3 wir Nullstellen in der Kac--Determinante aller
Verma--Moduln finden, die den Fockmoduln aus Satz \ref{Inter}
entsprechen.

\noindent{\bf Beweis der Determinantenformeln.}\\
Da die Methoden, um aus Satz \ref{Inter} die Determinantenformeln zu
beweisen, bekannt sind, werden wir die Beweise hier nur skizzieren.
Ausf"uhrlichere Beweise findet man in \cite{KR}, \cite{TK} oder
\cite{CdG}.\\
Zun"achst sind die S"atze \ref{kacdet1} und \ref{kacdet2} "aquivalent wegen
\ben\label{ka5}
\Phi_{r,s}(h,c)\Phi_{s,r}(h,c) = (h-h_{r,s})^2 (h-h_{s,r})^2 \mbox{ f"ur }
r\not= s
\een
und
\[
\Phi_{s,s}(h,c) = (h-h_{s,s})^2.
\]
Es gibt auch eine Relation zwischen Satz \ref{kacdet1} und Satz
\ref{fockdet}, es gilt
{\small\mathindent0mm\ben\label{ka6}
\Phi_{r,s}\left(h(\alpha,\beta),c(\beta)\right)
\Phi_{s,r}\left(h(\alpha,\beta),c(\beta)\right)=
\left(\Phi'_{r,s}(\alpha,\beta)\Phi'_{s,r}
(\alpha,\beta)\Phi''_{r,s}(\alpha,\beta)
\Phi''_{s,r}(\alpha,\beta)\right)^2.
\een}
Satz \ref{kacdet1} w"urde also aus Satz \ref{fockdet} folgen. Leider wird
aber Satz \ref{kacdet1} zum Beweis von Satz \ref{fockdet} ben"otigt.

\noindent{\bf Beweis von Satz \ref{kacdet2}.}\\
Zun"achst mu\3 man sich "uberzeugen, da\3 beide Seiten von (\ref{kacformel2})
den gleichen Grad als Polynom in $h$ und $c$ haben. Den Grad der rechten
Seite von (\ref{kacformel}) bzw. (\ref{kacformel2}) kann man unmittelbar
ablesen.

Die folgende "Uberlegung liefert die Grade der linken Seite: Die
Shapovalov--Form von zwei Basisvektoren $L_{-n_k} \ldots L_{-1}v_{h,c}$
und $L_{-m_k} \ldots L_{-1}v_{h,c}$ vom gleichen Grad in $V(h,c)$ ist der
Eigenwert von $L_1 \ldots L_{n_k} L_{-m_k} \ldots L_{-1} v_{h,c}$. Um den
Eigenwert zu berechnen, mu\3 man die Operatoren $L_k \; (k>0)$ mit
(\ref{virrel}) nach rechts bringen. Ein Faktor $h$ bzw. $c$ entsteht nur
aus $[L_{n},L_{-n}]=2n L_0 + \frac{n^3-n}{12} z$. Deshalb hat die
Shapovalov--Matrix auf der Diagonalen die Eintr"age mit dem h"ochsten Grad
in $h$ und $c$. Diese Grade kann man angeben, und eine einfache Rechnung
zeigt, da\3 sie mit den Graden der rechten Seite "ubereinstimmen.\\
Es reicht also zu zeigen, da\3 in (\ref{kacformel2}) die linke Seite durch
die rechte Seite teilbar ist.

Als n"achstes bestimmt man die Werte $h_{r,s}(c)$, die den Werten von
$\alpha$ und $\beta$ aus Satz \ref{Inter} entsprechen. Man erh"alt genau
die Ausdr"ucke (\ref{hrs}). Da dann ${\cal F}(\alpha,\beta)$ einen
singul"aren Vektor vom Grad $rs$ enth"alt, mu\3 nach Lemma \ref{Aequi} dies
auch f"ur $V(h_{rs}(c),c)$ gelten, und es mu\3
$\det S_{rs}(h=h_{r,s}(c),c)=0$ sein, d.h. es mu\3 $\det S_{rs}(hc)$ durch
$(h-h_{r,s}(c))$ teilbar sein. Der singul"are Vektor in $V(h_{r,s}(c),c)$
erzeugt einen Untermodul $W$ mit $\dim W_n=p(n-rs)$. Wegen (\ref{rad})
gilt $W \subset \mbox{ Rad } S(h_{r,s}(c),c)$, folglich mu\3 $\det
S_n(h,c)$ durch
$(h-h_{r,s}(c))^{p(n-rs)}$ teilbar sein (vgl. \cite{KR}, Lemma 8.4). Da
$r,s \in \NN$ beliebig waren, folgt schon die Behauptung.
\hfill$\Box$

\noindent{\bf Beweis von Satz \ref{fockdet}.}\\
Wir folgen hier \cite{TK}. W"ahle ein $\beta \in \CC$. $\gamma_{\pm}$
seien die L"osungen von $\beta= {\dst\frac{1}{\gamma} - \frac{\gamma}{2}}$
und es gelte $\gamma^2_{\pm} \notin \QQ$. Dann hat f"ur beliebige $\alpha
\in \QQ$ die Gleichung
\[
\alpha+\frac{r}{2} \gamma_+ +\frac{s}{2} \gamma_-=0
\]
h"ochstens eine L"osung $(r,s)\in \ZZ^2$. Sei nun $r,s \in \NN$ und setze
$\alpha_0 =\alpha_{r,s}(\gamma)= \frac{r}{2} \gamma_+ +\frac{s}{2}
\gamma_-$, $c=c(\beta)$ und $h_0=h(\alpha_0,\beta$. Dann besitzt $\det
S_{rs}(h=h_0,c)$ eine einfache Nullstelle und wegen der Vorbemerkung ist
damit schon klar, da\3 $M(h_0,c)$ nur von einem einzigen singul"aren Vektor
vom Grad $rs$ erzeugt wird.\\
Es gilt also $M(h_0,c)\simeq V(h_0+rs,c)$.

Aus Satz \ref{Inter} erhalten wir einen Intertwiner vom Grad $-rs$
\[
Q:{\cal F}(-\alpha_0,\beta) \lra {\cal F}(-\alpha'_0,\beta)
\]
mit $\alpha'_0=\alpha_{-r,s}(\gamma)$. Wir erhalten die Sequenz
\[
V(h,c) \stackrel{S'(-\alpha_0,\beta)}{\strich\strich\lra} {\cal
F}(-\alpha_0,\beta) \stackrel{Q}{\strich\strich\lra} {\cal
F}(-\alpha'_0,\beta),
\]
denn es gilt $\deg (Q \circ S'(-\alpha_0,\beta))=-rs$ und deswegen $Q
\circ S'(-\alpha_0,\beta)v_{h,c} \in {\cal F}(-\alpha_0,\beta)_{-rs}=\{ 0
\}$. Da $Q$ nichttrivial ist, kann $S'(-\alpha_0,\beta)$ kein
Isomorphismus sein, und es mu\3 $\ker S'(-\alpha_0,\beta)=M(h_0,c)$ gelten.
Es folgt
$\dim \ker S'(-\alpha_0,\beta)_n=p(n-rs)$
und folglich ist $\det S_n(\alpha,\beta)$ durch
$(\alpha+\alpha_0)^{p(n-rs)}=( \alpha +\frac{r}{2} \gamma_+ + \frac{s}{2}
\gamma_-)^{p(n-rs)}$ teilbar. Durch Dualisierung folgt, da\3 $\det
S''(\alpha,\beta)$ durch $(\alpha-\frac{r}{2} \gamma_+ -\frac{s}{2}
\gamma_-)^{p(n-rs)}$ teilbar ist, denn es gilt
\[
\left(S''(\alpha,\beta):{\cal F}(\alpha,\beta) \lra
 \overline{V}(h,c)\right)'=
S'(-\alpha,\beta):V(h,c) \lra {\cal F}(-\alpha,\beta).
\]
Wegen(\ref{ka6}) ist klar, da\3 wir damit schon alle Teiler gefunden haben
und Satz \ref{fockdet} ist bewiesen.
\hfill$\Box$\\
\chapter{Singul"are Vektoren in Virasoro--Moduln}
In diesem Abschnitt wollen wir die Determinantenformeln verwenden, um
tiefergehende Eigenschaften der Verma-- und  Fockmoduln zu beweisen.
Das Hauptresultat ist dabei die Klassifikation der Moduln bez"uglich
der Struktur der singul"aren bzw. kosingul"aren Vektoren. Dazu verwenden wir
Filtrationstechniken, die es erlauben, die Moduln immer weiter zu
reduzieren.

In H"ochstgewichts--Moduln der Virasoro--Algebra k"onnen entweder singul"are
oder kosingul"are Vektoren auftreten. Verma--Moduln enthalten nur singul"are
Vektoren und kontragrediente Verma--Moduln nur kosingul"are Vektoren. In
Fock--Moduln treten dagegen im allgemeinen beide Typen auf. Die Kenntnis
der Struktur der singul"aren Vektoren erlaubt zum einen die Konstruktion
der irreduziblen Moduln,  zum anderen die Klassifikation der
Homomorphismen
zwischen den Moduln.

\section{Allgemeines}
Zun"achst m"ochten wir einige allgemeine Aussagen "uber singul"are Vektoren
beweisen. Wir f"uhren  eine  Parametrisierung von
$(h,c)\in \CC^2$ ein: F"ur $r,s \in \NN$ und $t \in \CC^*$ sei
\ben\label{para}
 c(t)=6 t + 13 +6 t^{-1} \mbox{ und } h_{r,s}(t)=\frac{(1+t)^2-(s+t r )^2
}{4t}.
\een
Eine einfache Rechnung zeigt, da\3 $\Phi_{r,s}(h_{r,s}(t),c(t))=0$ f"ur $t
\in \CC^*$ gilt und da\3 umgekehrt alle Punkte $(h,c)$ mit
$\Phi_{r,s}(h,c)=0$ auf einer solchen Kurve liegen.
Wir werden gleich sehen, da\3 dann $V(h_{r,s}(t),c(t))$ einen singul"aren
Vektor vom Grad $rs$ enth"alt, was zun"achst nur klar ist, falls die
Shapovalov--Form in den Graden $n < rs$ nicht ausgeartet ist, d.~h. falls
$\det S_n(h_{r,s}(t),c(t)) \ne 0$ f"ur $n<rs$ gilt.

Zun"achst stellen wir uns eine andere Frage: Wann kann eine weitere Kurve
$(h_{r',s'}(t'),c(t')) \subset \CC^2$ mit $(r,s)\ne (r',s')$ und
\[
c(t_0) =c(t_0') \mbox{ und } h_{r,s}(t_0)=h_{r',s'}(t_0')
\]
existieren? Es ist leicht aus (\ref{para}) zu sehen, da\3 $t_0$ dann
rational sein mu\3.\\
Wir unterscheiden zwei F"alle: Sei $t=\pm q/p$ f"ur $p,q \in \NN$ und $q$
und $p$ relativ prim. Wir erhalten
{ \mathindent=0mm
\ben\label{ppq} (+)
 \quad c(q/p)= 25+6\frac{(p-q)^2}{pq},\quad
h_{r,s}(q/p)=\frac{(p+q)^2 -(rq+sp)^2}{4pq}=:h_{r,s}^+\;
\mbox{und}
\een
\ben\label{mpq}
(-) \quad c(-q/p)=1-6\frac{(p-q)^2}{pq}, \quad h_{r,s}(-q/p)=
\frac{(rq-sp)^2-(p-q)^2}{4pq}:=h_{r,s}^-,
\een}
d.h. es mu\3 dann insbesondere entweder $c>25$ oder $c<1$ gelten.
Bemerkenswerterweise treten in der zweiten Version der
Kac--Determinantenformel (\ref{kacformel2}) genau die Terme  $h_{r,s}^-$
auf, wenn man dort in $h_{r,s}(c)$\quad $c$ durch $c(-q/p)$ aus
(\ref{mpq}) ersetzt. Damit ist die Kac--Determinante von $V(h,c(-q/p))$
gegeben durch
\ben\label{detmini}
\det S_n\left(h,c(-q/p)\right)= K_n \prod_{rs \le n} \left(h -
h_{r,s}^-\right)^{p(n-rs)}.
\een
Genauso k"onnen wir die Kac--Determinante von $V(h,c(q/p))$ schreiben als
\ben\label{notdetmini}
\det S_n\left(h,c(q/p)\right)= K_n \prod_{rs \le n} \left(h -
h_{r,s}^+\right)^{p(n-rs)}.
\een
Dies wird im folgenden sehr n"utzlich sein, da wir nun die Nullstellen in
der Kac--Determinante f"ur Moduln mit mehr als einem singul"aren Vektor
explizit angeben k"onnen.

F"ur die Fock--Moduln gilt Analoges zu dem eben gesagten, auch hier k"onnen
wir eine Parametrisierung der singul"aren Vektoren einf"uhren, die aus der
entsprechenden Determinantenformel entspringt: F"ur $r,s\in \NN$ und $t \in
\CC^*$ sei
\ben\label{fockpara}
\alpha(t)=\frac{s}{2} t - r t^{-1},\qquad \beta(t)=\frac{1}{2}t + t^{-1}.
\een
Die Determinanten der kanonischen Abbildungen $S'(\alpha,\beta):V(h,c)\lra
{\cal F}(\alpha,\beta)$ und $S''(\alpha,\beta):{\cal F}(\alpha,\beta)\lra
\overline{V}(h,c)$ sind parametrisiert durch $\gamma$ mit $\beta=1/\gamma
-\gamma/2$. Es gilt $\Phi''(\alpha(t),\beta(t))=0$. Wir k"onnen durch eine
einfache Rechnung die Punkte $(\alpha,\beta)$ bestimmen, die im
Schnittpunkt mehrerer Kurven der Form (\ref{fockpara}) liegen. Wir
erhalten f"ur $\gamma$ die L"osungen $\gamma_+=i\sqrt{2p/q}$ und
$\gamma_-=\sqrt{2p/q}$ mit relativ primen $p,q \in \NN$ und v"ollig analog
zu obigen F"allen:
{\mathindent0mm
\ben
(+)\quad \alpha_{r,s}(\gamma_+)=i \frac{rq+sp}{\sqrt{2pq}}, \quad
\beta(\gamma_+)=i \frac{p-q}{\sqrt{2pq}},
\een
\ben\label{G28}
(-)\quad \alpha_{r,s}(\gamma_-)=\frac{rq-sp}{\sqrt{2pq}}, \quad
\beta(\gamma_-)=\frac{q-p}{\sqrt{2pq}}.
\een}
Diese F"alle entsprechen genau denen f"ur die Verma--Moduln mittels der
Beziehungen $c=1-12\beta^2$ und $h=\frac{1}{2}(\alpha^2-\beta^2)$.
Diese Nullstellen in den Determinanten entsprechen singul"aren Vektoren in
den Moduln.
\begin{lem}\label{exist}
\begin{itemize}
\item[(i)] Ist f"ur ein $(h,c)\in \CC^2$ und $r,s \in \NN$
$\Phi_{r,s}(h,c)=0$, so enth"alt $V(h,c)$ einen singul"aren Vektor vom Grad
$rs$.
\item[(ii)] Ist f"ur ein $(\alpha,\beta)\in \CC^2$ und $r,s \in \NN$
$\Phi''_{r,s}(\alpha,\beta)=0$, so enth"alt $\cal F(\alpha,\beta)$ einen
singul"aren Vektor vom Grad $rs$.
\end{itemize}
\end{lem}
{\bf Beweis.} (i):\\
Wir definieren $L_{h,c} : V(h,c) \lra V(h,c)$ durch $L_{h,c}= \sum_{k>0}
L_k(h,c)$. $L_{h,c}$ ist definiert auf $V(h,c)$, denn f"ur jedes $x \in
V(h,c)_n$ gilt $L_{h,c}x=\sum_{k=1}^n L_k(h,c)x$.

$v\in V(h,c)$ ist offensichtlich genau dann ein singul"arer Vektor, wenn $v
\ne 0$ und $v\in \ker L_{h,c}$ gilt. Weiter h"angt $L_{h,c}:V(h,c)_n \lra
\plus_{k=0}^{n-1} V(h,c)_k$ stetig (sogar polynominal) von $h$ und $c$
ab. Wir betrachten nun
\[
\left( L_{h_{r,s}(t),c(t)}\right )_{rs} = L_{h_{r,s}(t),c(t)}:
V(h_{r,s}(t),c(t))_{rs} \lra \plus_{k=0}^{rs-1} V(h_{r,s}(t),c(t))_k
\]
 mit der Parametrisierung (\ref{para}). Nach dem oben Gesagten gilt f"ur
alle $t$ mit $t\notin \QQ$
\[
\dim \ker (L_{h_{r,s}(t),c(t)})_{rs} \ge 1.
\]
Das mu\3 dann
aber auch f"ur $t=\pm q/p$ gelten, denn die Injektivit"at von
$L_{h_{r,s}(t),c(t)}$ f"ur ein $t=\pm q/p$
m"u\3te auf einer ganzen Umgebung von $t=\pm q/p$ gelten.

(ii) Der Beweis verl"auft im wesentlichen analog, wenn man Folgendes
ber"ucksichtigt: Wir verwenden die Faktorisierung der Shapovalov--Abbildung
"uber die Fock--Moduln
\[
S(h,c)=S''(\alpha,\beta) S'(\alpha,\beta),
\]
wobei  ${\cal F}(\alpha,\beta)$ ein
Modul vom Typ $(h,c)$ ist. Wir m"ussen  nach Nullstellen in $\det
S''(\alpha,\beta)$ suchen, um singul"are Vektoren in ${\cal
F}(\alpha,\beta)$ zu finden, denn in diesem Fall ist $\det S(h,c)=0$ und
$V(h,c)$ enth"alt nach (i) einen singul"aren Vektor, der durch
$S'(\alpha,\beta)$ auf einen singul"aren Vektor in ${\cal F}(\alpha,\beta)$
abgebildet wird.

Genau wie in (i) funktioniert das bis auf eine Ausnahmemenge, die auf der
reellen und komplexen Achse liegt, und mit der entsprechenden Abbildung
\[
L_{\alpha,\beta}=\sum_{k>0} L_k(\alpha,\beta)
\]
erschlagen wird.\hfill$\Box$

Interessanterweise gibt es in einem Fock-Modul nie zwei linear unabh"angige
singul"are Vektoren desselben Grades, es gilt genauer:
\begin{lem}[\cite{TK}]\label{algset}
${\frS}_d=\left\{(\alpha,\beta)\in \CC^2: \ker
\left(\left(L_{\alpha,\beta}\right)_d\right) \ne \{0\} \right\}$ ist eine
algebraische Menge der Dimension $\le 1$.
\end{lem}
{\bf Beweis.}\\
Wir verwenden die kanonische Basis $\{\Phi_{\delta}\}$ von ${\cal
F}(\alpha,\beta)$. Um die Wirkung von $L_{\alpha,\beta}$ auf der Basis
einfacher hinzuschreiben, f"uhren wir zun"achst in den Ausdr"ucken f"ur
$L_n(\alpha,\beta)$ eine Wick--Ordnung durch (d.h. Erzeugeroperatoren
 links von Vernichteroperatoren) und erhalten f"ur $n>0$
\ben
L_n(\alpha,\beta)=\sum_{k=1}^{\infty} a_{-k} a_{n+k} + \frac{1}{2}
\sum_{k=1}^{n-1} a_k a_{n-k} + (\alpha+n \beta)a_n.
\een
F"ur $n<0$ kann man einen "ahnlichen Ausdruck herleiten, den wir aber hier
nicht ben"otigen. Es folgt
{\mathindent5mm\bea
L_n(\alpha,\beta) \Phi_{\delta} &=& \sum_{k=1}^{\infty} \sqrt{k
(\delta_k+1)(n+k)\delta_{n+k}} \Phi_{\delta-e_{n+k}+e_k} \nonumber\\
&&+ \frac{1}{2}\sum_{k=1}^{n-1} \sqrt{k \delta_k(n-k)\delta_{n-k}}
\Phi_{\delta-e_{n-k}-e_k} + (\alpha+n \beta)\sqrt{n \delta_n}
\Phi_{\delta-e_n}.
\eea
Daraus folgt f"ur $L_{\alpha,\beta} \Phi_{\delta}$ mit $\|\delta\| =d$
\bea\label{Lop}
L_{\alpha,\beta}\Phi_{\delta} &=& \sum_{n=1}^d \bigg\{ \sum_{k=1}^{\infty}
\sqrt{k (\delta_k+1)(n+k)\delta_{n+k}} \Phi_{\delta-e_{n+k}+e_n}
\nonumber\\
&&+\frac{1}{2} \sum_{k=1}^{n-1} \sqrt{k \delta_k(n-k)\delta_{n-k}}
\Phi_{\delta-e_{n-k}-e_k} + (\alpha+n \beta)\sqrt{n \delta_n}
\Phi_{\delta-e_n}\bigg\},
\eea}
woran wir explizit sehen, da\3 ${\frS}_d$ eine algebraische Menge ist. Im
folgenden werden wir die Abh"angigkeit  von R"aumen und
Operatoren von $(\alpha,\beta)$ nicht mehr mitf"uhren.

Wir f"uhren die folgenden Bezeichnungen
ein: Sei ${\cal F}_d = \Lin \{ \Phi_{\delta} : \|\delta\| = d\}$,
${\cal F}_d^j=\Lin \{\Phi_{\delta} : \|\delta\|=d, \delta_1=j \}$,
$\overline{{\cal F}}_d = \plus_{l=0}^d {\cal F}_l$ und $\overline{\cal
F}_d^j=\plus_{l=0}^d {\cal F}_l^j.$

Es ist ${\cal F}_d^d=\CC\: \Phi_{(d,0,0,\ldots)}$, d.h $\dim {\cal
F}_d^d=1$. Au\3erdem ist $\dim {\cal F}_d^{d-1}=0$. Es reicht also zu
zeigen, da\3 $L\restrict{\plus_{j=0}^{d-2} {\cal F}_d^j}$ injektiv ist. Wie
man an (\ref{Lop}) und
\ben
L({\cal F}_d^k) \subset \plus_{j=k-2}^{k+1} \overline{\cal F}_{d-1}^j
\een
erkennt, reicht es sogar zu zeigen, da\3
\ben
L(k):{\cal F}_d^k \lra \overline{\cal F}_{d-1}^{k+1},\qquad L(k)=
\pr_{\overline{\cal F}_{d-1}^{k+1} } \circ L
\een
f"ur $k=0,1,\ldots,d-2$ injektiv ist:\\
Sei $\Psi=(\Psi_0,\Psi_1,\ldots,\Psi_{d-2}) \in \plus_{j=0}^{d-2} {\cal
F}_d^j$ mit $L\Psi=\sum_{j=0}^{d-2} L \Psi_j =0.$
$L$ ist injektiv als Abbildung ${\cal F}_d^{d-2} \lra \overline{\cal
F}_d^{d-1}$, denn dort stimmt sie genau mit $L(d-2)$ "uberein, und es folgt
$\Psi_{d-2}=0$. Wir erhalten also $L\Psi=\sum_{j=0}^{d-3} L\Psi_j$ und
$L\Psi_{d-3}\in {\cal F}_{d-1}^{d-4}\oplus {\cal
F}_{d-1}^{d-3}\oplus {\cal F}_{d-1}^{d-2}$. Die Abbildung in den dritten
 Summanden ist $L(d-3)$, aus deren Injektivit"at folgt damit
$\Psi_{d-3}=0$. Induktiv erhalten wir aus diesen Argumenten $\Psi=0$.

Es bleibt also die Injektivit"at von $L(k)$ zu zeigen. Sei $\Phi_{\delta}
\in {\cal F}_d^k$. Es gilt
\ben
L(k)\Phi_{\delta}=\sum_{n=1}^l\sqrt{(k+1)(n+1)\delta_{n+1}}\Phi_{
\delta-e_{n+1}+e_1}.
\een
Jeder Basisvektor in $\overline{\cal F}_{d-1}^{k+1}$ tritt h"ochstens im Bild
eines Basisvektors aus ${\cal F}_d^k$ auf, denn ist $\langle
\Phi_{\varepsilon},L(k),\Phi_{\delta} \rangle \ne 0$ und $\|\varepsilon\|
= d-l$, so gilt
$\varepsilon=(\varepsilon_1,\varepsilon_2,\ldots)=(k+1,\varepsilon_2,
\ldots)$ und
$\delta=(k,\varepsilon_2,\ldots,\varepsilon_l,
\varepsilon_{l+1}+1,\ldots).$ Damit mu\3 $L(k)$ injektiv sein. \hfill$\Box$

Die Aussage von Lemma \ref{algset} gilt in "aquivalenter Form auch f"ur die
Verma--Moduln, es gibt aber keinen so elementaren Beweis f"ur diese
Behauptung, da die Kommutatoren der Virasoro--Algebra viel komplizierter
sind als die der Heisenberg\-algebra. Deshalb folgt das erst aus der
Klassifizierung der Verma--Darstellungen.

Wir beschlie\3en diesen Abschnitt mit einer elementaren Dualit"atsaussage
f"ur Vir--Moduln und ihre kontragredienten Moduln:
\begin{lem}\label{kontra}
Sei $M$ ein $\ZZ$--graduierter Vir--Modul von endlichem Typ (d.h. $M=\plus
M_n$ und $\dim(M_n)< \infty$), und $\overline{M}$ sein kontragredienter
Modul.
Dann sind "aquivalent:
\begin{itemize}
\item[(i)] $M_n$ enth"alt einen singul"aren Vektor vom Grad $n$,
\item[(ii)] $\overline{M}_n$ enth"alt einen kosingul"aren Vektor vom  Grad
$n$.
\end{itemize}
\end{lem}
{\bf Beweis.}\\
Seien $L_k$ die Generatoren von Vir in der Darstellung auf $M$ und
$\overline{L}_k$ die entsprechenden Generatoren auf $\overline{M}$.
Sei
\[
L_+=\sum_{k=1}^n L_k : M_n \lra \plus_{l<n} M_l,
\]
und
\[
\overline{L}_-=\sum_{k=1}^n \overline{L}_{-k} : \plus_{l<n}
\overline{M}_l  \lra \overline{M}_n.
\]
Die Abbildungen $L_+$ und $\overline{L}_-$ sind zueinander dual, woraus
die Behauptung folgt.\hspace*{\fill}$\Box$

Aus Lemma \ref{kontra} folgt sofort, da\3 die kontragredienten
Verma--Moduln $\overline{V}(h,c)$ abgesehen von dem H"ochstgewichtsvektor
 keine singul"aren, sondern nur kosingul"are Vektoren enthalten, da
Verma--Moduln nur singul"are und keine kosingul"are Vektoren enthalten.
Insbesondere besitzen reduzible Moduln $\overline{V}(h,c)$ keinen
zyklischen H"ochstgewichtsvektor.
\section{Die Struktur der Verma--Moduln}
\subsection{Die Jantzen--Filtration in $V(h,c)$}
\label{jantzenfilter}
Wir verwenden hier \cite{RC1} und \cite{RCW}.
Sei $(.,.)_{h,c}=S_{h,c}(.,.)$ die Shapovalov Form auf $V(h,c)$ und sei
$(h,c)\in \CC^2$ fest gew"ahlt.
Sei $V={\frU}(N_-)={\frU}(\Lin\{L_n:n<0\})$
und seien $V_n$ die Elemente vom Grad $n$ in der kanonischen Graduierung
von $V$.
F"ur jedes $z \in \CC$ existiert ein Vektorraumisomorphismus
\bea i_z : V(h+z,c) &\lra& V   \nonumber\\
           w.v_{h+z,c} &\lmt& w,\nonumber
\eea
wobei $v_{h+z,c}$ der H"ochstgewichtsvektor von $V(h+z,c)$ ist. Wir
definieren eine Vir--Modulstruktur auf $V$ durch
\ben
\pi_z(X).w := i_z X. i_z ^{-1}w
\een
f"ur $X \in$ Vir, $w \in V$. Es ist  offensichtlich $(V,\pi_z)\simeq
V(h+z,c)$ als Vir--Moduln. Wir definieren die Shapovalov--Form auf $V$
durch
\[
B_z(w_1,w_2):= (i_z^{-1} w_1,i_z^{-1} w_2)_{h+z,c}
\]
f"ur $w_1,w_2 \in V$. $B_{z,n}$ sei die Einschr"ankung von $B_z$ auf $V_n$.

Sei ${\cal O}(V)$ die Menge aller Keime in Null analytischer Funktionen
mit Werten in einem endlichdimensionalen Teilraum von $V$ und
sei ${\cal O}(\CC)$ die Menge der Keime in Null analytischer Funktionen
mit Werten in $\CC$.

Wir definieren f"ur $k \in \NN_0$
\ben
{\cal O}_{k}(V) := \left\{ f \in {\cal O}(V) : B_z(f(z),w) \in z^k {\cal
O}(\CC) \mbox{ f"ur alle w }\in V \right\},
\een
\ben
V_{(k)}:=\left\{ f(0) : f \in {\cal O}_k(V)\right\}
\een
und $J_k:=i_0^{-1}(V_{(k)})$.
Es gilt $V_{h,c}=J_0 \supset J_1 \supset J_2 \supset \ldots$ und
$\bigcap\limits_{k \ge 0} J_k = \{ 0\}$, denn $B_{z,n}$ ist f"ur jedes $n$
ein Polynom in $h$ und $c$.\\
Au\3erdem ist jedes $J_k$ ein Untermodul, denn
ist $v \in J_k$, $X \in$ Vir, so existiert zu $v'=i_0 v$ ein Keim $f\in
{\cal O}_k(V)$ mit $f(0)=v'$, und wegen $B_z(X.f(z),w)=B_z(f(z),X.w) \in
z^k {\cal O}_k(\CC)$ f"ur alle $w \in V$  folgt $X.f \in {\cal O}_k(V)$
oder  $X.v \in J_k$.

$\{J_k : k\in \NN_0\}$ ist damit eine absteigende Filtration in $V(h,c)$,
die Jantzen--Filtration. Wir wollen  diese Filtration nun etwas genauer
untersuchen.
\subsection{Hilfsmittel}
Sei $V$ ein endlichdimensionaler Vektorraum "uber $\CC$ und $S_z \in {\cal
O}(\End(V))$. F"ur $k \in \NN_0$ definieren wir
\[
 {\cal O}_k^S(V):= \left\{f \in {\cal O}(V): S_z f(z) = z^k g(z), g \in
{\cal O}(V)\right\}
\]
als die Keime, die unter $S_z$ mit der Ordnung $\ge  k$ in Null
verschwinden, und
\[
V^S_k =\left\{f(0):f \in {\cal O}_k^S(V)\right\}.
\]
F"ur $f \in {\cal O}(\CC)$, $f(z)=\sum_{i \ge 0} f_i z^i$ sei $\ord_0
f:=\min\{i : f_i \ne 0\}$. Wir bestimmen zuerst eine Normalform f"ur $S_z$.
\begin{lem}\label{normal}
Sei $e_1,\ldots e_n$ eine Basis von $V$. Weiter sei $S_z \in {\cal
O}(\End(V))$ und es existiere ein $\varepsilon>0$ so, da\3 $S_z$ bijektiv
f"ur $z \in U_{\varepsilon}(0)\setminus\{0\}$ ist.  Dann gibt es
$L_z,R_z\in {\cal O}(\End V)$, so da\3 $\det L_z$ und $\det R_z$ konstant
und  ungleich Null sind und   $L_z S_z R_z$ bez"uglich der Basis $\{e_i\}$
durch die Diagonalmatrix
\[
D_z=\pmatrix{s_1(z)&\cdots&0 \cr
               \vdots&\ddots&\vdots\cr
               0&\cdots&s_n(z)\cr}
\]
mit $0\le \ord_0 s_1\le \ldots \le \ord_0 s_n $ gegeben ist.
\end{lem}
{\bf Beweis.}\\
Sei
$E_{i,j}=(\delta_{i,k} \delta_{j,l})_{k,l}\in M_{n,n}$,
$P_{i,j}=\mbox{Id}-E_{i,i}-E_{j,j}+E_{i,j}+E_{j,i}$ und
$Q_{i,j}(\beta(z))=\mbox{Id}+\beta(z)E_{i,j}$.
(Linksmultiplikation mit $P_{i,j}$ vertauscht $i$-te und $j$-te Zeile,
Rechtsmultiplikation mit $P_{i,j}$ die entsprechenden Spalten, es ist
$\det P_{i,j}=1$ und $\det Q_{i,j}(\beta)=1+\delta_{i,j}\beta$.)\\
Wir f"uhren den Beweis durch Induktion nach n: F"ur $n=1$ ist nichts zu
zeigen. Sei $n>1$ und seien $s_{i,j}(z)$ die Matrixelemente von $S_z$
bez"uglich der Basis $\{e_1,\ldots,e_n\}$. Es existieren $k,l$ mit
$\ord_0s_{k,l} \le \ord_0 s_{i,j}$ f"ur alle $i,j$. Die Matrix
$C_z:=P_{1,k} S_z P_{1,l}$ hat $s_{k,l}$ in der ersten Zeile und Spalte.
F"ur die Matrixelemente $c_{i,j}(z)$ von $C_z$ gilt
$\frac{c_{i,j}}{c_{1,1}}\in {\cal O}(\CC)$. Wir multiplizieren $C_z$
von links mit
\[
T_z=Q_{n,1}(-\frac{c_{n,1}(z)}{c_{1,1}(z)}) \cdots
Q_{2,1}(-\frac{c_{2,1}(z)}{c_{1,1}(z)})
\]
und von rechts mit
\[
U_z=Q_{1,2}(-\frac{c_{1,2}(z)}{c_{1,1}(z)}) \cdots Q_{1,n}(-\frac{
c_{1,n}(z)}{c_{1,1}(z)}).
\]
Die resultierende Matrix hat in der ersten Zeile und Spalte bis auf
$c_{1,1}$ nur Nullen als Eintr"age, wir k"onnen also auf die Abbildung
$\tilde{S}$, die durch Streichen der ersten Zeile und Spalte entsteht, die
Induktionsvoraussetzung anwenden, und erhalten
$\tilde{L_z}\tilde{S_z}\tilde{R_z}=\tilde{D_z}.$ Wir erhalten insgesamt
$(1\plus\tilde{L_z}) T_z P_{1,k} S_z P_{1,l} U_z (1\plus \tilde{R_z})=D_z$
als Normalform f"ur $S_z$, die alle gew"unschten Eigenschaften
hat.\hfill$\Box$\\
Damit k"onnen wir die folgende wichtige Beziehung beweisen:
\begin{lem}\label{dimension}
Sei $S_z\in {\cal O}(\End(V))$ und $\det S_z \ne 0$ f"ur $z\in
U_{\varepsilon}(0)\setminus\{0\}$.
Dann gilt
\ben
\sum_{k\ge 1} \dim V_k^S =\ord_0 \det S_z.
\een
\end{lem}
{\bf Beweis.}\\
Nach Lemma \ref{normal} existieren $L_z$ und $R_z$ mit
\[
D_z=L_z S_z R_z=\pmatrix{s_1(z)&&0\cr&\ddots&\cr 0&&s_n(z)\cr},
\]
und mit
$\alpha_i:=\ord_0s_i(z)$ gilt $\ord_0 \det S_z=\ord_0 D_z =\sum \alpha_i$.
Weiter ist
\bea
{\cal O}_k^D(V)&=& \left\{h\in {\cal O}(V):D_z h(z)=z^k g(z), g \in {\cal
O}(W) \right\}\nonumber \\
&=&\left\{h \in {\cal O}(V):L_z S_z R_zh(z)=z^k g(z), g \in {\cal
O}(W)\right\} \nonumber\\
&=& \left\{ R^{-1}f \in {\cal O}(V): L_z S_z f(z)=z^k g(z), g \in {\cal
O}(W) \right\} \nonumber \\
&=&\left\{ R^{-1} f \in {\cal O}(V): S_z f(z)=z^k \tilde{g}(z), \tilde{g}
\in {\cal O}(W)\right\}, \nonumber
\eea
denn $S_zf(z)=z^kg(z) \Longleftrightarrow L_z S_z f(z)=z^k L_z g(z) =:
z^k\tilde{g}(z)$. Wir haben damit gezeigt, da\3 $V_k^D=R_0^{-1} V_k^S$
gilt. Wir w"ahlen nun f"ur $j=1,\ldots,n$ $h_j \in {\cal O}(V)$ mit
$h(0)=e_j$. Es folgt $D_z h_j(z)=z^{\alpha_j}g(z)$ und
$V_k^D=\sum\limits_{\alpha_j \ge k} \CC e_j$. Wir erhalten $\dim
V_k^S=\dim V_k^D=\sharp\{j:\alpha_j \ge k\}$ und schlie\3lich
\[
\sum_{k \ge 1}\dim V_k^S = \sum_{k \ge 1} \sharp\{j:\alpha_j \ge
k\}=\sum_{k=1}^n \alpha_j = \ord_0 \det S_z.
\]
 \nopagebreak\hspace{\fill}$\Box$\newpage
Sei  weiter $(.,.)$ eine nichtdegenerierte, symmetrische Bilinearform auf
$V$ mit $(S_z v,w)=(v,S_z w)$ f"ur alle $v,w\in V$.
Wir m"ochten nun eine Form auf $V_k^S$ definieren. Dazu setzen wir
$T_z(v,w)=(S_z v,w)$. Es ist offensichtlich $T_z\in {\cal O}(\Hom(V\times
V,\CC))$.\\
F"ur $f,g\in {\cal O}(V)$ setzen wir $\langle f,g\rangle(z)\equiv\langle
f(z),g(z) \rangle:=T_z(f(z),g(z))$, und f"ur $f,g\in {\cal O}^S_k(V)$
setzen wir $\langle f,g\rangle_k(z):= z^{-k}\langle f,g\rangle (t)$.

$\langle f,g\rangle_k(0)$ h"angt nur von $f(0)$ und $g(0)$ ab: F"ur
$f,g\in{\cal O}^S_k(V)$ mit $S_zg(z)=z^k\tilde{g}(z)$, $f(0)=0$ und
deshalb $f(z)=zh(z)$ f"ur ein $h\in {\cal O}(V)$ folgt
\bea\label{kform}
\langle f,g\rangle_k(z)&=& z^{-k} \langle f,g\rangle(t)=z^{-k}
T_z(f(z),g(z)) \nonumber\\
&=& z^{-k} (f(z),S_z g(z))= z (h(z),\tilde{g}(z)) \stackrel{z\to 0}{\lra}
0.
\eea
Damit ist f"ur $v,w\in V_k^S$, $v=f(0)$ und $w=g(0)$
\ben\label{defform}
\langle v,w\rangle_k :=\langle f,g\rangle_k(0)
\een
wohldefiniert, und, wie man an (\ref{kform}) sieht,  gilt $\langle
V_k^S,V^S_{k+1}\rangle =0$. Konsequenterweise induziert $\langle
.,.\rangle$ eine symmetrische Bilinearform auf $V_k^S/V^S_{k+1}$, die wir
ebenfalls mit $\langle .,.\rangle_k$ bezeichnen.
\begin{lem}\label{lemma3} Sei $S_z$ wie vorher. Dann ist
$\langle.,.\rangle_k$ nichtdegeneriert auf
$\displaystyle\qquot{V_k^S}{V^S_{k+1}}$, d.h. es gilt
$\Rad\langle.,.\rangle_k=V_{k+1}^S$.
\end{lem}
{\bf Beweis.}
Wir verwenden dieselben Bezeichnungen wie im vorherigen Lemma.
Sei $v_i:=R_0 e_i$ und $f_i(z):=R_z e_i$. Dann ist
$V_k^S=\sum\limits_{\alpha_i \ge k} \CC v_i$.\\
Ist nun $\alpha_i=k$ und $\alpha_j \ge k$ f"ur ein $i \ne j$, dann gilt
\bea
\langle f_i,f_j\rangle_k(z)&=& z^{-k} (S_z R_z e_i,f_j(z))\nonumber\\
&=& z^{-k}(L_z^{-1} s_i(z) e_i,f_j(z)) \nonumber \\
&=& (L_z^{-1} \tilde{s}_i(z)e_i,f_j(z)),
\eea
wobei wir $s_i(z)=z^{k}\tilde{s}(z)$ gesetzt haben. Mit $w_i :=
L_0^{-1}\tilde{s}_i(z) e_i$ folgt $\langle v_i,v_j\rangle_k=(w_i,v_j)$,
und f"ur $\alpha_j >k$ ist $0=\langle v_i,v_j\rangle =(w_i,v_j)$. Da
$V=\sum \CC w_i$ ist folgt insgesamt, da\3 die Matrix $(w_i,v_j)_{i,j}$ die
Form
\[
\left(\matrix{\mbox{\fbox{$W_0$}}&&0\cr
         & \ddots& \cr
         *&&\mbox{\fbox{$W_l$}}\cr
         }\right)
\mbox{ mit  } W_k=\{ (w_i,v_j)_{i,j} : \alpha_i=\alpha_j=k \}
\]
hat. Da $(.,.)$ nicht ausgeartet ist, folgt $0 \ne \det\left(
(W_i,v_j)_{i,j}\right) = \prod \det W_k$ und deshalb $\det W_k \ne 0$,
falls $V_k^S \ne V_{k+1}^S$. Insgesamt gilt, wenn wir $\langle
.,.\rangle_k$ als Form auf $\qquot{V_k^S}{V_{k+1}^S}$ auf\/fassen, $\det
\langle .,.\rangle_k= \det W_k \ne 0$ falls $V_k^S \ne
V_{k+1}^S$.\hfill$\Box$

Kommen wir nun zur"uck zur Shapovalov-Form: F"ur jedes $n \in \NN_0$ ist
$B_{z,n}:V_n\times V_n \lra \CC$ eine symmetrische Bilinearform. Sei
$(.,.)_n$ das kanonische Skalarprodukt auf $V_n$. Es gibt genau ein
$\tilde{B}_{z,n} \in {\cal O}(\End V_n)$ mit
$B_{z,n}(v,w)=(\tilde{B}_{z,n} v,w)=(v,\tilde{B}_{z,n} w)$ f"ur alle $v,w
\in V_n$. Weiter haben wir ${\cal O}_k(V_n)={\cal
O}_k^{\tilde{B_{z,n}}}(V_n)$. Wir k"onnen also die Lemmata \ref{normal},
\ref{dimension} und \ref{lemma3} auf $\tilde{B_{z,n}}$ f"ur jedes $n$
anwenden und erhalten f"ur die Jantzen--Filtration von $V(h,c)$:
\begin{satz}\label{jantzen} Die Jantzen--Filtration $V(h,c)=J_0\supset J_1
\supset \ldots$ der Verma--Moduln hat folgende Eigenschaften:
\begin{itemize}
\item[(i)] Die Shapovalov--Form $S_{h,c}(.,.)$ definiert eine invariante,
nicht--degenerierte Form $\langle.,.\rangle_k$ auf den irreduziblen
Quotienten--Moduln $J_k/J_{k+1}$.
\item[(ii)] Sei $J_k = \plus_{n \ge 0} J_{k,n}$ und $S_{h,c}(.,.)=\plus_{n
\ge 0} S_{h,c,n}(.,.)$. Dann gilt f"ur alle $n\in \NN_0$
\ben\label{dimformel} \ord_{z=0}\left( \det S_{h+z,c,n}(.,.)
\right)=\sum_{k \ge 0} \dim J_{k,n}.
\een
\end{itemize}
\end{satz}
{\bf Beweis.}\\
Die Invarianz von $\langle.,.\rangle_k$ folgt sofort aus (\ref{defform})
und der Invarianz der Shapovalov--Form. Es bleibt zu zeigen, da\3
$J_k/J_{k+1}$ irreduzibel ist. Ist $J_k \ne J_{k+1}$, so folgt das genau
wie f"ur $V(h,c)$ und $S_{h,c}(.,.)$, wo man zeigt, da\3 der gr"o\3te echte
Untermodul genau $\Rad S_{h,c}(.,.)$ ist. Hier erhalten wir als gr"o\3ten
echten Untermodul von $J_k$ genau $J_{k+1}$. Die "ubrigen Aussagen folgen
direkt aus Lemma \ref{dimension} und Lemma \ref{lemma3}.\hfill$\Box$\\

\subsection{Der Klassifikationssatz f"ur Verma--Moduln}
Wir haben nun die n"otigen Hilfsmittel bereit, um die Verma--Moduln zu
klassifizieren. Dazu ben"otigen wir die folgende Fallunterscheidung, die
sich an \cite{FF} anlehnt. Seien $(h,c)\in \CC^2$ beliebig. Wir
unterscheiden:
\begin{itemize}
\item[$I:$] $\Phi_{r,s}(h,c)=0$ hat keine ganzzahlige L"osung $r,s$.
\item[$II:$] $\Phi_{r,s}(h,c)=0$ hat eine ganzzahlige L"osung $r,s$ mit $rs
>0$ ($II_+$) oder $rs \le 0$ ($II_-$).
\item[$III:$] $\Phi_{r,s}=0$ hat unendlich viele ganzzahlige L"osungen, d.h
es ist entweder $c<1$ ($III_-$) oder $c>25$ ($III_+$) und $c$ und $h$ sind
durch (\ref{para}) gegeben.\\
Wir unterscheiden weiter\\
\indent$III_-,III_+$: \qquad$r \not\equiv 0 \mod p$ und $s \not\equiv 0
\mod q$, \\
\indent$III_-^0, III_+^0$: \qquad$r \equiv 0 \mod p$ oder $s \equiv 0 \mod
q$.
\end{itemize}
Das Ziel dieses Abschnittes ist der folgende Satz:
\begin{satz}\label{klasse} Die Verma--Moduln sind im Fall $I$ und $II_-$
irreduzibel, und enthalten im Fall $II_+$ genau einen singul"aren Vektor,
der den gr"o\3ten echten Untermodul erzeugt. F"ur die anderen F"alle gilt:
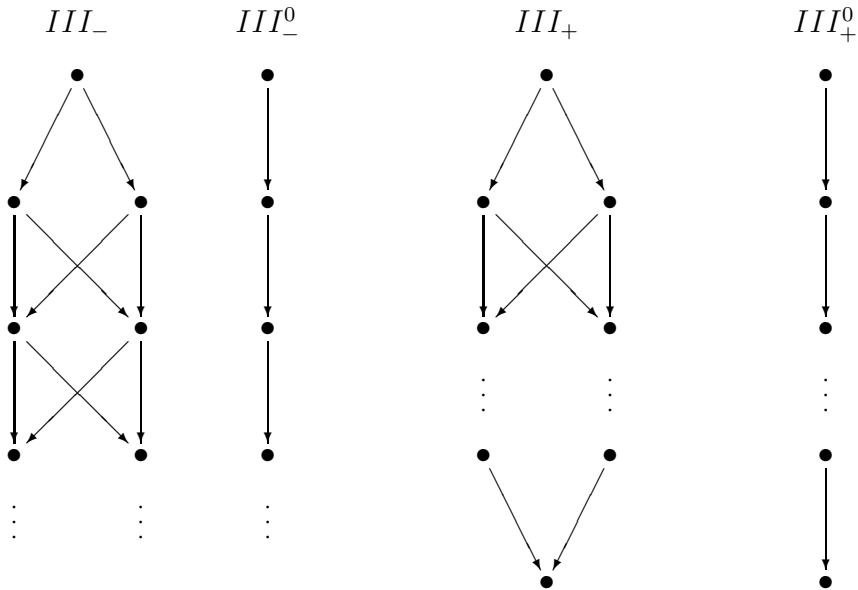
\begin{center}\setlength{\unitlength}{.008em}
\begin{picture}(3200,2200)
\put(250,2200){\makebox(0,0){$III_-$}}
\put(250,2000){\makebox(0,0){$\bullet$}}
\put(225,1950){\vector(-1,-2){200}}
\put(275,1950){\vector(1,-2){200}}
\put(0,1500){\makebox(0,0){$\bullet$}}
\put(50,1450){\vector(1,-1){400}}
\put(450,1450){\vector(-1,-1){400}}
\put(0,1450){\vector(0,-1){400}}
\put(0,1000){\makebox(0,0){$\bullet$}}
\put(50,950){\vector(1,-1){400}}
\put(450,950){\vector(-1,-1){400}}
\put(0,950){\vector(0,-1){400}}
\put(0,500){\makebox(0,0){$\bullet$}}
\multiput(0,300)(0,-60){3}{\circle*{1}}
\put(500,1500){\makebox(0,0){$\bullet$}}
\put(500,1450){\vector(0,-1){400}}
\put(500,1000){\makebox(0,0){$\bullet$}}
\put(500,950){\vector(0,-1){400}}
\put(500,500){\makebox(0,0){$\bullet$}}
\multiput(500,300)(0,-60){3}{\circle*{1}}

\put(1000,2200){\makebox(0,0){$III_-^0$}}
\put(1000,2000){\makebox(0,0){$\bullet$}}
\put(1000,1950){\vector(0,-1){400}}
\put(1000,1500){\makebox(0,0){$\bullet$}}
\put(1000,1450){\vector(0,-1){400}}
\put(1000,1000){\makebox(0,0){$\bullet$}}
\put(1000,950){\vector(0,-1){400}}
\put(1000,500){\makebox(0,0){$\bullet$}}
\multiput(1000,300)(0,-60){3}{\circle*{1}}
%
\put(2100,2200){\makebox(0,0){$III_+$}}
\put(2100,2000){\makebox(0,0){$\bullet$}}
\put(2075,1950){\vector(-1,-2){200}}
\put(2125,1950){\vector(1,-2){200}}
\put(1850,1500){\makebox(0,0){$\bullet$}}
\put(1900,1450){\vector(1,-1){400}}
\put(2300,1450){\vector(-1,-1){400}}
\put(1850,1450){\vector(0,-1){400}}
\put(1850,1000){\makebox(0,0){$\bullet$}}
\put(1850,500){\makebox(0,0){$\bullet$}}
\multiput(1850,800)(0,-60){3}{\circle*{1}}
\put(2350,1500){\makebox(0,0){$\bullet$}}
\put(2350,1450){\vector(0,-1){400}}
\put(2350,1000){\makebox(0,0){$\bullet$}}
\put(2350,500){\makebox(0,0){$\bullet$}}
\multiput(2350,800)(0,-60){3}{\circle*{1}}
\put(1875,450){\vector(1,-2){200}}
\put(2325,450){\vector(-1,-2){200}}
\put(2100,0){\makebox(0,0){$\bullet$}}
%
\put(3200,2200){\makebox(0,0){$III_+^0$}}
\put(3200,2000){\makebox(0,0){$\bullet$}}
\put(3200,1950){\vector(0,-1){400}}
\put(3200,1500){\makebox(0,0){$\bullet$}}
\put(3200,1450){\vector(0,-1){400}}
\put(3200,1000){\makebox(0,0){$\bullet$}}
\multiput(3200,800)(0,-60){3}{\circle*{1}}
\put(3200,500){\makebox(0,0){$\bullet$}}
\put(3200,450){\vector(0,-1){400}}
\put(3200,0){\makebox(0,0){$\bullet$}}
\end{picture}\setlength{\unitlength}{.01em}
\end{center}
Dabei entsprechen Punkte singul"aren Vektoren in dem Modul, und ein Pfeil
geht von einem Punkt zu einem zweiten Punkt, wenn der zweite Vektor in dem
vom ersten Vektor erzeugten Untermodul liegt. Der gr"o\3te echte Untermodul
ist im Fall $III_-^0$ und $III_+^0$ von einem, im Fall $III_-$ und $III_+$
von zwei singul"aren Vektoren erzeugt.
\end{satz}
\begin{bem} Im Fall  $III_-$ mit $0<r<p$ und $0<s<q$ kann man die Gewichte
und Grade der singul"aren Vektoren explizit angeben. Mit dem Gewicht eines
Vektors $w$ meinen wir den $L_0$--Eigenwert von $w$, den wir mit {\rm
wt}$(w)$ bezeichnen. Aus diesen Moduln ist der Zustandsraum der minimalen
Modelle der konformen Quantenfeldtheorie zusammengesetzt.\\[1mm]
\parbox{3cm}{\begin{center}
\setlength{\unitlength}{.005em}%
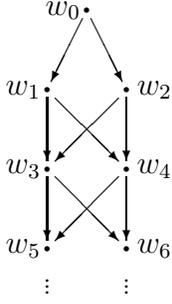
\begin{picture}(1000,2000)
\put(500,2000){\circle*{30}}
\put(475,1950){\vector(-1,-2){200}}
\put(525,1950){\vector(1,-2){200}}
\put(250,1500){\circle*{30}}
\put(300,1450){\vector(1,-1){400}}
\put(700,1450){\vector(-1,-1){400}}
\put(250,1450){\vector(0,-1){400}}
\put(250,1000){\circle*{30}}
\put(300,950){\vector(1,-1){400}}
\put(700,950){\vector(-1,-1){400}}
\put(250,950){\vector(0,-1){400}}
\put(250,500){\circle*{30}}
\multiput(250,300)(0,-40){3}{\circle*{1}}
\put(750,1500){\circle*{30}}
\put(750,1450){\vector(0,-1){400}}
\put(750,1000){\circle*{30}}
\put(750,950){\vector(0,-1){400}}
\put(750,500){\circle*{30}}
\multiput(750,300)(0,-40){3}{\circle*{1}}
\put(350,2000){\makebox(0,0){$w_0$}}
\put(100,1500){\makebox(0,0){$w_1$}}
\put(925,1500){\makebox(0,0){$w_2$}}
\put(100,1000){\makebox(0,0){$w_3$}}
\put(925,1000){\makebox(0,0){$w_4$}}
\put(100,500){\makebox(0,0){$w_5$}}
\put(925,500){\makebox(0,0){$w_6$}}
\end{picture}\end{center}
\setlength{\unitlength}{.01em}%
}
\hfill\parbox{12cm}{Die singul"aren Vektoren numerieren wir wie im Bild
angegeben. Wir erhalten dann f"ur $V(h_{r,s}(-q/p),c(-q/p))$ und $i \ge 0$:
{
\begin{eqnarray}\label{gewicht}
{\rm wt}(w_{4i})&=&h^-_{r,s+2iq}, \nonumber\\
{\rm wt}(w_{4i+1})&=&h^-_{r,-s+2(i+1)q}, \nonumber\\
{\rm wt}(w_{4i+2})&=&h^-_{r,-s-2iq},\\
{\rm wt}(w_{4i+3})&=&h^-_{r,s-2(i+1)q}.\nonumber
\end{eqnarray}
}}
Die Grade sind gegeben durch
\begin{eqnarray}\label{grade}
\deg(w_{4i})&=&\big(ip-r\big)\big(iq+s\big) + rs,\nonumber\\
\deg(w_{4i+1})&=&\big((i+1)p-r\big)\big((i+1)q-s\big),\nonumber\\
\deg(w_{4i+2})&=&\big(ip+r\big)\big(iq+s\big),\\
\deg(w_{4i+3})&=&\big((i+1)p+r\big)\big((i+1)q-s\big) + rs.\nonumber
\end{eqnarray}
\end{bem}
Wir bezeichnen mit $[w_i]$ den von $w_i$ erzeugten Untermodul. Aus den
Gleichungen (\ref{gewicht}) folgt $[w_{4i}] \simeq
V(h_{r,s+2iq}^-,c(-q/p))$ (analog f"ur die anderen singul"aren Vektoren),
denn jeder durch einen singul"aren Vektor erzeugte Untermodul ist frei "uber
$\Vir_+$ und somit wieder ein Verma--Modul. An Gl. (\ref{grade}) erkennen
wir, da\3 die Vektoren $w_{4i}$ f"ur $i >0$ und $w_{4i+3}$ f"ur $i \ge 0$
nicht auf Kurven der Form (\ref{para}) durch $V(h_{r,s}^-,c(-q/p))$
liegen, da die entsprechenden Grade nicht faktorisieren. Diese Vektoren
entstehen als singul"are Vektoren von Untermoduln, z.B. liegen sie auf
Kurven der Form (\ref{para}) durch $V(h_{r,s}^- + rs ,c(-q/p))\simeq
[w_2]$.\\
Feigin und Fuks haben in \cite{FF} einen Beweis f"ur Satz \ref{klasse}
angegeben, der Methoden der algebraischen Geometrie verwendet. Wir werden
im folgenden sehen, da\3 man Satz \ref{klasse} auch mit elementaren
Methoden beweisen kann, wenn man konkrete Eigenschaften der $h_{r,s}$ und
die Aussagen "uber die Jantzen--Filtration von Satz \ref{jantzen}
verwendet. Da die F"alle $I$ und $II$ trivial sind, k"onnen wir in den
verbleibenden F"allen die Determinantenformeln (\ref{detmini}) und
(\ref{notdetmini}) verwenden. Damit sind die Aussagen "uber singul"are
Vektoren auf einfache Aussagen "uber $h_{r,s}^{\pm}$ zur"uckgespielt. Wir
m"ussen dazu einfach feststellen, auf welchen Levels in der
Kac--Determinante neue Nullstellen auftauchen. Anders gesagt, wir m"ussen
 bestimmen, wann $h_{r,s}^{\pm}=h_{m,n}^{\pm}$ gilt.
\begin{lem}\label{lemma6}
\ben h_{r,s}^+=h_{m,n}^+ \Longleftrightarrow \cases{ m=r - k p,\; n=s
+kq&f"ur $k \in \ZZ$ \cr
\mbox{oder}\cr
m=-r-kp,\; n=-s+kq & f"ur $k \in \ZZ$\cr}
\een
\ben h_{r,s}^-=h_{m,n}^- \Longleftrightarrow \cases{ m=r+kp,\; n=s+kq&f"ur
$k \in \ZZ$ \cr
\mbox{oder}\cr
m=-r+kp,\; n=-s+kq & f"ur $k \in \ZZ$\cr}
\een
\end{lem}
Der Beweis ist eine triviale Rechnung.\hfill$\Box$\\[3mm]
{\bf Beweis von Satz \ref{klasse}}\\
Da $h_{r,s}^-=h_{-r,-s}^-=h_{r+p,s+q}^-$ gilt, k"onnen wir uns im Fall
$III_-$ o.B.d.A auf $0<r<p$ und $
jq<s<(j+1)q$ f"ur ein $j \in \NN_0$ beschr"anken. Weiter sehen  wir, da\3
wegen (\ref{gewicht}) die Moduln $V(h_{r,s+jq},c)$ mit $0<r<p$ und $0<s<q$
 als Untermoduln von $V(h_{r,s},c)$ auftreten, es reicht also den Fall
$III_-$ f"ur diese Werte von $r$ und $s$ f"ur beliebige, relativ prime $p$
und $q$ zu beweisen.

Ist z.B $r \equiv 0 \mod p$, was wir wegen $h_{r,s}^-=h_{r-p,s-q}^-$ auf
$r=0$ und $s$ beliebig zur"uckf"uhren k"onnen,  sieht man leicht, da\3
(zun"achst f"ur $0<s<q$) nur die Vektoren $w_0,w_1,w_4,w_5,w_8,\ldots$ des
Diagrammes f"ur den Fall $III_-$ "ubrigbleiben, die ein Diagramm vom Typ
$III_-^0$ bilden, die "ubrigen Nullstellen verschwinden (d.~h. sie haben
Grad $\le 0$ in den entsprechenden Moduln).\\
Auch hier treten die F"alle mit $jq<s<(j+1)q$ als Untermoduln auf.
Ein interessanter Spezialfall ist $r\equiv 0 \mod p$ und $s\equiv 0 \mod
q$. Hier erh"alt man durch Entartung doppelte Nullstellen, d.~h. die
entsprechenden singul"aren Vektoren liegen auf zwei Kurven der Form
(\ref{para}). F"ur $r=0$ und $s=0$ erh"alt man wt$(w_0)={\rm wt}(w_2)$ und
${\rm wt}(w_{4k+1})={\rm wt}(w_{4k+3})={\rm wt}(w_{4k+4})={\rm
wt}(w_{4k+6})$. Die Vektoren $w_{4k+3}$ und $w_{4k+4}$ liegen auf keiner
Kurve der Form (\ref{para}), deshalb liegen $w_{4k+1}=w_{4k+6}$ nur auf
dem Schnittpunkt zweier Kurven.\\ Eine analoge "Uberlegung kann man f"ur die
F"alle $III_+$ und $III_+^0$ anstellen, dazu aber sp"ater.

\noindent Beweis f"ur den Fall $III_-$ mit $0<r<p$ und $0<s<q$:\\
Sei kurz $V=V\left(h_{r,s}^-,c(-q/p)\right)$. Wir wollen (\ref{gewicht})
durch Induktion beweisen.
Wenn wir Lemma \ref{lemma6} anwenden, erhalten wir einfache Nullstellen in
der Kac--Determinante von V f"ur $(n,m)=(r+kp,s+kq), \;(k\ge0)$ und
$(n,m)=(-r+kp,-s+kq),\;(k\ge 1)$. Diese Nullstellen erzeugen nach Lemma
\ref{exist} singul"are Vektoren  $w_{4k+2}$ und $w_{4(k-1)+1}$. Man
"uberzeugt sich leicht, da\3 weder $w_1 \in [w_2]$ noch $w_2 \in [w_1]$
gilt, da wir sonst Nullstellen in den entsprechenden Determinanten finden
m"u\3ten. Wir haben also das Diagramm
\ben\begin{array}{c}
\setlength{\unitlength}{.005em}%
\begin{picture}(1000,500)
\put(500,500){\circle*{30}}
\put(475,450){\vector(-1,-2){200}}
\put(525,450){\vector(1,-2){200}}
\put(250,0){\circle*{30}}
\put(750,0){\circle*{30}}
\put(350,500){\makebox(0,0){$w_0$}}
\put(100,0){\makebox(0,0){$w_1$}}
\put(925,0){\makebox(0,0){$w_2$}}
\end{picture}
\setlength{\unitlength}{.01em}%
\end{array}\een
bewiesen.
Das ist auch der Induktionsanfang. Wir untersuchen nun $[w_2]\simeq
V\left(h^-_{r,-s},c(-q/p)\right)$. $w_2$ hat in $V$ den Grad $rs$, und die
Kac--Determinante von $[w_2]$ liefert singul"are Vektoren der Grade
$d_k^+=(r+kp)(-s+kq)$ bzw. $d_k^-=(-r+kp)(s+kq)$ f"ur $k\ge 1$. Diese
Vektoren haben in $V$ den Grad $d_k^{\pm} + rs$, was genau $\deg
w_{4(k-1)+3}$ bzw. $\deg w_{4k}$ entspricht. Damit haben wir, wie sich
sp"ater zeigen wird, bereits alle singul"aren Vektoren in $V$ gefunden. In
$[w_1]$ finden wir singul"are Vektoren der Grade
$\tilde{d}_k^+=(r+(k-1)p)(-s+(k+1)q)$ und $\tilde{d}_k^-=(-r +
(k+1)p)(s+(k-1)q)$ f"ur $k\ge 1$. Es gilt $\tilde{d}_k^{\pm}
+(p-r)(q-s)=d_k^{\pm} + rs$, d.h. die singul"aren Vektoren von $[w_1]$ und
$[w_2]$ haben in $V$ denselben Grad.

Wir f"uhren folgende Sprechweise ein: Zwei Teilr"aume $U=\plus U_n, W=\plus
W_n$ von $V$ hei\3en bis zum Grad $n$ gleich, falls f"ur $i=0,\ldots,n$
$U_i=W_i$ gilt. Wir haben bis jetzt gezeigt, da\3 $J_1=[w_1]+[w_2]$ bis
unterhalb des Grades des n"achsten singul"aren Vektors gilt, der z.~B. $w_3$
sei.

$V$ k"onnte im Prinzip zwei linear unabh"angige singul"are Vektoren vom Grad
$\deg w_3$ enthalten, denn sowohl $[w_1]$ als auch $[w_2]$ enthalten einen
solchen Vektor. Diese Vektoren fallen in $V$ aber zusammen: Sei $\deg
w_3=d$, $w_3$ und $w_3'$ seien die von $[w_1]$ und $[w_2]$ induzierten
singul"aren Vektoren in $V$ vom Grad $d$. Aus Satz \ref{jantzen} folgt, da\3
$w_3,w_3' \in J_2$ gelten mu\3. Aus der Dimensionsformel von Satz
\ref{jantzen} folgt
{\mathindent0mm\[
 \dim J_{1,d}+\dim J_{2,d}=\ord_0 \det S_d(h_{r,s},c(-q/p))= p(d-\deg
w_1)+p(d-\deg w_2) + 1.
\]}
W"aren $w_3$ und $w_3'$ linear unabh"angig, so w"are die Summe $[w_1]+[w_2]$
bis zum Grad $d$ direkt, und wir w"urden  $\dim J_{1,d}+\dim
J_{2,d}=p(d-\deg w_1) + p(d-\deg w_2)+2$ erhalten, was einen Widerspruch
ergibt. Es folgt also bis zum Grad $d$: $[w_1]\cap[w_2]=[w_3]=[w_3']$,
$J_1=[w_1]+[w_2]$ und $J_2=[w_3]$. Analog zeigt man $w_4=w_4'$, d.h
$J_2=[w_3]+[w_4]$ bis zum Grad $\deg w_4$. Wir haben damit bis jetzt das
Diagramm
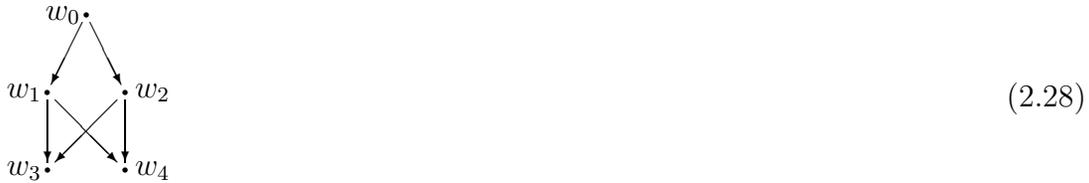
\ben\begin{array}{c}
\setlength{\unitlength}{.005em}%
\begin{picture}(1000,1000)
\put(500,1000){\circle*{30}}
\put(475,950){\vector(-1,-2){200}}
\put(525,950){\vector(1,-2){200}}
\put(250,500){\circle*{30}}
\put(300,450){\vector(1,-1){400}}
\put(700,450){\vector(-1,-1){400}}
\put(250,450){\vector(0,-1){400}}
\put(250,0){\circle*{30}}
\put(750,500){\circle*{30}}
\put(750,450){\vector(0,-1){400}}
\put(750,0){\circle*{30}}
\put(350,1000){\makebox(0,0){$w_0$}}
\put(100,500){\makebox(0,0){$w_1$}}
\put(925,500){\makebox(0,0){$w_2$}}
\put(100,0){\makebox(0,0){$w_3$}}
\put(925,0){\makebox(0,0){$w_4$}}
\end{picture}
\setlength{\unitlength}{.01em}%
\end{array}\een
bewiesen.\\
Der Induktionsschritt: Wir nehmen an, da\3 wir das folgende Diagramm
bewiesen haben,
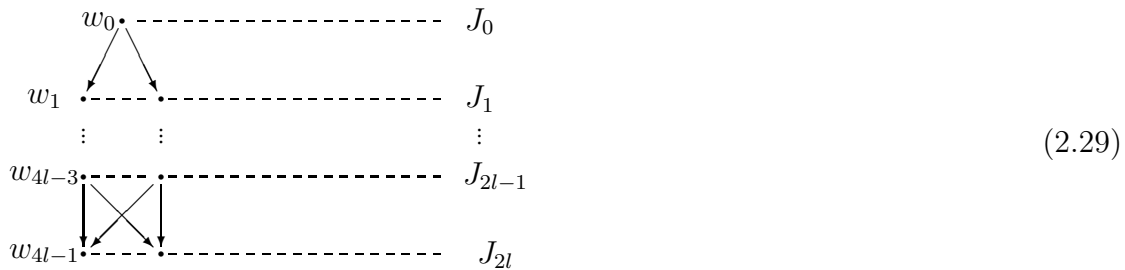
\ben\begin{array}{c}
\setlength{\unitlength}{.005em}%
\begin{picture}(3000,1500)
\put(500,1500){\circle*{30}}
\put(475,1450){\vector(-1,-2){200}}
\put(525,1450){\vector(1,-2){200}}
\put(250,1000){\circle*{30}}
\put(250,500){\circle*{30}}
\put(300,450){\vector(1,-1){400}}
\put(700,450){\vector(-1,-1){400}}
\put(250,450){\vector(0,-1){400}}
\put(250,0){\circle*{30}}
\multiput(250,800)(0,-40){3}{\circle*{1}}
\put(750,1000){\circle*{30}}
\put(750,500){\circle*{30}}
\put(750,450){\vector(0,-1){400}}
\put(750,0){\circle*{30}}
\multiput(750,800)(0,-40){3}{\circle*{1}}
\put(350,1500){\makebox(0,0){$w_0$}}
\put(0,1000){\makebox(0,0){$w_1$}}
\put(0,0){\makebox(0,0){$w_{4l-1}$}}
\put(0,500){\makebox(0,0){$w_{4l-3}$}}
\multiput(600,1500)(100,0){20}{\line(1,0){50}}
\put(2800,1500){\makebox(0,0){$J_0$}}
\multiput(300,1000)(100,0){4}{\line(1,0){50}}
\multiput(800,1000)(100,0){18}{\line(1,0){50}}
\multiput(300,500)(100,0){4}{\line(1,0){50}}
\multiput(800,500)(100,0){18}{\line(1,0){50}}
\multiput(300,0)(100,0){4}{\line(1,0){50}}
\multiput(800,0)(100,0){18}{\line(1,0){50}}
\put(2800,1000){\makebox(0,0){$J_1$}}
\multiput(2800,800)(0,-40){3}{\circle*{1}}
\put(2900,500){\makebox(0,0){$J_{2l-1}$}}
\put(2900,0){\makebox(0,0){$J_{2l}$}}
\end{picture}
\setlength{\unitlength}{.01em}%
\end{array}
\een
wobei die Aussagen "uber die Jantzen--Filtration bis zum h"ochsten Grad der
im Bild auftretenden singul"aren Vektoren gelten sollen.

Lemma \ref{lemma6} liefert f"ur $[w_{4l-1}]=[w_{4(l-1)+3}]\simeq
V(h^-_{r,s-2(l-1)q},c(-q/p))$ und $[w_{4l}]\simeq
V(h^-_{r,s+2lq},c(-q/p))$ singul"are Vektoren der Grade $\deg w_{4l+1},\deg
w_{4l+2}, \deg w_{4l+5}, \ldots$.
Jedes der Untermoduln $w_0$ und $[w_{4r-1}],[w_{4r}], r=1,\ldots,l$
enth"alt einen singul"aren Vektor vom Grad $\deg w_{4l+1}=:d$. F"ur $r=l$
folgt ganz analog zum Induktionsanfang aus (\ref{dimformel}), da\3
 $w''_{4l+1}$ und $w'_{4l+1}$ in $[w_{4l-1}]$ und $[w_{4l}]$ linear
abh"angig sind.  Dies gilt dann nat"urlich auch in $V$. Falls ein $r$
existiert, so da\3 $[w_{4r-1}]$ oder $[w_{4r}]$ zwei linear unabh"angige
singul"are Vektoren vom Grad $d$ enth"alt, wenden wir auf diesen Modul die
Dimensionsformel an. Durch Umnumerierung k"onnen wir erreichen, da\3 dieser
Modul durch $w_0$ erzeugt wird, d.h. wir erhalten genau das Bild aus der
Induktionsannahme, wobei $V$ auf dem Level $d$ zwei linear unabh"angige
singul"are Vektoren enth"alt. Wir erhalten aus (\ref{detmini})
{\mathindent2mm\bea\label{ordnungen}
\ord_0\det S_d(h_{r',s'},c(-q/p))&=& p(d-\deg w_1)+p(d-\deg w_2)+p(d-\deg
w_5) + \ldots \nonumber \\
&&+ p(d-\deg w_{4l-3})+ p(d-\deg w_{4l-2}) \nonumber\\
&&+ p(d-\deg w_{4l+1})+ p(d-\deg w_{4l+2}),
\eea}
wobei wir $p(-n)=0$ f"ur $n\in \NN$ gesetzt haben. Andererseits wissen wir
"uber die Dimensionen der Jantzen--Filtration gem"a\3 Induktionsanahme
\bea
\dim J_{1,d}&=&p(d-\deg w_1)+p(d-\deg w_2)\nonumber \\
&&\mbox{}-p(d-\deg w_3)-p(d-\deg w_4)\nonumber \\
&&\mbox{}+p(d-\deg w_5)+p(d-\deg w_6)\nonumber \\
&&\mbox{}\;\vdots\nonumber \\
&&\mbox{}-p(d-\deg w_{4l-1})-p(d-\deg w_{4l})\nonumber \\
&&\mbox{}+2p(d-\deg w_{4l+1}).
\eea
Analog k"onnen wir die Dimensionen von $J_k,d$, $k=2,\ldots, 2l$ als
alternierende Summen angeben. Dort tritt der Faktor $2$ bei $p(d-\deg
w_{4l+1})$ wegen der  Wahl von $r$ nicht auf. Es folgt
\bea
\sum_{k=1}^{2l} \dim J_{k,d} &=& p(d-\deg w_1)+p(d-\deg w_2) \nonumber \\
&&\mbox{}+p(d-\deg w_5)+p(d-\deg w_6)\nonumber \\
&&\mbox{}\;\vdots \nonumber\\
&&\mbox{}+p(d-\deg w_{4l-3})+p(d-\deg w_{4l-2})\nonumber \\
&&\mbox{}+2p(d-\deg w_{4l+1})+p(d-\deg w_{4l+2}),
\eea
was im Widerspruch zu (\ref{ordnungen}) steht. Genauso folgt das f"ur die
singul"aren Vektoren vom Grad $\deg w_{4l+2}$, es folgt damit insgesamt
$J_{2l+1}=[w_{4l+1}]+[w_{4l+2}]$. Der gleiche Beweisschritt, angewendet
auf $[w_1]$ und $[w_2]$, liefert die Behauptung f"ur $w_{4l+3}$ und
$w_{4l+4}$, und damit folgt der Induktionsschritt. Damit ist der Fall
$III_-$ bewiesen.

\noindent Beweis f"ur den Fall $III_+$:\\
Wir werden den Beweis hier nur skizzieren, da er im wesentlichen aus
denselben Methoden wie im Fall $III_-$ folgt.
Wir haben die Symmetrien $h_{r,s}^+=h_{-r,-s}^+ =h_{r+p,s-q}^+$ und k"onnen
uns deswegen wieder auf die F"alle $0<r<p$ und $jq<s<(j+1)q$ f"ur ein $j\in
\NN_0$ beschr"anken. Wir behaupten f"ur diese Moduln das folgende
Diagramm:\\
\parbox{7cm}{
\begin{center}
\setlength{\unitlength}{.005em}%
\begin{picture}(1000,2000)
\multiput(250,1800)(0,-30){3}{\circle*{1}}
\multiput(750,1800)(0,-30){3}{\circle*{1}}
\put(250,1500){\circle*{30}}
\put(750,1500){\circle*{30}}
\put(300,1450){\vector(1,-1){400}}
\put(700,1450){\vector(-1,-1){400}}
\put(250,1450){\vector(0,-1){400}}
\put(750,1450){\vector(0,-1){400}}
\put(250,1000){\circle*{30}}
\put(750,1000){\circle*{30}}
\put(300,950){\vector(1,-1){400}}
\put(700,950){\vector(-1,-1){400}}
\put(250,950){\vector(0,-1){400}}
\put(750,950){\vector(0,-1){400}}
\put(250,500){\circle*{30}}
\put(750,500){\circle*{30}}
\put(500,0){\circle*{30}}
\put(275,450){\vector(1,-2){200}}
\put(725,450){\vector(-1,-2){200}}
\put(350,0){\makebox(0,0){$w_0$}}
\put(100,500){\makebox(0,0){$w_1$}}
\put(925,500){\makebox(0,0){$w_2$}}
\put(100,1000){\makebox(0,0){$w_3$}}
\put(925,1000){\makebox(0,0){$w_4$}}
\put(100,1500){\makebox(0,0){$w_5$}}
\put(925,1500){\makebox(0,0){$w_6$}}
\end{picture}
\end{center}
\setlength{\unitlength}{.01em}%
}
\hfill
\parbox{8cm}{
\bea
{\rm wt}(w_{4i})&=& h_{r,-s+2iq}^+\nonumber\\
{\rm wt}(w_{4i+1})&=& h_{r,s-2(i+1)q}^+\nonumber \\
{\rm wt}(w_{4i+2})&=& h_{r,s+2iq}^+\\
{\rm wt}(w_{4i+3})&=& h_{r,-s-2iq}^+\nonumber
\eea}\\
Alle oben genannten Moduln tauchen in diesem Diagramm auf. Dieses Diagramm
k"onnen wir durch Induktion ``von unten nach oben'' beweisen. Zun"achst
stellt man fest, da\3 die Kac--Determinante von $[w_0]\simeq
V(h_{r,-s}^+,c(q/p))$ keine Nullstelle hat, was den Induktionsanfang
darstellt.\\
Nun untersucht man $[w_1]$ und $[w_2]$ und stellt fest, da\3 beide genau
einen singul"aren Vektor, n"amlich $w_0$ enthalten. So f"ahrt man fort, wobei
man wieder zeigen mu\3, da\3 keine verschiedenen singul"aren Vektoren des
gleichen Gewichtes auftauchen, was genau wie im Fall $III_-$ aus Formel
(\ref{dimformel}) f"ur die Jantzen--Filtration folgt.
\hfill$\Box$

Wir k"onnen nun mit Hilfe von Satz \ref{klasse} und Lemma \ref{kontra} auch
die kontragredienten Verma--Moduln klassifizieren. Wir erhalten die
gleichen Typen von Darstellungen $I,II_{\pm},III_{\pm},III_{\pm}^0$. Jeder
singul"are Vektor ungleich dem H"ochstgewichtsvektor in einem Verma--Modul
entspricht einem kosingul"aren Vektor in dem kontragredienten Modul. Die
Diagramme f"ur $\overline{V}(h,c)$ erhalten wir einfach aus den
 entsprechenden Diagrammen f"ur $V(h,c)$, indem wir alle Pfeile umdrehen,
denn dies bedeutet genau, da\3 der Punkt von dem der Pfeil ausgeht, einem
kosingul"aren Vektor entspricht.
\begin{kor} Die kontragredienten Verma--Moduln $\overline{V}(h,c)$ sind
 im Fall $I,II_-$ irreduzibel. Im Fall $II_+$ enth"alt $\overline{V}(h,c)$
genau einen kosingul"aren Vektor. F"ur die F"alle $III_{\pm}, III^0_{\pm}$
erhalten wir die Diagramme der kosingul"aren Vektoren nach obiger
Vorschrift. In jedem Fall erzeugt der H"ochstgewichtsvektor einen
irreduziblen Untermodul.
\end{kor}
F"ur zwei Vir--Moduln $V_1,V_2$ sei $\Hom(V_1,V_2)$ der Raum der
Intertwiner zwischen $V_1$ und $V_2$. Eine direkte Folgerung aus Satz
\ref{klasse} ist, da\3 f"ur zwei Verma--Moduln stets $\dim \Hom(V_1,V_2) \le
1$ gilt. Insbesondere gibt es genau dann einen nichttrivialen Intertwiner
$T:V_1\lra V_2$, wenn $V_2$ einen singul"aren Vektor mit dem Gewicht des
H"ochstgewichtsvektors von $V_1$ enth"alt. $T$ ist dann sogar eine
Einbettung.

Im Fall der kontragredienten Verma--Moduln ist die Situation noch
einfacher, es gilt, falls $V_1\not\simeq V_2$, immer
$\Hom(V_1,V_2)=\{0\}$.
\section{Die Struktur der Fock--Moduln}
Wir werden uns nun den Fock--Moduln zuwenden und ein Analogon zu Theorem
\ref{klasse} f"ur die Fock--Moduln beweisen. Zun"achst m"ussen wir aber die
Begriffe aus Abschnitt \ref{jantzenfilter} verallgemeinern.
\subsection{Die Jantzen--Filtrationen von Fock--Moduln}
Sei ${\cal F}(\alpha,\beta)$ ein Fock--Modul vom Typ $(h,c)$. Dann haben
wir kanonische Vir--Homomorphismen
\ben
V(h,c)\stackrel{S'(\alpha,\beta)}{\strich\lra}{\cal
F}(\alpha,\beta)\stackrel{S''(\alpha,\beta)}{\strich\lra}
 \overline{V}(h,c),
\een
deren Zusammensetzung gerade die Shapovalov--Abbildung ist.\\
Wir haben f"ur $z\in \CC$ Vektorraumisomorphismen
\bea
V(h+z,c)& \stackrel{i_z}{\lra}&\frU(N_-)=:V,\\
{\cal F}(\alpha(z),\beta(z))&\stackrel{j_z}{\lra}& \frU\left(
\Lin\{a_{-1},a_{-2},\ldots\}\right) =:{\cal F}\mbox{ und}\\
\overline{V}(h+z,c)&\stackrel{\bar{i}_z}{\lra}& \overline{V},
\eea
und erhalten folgendes kommutative Diagramm:
\begin{center}
\xext=3000\yext=1000
\begin{picture}(\xext,\yext)(\xoff,\yoff)
\setsqparms[1`1`1`1;1500`500]
\putsquare(0,250)[V(h+z,c)`{\cal F}(\alpha(z),\beta(z))`V`{\cal F};
S'(\alpha(z),\beta(z))`i_z`j_z`S'_z]
\setsqparms[1`0`1`1;1500`500]
\putsquare(1500,250)[\phantom{{\cal F}(\alpha(z),\beta(z))}
`\overline{V}(h+z,c)`\phantom{\cal
F}`\overline{V};S''(\alpha(z),\beta(z))``\bar{i}_z`S''_z]
\put(1500,850){\oval(3000,150)[t]}
\put(3000,850){\vector(0,-1){1}}
\put(1500,1000){\makebox(0,0){$S(h+z,c)$}}
\put(1500,150){\oval(3000,150)[b]}
\put(3000,150){\vector(0,1){1}}
\put(1500,0){\makebox(0,0){$S_z$}}
\end{picture}
\end{center}
$V$ haben wir bereits in Abschnitt \ref{jantzenfilter} auf nat"urliche
Weise zu einem Vir--Modul gemacht, und analog setzen wir
\bea
\tilde{\pi}_z(X).f &:=& j_z X.j_z^{-1}f\qquad (f\in {\cal F}),\nonumber\\
\bar{\pi}_z(X).\bar{v} &:=& \bar{i}_z X.\bar{i}_z^{-1}\bar{v}\qquad
(\bar{v} \in \overline{V}).
\nonumber\eea
Es gilt ${\cal F}(\alpha(z),\beta(z))\simeq({\cal F},\tilde{\pi}_z)$ und
$\overline{V}(h+z,c)\simeq(\overline{V},\bar{\pi}_z)$. Die induzierten
 Abbildungen $S'_z$ und $S''_z$ sind Intertwiner, denn
\[
S'_z \pi(X)v=j_z S'(h+z,c)i_z^{-1} i_z Xi_z^{-1} v=j_zX j_z^{-1} j_z
S'(h+z,c)i_z^{-1}v=\tilde{\pi}_z(X)S'_zv.
\]
Per Definition gilt$S_z=S''_z \circ S'_z$, d.h f"ur alle $f\in{\cal O}(V)$
gilt $S_z f(z)=S''_z \circ S'_z f(z)$ f"ur  $z\in U_f$ und daher
faktorisiert $S_z$ auch als Abbildung ${\cal O}(V)\lra {\cal O}
(\overline{V})$ "uber ${\cal O}({\cal F})$.
 Sei
\ben
{\cal O}_k(S'):=\left\{ f\in {\cal O}(V)\;:\;S'f \in {\cal O}_k({\cal
F})\right\}
\een
(analog f"ur $S''$) und $V_k:=\{f(0):f\in{\cal O}_k(V)\}$. Es gilt
\begin{bem} \label{faktor}Ist $Sf\in {\cal O}_n(\overline{V})$ und
 $h=S'f\in {\cal O}_l({\cal F})$, dann gilt mit $\wt{h}(z)=h(z)z^{-l}$,
da\3 $S''\wt{h}\in {\cal O}_{n-l}(\overline{V})$ ist.
\end{bem}
 Wir wollen nun Abbildungen $S'_k:V_k\lra ??$ definieren, wobei der
Bildraum noch zu bestimmen ist;  die einfachste Wahl ${\cal F}$ liefert
i.~allg. keine wohldefinierten Abbildungen. Dazu sei $v\in V_k$. Es
existiert ein $f\in {\cal O}_k(V)$ mit $f(0)=v$. Wir definieren
\[\wt{S}'_k v =\lim_{z\to 0} \frac{1}{z^k} S'_z f(z).
\]
$\wt{S}'_k v$ kann aber von der Wahl von $f\in {\cal O}_k(V)$ abh"angen,
was wir nun induktiv bereinigen.
\begin{itemize}
\item[(i)] $S'_0 =\wt{S}'_0: V \lra {\cal F}$ ist wohldefiniert.
\item[(ii)] Sei $f \in{\cal O}_1(S')$ mit $f(0)=0$. Wir k"onnen also
$f(z)=zh(z)$ f"ur ein $h\in {\cal O}(V)$ schreiben. Es folgt
\[
\frac{1}{z} S'_z f(z) = \frac{1}{z} S'_z z h(z) = S'_z h(z)
\]
und
\[
\lim_{z\to 0} \frac{1}{z}S'_z f(z)=S_0 h(0) \in \im S_0.
\]
Damit ist
\[
S'_1 : V_1 \lra \quot{\cal F}{\im S'_0} = \coker S'_0
\]
wohldefiniert.
\item[(iii)] Sei $S'_{k-1}:V_{k-1} \lra \coker S'_{k-2}$ wohldefiniert und
$f \in {\cal O}_k(S')$ mit $f(z)=z h(z)$. Wir erhalten $1/z^k S'_z
f(z)=1/z^{k-1} S'_z h(z)$. Da der Grenzwert auf der linken Seite
existiert, ist $h \in {\cal O}_{k-1}(S')$ und es folgt weiter, wenn wir im
Bild auf $\coker S'_{k-2}$ projizieren,
\[
\lim_{z\to 0} \frac{1}{z^k}S'_z f(z)=\lim_{z\to 0} \frac{1}{z^{k-1}}S'_z
h(z) = S'_{k-1} h(0) \in \im S_{k-1}.
\]
Folglich ist
\[
S'_k : V_k \lra \quot{\coker S'_{k-2}}{\im S'_{k-1}} =\coker S'_{k-1}
\]
wohldefiniert.
\end{itemize}
Wenn wir die gleiche Prozedur auf $S_z$ und $S''_z$ anwenden, erhalten wir
drei Folgen von Abbildungen
\bea\label{abb0}
S_0:V \lra \overline{V}&,&\quad S_k : \ker S_{k-1} \lra \coker S_{k-1},\\
\label{abb1}
S'_0 : V \lra {\cal F}&,&\quad S'_k : \ker S'_{k-1} \lra \coker S'_{k-1}
\mbox{ und}
\\ \label{abb2}
S''_0:{\cal F}\lra \overline{V}&,&\quad S''_k : \ker S''_{k-1}\lra \coker
S''_{k-1}.
\eea
Sei $\pr_k:{\cal F}\lra \coker S'_k$ die kanonische Projektion. Wir
definieren zwei Filtrationen in ${\cal F}$ durch
\bea\label{filter1}
J'_k &:=& \ker \{ \pr_k : {\cal F} \lra \coker S'_k \},\\
\label{filter2}
J''_0 &:=& {\cal F}, \quad J''_{k} := \ker S''_{k-1}\; (k>0).
\eea
\begin{lem} $J'_k$ und $J''_k$ sind Untermoduln von ${\cal F}$. Es gilt
\begin{itemize}
\item[(i)]
$J'_k \subset J'_{k+1}$ und $\cup_k J'_k = {\cal F}$,
\item[(ii)] $J''_k \supset J''_{k+1}$ und $\cap_k J''_k = \{0\}$.
\end{itemize}
\end{lem}
{\bf Beweis.}\\
 Man sieht wie in Abschnitt \ref{jantzenfilter}, da\3  $J''_{k+1}\subset
J''_k$ und $\cap_{k} J''_k=\{0\}$ gilt, da\3 also  $\{J''_k\}$  eine
absteigende Filtration von ${\cal F}$ ist. Wir wollen nun zeigen, da\3 $\{
J'_k\}$ eine aufsteigende Filtration von ${\cal F}$ ist. Zun"achst folgt
wegen
\ben\label{quoti}
\coker S'_k \simeq \quot{{\cal F}}{\im \wt{S}'_0 + \ldots + \im
\wt{S}'_k},
\een
da\3 $J'_k \simeq \im \wt{S}'_0 + \ldots \im\wt{S}'_k$ (die Summe ist
wohldefiniert, da $S'_i$ wohldefiniert modulo $\im S'_{i-1}$ ist), und da\3
$J'_k \subset J'_{k+1}$ gilt. Zu zeigen bleibt $\cup_k J'_k={\cal F}$.
Seien $V_n$ die homogenen Elemente vom Grad $n$ in $V$, $(J_k)_n=(\ker
S'_k)_n$. Wegen $\cap J_k =\{0\}$ folgt, da\3 $a<b \in \NN$ existieren mit
$(J_a)_n=V_n$ und $(J_{b})_n=\{0\}$. Wir erhalten als Isomorphie der
Vektorr"aume
\[
V_n \simeq \quot{(J_a)_n}{(J_{a+1})_n}
\oplus\quot{(J_{a+1})_n}{(J_{a+2})_n} \oplus \cdots \oplus
\quot{(J_{b-1})_n}{(J_{b})_n}.
\]
Au\3erdem gilt wegen (\ref{quoti})
\[
S'_k : \quot{(J_{k-1})_n}{(J_{k})_n} \stackrel{\simeq}{\lra} \quot{(\coker
S'_{k-1})_n}{(\coker S'_k)_n}\simeq \quot{(J'_k)_n}{(J'_{k-1})_n}.
\]
Wir erhalten insgesamt:
\begin{center}\xext=3000\yext=700
\begin{picture}(\xext,\yext)(\xoff,\yoff)
\putmorphism(700,700)(0,-1)[\qquot{(J_a)_n}{(J_{a+1})_n}`\qquot{(J'_{a+1})_
n}{(J'_a)_n}`S'_{a+1}]{700}1r
\put(600,350){\makebox(0,0){$\simeq$}}
\putmorphism(1400,700)(0,-1)[\qquot{(J_{a+1})_n}{(J_{a+2})_n}`\qquot{(J'_{a
+2})_n}{(J'_{a+1})_n}`S'_{a+2}]{700}1r
\put(1300,350){\makebox(0,0){$\simeq$}}
\putmorphism(2600,700)(0,-1)[\qquot{(J_{b-1})_n}{(J_{b})_n}`\qquot{(J'_{b})
_n}{(J'_{b-1})_n}`S'_{b}]{700}1r
\put(2500,350){\makebox(0,0){$\simeq$}}
\put(200,700){\makebox(0,0){$V_n \simeq$}}
\put(200,0){\makebox(0,0){$\qquot{(J'_{b})_n}{(J'_a)_n}\simeq$}}
\put(1050,700){\makebox(0,0){$\oplus$}}
\put(1050,0){\makebox(0,0){$\oplus$}}
\put(2000,700){\makebox(0,0){$\oplus\cdots\oplus$}}
\put(2000,0){\makebox(0,0){$\oplus\cdots\oplus$}}
\end{picture}
\end{center}
$(J'_a)_n$ ist trivial, denn da $(\ker S'_a)_n = V_n$ ist, ist $(\coker
S'_a)_n={\cal F}_n$ und deswegen $(J'_a)_n=\{0\}$. Es folgt also $(J'_b)_n
={\cal F}_n$, was $\cup_n J'_n ={\cal F}$ beweist.\hfill $\Box$

Mit den Isomorphismen $i_z,j_z,\bar{i}_z$ folgt alles soeben Bewiesene
auch f"ur die Abbildungen zwischen den (kontragredienten) Verma--Moduln und
den Fock--Moduln, wir erhalten somit drei Folgen von Abbildungen
(\ref{abb0}), (\ref{abb1}) und (\ref{abb2}) sowie zwei Filtrationen von
${\cal F}(\alpha,\beta)$, die durch (\ref{filter1}) und (\ref{filter2})
gegeben sind.
\subsection{Der Klassifikationssatz f"ur Fock--Moduln}
Wir wollen nun unser Wissen "uber die Struktur der Verma--Moduln verwenden,
um die Struktur der Fock--Moduln zu beweisen. Wir haben aus $S(h,c)$,
$S'(\alpha,\beta)$ und $S''(\alpha,\beta)$ drei Folgen von Abbildungen
$S_k$, $S'_k$ und $S''_k$ konstruiert.  Den Zusammenhang zwischen diesen
Abbildungen kl"art das n"achste Lemma.
\begin{lem}[\cite{FF}]\label{fklem}
Sei $v\in \ker S_{k+l-1}\cap S'_{k-1} (=D(S_{k+l})\cap D(S'_k))$ und $S'_k
v \ne 0$. Dann gilt:
\begin{itemize}
\item[(i)] Es existiert $v\in \ker S''_{l-1}\subset{\cal F}(\alpha,\beta)$
mit $S'_k v= \pr_{k-1} w$, wobei $\pr_{k-1}$ die kanonische Projektion
${\cal F}(\alpha,\beta)\lra \coker S'_{k-1}$ ist.
\item[(ii)] Ist $S'_l w=0$, dann ist auch $S_{k+l}v=0$.
\end{itemize}
\end{lem}
{\bf Beweis.}\\
Es existiert ein $f\in {\cal O}(V)$ mit $f(0)=v$ und $Sf\in{\cal O}_{k+l}
(\overline{V})$, $S'f\in {\cal O}_k({\cal F})$. Wir setzen
\[w:=\lim_{z\to 0} \frac{1}{z^k} S'_z f(z),
\]
womit $S'_k v =\pr_{k-1} w$ folgt. Es gilt $w\in \ker S''_{l-1}$, denn
\[
\lim_{z\to 0} \frac{1}{z^{l-1}} S''_z S'_z \frac{1}{z^k} f(z)=\lim_{z \to
0} \frac{1}{z^{k+l-1} } S_z f(z)=0.
\]
Ist $w=0$, so folgt $f\in {\cal O}_{k+1}(S')$, und wegen $S'f\in {\cal
O}_l(S'')$ ist $Sf=S''S'f \in {\cal O}_{k+l+1}(\overline{V})$, und
 deshalb ist $\lim_{z \to 0} 1/z^{k+l} S_z f(z)=0$. \hfill$\Box$

Bei der Klassifikation der Fock--Moduln tritt eine weitere Aufspaltung der
bei der Klassifikation der Verma--Moduln aufgetretenen F"alle auf. Zum
einen r"uhrt das daher, da\3 die kontragredienten Fock--Moduln wieder
Fock--Moduln sind. Wegen der Beziehungen $h=1/2(\alpha^2-\beta^2)$ und
$c=1-12 \beta^2$ gibt es vier Fock--Moduln vom Typ $(h,c)$. Wie man aber
leicht an (\ref{fockrep}) sieht, sind ${\cal F}(\alpha,\beta)$ und ${\cal
F}(-\alpha,-\beta)$ isomorph, und es gilt ${\cal
F}(-\alpha,\beta)=\overline{{\cal F}}(\alpha,\beta)$. In den Diagrammen
unterscheiden sich diese Moduln einfach durch die Vertauschung von
singul"aren und kosingul"aren Vektoren und der entsprechenden Pfeile. Wir
erhalten deshalb die F"alle
$\{I,\overline{I},II_{\pm},\overline{II}_{\pm},III_{\pm},
\overline{III}_{\pm},III_{\pm}^0,\overline{III}_{\pm}^0 \}$. Es tritt aber
noch ein zus"atzlicher Fall auf, der aus $III_{-}^0$ entsteht, wenn $r
=0\mod p$ und $s=0\mod q$ gilt. Die doppelten Nullstellen in der
Kac--Determinante f"uhren zu gleichzeitigen Nullstellen sowohl in $\det S'$
als auch in $\det S''$, im Unterschied zu den F"allen mit einfachen
Nullstellen in der Kac--Determinante (vergl. Beweis von Satz
\ref{klasse}). Wir bezeichnen diesen Fall mit $III_{-}^{00}$ (bzw.
$\overline{III}_-^{00}$, aber diese Moduln sind isomorph).

Im folgenden sei f"ur einen Vektor $w\in {\cal F}$ mit $[w]$ der von $w$
erzeugte Untermodul in ${\cal F}$ bezeichnet.
Das folgende Theorem wurde bis auf einen Fehler in der Klassifizierung der
Moduln vom Typ ``+'' zuerst von Feigin und Fuks in \cite{FF} bewiesen.
\begin{satz}\label{fockklasse} Die Fock--Moduln vom Typ
$I,\overline{I},II_-,\overline{II}_-$ sind irreduzibel. Im Fall $II_+$
enth"alt ${\cal F}(\alpha,\beta)$ genau einen singul"aren Vektor mit
irreduziblem Quotienten (es gilt ${\cal F}(\alpha,\beta)\simeq V(h,c)$),
im Fall $\overline{II}_+$ enth"alt ${\cal F}(\alpha,\beta)$ genau einen
kosingul"aren Vektor, und es gilt ${\cal F}(\alpha,\beta)\simeq
\overline{V}(h,c)$.

Die Moduln vom Typ $III_{\pm}^{00}$ enthalten  endlich oder unendlich
viele  singul"are (und gleichzeitig kosingul"are) Vektoren $w_1,w_2,\ldots$,
und es gilt ${\cal F}(\alpha,\beta)\simeq [w_1]\oplus [w_2]\oplus \ldots$,
wobei $[w_i]$ irreduzible Moduln sind.

Die Moduln vom Typ $III_+$ und $III_+^0$ sind isomorph zu den
entsprechenden Verma--Moduln, die vom Typ $\overline{III}_+$ und
$\overline{III}_+^0$ sind isomorph zu den entsprechenden kontragredienten
Verma--Moduln.\\
Wir erhalten f"ur die F"alle $III$  die folgenden Diagramme:
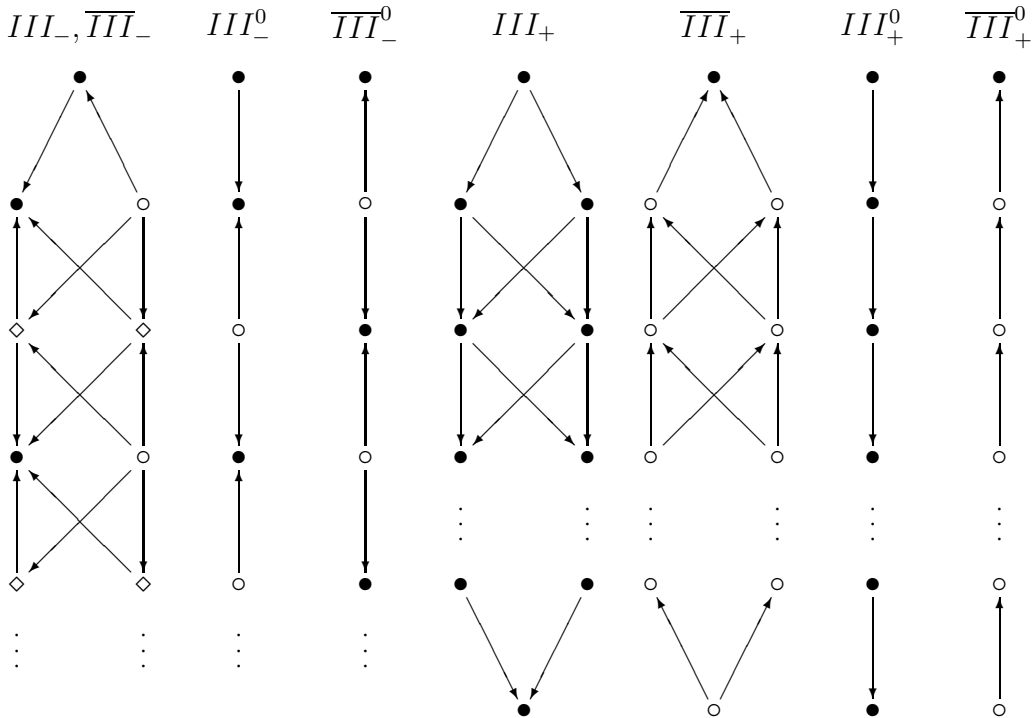
\begin{center}
\setlength{\unitlength}{.008em}%
\begin{picture}(4000,2700)
\put(250,2700){\makebox(0,0){$III_-,\overline{III}_-$}}
\put(250,2500){\makebox(0,0){$\bullet$}}
\put(225,2450){\vector(-1,-2){200}}
\put(475,2050){\vector(-1,2){200}}
\put(0,2000){\makebox(0,0){$\bullet$}}
\put(500,2000){\makebox(0,0){$\circ$}}
\put(450,1550){\vector(-1,1){400}}
\put(450,1950){\vector(-1,-1){400}}
\put(0,1550){\vector(0,1){400}}
\put(500,1950){\vector(0,-1){400}}
\put(0,1500){\makebox(0,0){$\diamond$}}
\put(500,1500){\makebox(0,0){$\diamond$}}
\put(450,1050){\vector(-1,1){400}}
\put(450,1450){\vector(-1,-1){400}}
\put(0,1450){\vector(0,-1){400}}
\put(500,1050){\vector(0,1){400}}
\put(0,1000){\makebox(0,0){$\bullet$}}
\put(500,1000){\makebox(0,0){$\circ$}}
\put(450,550){\vector(-1,1){400}}
\put(450,950){\vector(-1,-1){400}}
\put(0,550){\vector(0,1){400}}
\put(500,950){\vector(0,-1){400}}
\put(0,500){\makebox(0,0){$\diamond$}}
\put(500,500){\makebox(0,0){$\diamond$}}
\multiput(0,300)(0,-60){3}{\circle*{1}}
\multiput(500,300)(0,-60){3}{\circle*{1}}
%
\put(875,2700){\makebox(0,0){$III_-^0$}}
\put(875,2500){\makebox(0,0){$\bullet$}}
\put(875,2450){\vector(0,-1){400}}
\put(875,2000){\makebox(0,0){$\bullet$}}
\put(875,1550){\vector(0,1){400}}
\put(875,1500){\makebox(0,0){$\circ$}}
\put(875,1450){\vector(0,-1){400}}
\put(875,1000){\makebox(0,0){$\bullet$}}
\put(875,550){\vector(0,1){400}}
\put(875,500){\makebox(0,0){$\circ$}}
\multiput(875,300)(0,-60){3}{\circle*{1}}
\put(1375,2700){\makebox(0,0){$\overline{III}_-^0$}}
\put(1375,2500){\makebox(0,0){$\bullet$}}
\put(1375,2050){\vector(0,1){400}}
\put(1375,2000){\makebox(0,0){$\circ$}}
\put(1375,1950){\vector(0,-1){400}}
\put(1375,1500){\makebox(0,0){$\bullet$}}
\put(1375,1050){\vector(0,1){400}}
\put(1375,1000){\makebox(0,0){$\circ$}}
\put(1375,950){\vector(0,-1){400}}
\put(1375,500){\makebox(0,0){$\bullet$}}
\multiput(1375,300)(0,-60){3}{\circle*{1}}
%
\put(2000,2700){\makebox(0,0){$III_+$}}
\put(2000,2500){\makebox(0,0){$\bullet$}}
\put(1975,2450){\vector(-1,-2){200}}
\put(2025,2450){\vector(1,-2){200}}
\put(2250,2000){\makebox(0,0){$\bullet$}}
\put(1750,2000){\makebox(0,0){$\bullet$}}
%
\multiput(1750,1500)(500,0){2}{\makebox(0,0){$\bullet$}}
\multiput(1800,1450)(0,500){2}{\vector(1,-1){400}}
\multiput(1750,1450)(0,500){2}{\vector(0,-1){400}}
\multiput(2200,1450)(0,500){2}{\vector(-1,-1){400}}
\multiput(2250,1450)(0,500){2}{\vector(0,-1){400}}
\multiput(1750,1000)(500,0){2}{\makebox(0,0){$\bullet$}}
\multiput(1750,800)(0,-60){3}{\circle*{1}}
\multiput(2250,800)(0,-60){3}{\circle*{1}}
\multiput(1750,500)(500,0){2}{\makebox(0,0){$\bullet$}}
\put(1775,450){\vector(1,-2){200}}
\put(2225,450){\vector(-1,-2){200}}
\put(2000,0){\makebox(0,0){$\bullet$}}
%
%
\put(2750,2700){\makebox(0,0){$\overline{III}_+$}}
\put(2750,2500){\makebox(0,0){$\bullet$}}
\put(2525,2050){\vector(1,2){200}}
\put(2975,2050){\vector(-1,2){200}}
\multiput(2500,2000)(500,0){2}{\makebox(0,0){$\circ$}}
%
\multiput(2500,1500)(500,0){2}{\makebox(0,0){$\circ$}}
\multiput(2550,1050)(0,500){2}{\vector(1,1){400}}
\multiput(2500,1050)(0,500){2}{\vector(0,1){400}}
\multiput(2950,1050)(0,500){2}{\vector(-1,1){400}}
\multiput(3000,1050)(0,500){2}{\vector(0,1){400}}
\multiput(2500,1000)(500,0){2}{\makebox(0,0){$\circ$}}
\multiput(2500,800)(0,-60){3}{\circle*{1}}
\multiput(3000,800)(0,-60){3}{\circle*{1}}
\multiput(2500,500)(500,0){2}{\makebox(0,0){$\circ$}}
\put(2775,50){\vector(1,2){200}}
\put(2725,50){\vector(-1,2){200}}
\put(2750,0){\makebox(0,0){$\circ$}}
%
%
\put(3375,2700){\makebox(0,0){$III_+^0$}}
\multiput(3375,2500)(0,-500){6}{\makebox(0,0){$\bullet$}}
\multiput(3375,2450)(0,-500){3}{\vector(0,-1){400}}
\multiput(3375,800)(0,-60){3}{\circle*{1}}
\put(3375,450){\vector(0,-1){400}}
%
%
\put(3875,2700){\makebox(0,0){$\overline{III}_+^0$}}
\put(3875,2500){\makebox(0,0){$\bullet$}}
\multiput(3875,2000)(0,-500){5}{\makebox(0,0){$\circ$}}
\multiput(3875,2050)(0,-500){3}{\vector(0,1){400}}
\multiput(3875,800)(0,-60){3}{\circle*{1}}
\put(3875,50){\vector(0,1){400}}
%
%
\end{picture}\end{center}
\setlength{\unitlength}{.01em}%

Dabei entsprechen $\bullet$ singul"aren Vektoren, $\circ$ kosingul"aren
Vektoren und $\diamond$ Vektoren, die nach der Faktorisierung des Moduls
durch den von den singul"aren Vektoren erzeugten Untermodul singul"ar
werden.
Im Fall $III_-$ erhalten wir folgende Struktur: ${\cal F}(\alpha,\beta)$
enth"alt unendlich viele singul"are Vektoren (``$\bullet$''), genauer die
direkte Summe der durch ``$\bullet$'' erzeugten Untermoduln. Der
Quotientenmodul von ${\cal F}(\alpha,\beta)$ mit dieser direkten Summe
enth"alt die singul"aren Vektoren ``$\diamond$''. Wenn wir wieder den
Quotienten durch den von diesen singul"aren Vektoren erzeugten Untermodul
bilden, erhalten wir einen Modul, der die direkte Summe von durch die
singul"aren Vektoren ``$\circ$'' erzeugten irreduziblen Moduln ist.
\end{satz}
Die Strukturbeweise in den F"allen $III_-$ sind f"ur die Fock--Moduln
deutlich komplizierter als f"ur die Verma--Moduln, denn dort waren alle
Untermoduln wieder (Summen von) Verma--Moduln. Dies ist im Fall der
Fock--Moduln nicht so einfach, die singul"aren Vektoren in einem
Fock--Modul erzeugen i.~allg. keinen Fock--Modul.

Eine weitere Schwierigkeit liegt darin, da\3 die Abbildungen $S'_k$ nur
projektiv in die Fock--Moduln abbilden. Diese Schwierigkeit wird nur
teilweise von Lemma \ref{fklem} behoben.

\noindent{\bf Beweis.}\\
Die F"alle $I$ und $II_-$ sind klar. Im Fall $II_+$ bzw. $\overline{II}_+$
hat entweder $\det \Phi''$ ($II_+$) oder $\det \Phi'$ ($\overline{II}_+$)
genau eine Nullstelle. Es folgt aus Lemma \ref{exist}, da\3 $\cal F$ (bzw.
$\overline{\cal F}$) genau einen singul"aren und $\overline{\cal F}$ (bzw.
$\cal F$) genau einen kosingul"aren Vektor enth"alt.

Wir werden den Fall $III_-$ ausf"uhrlich beweisen, da dieser Fall der
komplizierteste und der im Weiteren  f"ur uns interessanteste ist. Im Fall
$III_-$ haben die Verma--Moduln die folgende Struktur:
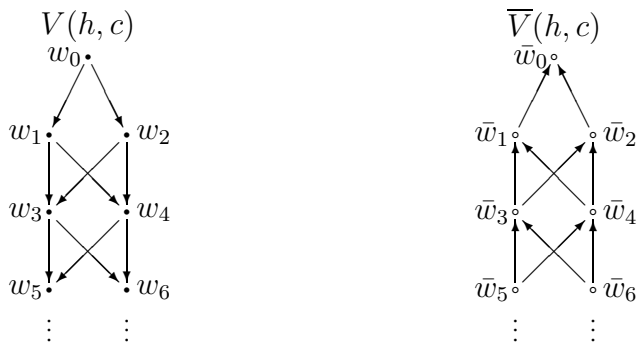
\begin{center}
\setlength{\unitlength}{.005em}%
\begin{picture}(4000,2200)
\put(500,2200){\makebox(0,0){$V(h,c)$}}
\put(500,2000){\circle*{30}}
\put(475,1950){\vector(-1,-2){200}}
\put(525,1950){\vector(1,-2){200}}
\put(250,1500){\circle*{30}}
\put(750,1500){\circle*{30}}
\put(300,1450){\vector(1,-1){400}}
\put(700,1450){\vector(-1,-1){400}}
\put(250,1450){\vector(0,-1){400}}
\put(750,1450){\vector(0,-1){400}}
\put(250,1000){\circle*{30}}
\put(750,1000){\circle*{30}}
\put(300,950){\vector(1,-1){400}}
\put(700,950){\vector(-1,-1){400}}
\put(250,950){\vector(0,-1){400}}
\put(750,950){\vector(0,-1){400}}
\put(250,500){\circle*{30}}
\put(750,500){\circle*{30}}
\multiput(250,300)(0,-60){3}{\circle*{1}}
\multiput(750,300)(0,-60){3}{\circle*{1}}
\put(350,2000){\makebox(0,0){$w_0$}}
\put(100,1500){\makebox(0,0){$w_1$}}
\put(925,1500){\makebox(0,0){$w_2$}}
\put(100,1000){\makebox(0,0){$w_3$}}
\put(925,1000){\makebox(0,0){$w_4$}}
\put(100,500){\makebox(0,0){$w_5$}}
\put(925,500){\makebox(0,0){$w_6$}}
\put(3500,2200){\makebox(0,0){$\overline{V}(h,c)$}}
\put(3500,2000){\circle{30}}
\put(3275,1550){\vector(1,2){200}}
\put(3725,1550){\vector(-1,2){200}}
\put(3250,1500){\circle{30}}
\put(3750,1500){\circle{30}}
\put(3300,1050){\vector(1,1){400}}
\put(3700,1050){\vector(-1,1){400}}
\put(3250,1050){\vector(0,1){400}}
\put(3750,1050){\vector(0,1){400}}
\put(3250,1000){\circle{30}}
\put(3750,1000){\circle{30}}
\put(3300,550){\vector(1,1){400}}
\put(3750,550){\vector(0,1){400}}
\put(3700,550){\vector(-1,1){400}}
\put(3250,550){\vector(0,1){400}}
\put(3250,500){\circle{30}}
\put(3750,500){\circle{30}}
\multiput(3250,300)(0,-60){3}{\circle*{1}}
\multiput(3750,300)(0,-60){3}{\circle*{1}}
\put(3350,2000){\makebox(0,0){$\bar{w}_0$}}
\put(3100,1500){\makebox(0,0){$\bar{w}_1$}}
\put(3925,1500){\makebox(0,0){$\bar{w}_2$}}
\put(3100,1000){\makebox(0,0){$\bar{w}_3$}}
\put(3925,1000){\makebox(0,0){$\bar{w}_4$}}
\put(3100,500){\makebox(0,0){$\bar{w}_5$}}
\put(3925,500){\makebox(0,0){$\bar{w}_6$}}
\end{picture}\end{center}
\setlength{\unitlength}{.01em}%
Im folgenden werden $V(h,c)$ kurz mit $V$ und $\overline{V}(h,c)$  mit
$\overline{V}$ bezeichnet. F"ur einen  kosingul"aren Vektor $\bar{v}\in
\overline{V}$ sei $[\bar{v}]:= \overline{[v]}$, wobei $v$ der zu $\bar{v}$
duale singul"are Vektor in $V$ ist. $[\bar{v}]$ ist kein Untermodul,
sondern ein Quotientenmodul von $\overline{V}$.

Eine einfache Folgerung aus Satz \ref{klasse} zeigt, da\3 f"ur $S_k:\ker
S_{k-1}\lra\coker S_{k-1}$
\[ \ker S_k=[w_{2k-1}]+[w_{2k}],\qquad \coker
S_k=[\bar{w}_{2k+1}]+[\bar{w}_{2k+2}]
\]
gilt. Es folgt daher
\[
S_k w_{2k-1}=\bar{w}_{2k-1},\qquad S_k w_{2k}=\bar{w}_{2k},\qquad S_k
w_j=0 \mbox{ f"ur } j >2k.
\]

Wir werden nun die Kerne von $S'_k$ in $V$ und die Kokerne von $S''_k$ in
$\overline{V}$ bestimmen. Das wird den Beweis des entsprechenden Teils
 von Satz \ref{fockklasse} erlauben.
\begin{lem} \label{grausam}F"ur die Abbildungen $S'_k : \ker S'_{k-1} \lra
\coker S'_{k-1}$ und $S''_k:\ker S''_{k-1}\lra \coker S''_{k-1}$ gilt:
\begin{itemize}
\item[(i)] $\ker S'_k=[w_{4k+2}]$ und damit
\bea
S'_0:\quot{V}{[w_2]}&\stackrel{\simeq}{\strich\lra}& \quot{\cal F}{J'_0},
\nonumber \\
S'_k : \quot{[w_{4k-2}]}{[w_{4k+2}]}&\stackrel{\simeq}{\strich\lra}&
\quot{J'_{k-1}}{J'_k}.\nonumber
\eea
\item[(ii)] $\coker S''_k=[\bar{w}_{4k+1}]$ und deshalb
\bea
S''_0:\quot{\cal F}{J''_1}&\stackrel{\simeq}{\strich\lra}&\ker \{
\pr:\overline{V}\lra [\bar{w}_1]\},
\nonumber\\
S''_k:\quot{J''_k}{J''_{k+1}}&\stackrel{\simeq}{\strich\lra}& \ker \{\pr :
[\bar{w}_{4k-3}] \lra [\bar{w}_{4k+1}]\}.\nonumber
\eea
\end{itemize}
\end{lem}
Die Aussage von Lemma \ref{grausam} ist "aquivalent zu dem folgenden
Diagramm:
\begin{center}
\setlength{\unitlength}{.009em}%
\begin{picture}(4000,2200)
\put(500,2200){\makebox(0,0){$V$}}
\put(500,2000){\makebox(0,0){$\bullet$}}
\put(475,1950){\vector(-1,-2){200}}
\put(525,1950){\vector(1,-2){200}}
\put(250,1500){\makebox(0,0){$\bullet$}}
\put(750,1500){\makebox(0,0){$\bullet$}}
\put(300,1450){\vector(1,-1){400}}
\put(700,1450){\vector(-1,-1){400}}
\put(250,1450){\vector(0,-1){400}}
\put(750,1450){\vector(0,-1){400}}
\put(250,1000){\makebox(0,0){$\bullet$}}
\put(750,1000){\makebox(0,0){$\bullet$}}
\put(300,950){\vector(1,-1){400}}
\put(700,950){\vector(-1,-1){400}}
\put(250,950){\vector(0,-1){400}}
\put(750,950){\vector(0,-1){400}}
\put(250,500){\makebox(0,0){$\bullet$}}
\put(750,500){\makebox(0,0){$\bullet$}}
\multiput(250,300)(0,-60){3}{\circle*{1}}
\multiput(750,300)(0,-60){3}{\circle*{1}}
\put(400,2000){\makebox(0,0){$w_0$}}
\put(150,1500){\makebox(0,0){$w_1$}}
\put(875,1500){\makebox(0,0){$w_2$}}
\put(150,1000){\makebox(0,0){$w_3$}}
\put(875,1000){\makebox(0,0){$w_4$}}
\put(150,500){\makebox(0,0){$w_5$}}
\put(875,500){\makebox(0,0){$w_6$}}
\put(2000,2200){\makebox(0,0){$\cal F$}}
\put(2000,2000){\makebox(0,0){$\bullet$}}
\put(1975,1950){\vector(-1,-2){200}}
\put(2225,1550){\vector(-1,2){200}}
\put(1750,1500){\makebox(0,0){$\bullet$}}
\put(2250,1500){\makebox(0,0){$\circ$}}
\put(2200,1050){\vector(-1,1){400}}
\put(2200,1450){\vector(-1,-1){400}}
\put(1750,1050){\vector(0,1){400}}
\put(2250,1450){\vector(0,-1){400}}
\put(1750,1000){\makebox(0,0){$\diamond$}}
\put(2250,1000){\makebox(0,0){$\diamond$}}
\put(2200,550){\vector(-1,1){400}}
\put(2200,950){\vector(-1,-1){400}}
\put(1750,950){\vector(0,-1){400}}
\put(2250,550){\vector(0,1){400}}
\put(1750,500){\makebox(0,0){$\bullet$}}
\put(2250,500){\makebox(0,0){$\circ$}}
\multiput(1750,300)(0,-60){3}{\circle*{1}}
\multiput(2250,300)(0,-60){3}{\circle*{1}}
%
\put(3500,2200){\makebox(0,0){$\bar{V}$}}
\put(3500,2000){\makebox(0,0){$\bullet$}}
\put(3275,1550){\vector(1,2){200}}
\put(3725,1550){\vector(-1,2){200}}
\put(3250,1500){\makebox(0,0){$\bullet$}}
\put(3750,1500){\makebox(0,0){$\bullet$}}
\put(3300,1050){\vector(1,1){400}}
\put(3700,1050){\vector(-1,1){400}}
\put(3250,1050){\vector(0,1){400}}
\put(3750,1050){\vector(0,1){400}}
\put(3250,1000){\makebox(0,0){$\bullet$}}
\put(3750,1000){\makebox(0,0){$\bullet$}}
\put(3300,550){\vector(1,1){400}}
\put(3750,550){\vector(0,1){400}}
\put(3700,550){\vector(-1,1){400}}
\put(3250,550){\vector(0,1){400}}
\put(3250,500){\makebox(0,0){$\bullet$}}
\put(3750,500){\makebox(0,0){$\bullet$}}
\multiput(3250,300)(0,-60){3}{\circle*{1}}
\multiput(3750,300)(0,-60){3}{\circle*{1}}
\put(3600,2000){\makebox(0,0){$\bar{w}_0$}}
\put(3150,1500){\makebox(0,0){$\bar{w}_1$}}
\put(3875,1500){\makebox(0,0){$\bar{w}_2$}}
\put(3150,1000){\makebox(0,0){$\bar{w}_3$}}
\put(3875,1000){\makebox(0,0){$\bar{w}_4$}}
\put(3150,500){\makebox(0,0){$\bar{w}_5$}}
\put(3875,500){\makebox(0,0){$\bar{w}_6$}}
%
%
\put(550,2000){\vector(1,0){1400}}
\put(1250,2100){\makebox(0,0){$S'_0$}}
\put(2050,2000){\vector(1,0){1400}}
\put(2750,2100){\makebox(0,0){$S''_0$}}
\put(1000,1550){\oval(1475,200)[t]}
\put(1737,1550){\vector(0,-1){1}}
\put(1100,1550){\makebox(0,0){$S'_0$}}
\multiput(1000,1050)(0,-500){2}{\oval(1450,200)[t]}
\multiput(1725,1050)(0,-500){2}{\vector(0,-1){1}}
\put(1100,1050){\makebox(0,0){$S'_1$}}
\put(1100,550){\makebox(0,0){$S'_1$}}
\put(1500,1550){\oval(1475,300)[t]}
\put(2237,1550){\vector(0,-1){1}}
\put(1500,1800){\makebox(0,0){$S'_1$}}
\multiput(1500,1050)(0,-500){2}{\oval(1450,300)[t]}
\multiput(2225,1050)(0,-500){2}{\vector(0,-1){1}}
\put(1500,1300){\makebox(0,0){$S'_1$}}
\put(1500,800){\makebox(0,0){$S'_2$}}
\put(2500,1550){\oval(1475,200)[t]}
\put(3237,1550){\vector(0,-1){1}}
\put(2500,1550){\makebox(0,0){$S''_1$}}
\multiput(2500,1050)(0,-500){2}{\oval(1450,200)[t]}
\multiput(3225,1050)(0,-500){2}{\vector(0,-1){1}}
\put(2500,1050){\makebox(0,0){$S''_1$}}
\put(2500,550){\makebox(0,0){$S''_2$}}
\put(3000,1550){\oval(1475,300)[t]}
\put(3737,1550){\vector(0,-1){1}}
\put(3000,1800){\makebox(0,0){$S''_0$}}
\multiput(3000,1050)(0,-500){2}{\oval(1450,300)[t]}
\multiput(3725,1050)(0,-500){2}{\vector(0,-1){1}}
\put(3000,1300){\makebox(0,0){$S''_1$}}
\put(3000,800){\makebox(0,0){$S''_1$}}
\end{picture}\end{center}
\setlength{\unitlength}{.01em}%

Dabei sind die Vektoren in $\cal F$, die in dem Diagramm auftauchen, nach
Lemma \ref{fklem} nur bis auf "Aquivalenz bez"uglich der bereits erw"ahnten
Projektionen festgelegt. Die Abbildungen $S'_k$ und $S''_k$, die im Bild
eingezeichnet sind, sind nat"urlich erst nach der entsprechenden Projektion
wohldefiniert, z.B. mu\3 man  $S'_1$ als Abbildung $\ker S'_0\lra {\cal F}
/\im S'_0={\cal F}/[S'_0 w_0]$ auf\/fassen. Aus der G"ultigkeit dieses
Diagrammes folgt die Behauptung des Satzes \ref{fockklasse} f"ur den Fall
$III_-$.

\noindent{\bf Beweis von Lemma \ref{grausam}.}

Wir wissen aus Lemma \ref{exist} und Lemma \ref{kontra}, da\3 ${\cal F}$
singul"are Vektoren der Grade $\deg w_{4k+1}$ und kosingul"are Vektoren der
Grade $\deg w_{4k+2}$ f"ur $k \ge 0$ enth"alt. Aus der Determinantenformel
von $S'(\alpha,\beta)$ folgt, da\3 $\ker S'_0=\ker S'(\alpha,\beta)=[w_2]$
gilt, d.h. wir  $S'_0:V/[w_2] \stackrel{\simeq}{\lra} \im S'_0$ haben. Da
$S_1 w_2 \ne 0 (=\bar{w}_2)$, mu\3 $S'_1 w_2 \ne 0$ sein (sonst w"are
$S'w_2$ von der Ordnung $z^2$ und $S_1 w_2=0$). Wegen Lemma \ref{fklem}
mu\3 ein $\tilde{w}_2$  mit $\pr_0 \tilde{w}_2 = S'_1 w_2$ und  $S''_0
\tilde{w}_2 =\bar{w}_2$ existieren. Au\3erdem gilt $S''_0 S'_0 w_1 =0$
wegen $S_0 w_1=0$ und $S''_1 S'_0 w_1=\bar{w}_1$ in $\coker S''_0$, da
$S_1 w_1=\bar{w}_1$ gilt. Es folgt $\coker S''_0 = \overline{V}/\im S''_0
=[\bar{w}_1]$. Das liefert den Induktionsanfang.

Sei die Behauptung f"ur alle $i<i_0$ bewiesen. Wir haben
\[ \ker S'_{i_0-2}=[w_{4i_0-6}],\quad \ker S'_{i_0-1}=[w_{4i_0-2}],\]
\[ \coker S''_{i_0-2} = [w_{4i_0-7}] \mbox{ und } \coker
S''_{i_0-1}=[w_{4i_0-3}]. \]
Weiter gilt
\bea
S''_{i_0-2} : \quot{J''_{i_0-2}}{J''_{i_0}}
&\stackrel{\simeq}{\strich\lra}& \ker \{ \pr : [\bar{w}_{4i_0-11}]\lra
[\bar{w}_{4i_0-7}] \}, \nonumber \\
S''_{i_0-1} : \quot{J''_{i_0-1}}{J''_{i_0}}
&\stackrel{\simeq}{\strich\lra}&\ker \{ \pr :[\bar{w}_{4i_0-7}] \lra
[\bar{w}_{4i_0-3}]\}. \nonumber
\eea
Wir m"ussen die  folgende Aussagen "uber $S'_{i_0}$, $S''_{i_0}$,
$S'_{i_0+1}$ und $S''_{i_0+1}$  beweisen.
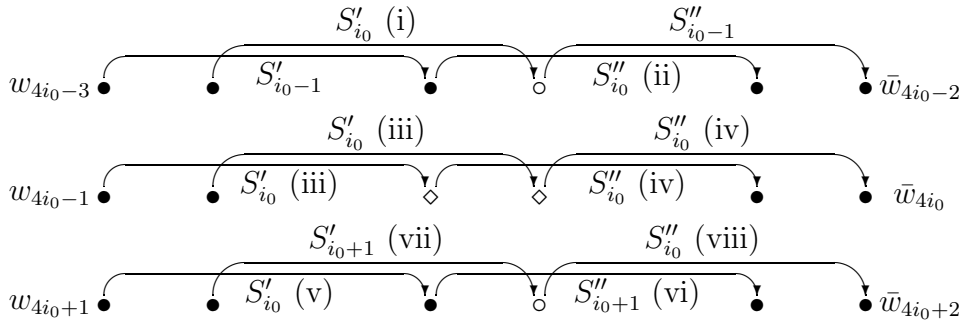
\begin{center}
\setlength{\unitlength}{.007em}%
\begin{picture}(4000,1300)
\put(250,1000){\makebox(0,0){$\bullet$}}
\put(750,1000){\makebox(0,0){$\bullet$}}
\put(250,500){\makebox(0,0){$\bullet$}}
\put(750,500){\makebox(0,0){$\bullet$}}
\put(250,0){\makebox(0,0){$\bullet$}}
\put(750,0){\makebox(0,0){$\bullet$}}
\put(0,1000){\makebox(0,0){$w_{4i_0-3}$}}
\put(0,500){\makebox(0,0){$w_{4i_0-1}$}}
\put(0,0){\makebox(0,0){$w_{4i_0+1}$}}
%
\put(1750,1000){\makebox(0,0){$\bullet$}}
\put(2250,1000){\makebox(0,0){$\circ$}}
\put(1750,500){\makebox(0,0){$\diamond$}}
\put(2250,500){\makebox(0,0){$\diamond$}}
\put(1750,0){\makebox(0,0){$\bullet$}}
\put(2250,0){\makebox(0,0){$\circ$}}
%
\put(3250,1000){\makebox(0,0){$\bullet$}}
\put(3750,1000){\makebox(0,0){$\bullet$}}
\put(3250,500){\makebox(0,0){$\bullet$}}
\put(3750,500){\makebox(0,0){$\bullet$}}
\put(3250,0){\makebox(0,0){$\bullet$}}
\put(3750,0){\makebox(0,0){$\bullet$}}
\put(4000,1000){\makebox(0,0){$\bar{w}_{4i_0-2}$}}
\put(4000,500){\makebox(0,0){$\bar{w}_{4i_0}$}}
\put(4000,0){\makebox(0,0){$\bar{w}_{4i_0+2}$}}
%
%
%
\put(1100,1050){\makebox(0,0){$S'_{i_0-1}$}}
\multiput(987,1050)(0,-500){3}{\oval(1475,200)[t]}
\multiput(1725,1050)(0,-500){3}{\vector(0,-1){1}}
\put(1100,550){\makebox(0,0){$S'_{i_0}$ (iii)}}
\put(1100,050){\makebox(0,0){$S'_{i_0}$ (v)}}
\put(1500,1300){\makebox(0,0){$S'_{i_0}$ (i)}}
\multiput(1487,1050)(0,-500){3}{\oval(1475,300)[t]}
\multiput(2225,1050)(0,-500){3}{\vector(0,-1){1}}
\put(1500,800){\makebox(0,0){$S'_{i_0}$ (iii)}}
\put(1500,300){\makebox(0,0){$S'_{i_0+1}$ (vii)}}
\put(2700,1050){\makebox(0,0){$S''_{i_0}$ (ii)}}
\multiput(2513,1050)(0,-500){3}{\oval(1475,200)[t]}
\multiput(3250,1050)(0,-500){3}{\vector(0,-1){1}}
\put(2700,550){\makebox(0,0){$S''_{i_0}$ (iv)}}
\put(2700,50){\makebox(0,0){$S''_{i_0+1}$ (vi)}}
\put(3000,1300){\makebox(0,0){$S''_{i_0-1}$}}
\multiput(3013,1050)(0,-500){3}{\oval(1475,300)[t]}
\multiput(3750,1050)(0,-500){3}{\vector(0,-1){1}}
\put(3000,800){\makebox(0,0){$S''_{i_0}$ (iv)}}
\put(3000,300){\makebox(0,0){$S''_{i_0}$ (viii)}}
\end{picture}\end{center}
\setlength{\unitlength}{.01em}%

\begin{itemize}
\item[(i)] Da $S_{2i_0-1} w_{4i_0-2} \ne 0$, mu\3 $S'_{i_0} w_{4i_0-2} \ne
0$ sein.
\item[(ii)] Da $S_{2i_0-1} w_{4i_0-3} \ne 0$ und $\bar{w}_{4i_0-3} \in
\coker S''_{i_0-1}$, mu\3 $\bar{w}_{4i_0-3} \in \im S''_{i_0}$ gelten.
\item[(iii)] Es kann nicht $S'_{i_0} w_{4i_0-1} =0$ gelten, denn sonst
m"u\3te ein $x \in \ker S''_{j-1}$ f"ur ein $j<i_0-1$  mit $S''_j x
=\bar{w}_{4i_o-1}$ existieren. Das ist aber nach der Induktionsannahme
ausgeschlossen. Das gleiche Argument funktioniert f"ur $w_{4i_0}$.
\item[(iv)] Es mu\3 $S''_{i_0} S'_{i_0} w_{4i_0-1} = \bar{w}_{4i_0-1}$ und
$S''_{i_0} S'_{i_0} w_{4i_0}=\bar{w}_{4i_0}$ gelten.
\item[(v)] $J'_{i_0-1}$ enth"alt keinen singul"aren Vektor vom Grad $\deg
w_{4i_0+1}$: \\
Es m"u\3te sonst ein $j<i_0$ existieren, so da\3 $S'_j w_{4i_0+1}$ gleich
diesem singul"aren Vektor ist. Es gilt aber $w_{4i_0+1}\in \ker S'_j$ f"ur
alle $j<i_0$.\\
Da ${\cal F}$ aber einen singul"aren Vektor diesen Grades enth"alt, folgt,
da\3 auch ${\cal F}/J'_{i_0-1}\simeq \coker S'_{i_0-1}$ einen singul"aren
Vektor diesen Grades enth"alt.\\
 $S'_{i_0} w_{4i_0+1}$ mu\3 gleich diesem Vektor sein:\\
Sonst m"u\3te das f"ur ein $j=i_0+k>i_0$ gelten, und damit w"urde folgen, da\3
$S'_{j} w_{4i_0+1}\ne 0$ ist und $S''_{i_0-k+1} S'_{i_0+k} w_{4i_0+1}$ ein
singul"arer Vektor in $\coker S'_{i_0-k}$ ist. Es gibt dort aber keinen
singul"aren Vektor diesen Grades.
\item[(vi)] Es mu\3 $S''_{i_0} S'_{i_0} w_{4i_0+1}=0 $ sein, d.h.
$S'_{i_0}\in \ker S''_{i_0}$. Weiter mu\3 wegen $S_{2i_0+1}
w_{4i_0+1}=\bar{w}_{4i_0+1}$ $S''_{i_0+1} S'_{i_0} w_{4i_0+1} \ne 0$
gelten.
\item[(vii)] Es mu\3 $S'_{i_0} w_{4i_0+2}=0$ gelten, denn ansonsten w"urde
$\coker S'_{i_0-1}$ einen singul"aren Vektor diesen Grades enthalten, der
sogar schon in ${\cal F}$ singul"ar sein m"u\3te. ${\cal F}$ enth"alt aber
einen kosingul"aren Vektor diesen Grades, der dann f"ur keine $j$ in $\im
S'_j$  liegen w"urde, was einen Widerspruch bedeuten w"urde. Es folgt damit
$\ker S'_{i_0}=[w_{4i_0+2}]$.
\item[(viii)] Aus (vii) folgt weiter $S'_{i_0+1} w_{4i_0+2} \ne 0$ und
$S''_{i_0} S'_{i_0+1} w_{4i_0+2}$. Insgesamt mu\3 $\im
S''_{i_0}=[\bar{w}_{4i_0+2}]/[\bar{w}_{4i_0-2}]$ gelten oder "aquivalent
$\coker S''_{i_0}=[\bar{w}_{4i_0-3}]/[\bar{w}_{4i_0+1}]$.
\end{itemize}
Lemma \ref{grausam} ist bewiesen.\hfill$\Box$

Der Beweis von Fall $III_-$ folgt nun aus folgender "Uberlegung: Wir
erfahren aus Lemma \ref{grausam}, da\3 $S'_k$ Isomorphismen von
$[w_{4k-2}]/[w_{4k+2}]$ nach $J'_{k-1}/J'_k$ sind. Das Urbild ist der  von
dem Vektor $w_{4k-2}$ erzeugte Modul mit drei singul"aren Vektoren und der
Struktur:
\ben\label{diag1}\begin{array}{c}
\setlength{\unitlength}{.005em}%
\begin{picture}(500,1000)
\multiput(0,500)(0,-500){2}{\circle*{30}}
\multiput(500,1000)(0,-500){2}{\circle*{30}}
\multiput(450,950)(0,-500){2}{\vector(-1,-1){400}}
\multiput(500,950)(-500,-500){2}{\vector(0,-1){400}}
\end{picture}
\setlength{\unitlength}{.01em}
\end{array}\een
 Dasselbe gilt nat"urlich dann auch f"ur das Bild. Da wir die Abbildungen
$S_k$ aber nach Lemma \ref{fklem} ``liften'' k"onnen, w"ahlen wir beliebige
Elemente in ${\cal F}$, die unter der Projektion auf das Bild auf die
entsprechenden Vektoren in $J'_{k-1}/J'_k$ abgebildet werden. Diese
Vektoren sind (bis auf das Bild von $w_{4k+1}$) sicherlich nicht mehr
singul"ar, aber dennoch "ubertr"agt sich die obige Struktur auf diese
Vektoren, d.h. wir erhalten in ${\cal F}$ Vektoren mit der obigen
Struktur. Analog sieht man, da\3  aus den Aussagen "uber $S''_k$ Diagramme
der Form
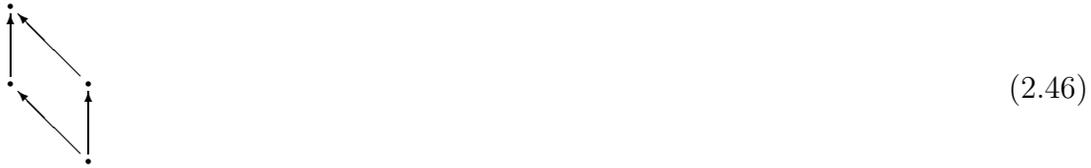
\ben\label{diag2}
\begin{array}{c}
\setlength{\unitlength}{.005em}%
\begin{picture}(500,1000)
\multiput(0,1000)(0,-500){2}{\circle*{30}}
\multiput(500,500)(0,-500){2}{\circle*{30}}
\multiput(450,550)(0,-500){2}{\vector(-1,1){400}}
\multiput(0,550)(500,-500){2}{\vector(0,1){400}}
\end{picture}
\setlength{\unitlength}{.01em}
\end{array}
\een
 in ${\cal F}$ folgen. Diese beiden Diagramme zusammengef"ugt ergeben das
Diagramm von ${\cal F}$. Die verschiedenen Typen von Vektoren in ${\cal
F}$ ergeben sich direkt aus der Lage der singul"aren Vektoren in ${\cal F}$
und den Richtungen der Pfeile. Damit erhalten wir insgesamt die behauptete
Struktur von ${\cal F}$. Der Fall $III_-$ ist bewiesen.

Einige Bemerkungen zum Beweis der restlichen F"alle:\\
Die aufmerksame Leserin hat im Beweis von Lemma \ref{grausam} bemerkt, da\3
es letztendlich entscheidend war, wo singul"are und kosingul"are Vektoren in
den Fock--Moduln liegen; sie markieren die Extremalpunkte in den
Diagrammen (\ref{diag1}) und (\ref{diag2}). Man mu\3 also wieder einige
kleine Rechnungen mit den entsprechenden Determinantenformeln ausf"uhren,
um diese Vektoren zu finden. In den F"allen ``+'', d.h $III_+, III_+^0,
\overline{III}_+, \overline{III}_+^0$ erhalten wir als Ergebnis, da\3 alle
Nullstellen von $\det S$ entweder ausschlie\3lich in $\det S'$ oder $\det
S''$ liegen. Damit ist entweder $S'$ oder $S''$ ein Isomorphismus, und die
Struktur des Moduls entspricht einem (kontragredienten) Verma--Modul.

Im Fall $III_-^0$ erhalten wir abwechselnd Nullstellen in $\det S'$ und
$\det S''$, die abwechselnd singul"are und kosingul"are Vektoren in $\cal F$
erzeugen. Im Fall $III_-^{00}$ schlie\3lich erhalten wir Nullstellen
gleichzeitig in $\det S'$ und $\det S''$, was zu zugleich singul"aren und
kosingul"aren Vektoren f"uhrt.
Man kann leicht die entsprechende Form von Lemma \ref{grausam} aus diesen
"Uberlegungen extrahieren und einen Beweis analog zu dem Beweis von Lemma
\ref{grausam} f"uhren. Dies liefert dann den Beweis der entsprechenden
Behauptung von Satz \ref{fockklasse}.\hfill$\Box$
\chapter{Vertex--Operatoren}
In diesem Kapitel werden wir die Vertex--Operatoren einf"uhren. Sie sind
die wichtigsten Objekte zur Konstruktion von Modellen der konformen
Quantenfeldtheorie und zum Beweis der Determinantenformeln. Wir werden
sie als Operatoren in den Hilbertr"aumen ${\cal H} (\alpha,\beta)$
definieren und nicht nur als formale Exponentiale, wie es "ublicherweise
in der konformen Quantenfeldtheorie  gemacht wird. Wir werden dann
Produkte von Vertex--Operatoren untersuchen und zeigen, da\3 sie als
Operatorprodukte auf einem dichten Teilraum existieren.\\
Die Ber"uhrung mit der Virasoro--Algebra kommt durch die Tatsache, da\3 wir
trotz der komplizierten Struktur der Vertex--Operatoren ihre Kommutatoren
mit der Virasoro--Algebra berechnen k"onnen. Dies wird der erste Schritt
sein, um die im Satz \ref{Inter} beschriebenen Intertwiner zu
konstruieren, und ebenso auf dem Weg zur Konstruktion der prim"aren Felder
f"ur die minimalen Modelle der konformen Quantenfeldtheorie.\\
Wir werden danach die Vertex--Operatoren noch etwas eingehender
untersuchen und eine Faktorisierung in einen Hilbert--Schmidt und einen
Diagonal--Operator beweisen. Wir werden aber auch beweisen, da\3 die
Vertex--Operatoren, wie sie in der konformen Quantenfeldtheorie benutzt
werden, nicht abschlie\3bare
Operatoren in ${\cal H}(\alpha,\beta)$ sind.
\section{Das freie Feld}
Vertex--Operatoren entstehen formal als Wick--geordnete Exponentiale des
freien Feldes, genauer des euklidischen, freien, masselosen Feldes, da\3
in radialer Quantisierung in ${\cal H}(\alpha,\beta)$ folgenderma\3en
gegeben
ist (vgl. \cite{GSW}). F"ur $0 \neq z \in \CC$ sei
\ben\label{01}
\Phi (z) = q-ia_0 \ln z +i \sum_{n \neq 0} \frac{z^{-n}}{n} a_n,
\een
wobei $[ q, a_n]= i \delta_{n,0}$ gilt. ($q$ und $a_0$ sind also ein
Schr"odinger--Paar, sie repr"asentieren Ort und Impuls des freien Feldes.)
Die Variable $z=re^{i \varphi}$ beschreibt die Raum--Zeit, der Winkel
$\varphi$ ist die Ortskoordinate und $\ln r$ ist die Zeitkoordinate. Der
Einheitskreis in $\CC$ ist also die ``Welt'' zur Zeit $t=\ln r=0$.
Mit $\Phi_+(z):=-ia_0-i\sum_{n>0} \frac{z^n}{n}a_{-n}$ und $\Phi_-(z):=
q+i\sum_{n>0} \frac{z^{-n}}{n} a_n$ gilt
\ben\label{korrfr1}
[\Phi_-(z_1),\Phi_-(z_2)]=[\Phi_+(z_1),\Phi_+(z_2)]=0
\een
und f"ur $|z_1|>|z_2|$
\ben\label{korrfr2}
[\Phi_-(z_1),\Phi_+(z_2)]=-\log(z_1-z_2).
\een
Aus (\ref{korrfr1}) und (\ref{korrfr2}) folgt unmittelbar f"ur
$|z_1|>|z_2|$
\ben\label{frfeld}
\langle v_{0,0},\Phi(z_1)\Phi(z_2) v_{0,0} \rangle =\log(z_1-z_2).
\een
Gleichung (\ref{frfeld}) rechtfertigt die Bezeichnung ``freies Feld''  f"ur
$\Phi$.
$\Phi(z)$ erf"ullt formal
\ben\label{02}
\Phi(z)^*= \Phi(\bar{z}^{-1}),
\een
insbesondere gilt also $\Phi(z)^*=\Phi(z)$ f"ur $|z|=1$. Es
gibt noch eine weitere interessante formale Relation
\ben\label{03}
\Phi(z) =|z|^N \Phi(\frac{z}{|\bar{z}|}) |z|^{-N}= e^{\ln |z|N} \Phi
(\frac{z}{|\bar{z}|})e^{-\ln |z|N},
\een
die das euklidische Heisenberg--Bild darstellt. $N$ kann also als
Hamiltonoperator interpretiert werden, dessen Halbgruppe $e^{-tN}$ sowie
$e^{tN}$ die Zeitentwicklung liefern. Man sieht hier schon ein Handikap
 der euklidischen Felder und eine der Ursachen f"ur unsere Schwierigkeiten,
die Zeitentwicklung wird nicht durch unit"are Operatoren beschrieben.\\
Die Gleichheit in (\ref{02}) kann man f"ur die Operatoren nat"urlich nicht
erwarten, man verlangt hier nur ``$\supset$''. Wir werden sehen, da\3 auch
diese Relation nur trivial erf"ullt ist, denn wir werden beweisen, da\3
$D(\Phi(z))=D \left( \Phi(\bar{z}^{-1})^* \right) =\{0\}$ f"ur $|z|>1$
gilt, da\3 also  das freie Feld f"ur $|z|>1$ nicht definierbar ist. F"ur
$|z|<1$ sieht man dagegen schnell, da\3 ${\cal F}(\alpha,\beta)$ im
Definitionsbereich von $\Phi(z)$ liegt, (\ref{01}) definiert also f"ur
$|z|<1$ immerhin einen dicht definierten Operator.

Bevor wir weiter fortfahren, noch einige Bemerkungen zu den bereits
erw"ahnten Schwierigkeiten mit dem freien Feld und den Vertex--Operatoren.
Es gibt zumindest zwei Gr"unde, weshalb die zu definierenden Operatoren
schlechte Eigenschaften haben k"onnen: Zum einen haben wir eine masselose
Theorie, die bekannterma\3en singul"arer als eine massive Theorie ist. Zum
anderen sind Quantenfelder oft nur nach einer ``Verschmierung''
wohldefiniert, d.h. wir m"ussen die Felder als operatorwertige
Distribution
"uber einem geeigneten Testraum interpretieren.

Es ist bekannt, da\3 man die freien Felder in $t$ lokalisieren kann
\cite{RSII}, es reicht also, im Ortsraum, d.h. in $S^1$ zu verschmieren.

Sei $f\in C^{\infty}(S^1)$ und $f_{-k} :=1/i \int_{S^1} f(w) w^{-k-1} {\rm
d}w$, so
da\3 $|f_k|\le C_1 \exp(-C_2 |k|^{1+\varepsilon})$ f"ur ein $\varepsilon >0$
und Konstanten $C_1,C_2>0$. Den linearen Raum aller $f$ mit dieser
Eigenschaft bezeichnen wir mit $\cal A$. Wegen $f(z)=\frac{1}{2\pi}
\sum_{k\in \ZZ} f_k z^k$ f"ur $|z|=1$ und der Bedingung an $\{f_k\}$
entstehen die Elemente von $\cal A$ als Einschr"ankung in $\CC \setminus
\{0\}$ analytischer Funktionen auf $S^1$. Deshalb enth"alt $\cal A$
nat"urlich keine nichttrivialen Funktionen mit kompaktem Tr"ager. Trotzdem
k"onnen wir in $\cal A$ ``lokalisieren'', d.h. wir finden Funktionen, die
ihren Tr"ager bis auf einen beliebig kleinen Rest in einer beliebig kleinen
Umgebung eines Punktes $\varphi_0\in S^1$ haben. Als Beispiel k"onnen wir
an Partialsummen $\sum_{k\le N} e^{ik(\varphi-\varphi_0)}$ denken, wegen
der Poisson--Formel $\delta(\varphi-\varphi_0)=\frac{1}{2\pi}
\sum_{k\in\ZZ} e^{ik(\varphi-\varphi_0)}$ haben diese Partialsummen die
gew"unschte Eigenschaft und liegen nat"urlich in $\cal A$.
Wir definieren f"ur $f\in \cal A$
\ben
\Phi(f)=q -ia_0 \log f_0 + i \sum_{n\ne 0} \frac{f_{-n}}{n} a_n
\een
als das verschmierte freie Feld zur Zeit $t=0$. Mittels (\ref{03})
definieren wir $\Phi$ f"ur beliebige Zeiten $r=\ln t$ durch
\ben\label{schmierfeld}
\Phi(r,f)=q-ia_0 \log f_0 +i \sum_{n\ne 0}\frac{r^{-n} f_{-n}}{n} a_n.
\een
Im Sinne der Poisson--Formel ist eine m"ogliche Verschmierung (d.h.
Regularisierung) des freien Feldes durch
\ben
\Phi_N (z) = q-ia_0 \ln z +i \sum_{n \neq 0, n\le N} \frac{z^{-n}}{n} a_n
\een
gegeben.
Aus unseren Voraussetzungen an $f_k$ folgt, da\3 $\sum_{n\in \ZZ} | r^n
f_n|$ existiert und damit folgt schnell, da\3 $\Phi(r,f)$ dicht definiert
ist.
\section{Definition der Vertex--Operatoren}
 Formal sind Vertex--Operatoren definiert durch
\ben\label{04}
\tilde{V}(\gamma ,z)=\; : \exp (i \gamma \Phi(z)):,
\een
wobei $\gamma\in \CC$, $0 \ne z\in \CC$ und $:\; :$ die Wick--Ordnung
(Erzeuger links von
Vernichteroperatoren) bezeichnet. Die Wick--Ordnung l"a\3t sich leicht
ausf"uhren, wir erhalten
\ben\label{05}
\tilde{V}(\gamma, z)=\exp \left( \gamma \sum_{n=1}^{\infty}
\frac{z^n}{n}a_{-n}
\right) \exp(i \gamma q)z^{\gamma a_0} \exp \left(- \gamma
\sum_{n=1}^{\infty} \frac{z^{-n}}{n}a_n \right).
\een
Der Operator $\exp(i\gamma q)$ bewirkt wegen $[a_0,\exp(i \gamma
q)]=\gamma \exp (i \gamma q)$, da\3 $\tilde{V}(\gamma, z): {\cal H}
(\alpha,\beta)\longrightarrow {\cal H}(\alpha+\gamma,\beta)$ gilt. Wir
werden kurz $T_{\gamma}$ f"ur $\exp(i \gamma q)$ schreiben.
$\tilde{V}(\gamma,z)$ erf"ullt analoge Relationen zu (\ref{02}) und
(\ref{03}), wir haben
\bea
\tilde{V}(\gamma,z)^*&=& \tilde{V}(-\bar{\gamma},\frac{1}{\bar{z}})
\mbox{ und }\\
\tilde{V}(\gamma,z)&=&|z|^N \tilde{V}(\gamma,\frac{z}{|z|}) |z|^{-N}.
\eea
Wir setzen $V(\gamma,z)=\exp(\gamma\sum_{n=1}^{\infty} \frac{z^n}{n}
a_{-n}) \exp(-\gamma \sum_{n=1}^{\infty} \frac{z^{-n}}{n} a_n)$, wir
haben also $\tilde{V}(\gamma,z)=T_{\gamma}z^{\gamma a_0} V(\gamma,z)$.
$V(\gamma,z)$ und $\tilde{V}(\gamma,z)$ sind als Operatoren "aquivalent,
wir werden im folgenden oft die zun"achst unwichtigen Faktoren
$T_{\gamma}z^{\gamma a_0}$ ignorieren, beziehungsweise werden wir
gegebenenfalls auch andere Potenzen von $z$ w"ahlen.

 Formal k"onnen wir $\tilde{V}(\gamma,z)$ auch als unendliches Produkt
schreiben (da $a_n$ und $a_m$ f"ur $n\neq -m$
vertauschen), wir erhalten
\ben\label{06}
\tilde{V}(\gamma,z)=\prod_{n=1}^{\infty} \exp \left( \gamma
 \frac{z^n}{n}
a_{-n}\right) T_{\gamma} z^{\gamma a_0} \prod_{n=1}^{\infty} \exp
\left(- \gamma \frac{z^{-n}}{n} a_n\right).
\een
Unsere erste Aufgabe wird es sein, zu zeigen, da\3 (\ref{05}) bzw.
(\ref{06}) f"ur $|z|<1$
einen dicht definierten Operator definiert. Zur Definition von
$V(\gamma,z)$ wollen wir
Ma\-trix\-ele\-men\-te
$\langle \Phi_{\eta},V(\gamma,z)\Phi_{\nu}\rangle$ berechnen. Die lassen
sich einfach berechnen, wenn wir $[a_k,V(\gamma,z)]$ kennen. Es reicht
dazu den Kommutator mit dem entsprechenden Faktor in (\ref{06}) zu
berechnen. Es gilt f"ur $k>0$ (nat"urlich immer noch formal)
\begin{eqnarray*}
[a_k,\exp(\gamma \frac{z^k}{k}a_{-k})]&=&\sum_{n=0}^{\infty}\frac{1}{n!}
\left(\frac{\gamma z^k}{k}\right)^n [a_k,a_{-k}^n]\\
&=& \sum_{n=1}^{\infty} \frac{1}{n!} \left(\frac{\gamma z^k}{k}\right)^n
nka_{-k}^{n-1}=\gamma z^k \exp(\gamma \frac{z^k}{k} a_{-k}).
\end{eqnarray*}
Analog erh"alt man $[a_{-k},\exp(\gamma \frac{z^{-k}}{k} a_k)]=\gamma
z^{-k} \exp(\gamma \frac{z^{-k}}{k}a_k)$. Weiter gilt
$[a_0,\tilde{V}(\gamma,z)]=\gamma V(\gamma,z)$. Es folgt
insgesamt f"ur $k\in \ZZ$
\ben\label{07}
[a_k,\tilde{V}(\gamma,z)]=\gamma z^k \tilde{V}(\gamma,z).
\een
Der einzige Unterschied zu $V(\gamma,z)$ ist, da\3 $[a_0,V(\gamma,z)]=0$
gilt.
Nun ist es eine leichte "Ubung, die Matrixelemente von $V(\gamma,z)$
anzugeben, es folgt
{\mathindent0mm\bea\label{08}
\langle \Phi_{\eta},V(\gamma,z)\Phi_{\nu}\rangle &=&
\frac{1}{\sqrt{\eta!\nu!}}\prod_{i=1}^{\infty} \left\{
\sum_{j=0}^{ \min(\eta_i,\nu_i)} {\eta_i \choose j}{\nu_i \choose j} j!
\left( \frac{\gamma}{\sqrt{i}} z^i\right)^{\eta_i-j} \left( -
\frac{\gamma}{\sqrt{i}} z^{-i}\right)^{\nu_i-j} \right\}\nonumber\\
&=& \frac{1}{\sqrt{\eta!\nu!}}\prod_{i=1}^{\infty}m_{\eta_i,\nu_i}
\left(\frac{\gamma}{\sqrt{i}} z^i,-\frac{\gamma}{\sqrt{i}}
z^{-i}\right)\nonumber\\
&=:& v_{\eta,\nu}(\gamma,z)
\eea}
mit
\ben
m_{n,m}(x,y):=\sum_{j=0}^{\min(n,m)} {n\choose j}{m\choose j}j! x^{n-j}
y^{m-j}.
\een
Das unendliche Produkt in (\ref{08}) entspricht dem Produkt in (\ref{06})
und macht keine Probleme, da $\eta_i\neq 0$ und $\nu_i\neq 0$ nur f"ur
endlich viele $i$ gilt und $m_{0,0}(x,y)=1$ ist.

Nat"urlich l"a\3t sich $m_{n,m}(x,y)$ durch bekannte Polynome ausdr"ucken,
z.B. "uber Charlier--Polynome, die durch
\[
C_n^{(a)}(x)=\sum_{l=0}^{n} {n \choose l}{x\choose l} l! (-a)^{n-l}
\]
definiert sind, es gilt  $m_{n,m}(x,y)=y^{m-n} C_n^{(-xy)} (m)$
\cite{Chi}.

Analog zu (\ref{schmierfeld}) k"onnen wir verschmierte Vertex--Operatoren
definieren, f"ur $f\in\cal A$  sei
\ben\label{14}
V(\gamma,r,f)=\exp \left( \gamma \sum_{n=1}^{\infty} \frac{r^n
f_n}{n}a^{-n} \right)  \exp \left(- \gamma
\sum_{n=1}^{\infty} \frac{r^{-n}f_{-n}}{n}a_n \right).
\een
Die Matrixelemente von $V(\gamma,r,f)$ sind gegeben durch
\ben\label{15}
\langle\Phi_{\eta},V(\gamma,r,f)\Phi_{\nu}\rangle =\frac{1}{ \sqrt{
\eta!\nu!}}
\prod_{i=1}^{\infty}m_{\eta_i,\nu_i}\left(\frac{\gamma}{\sqrt{i}}
r^i f_i,-\frac{\gamma}{\sqrt{i}} r^{-i}f_{-i}\right).
\een
Es wird sich zeigen, da\3 die Verschmierung leider entscheidende
Eigenschaften der Vertex--Operatoren zerst"ort, verschmierte
Vertex--Operatoren sind keine ``prim"aren Felder''. Deshalb sind sie f"ur
die konforme Quantenfeldtheorie nicht von gro\3em Nutzen. Zun"achst k"onnen
wir jedoch die verschmierten Vertices mitbehandeln. Dazu untersuchen wir
im folgenden allgemeiner Operatoren
\ben\label{verallg}
V(\omega)=\exp\left( \sum_{n=1}^{\infty} \frac{\omega_n}{\sqrt{n}}
a_{-n}\right) \exp \left(
-\sum_{n=1}^{\infty}\frac{\omega_{-n}}{\sqrt{n}} a_n\right),
\een
 f"ur
$\omega=\{\omega_i\}_{i\in \ZZ\setminus\{0\} }\subset \CC$, $V(\omega)$ ist
durch die Matrixelemente
\ben\label{16}
v_{\eta,\nu}(\omega)=\frac{1}{\sqrt{\eta!\nu!}}\prod_{i=1}^{\infty}
m_{\eta_i,\nu_i}(\omega_i,-\omega_{-i})
\een
definiert. Wir haben also die Beziehungen
\ben\label{17}
V(\gamma,r,f)=
V(\{\frac{\gamma}{\sqrt{i}} r^{\pm i} f_{\pm i}\}_{i\in\NN})
\een
und
\ben\label{18}
V(\gamma,z)= V(\{\frac{\gamma}{\sqrt{i}}
z^{\pm i}\}_{i\in\NN}).
\een
Im Weiteren werden wir, wenn die Abh"angigkeit der auftretetenden
Operatoren  von $\alpha$ und $\beta$ trivial ist, oft kurz $\cal H$ f"ur
${\cal H}(\alpha,\beta)$ und $\cal F$ f"ur ${\cal F}(\alpha,\beta)$
 schreiben.
Mit Hilfe der Matrixelemente k"onnen wir nun einen Operator in ${\cal H}
$ definieren durch \cite{Wd}
\bea\label{10}
D(V(\omega))&=&\Bigg\{ \Psi \in {\cal H}\;:\;\lim_{k\to\infty}
\sum_{\|\nu\|
\le k} v_{\eta,\nu}(\omega) \langle \Phi_v,\Psi\rangle \mbox{ existiert
f"ur alle } \eta \mbox{ und}\nonumber \\
&&\sum_{\eta}\left| \sum_{\nu} v_{\eta,\nu}(\omega)\langle
\Phi_{\nu},\Psi \rangle \right|^2 < \infty \Bigg\},
\eea
und
\ben\label{11}
V(\omega)\Psi := \sum_{\eta} \sum_{\nu} v_{\eta,\nu}(\omega)\langle
\Phi_{\nu } ,\Psi\rangle \mbox{ f"ur }\Psi \in D\left(V(\omega)\right).
\een
Wir m"ochten nun untersuchen, wann $V(\omega)$ dicht definiert ist. Dazu
verwenden wir die folgende Identit"at, die man als Verallgemeinerung der
Orthogonalit"atsrelation f"ur die Charlier--Polynome interpretieren kann
(\cite{Chi}, Ch. VI.1).
\begin{lem} \label{Identitaet} F"ur $x,y,z,w\in \CC$ und $i,j\in \NN_0$
gilt
\ben
\sum_{k=0}^{\infty} \frac{1}{k!} m_{i,k}(x,y) m_{k,j}(z,w)=m_{i,j}(x+z,
y+w ) e^{zy}.
\een
\end{lem}
Einen Beweis von Lemma \ref{Identitaet} findet man in \cite{BC}. Wir
k"onnen nun den folgenden Satz beweisen.
\begin{satz}\label{dicht}\begin{itemize}
\item[(i)] Sei $\sum_{i=1}^{\infty} |\omega_i|^2<\infty$. Dann gilt
${\cal F}(\alpha,\beta)\subset D(V(\omega))$, $V(\omega)$ ist also
insbesondere dicht definiert.
\item[(ii)] Sei $\sum_{i=1}^{\infty} |\omega_{-i}|^2 < \infty$. Dann ist
$V(\omega)$ abgeschlossen und $V(\omega)^*$ dicht definiert.
\end{itemize}
\end{satz}
{\bf Beweis.}\\
(i): Es gilt $\Phi_{\nu}\in D(V(\omega))$, falls $\sum_{\eta} |
v_{\eta,\nu}(\omega) |^2 <\infty$, es reicht also, die Konvergenz dieser
Summe f"ur alle $\nu$ zu zeigen. Wegen $|m_{i,j}(x,y)|\le m_{i,j}(|x|,
|y|)$ und $m_{i,j}(x,y)=m_{j,i}(y,x)$ k"onnen wir mit Lemma
\ref{Identitaet} folgenderma\3en absch"atzen:
\bea\label{dichtbound}
\sum_{\eta}|v_{\eta,\nu}(\omega)|^2&=&\frac{1}{\nu!}\sum_{\eta}
\frac{1}{\eta!} \prod_{i=1}^{\infty} |
m_{\eta_i,\nu_i}(\omega_i,-\omega_{-i}) |^2\nonumber\\
&\le& \frac{1}{\nu!} \prod_{i=1}^{\infty} \sum_{\eta_i=0}^{\infty}
\frac{1}{\eta_i!} m_{\nu_i,\eta_i}(|\omega_{-i}|,|\omega_i|)
m_{\eta_i,\nu_i} (|\omega_i|,|\omega_{-i}|) \nonumber\\
&=&\frac{1}{\nu!} \prod_{i=1}^{\infty} \left( m_{\nu_i,\nu_i}(|\omega_i|
+ |\omega_{-i}|,|\omega_i|+|\omega_{-i}| )\exp(|\omega_i|^2)\right)
\nonumber\\
&=&\frac{1}{\nu!}\prod_{i=1}^{\infty}\left(
m_{\nu_i,\nu_i}(|\omega_i|+|\omega_{-i} |,|\omega_i|+|\omega_{-i}|)
\right) \exp(\sum_{i=1}^{\infty} |\omega_i|^2)\\
&<& \infty,\nonumber
\eea
falls $\sum_{i=1}^{\infty} | \omega_i|^2<\infty$, da das "ubrigbleibende
Produkt f"ur jeden Multiindex $\nu$ mit $\|\nu\|<\infty$ endlich ist.\\
(ii): V"ollig analog zu (i) k"onnen wir $\sum_{\nu} |
v_{\eta,\nu}(\omega)|^2 )$  durch
\ben
\sum_{\nu}|v_{\eta,\nu}(\omega)|^2 \le \frac{1}{\mu!}\prod_{i=1}^{\infty}
m_{\mu_i,\mu_i}(|\omega_i|+|\omega_{-i}|, | \omega_i|+|\omega_{-i}|)
\exp(\sum_{i=1}^{\infty} |\omega_{-i}|^2)
\een
absch"atzen. Damit folgt die Behauptung unmittelbar aus Satz 6.20,
\cite{Wd}.
\hfill$\Box$\\[3mm]
F"ur die verschiedenen Typen von Vertex--Operatoren bedeutet das: Die
verschmierten Vertex--Operatoren sind abgeschlossene
Operatoren in $\cal H$. F"ur die nichtverschmierten Vertex--Operatoren
erhalten wir aus Satz \ref{dicht}: Ist $|z|<1$, so ist $V(\gamma,z)$ dicht
definiert. Ist $|z|>1$, so ist $V(\gamma,z)$ abgeschlossen. F"ur $|z|=1$
ist weder (i) noch (ii) aus Satz \ref{dicht} anwendbar.
\section{Produkte von Vertex--Operatoren}
Als n"achsten Schritt m"ochten wir Produkte von Vertex--Operatoren
einf"uhren. Da wir keinen unter den Vertex--Operatoren invarianten Teilraum
zur Verf"ugung haben, auf dem Produkte der Vertex--Operatoren existieren,
m"ussen wir uns anderer Methoden bedienen. Die Rettung
wird eine explizite Konstruktion der Produkte sein. Sei
\ben
V_+(\omega)=\exp\left(-\sum_{n=1}^{\infty} \frac{\omega_{-n}}{\sqrt{n}}
a_n\right)
\een
und
\ben
V_-(\omega)=\exp\left(\sum_{n=1}^{\infty} \frac{\omega_{n}}{\sqrt{n}}
a_{-n}\right).
\een
Mit Hilfe der Campbell--Baker--Hausdorff--Formel
\[
\exp(A) \exp(B) = \exp(B)\exp(A) \exp([A,B])
\]
falls $[A,[A,B]]=[B,[A,B]]=0$ erhalten wir formal
\bea
V_+(\omega^1) V_-(\omega^2)&=& V_-(\omega^2)V_+(\omega^1)
\exp\left(\left[-\sum_n \frac{\omega^1_{-n}}{\sqrt{n}} a_n,\sum_m \frac{
\omega^2_{m}}{ \sqrt{m}} a_{-m}\right]\right)\nonumber\\
&=&
V_-(\omega^2)V_+(\omega_1) \exp(-\sum_{n=1}^{\infty} \omega^1_{-n}
\omega^2_n).
\eea
Damit erhalten wir die folgende Funktionalgleichung f"ur die Produkte von
Vertex--Operatoren:
\bea\label{322}
V(\omega^1)V(\omega^2)&=& V_-(\omega^1)V_+(\omega^1)V_-(\omega^2)
V_+(\omega^2) \nonumber \\
&=&
\exp(-\sum_{n=1}^{\infty} \omega_{-n}^1 \omega_n^2)
V_-(\omega^1)V_-(\omega^2)V_+(\omega^1) V_+(\omega^2)\nonumber\\
&=&
\exp(-\sum_{n=1}^{\infty} \omega_{-n}^1 \omega_n^2) V_-(\omega^1+\omega^2)
V_+(\omega^1+\omega^2)\nonumber \\
&=&
\exp(-\sum_{n=1}^{\infty} \omega_{-n}^1 \omega_n^2) V(\omega^1+\omega^2).
\eea
Wir k"onnen also, falls $\sum_{n=1}^{\infty} \omega_{-n}^1 \omega_n^2$
endlich ist,  das Produkt von zwei Vertex--Operatoren $V(\omega^1)$ und
$V(\omega^2)$ einfach durch die rechte Seite von (\ref{322}) definieren.
Dies ist f"ur die verschmierten Operatoren immer der Fall, da
$(\omega^{1,2}_i)_{i\in\ZZ}$ dann schnellfallend sind. F"ur die
lokalisierten
Operatoren mit $\omega_i^{1,2}=\gamma_{1,2} \frac{z^i_{1,2}}{\sqrt{i}}$
ist
$-\sum_{n=1}^{\infty} \omega_{-n}^1\omega_n^2=-\gamma_1\gamma_2
\sum_{n=1}^{\infty} \frac{1}{n}(\frac{z_2}{z_1})^n=\gamma_1\gamma_2
\log(1-\frac{z_2}{z_1})$, falls $|z_1|>|z_2|$ gilt. Diese Einschr"ankung
ist ein aus der euklidischen Quantenfeldtheorie bekannter Effekt, nur
``zeitgeordnete'' Produkte von Feldern sind definiert.
F"ur das Produkt $\tilde{V}(\gamma_1,z_1)\tilde{V}(\gamma_2,z_2)$ m"ussen
wir die Relation
$\tilde{V}(\gamma,z)=V(\{\gamma\frac{z^i}{\sqrt{i}} \})
T_{\gamma}z^{\gamma
a_0 }$ beachten. Die zus"atzlichen Terme liefern bei der Produktbildung,
da $a_0$ in der Wick--Ordnung rechts von $q$ stehen soll,
\bea
T_{\gamma_1} z_1^{\gamma_1 a_0} T_{\gamma_2} z_2^{\gamma_2
a_0} &=& T_{\gamma_1} T_{\gamma_2}
z_1^{\gamma_1(a_0+\gamma_2)} z_2^{a_0} \nonumber \\
&=& T_{\gamma_1+\gamma_2} z_1^{\gamma_1 a_0} z_2^{\gamma_2 a_0}
z_1^{\gamma_1 \gamma_2}.
\eea
Damit erhalten wir in diesem Fall aus (\ref{322}), falls $|z_1|>|z_2|$,
{\mathindent0mm\small\bea\label{3221}
V(\gamma_1,z_1)V(\gamma_2,z_2)&=& (1-\frac{z_2}{z_1})^{\gamma_1 \gamma_2}
\exp
\left({\textstyle \sum\limits_{n=1}^{\infty} \frac{\gamma_1 z_1^n+\gamma_2
z_2^n}{n}
a_{-n} }\right) \exp\left({\textstyle \sum\limits_{n=1}^{\infty}
\frac{\gamma_1
z_1^{-n} +\gamma_2 z_2^{-n}}{n} a_n}\right)\nonumber\\
&=:& (1-\frac{z_2}{z_1})^{\gamma_1 \gamma_2} V(\gamma_1,\gamma_2;z_1,z_2)
\eea
\normalsize und
\ben\label{3222}
\tilde{V}(\gamma_1,z_2)\tilde{V}(\gamma_2,z_2) = (z_1-z_2)^{\gamma_1
\gamma_2} T_{\gamma_1+\gamma_2} z_1^{\gamma_1 a_0} z_2^{\gamma_2 a_0}
V(\gamma_1,\gamma_2;z_1,z_2).
\een}

Interessanterweise spielt auf der rechten Seite von (\ref{3221}) und
(\ref{3222}) die Zeitordnung keine Rolle mehr, sie ist f"ur beliebige
$z_i$ mit $|z_i|<1$
definiert. Die rechte Seite stellt also  eine analytische
Fortsetzung der linken Seite dar. Daf"ur haben wir uns allerdings eine
Mehrdeutigkeit durch die Funktion $(z_1-z_2)^{\gamma_1\gamma_2}$
eingehandelt. Wir werden die folgende Konvention verwenden: F"ur
$0<z_2<z_1$ soll $\log(z_1-z_2)$ reell sein.

Damit haben wir eine M"oglichkeit gefunden, Produkte von
Vertex--Operatoren wieder durch Vertex--Operatoren darzustellen. Wir
k"onnen also das Produkt einfach durch die rechte Seite von (\ref{322})
bzw. (\ref{3221}) definieren. F"ur die auf der rechten Seite stehenden
Operatoren ist Satz \ref{dicht} anwendbar (falls er es f"ur die Faktoren
war), und wir erhalten wieder dicht definierte Operatoren. Wir werden
aber zeigen, da\3 (\ref{322}) und (\ref{3221}) mehr als formale
Relationen sind, eine Anwendung von Lemma \ref{Identitaet} wird zeigen,
da\3
diese Gleichungen sogar stark auf $\cal F$ gelten.
\begin{satz}\label{produkt} Sei
$\sum_{i=1}^{\infty}|\omega_i^{r}|^2<\infty$ (r=1,2), und weiter
existiere
$\sum_{i=1}^{\infty} \omega_{-i}^1 \omega_i^2$. Dann gilt
\begin{itemize}
\item[(i)]
\ben
V(\omega^2)({\cal F})\subset D(V(\omega^1))\een
 und
\item[(ii)]
\ben
V(\omega^1)V(\omega^2)\restrict{\cal F} =\exp(\sum_{i=1}^{\infty}
\omega_{-i}^1 \omega_i^2) V(\omega^1+\omega^2)\restrict{\cal F}.
\een
\end{itemize}
\end{satz}
{\bf Beweis.}\\
Nehmen wir an, wir  k"onnten zeigen, da\3 unter den Voraussetzungen des
Satzes
\ben\label{toshow}
\sum_{\mu} v_{\eta,\mu}(\omega^1) v_{\mu,\nu}(\omega^2)=\exp\left(
\sum_{i=1}^{\infty} \omega_{-i}^1 \omega_i^2 \right)
v_{\eta,\mu}(\omega^1 + \omega^2)
\een
gilt. (Dabei und im folgenden soll $\sum_{\mu}$ immer die Summe "uber alle
Indizes endlicher L"ange bedeuten.) Dann folgt mit $\Psi_{\nu}= \sum_{\mu}
v_{\mu,\nu}(\omega^2) \Phi_{\mu}$, da\3 $\Psi_{\nu}\in D(V(\omega^1))$
gilt, denn aus (\ref{toshow}) folgt schnell, da\3 $\Psi_{\nu}$ die
Bedingungen in (\ref{10}) erf"ullt, und weiter folgt aus (\ref{11})
{\mathindent8mm\bea
V(\omega^1)V(\omega^2)\Phi_{\nu} &=& V(\omega^1) \Psi_{\nu}=\sum_{\eta}
\sum_{\mu} v_{\eta,\mu}(\omega^1) \langle \Phi_{\mu},\Psi_{\nu}\rangle
\nonumber\\
&=&\sum_{\eta}
\sum_{\mu}  v_{\eta,\mu}(\omega^1) v_{\mu,\nu}(\omega^2)=
\exp(-\sum_{i=1}^{\infty} \omega_{-i}^1 \omega_i^2)\sum_{\eta}
v_{\eta,\nu} (\omega^1+\omega^2)\nonumber\\
&=& \exp(-\sum_{i=1}^{\infty} \omega_{-i}^1 \omega_i^2) V(\omega^1 +
\omega^2) \Phi_{\nu}.
\eea
Es reicht also, (\ref{toshow}) zu zeigen. Wir haben
\ben\label{327}
\sum_{\mu} v_{\eta,\mu}(\omega^1) v_{\mu,\nu}(\omega^2)=
\frac{1}{\sqrt{\eta!\nu!}} \sum_{\mu} \prod_{i=1}^{\infty} \left\{
\frac{1}{\mu_i!} m_{\eta_i,-\mu_i}(\omega_i^1,\omega_{-i}^1)
m_{\mu_i,-\nu_i} (\omega_i^2,\omega_{-i}^2) \right\}.
\een}
Wir m"ochten nun in (\ref{327}) die Summe und das Produkt vertauschen. Als
Absch"atzung haben wir diesen Schritt bereits im Beweis von Satz
\ref{dicht} durchgef"uhrt. Diese Vertauschung entspricht der Gleichung
\ben\label{328}
\sum_{n_1,\ldots,n_N=0}^{\infty} \prod_{i=1}^{N} X_{i,n_i}=
\prod_{i=1}^{N} \left(\sum_{n_i=0}^{\infty} X_{i,n_i}\right)
\een
f"ur $N\to \infty$. Im Grenzfall stehen aber auf der linken Seite von
(\ref{328}) weniger Summanden als auf der rechten Seite,  da wir nur "uber
Multiindizes endlicher L"ange summieren. Da $\eta$ und $\nu$ fest sind,
gilt ab einem $i_0$ f"ur alle $i\ge i_0$ $\eta_i=\nu_i=0$. Dann ist
$m_{\eta_i,\mu_i}(\omega_i^1,-\omega_{-i}^1) m_{\mu_i,\nu_i}
(\omega_i^2,-\omega_{-i}^2)=(-\omega_{-i}^1\omega_i^2)^{\mu_i} =:
X_i^{\mu_i}$.
F"ur die endlich vielen Faktoren $i<i_0$ k"onnen wir (\ref{328}) anwenden
und es bleibt zu zeigen, da\3
\ben\label{329}
\sum_{\mu} \prod_{i=1}^{\infty} \frac{X_i^{\mu_i}}{\mu_i!} =
\prod_{i=1}^{\infty} \sum_{\mu_i=0}^{\infty} \frac{X_i^{\mu_i}}{\mu_i!} =
\prod_{i=1}^{\infty} \exp(X_i)
\een
gilt.
Wegen $\sum_{i=1}^{\infty} X_i<\infty$ folgt $\prod_{i=1}^{\infty} X_i=0$,
also haben wir f"ur jeden Multiindex $\mu$ mit $\|\mu\| = \infty$, da\3
$\prod_{i=1}^{\infty}\frac{X_i^{\mu_i}}{\mu_i!} =0$. Also ist
\bea
\sum_{\mu:\|\mu\|<\infty} \prod_{i=1}^{\infty}
\frac{X_i^{\mu_i}}{\mu_i!}&=& \sum_{\mu: \|\mu\|\le\infty}
\prod_{i=1}^{\infty} \frac{X_i^{\mu_i}}{\mu_i!}= \lim_{N\to\infty}
\sum_{\mu_1,\ldots,\mu_N=0}^{\infty} \prod_{i=1}^{N
}
\frac{X_i^{\mu_i}}{\mu_i!}
\nonumber\\
&=&\lim_{N\to\infty}\prod_{i=1}^{N
}\sum_{\mu_i}^{\infty}\frac{X_i^{\mu_i}}{\mu_i!}= \prod_{i=1}^{\infty}
\exp(X_i).
\eea
Wir k"onnen also in (\ref{327}) weiterschreiben und erhalten unter
Anwendung von Lemma \ref{Identitaet}
\bea
\sum_{\mu}v_{\eta,\mu}(\omega^1)v_{\mu,\nu}(\omega^2)&=&
\frac{1}{\sqrt{\eta!\nu!}} \prod_{i=1}^{\infty} \left\{
\sum_{\mu_i=0}^{\infty} \frac{1}{\mu_i!} m_{\eta_i,-\mu_i}(\omega_i^1,
\omega_{-i}^1) m_{\mu_i,\nu_i} (\omega_i^2,-\omega_{-i}^2) \right\}
\nonumber\\
&=& \frac{1}{\sqrt{\eta!\nu!}} \prod_{i=1}^{\infty} \Bigg\{
 m_{\eta_i,\nu_i}(\omega_i^1+ \omega_{i}^2,-\omega_{-i}^1-\omega_{-i}^2)
\exp(-\omega_{-i}^1\omega_i^2)\Bigg\} \nonumber\\
&=& \exp(-\sum_{i=1}^{\infty} \omega_{-i}^1\omega_i^2)
v_{\eta,\nu}(\omega^1 +\omega^2).
\eea
Satz \ref{produkt} ist bewiesen.\hfill$\Box$\\[3mm]
Eine triviale Verallgemeinerung von Satz \ref{produkt} erlaubt es,
beliebige Produkte von Vertex--Operatoren zu bilden, wir erhalten dann
\ben\label{331}
V(\omega^1)V(\omega^2)\cdots V(\omega^r)=\prod_{1\le i<j\le r}
\exp(-\sum_{l=1}^{\infty}\omega_{-l}^i\omega_l^j) V(\omega^1+\omega^2 +
\ldots \omega^r).
\een
Diese  Gleichung gilt ebenfalls stark auf $\cal F$. F"ur das Produkt der
Operatoren $V(\gamma_i,z_i)$ erhalten wir aus Satz \ref{produkt} unter der
Voraussetzung $|z_r|<|z_{r-1}|<\ldots<|z_1|<1$
{\small\mathindent0mm\bea\label{332}
\lefteqn{V(\gamma_1,z_1)V(\gamma_2,z_2)\cdots V(\gamma_r,z_r)=}\nonumber\\
&=& \prod_{1\le i<j\le r}(1-\frac{z_j}{z_i})^{\gamma_i\gamma_j}  \exp
\left( \sum_{n=1}^{\infty} \frac{\sum_{i=1}^r \gamma_i z_i^n}{n}
a_{-n}\right)T_{\Sigma \gamma_i}\prod_{i=1}^r z_i^{\gamma_i a_0} \exp
\left( \sum_{n=1}^{\infty} \frac{\sum_{i=1}^r \gamma_i z_i^{-n} }{n}
a_{n}\right)\nonumber\\
&=:&\prod_{1\le i<j\le r} (1-\frac{z_j}{z_i})^{\gamma_i\gamma_j}
V(\gamma_1,\ldots,\gamma_r;z_1,\ldots,z_r),
\eea}
(vgl. (\ref{3221}), nat"urlich gilt auch eine Verallgemeinerung von
(\ref{3222})). Auch diese Gleichung gilt unter den
angegebenen Bedingungen stark auf $\cal F$.
\section{Kommutatoren}
Wir haben bis jetzt gesehen, da\3 Vertex--Operatoren unter bestimmten
Bedingungen dicht definierte Operatoren sind. Zur Definition der
Operatoren haben wir eine formale Kommutatorrelation zwischen
Vertex--Operatoren und der Heisenbergalgebra verwendet (siehe
(\ref{07})). Wir m"ochten nun als erstes beweisen, da\3 die
Vertex--Operatoren  Gl. (\ref{07}) stark auf ${\cal F}$ erf"ullen.
\begin{lem}\label{hdef}Sei $\sum_{i=1}^{\infty} |\omega_i|^2<\infty$.
Dann gilt:
\begin{itemize}
\item[(i)]
F"ur jedes Monom $P(a_{i_1}, \ldots , a_{i_k})$
ist $V(\omega)({\cal F}) \subset D(P(a_{i_1}, \ldots ,a_{i_k}))$.
\item [(ii)]
F"ur alle $\Phi \in {\cal F} $ und $k > 0$
\bea\label{33}
a_k V(\omega) \Phi  &=& V(\omega)(a_k + \sqrt{k} \omega_k) \Phi,
\nonumber\\
a_{-k} V(\omega) \Phi &=& V(\omega) (a_{-k} + \sqrt{k} \omega_{-k}) \Phi.
\eea
\item[(iii)]  $\| a_{\pm k} V(\omega) \Phi_{\nu}\| \le C(\nu) k^{1/2}$
 \quad f"ur $k>0$.
\end{itemize}
\end{lem}
{\bf Beweis.}\\
(ii):
Da wir zur formalen Berechnung der Matrixelemente von $V(\omega)$
(\ref{07}) bzw.
(\ref{33}) verwendet haben, ist klar, da\3 (\ref{33}) schwach auf ${\cal F}
\times {\cal F}$ gilt. Damit reicht es (i) zu beweisen.\\
(i):
Wir zeigen $V(\omega) ({\cal F}) \subset D(a_k)$ f"ur beliebige $k>0$. (i)
ist dann eine offensichtliche Verallgemeinerung. Es gilt f"ur $k>0$
{\small\mathindent0mm\bea\label{34}
||a_k V(\omega) \Phi_{\nu}||^2 &=& \sum_{\mu} |\langle \Phi_{\mu}, a_k
V(\omega) \Phi_{\nu} \rangle |^2
= \sum_{\mu} | \langle a_{-k} \Phi_{\mu}, V(\omega) \Phi_{\nu} \rangle|^2
\nonumber\\
&=& \sum_{\mu} k(\mu_k +1) |\langle \Phi_{\mu +e_k}, V(\omega)
\Phi_{\nu}\rangle |^2
 \leq k \sum_{\mu} (\mu_k +2) |\langle \Phi_{\mu}, V(\omega)
\Phi_{\nu}\rangle |^2 \nonumber\\
& \leq & \frac{k}{\nu !} \prod_{i \not= k} \left(
\sum_{\mu_i=0}^{\infty}\frac{1}{\mu_i} | m_{\mu_i , \nu_i} (\omega) |^2
\right) \left(
\sum_{\mu_k=0}^{\infty} \frac{1}{\mu_k !} (\mu_k+2) | m_{\mu_k , \nu_k}
(\omega) |^2 \right).
\eea}
Eine einfache "Uberlegung zeigt
\ben\label{35}
|m_{i,j} (x,y)|^2 \leq C(j) i^2 (\max (1,x))^{2i},
\een
so da\3 die Summe "uber $\mu_k$ existiert. Die "ubrigen Faktoren sind genau
die aus dem Beweis von Satz \ref{dicht}. Es folgt (da $a_k$ abgeschlossen
ist) $V(\omega) \Phi_{\nu} \in D(a_k)$. Der Beweis f"ur $a_{-k}$ geht
analog.\\
(iii) folgt unmittelbar aus (\ref{34}), das "ubrigbleibende Produkt "uber
$i\ne k$ liefert die $\nu$-abh"angige Konstante. \hfill$\Box$\\[3mm]
Wir wollen uns nun die Virasoro--Operatoren $L_n$ etwas genauer anschauen.

Wir waren bis jetzt damit zufrieden, da\3 $L_n$ auf ${\cal F}$ definiert
sind. Da wir aber Kommutatoren $[L_n,V(\omega)]$ berechnen wollen und die
Bilder $V(\omega)({\cal F})$ sicher nicht mehr in $\cal F$ liegen, m"ussen
wir den Definitionsbereich von $L_n$ vergr"o\3ern. Zun"achst ist wegen dem
folgenden Lemma klar, da\3 $L_n$ abschlie\3bare Operatoren sind.
\begin{lem}\label{seppel}
In einem separablen Hilbertraum ist die direkte Summe von
endlichdimensionalen Operatoren abschlie\3bar.
\end{lem}
{\bf Beweis.} Klar.\hfill$\Box$\\[3mm]
$L_n$ k"onnen wir als direkte Summe schreiben, da $L_n : {\cal F}_k \lra
{\cal F}_{k-n}$, es gilt $L_n=\oplus_{k=0}^{\infty} L_n|_{{\cal F}_k}$.
Ab sofort seien also $L_n(\alpha,\beta)$ als der Abschlu\3 von
$L_n(\alpha,\beta)|_{\cal F}$ definiert. Der Defini\-tions\-bereich der
$L_n$ ist gen"ugend gro\3:
\begin{lem}
\begin{itemize}
\item[(i)] $D(L_0)=\left\{ \Phi=\sum_{\mu} c_{\mu} \Phi_{\mu}\; : \;
\sum_{\mu}
|c_{\mu}|^2 \|\mu\|^2 < \infty \right\}.$
\item[(ii)] $D(L_0)\subset D(L_n)$ f"ur alle $n\in\ZZ$.
\item[(iii)] Ist $\sum_{k=1}^{\infty}k^2 |\omega_k|^2 < \infty$, so gilt
$V(\omega)({\cal F}) \subset D(L_0).$
\end{itemize}
\end{lem}
{\bf Beweis.}\\
(i) ist offensichtlich, da $L_0\Phi_{\mu} = \|\mu\| \Phi_{\mu}$.\\
(ii): Wir schreiben $L_n$ f"ur $n\ne 0$ in der Wick--Ordnung, dann gilt
f"ur $n>0$
\ben\label{301}
L_n(\alpha,\beta)=\sum_{k=1}^{\infty} a_{-k} a_{n+k} +\frac{1}{2}
\sum_{k=1}^{n-1}a_k a_{n-k} + (\alpha+n\beta)a_n
\een
und
\ben\label{302}
L_{-n}(\alpha,\beta)=\sum_{k=1}^{\infty} a_{-n-k} a_{k} +\frac{1}{2}
\sum_{k=1}^{n-1}a_{-k} a_{-n+k} + (\alpha-n\beta)a_{-n}.
\een
Die endlichen Summen in (\ref{301}) und (\ref{302}) sind
auf $D(L_0)$ definiert, da $D(L_0)\subset D(a_{\pm k})$ gilt, es reicht
 also, jeweils die Anteile
\ben\label{303}
\tilde{L}_n=\sum_{k=1}^{\infty} a_{-k}a_{n+k} \mbox{ und } \tilde{L}_{-n}
= \sum_{k=1}^{\infty} a_{-n-k} a_k
\een
zu untersuchen. Es gilt mit $N(\mu)=\max \{ i:\mu_i\ne 0\} $ und f"ur $n>0$
\ben\label{304}
\tilde{L}_n \Phi_{\mu} = \sum_{k=1}^{N(\mu)-n} a_{-k} a_{n+k} \Phi_{\mu}
=\sum_{k=1}^{N(\mu)-n} \sqrt{ k (\mu_k+1) (n+k) \mu_{n+k}}
\Phi_{\mu-e_{n+k}+e_k}
\een
und deshalb
\bea\label{305}
\| \tilde{L}_n \Phi_{\mu}\|^2 &=& \sum_{k=1}^{N(\mu)-n}
k(\mu_k+1)(n+k)\mu_{n+k}\nonumber\\
&=& \sum_{k=1}^{N(\mu)-n} k \mu_k (n+k)\mu_{n+k}
+k(n+k)\mu_{n+k}\nonumber\\
&\le& \sum_{k=1}^{N(\mu)} k^2 \mu_k^2 + k^2 \mu_k \le 2 \|\mu\|^2.
\eea
Analog k"onnen wir $\|\tilde{L}_{-n} \Phi_{\mu}\|$  durch
\ben\label{306}
\|\tilde{L}_{-n} \Phi_{\mu}\|^2 \le (n+2) \|\mu\|^2
\een
absch"atzen. Damit folgt, da\3 $\tilde{L}_n$ und damit auch $L_n$ auf
$D(L_0)$ definiert sind.\\
(iii): Zun"achst zeigen wir, da\3 $V(\omega)\Phi_0\in D(L_0)$ gilt. Aus der
Form der Matrixelemente von $V(\omega)$ (\ref{16}) folgt, da\3
\ben\label{307}
V(\omega)\Phi_0=\sum_{\mu} \bigg(\prod_{i=1}^{\infty}
\frac{\omega_i^{\mu_i}}{\sqrt{\mu_i!}}\bigg) \Phi_{\mu} =: \sum_{\mu}
v_{\mu} \Phi_{\mu}
\een
gilt. Wir m"ussen also $\sum_{\mu} |v_{\mu}|^2 \|\mu\|^2$ absch"atzen.\\
Wegen $(\sum_{k=1}^{\infty} k \mu_k)^2 \le 2 \prod_{k=1}^{\infty} (k^2
\mu_k^2+1)$ folgt
\bea\label{308}
\sum_{\mu} | v_{\mu}|^2 (\sum_{k=1}^{\infty} k \mu_k)^2 &\le& 2
\sum_{\mu}\prod_{k=1}^{\infty}
\frac{|\omega_k|^{2\mu_k}}{\mu_k!}(1+\mu_k^2
k^2)\nonumber\\
&=& 2 \prod_{k=1}^{\infty}
\sum_{\mu_k=0}^{\infty}\frac{|\omega_k|^{2\mu_k}}{\mu_k!}(1+\mu_k^2
k^2)\nonumber\\
&=& 2 \prod_{k=1}^{\infty}(1+k^2 |\omega_k|^2 + k^2 | \omega_k|^4)
\exp(|\omega_k|^2) <\infty,
\eea
denn das Produkt existiert nach Voraussetzung. Wir wissen also, da\3
$V(\omega)\Phi_0\in D(L_0)$ gilt. Es reicht nun zu zeigen, da\3 f"ur jedes
Monom $P(a_{-1},\ldots,a_{-l})$ gilt $P(a_{-1},\ldots,a_{-l})
V(\omega)\Phi_0\in D(L_0)$, denn wegen (\ref{33}) k"onnen wir
$V(\omega)\Phi_{\mu}$ als Linearkombination solcher Ausdr"ucke schreiben.
Das folgt aber schnell, nach Lemma \ref{hdef} ist  z.B.
\ben\label{309}
a_{-k} V(\omega)\Phi_0=a_{-k} \sum_{\nu} v_{\nu} \Phi_{\nu}=\sum_{\nu}
v_{\nu} \sqrt{k(\nu_k+1)} \Phi_{\nu+e_k}.
\een
Analog k"onnen wir die Wirkung beliebiger Monome auf $V(\omega)\Phi_0$
hinschreiben. Mit den Koeffizienten aus (\ref{309}) k"onnen wir auf die
gleiche Weise wie in (\ref{308}) verfahren, nur endlich viele der
 Faktoren aus (\ref{307}) werden gest"ort, die Konvergenz des Produktes
bleibt dabei unber"uhrt. \hfill$\Box$\\[3mm]
Wir kommen nun zum wichtigsten Punkt in diesem Abschnitt, den
Kommutatoren zwischen den Virasoro--Operatoren und den
Vertex--Operatoren. Zun"achst gilt ganz allgemein:
\begin{lem}\label{kom01} Sei $\sum_{k=1}^{\infty} k |\omega_k| < \infty$.
Setze $\theta_{\pm k}
=\sqrt{k} \omega_{\pm k}$, $\theta_0=\gamma$. Dann gilt stark auf $\cal
F$:
\bea\label{k001}
\lefteqn{L_n(\alpha+\gamma,\beta) V(\omega) - V(\omega)L_n(\alpha,\beta)=
\sum_{k=1}^{\infty} \theta_{n+k} a_{-k} V(\omega) + \sum_{k=1}^{\infty}
\theta_{n-k} V(\omega)a_k} \nonumber\\
&&+ \left(\frac{1}{2} \sum_{k=1}^{n-1} \theta_k \theta_{n-k} +
\theta_n(\alpha+\gamma+n\beta)\right) V(\omega).
\eea
\end{lem}
{\bf Beweis.}\\
Sei $n > 0$. Weiter sei  $L_n^N(\alpha,\beta) = \sum_{k=1}^{N} a_{-k}
a_{n+k} +\frac{1}{2} \sum_{k=1}^{n-1} a_k a_{n-k} +(\alpha+ n\beta)a_n$.
Aus Lemma \ref{hdef} folgt stark auf $\cal F$
{\mathindent0mm\bea\label{h001}
L^N_n (\alpha+\gamma,\beta)V(\omega) &=& \sum_{k=1}^N a_{-k} V(\omega)
(a_{n+k} + \theta_{n+k})+ \frac{1}{2} \sum^{n-1}_{k=1} V(\omega)(a_k
+\theta_k)(a_{n-k}+\theta_{n-k})
\nonumber\\&&+
(\alpha+\gamma+n\beta)V(\omega)(a_n+ \theta_n)
\eea}
und
{\mathindent0mm\bea\label{h002}
V(\omega) L^N_n(\alpha,\beta) &=& \sum^N_{k=1} (a_{-k}
-\theta_{-k})V(\omega)a_{n+k}
+\frac{1}{2} \sum_{k=1}^{n-1} V(\omega)a_k a_{n-k}\nonumber\\
&&+ (\alpha+n\beta) V(\omega)a_n.
\eea}
F"ur den Kommutator, also die Differenz von (\ref{h001}) und (\ref{h002})
ergibt sich
{\mathindent0mm\bea
[\qquad] &=& \sum^N_{k=1} a_{-k} \theta_{n+k} V(\omega) + \frac{1}{2}
\sum_{k=1}^{n-1} V(\omega) a_k \theta_{n-k} + \frac{1}{2} \sum^{n-1}_{k=1}
V(\omega)a_{n-k} \theta_k +
\nonumber\\
&&+ \gamma V(\omega) a_n + \theta_n (\alpha +\gamma + n\beta) V(\omega) +
\sum_{k=1}^N \theta_{-k} V(\omega) a_{n+k}
\nonumber\\
&=& \sum_{k=1}^N \theta_{n+k} a_{-k} V(\omega) + \sum_{k=1}^{n-1}
\theta_{n-k} V(\omega) a_k + \sum_{k=n+1}^{N+n} \theta_{n-k} V(\omega)
a_k\nonumber\\
&&+ \gamma V(\omega) a_n + \theta_n ( \alpha + \gamma + n\beta)
V(\omega).
\eea}
Zu zeigen bleibt nur die Konvergenz des ersten Terms. F"ur $\Phi \in {\cal
F}$ gilt nach Lemma \ref{hdef}, (iii)
\[ \left\|\sum_{k=1}^N \theta_{n+k} a_{-k} V(\omega) \Phi\right\| \leq C
\sum_{k=1}^N \sqrt{n+k} \sqrt{k} |\omega_k| < C_1(n)<\infty\]
f"ur beliebige $n\in \NN$, falls $ \sum_{k=1}^{\infty} k |\omega_k| <
 \infty$ gilt. Die F"alle $n=0$ bzw. $n<0$ gehen analog unter Verwendung
von (\ref{fockrep}) bzw. (\ref{302}). \hfill$\Box$\\[3mm]
Nun werden wir zum ersten Male sehen, wieso die nichtverschmierten
Operatoren f"ur uns interessanter sind. F"ur $V(\gamma,z)$ bzw.
$V(\gamma_1,\ldots,\gamma_r;z_1,\ldots,\gamma_r)$ gilt:
\begin{lem}\label{k002}Sei $|z_i|<1 \; (i=1,\ldots ,r)$. Dann ist f"ur
$\Phi \in {\cal F}$ $V(\gamma_1,\ldots,\gamma_r;z_1,\ldots,z_r)\Phi$
holomorph in $z_1,\ldots,z_r$ und es gilt
{\mathindent0mm\bea\label{w01}
z_j \frac{\partial}{\partial z_j} V(\gamma_1, \ldots ,\gamma_r ; z_1,
 \ldots ,z_r)\Phi &=& \gamma_j \sum_{k=1}^{\infty} z_j^k a_{-k}
V(\gamma_1,\ldots,\gamma_r;z_1,\ldots,z_r) \Phi \nonumber\\
&&+ \gamma_j \sum_{k=1}^{\infty}z^{-k}_j
V(\gamma_1,\ldots,\gamma_r,z_1,\ldots,z_r) a_k \Phi.
\eea}
\end{lem}
F"ur die Aussage von Lemma \ref{k002} ben"otigen wir wesentlich die Form von
$\theta_k=C z^k$, wodurch $V(\omega)$ von der Form $V(\gamma,z)$ sein
mu\3.\\
{\bf Beweis.}\\
Wir beweisen nur den Fall $r=1$, die Verallgemeinerung auf beliebiges
 $r>1$ ist offensichtlich.
Sei $F(z):= V( \gamma,z) \Phi_{\nu}$. Nach Satz \ref{dicht},
(\ref{dichtbound}) ist $\|F(z)\|$ lokal gleichm"a\3ig beschr"ankt in dem
gelochten Einheitskreis. Deshalb reicht es, die  Differenzierbarkeit von
$\langle \Phi,F(z)\rangle$ f"ur $\Phi\in \cal F$ nachzuweisen, denn
f"ur $\Phi=\sum_{\mu} c_{\mu}\Phi_{\mu} \in {\cal H}$ gilt dann
\begin{eqnarray*}
\left|\sum_{\mu} \overline{c}_{\mu} \langle \Phi_{\mu}, F(z) \rangle -
\sum_{\|\mu\| \leq N} \overline{c}_{\mu} \langle \Phi_{\mu} , F(z) \rangle
\right|
&=& \left| \sum_{\| \mu \| >N} \overline{c}_{\mu} \langle \Phi_{\mu}, F(z)
\rangle \right| \\
 \leq  \left( \sum_{\| \mu \| >N} |c_{\mu}|^2 \sum_{\mu} \left| \langle
\Phi_{\mu},F(z) \rangle \right| \right)^{\frac{1}{2}}
& \leq & \left( \sum_{\| \mu \| >N} | c_{\mu}|^2 \right)^{\frac{1}{2}}
\left\| F(z) \right\| \stackrel{N\to\infty}{\strich\lra} 0
\end{eqnarray*}
lokal gleichm"a\3ig. $F(z)$ ist dann folglich schwach holomorph und damit
auch (stark) holomorph. Es reicht demnach, (\ref{w01}) schwach auf
${\cal F} \times {\cal F}$ nachzuweisen, da\3 hei\3t f"ur die Matrixelemente
von $V(\gamma ,z)$. Wir erhalten aus (\ref{08})
{\mathindent0mm\begin{eqnarray*}
\lefteqn{z\frac{\partial}{\partial z} \langle \Phi_{\mu} ,V(\gamma,z)
\Phi_{\nu}\rangle
= z\frac{\partial}{\partial z} \frac{1}{\sqrt{\mu ! \nu!}}
\prod_{i=1}^{\infty} m_{\mu_i ,\nu_i} \left( \gamma \frac{z^i}{\sqrt{i}},-
\gamma \frac{z^{-i}}{\sqrt{i}}\right)}\\
&=& \sum_{k=1}^{\infty} \left(\prod_{i\neq k} \frac{1}{\sqrt{\mu_i ! \nu_i
!}} m_{\mu_i \nu_i} \left( \gamma \frac{z^i}{\sqrt{i}},-\gamma
\frac{z^{-i}}{\sqrt{i}} \right)\right)\left( z\frac{\partial}{\partial z}
\frac{1}{\sqrt{\mu_k ! \nu_k !}} m_{\mu_k, \nu_k} \left( \gamma
\frac{z^k}{\sqrt{k}}, -\gamma \frac{z^{-k}}{\sqrt{k}} \right)\right).
\end{eqnarray*}}
Wir haben dabei die gew"ohnliche Produktregel angewendet, denn alle
auftretenden Produkte und Summen brechen nach endlich vielen Termen ab.
Durch eine einfache Rechnung erh"alt man die folgende Identit"at:
{\mathindent0mm\small\begin{eqnarray*}
\lefteqn{z\frac{\partial}{\partial z}\frac{1}{\sqrt{\mu_k! \nu_k!}}
m_{\mu_k,\nu_k}\left(\gamma\frac{z^k}{\sqrt{k}},-\gamma\frac{z^{-k}}
{\sqrt{k}}  \right)}\\
&=&
z \frac{\partial}{\partial z} \frac{1}{\sqrt{\mu_k! \nu_k!}}
 \sum_{j=0}^{\min (\mu_k, \nu_k)} { \mu_k \choose j} {\nu_k \choose j} j!
\left( \gamma \frac{z^k}{\sqrt{k}} \right)^{\mu_k -j} \left( -\gamma
\frac{z^{-k}}{\sqrt{k}} \right)^{v_k -j}\\
&=&\frac{\gamma z^k \sqrt{k \mu_k}}{\sqrt{\left(\mu_k -1\right)! v_k!}}
m_{\mu_k -1, v_k} \left( \gamma \frac{z^k}{\sqrt{k}}, -\gamma
\frac{z^{-k}}{\sqrt{k}} \right) + \frac{\gamma z^{-k}\sqrt{k
\nu_k}}{\sqrt{\mu_k!\left( \nu_k-1\right)!}} m_{\mu_k, \nu_{k-1}}\left(
\gamma \frac{z^k}{\sqrt{k}},-\gamma \frac{z^{-k}}{\sqrt{k}}\right).
\end{eqnarray*}}
Es folgt also insgesamt
{\mathindent0mm\begin{eqnarray*}
\lefteqn{z \frac{\partial}{\partial z} \left\langle \Phi_{\mu}, V\left(
\gamma,z\right) \Phi_{\nu}\right\rangle}\\
&=& \sum_{k=1}^{\infty}\gamma z^k \sqrt{k \mu_k} \left\langle \Phi_{\mu
-e_k} ,V\left( \gamma,z \right) \Phi_{\nu} \right\rangle
+ \sum_{k=1}^{\infty}\gamma z^{-k} \sqrt{ k \nu_k} \left\langle
\Phi_{\mu}, V \left( \gamma,z \right)\Phi_{\nu -e_k} \right\rangle\\
&=& \sum_{k=1}^{\infty} \gamma z^k \left\langle \Phi_{\mu},a_{-k} v\left(
\gamma,z \right) \Phi_{\nu}\right\rangle + \sum_{k=1}^{\infty}\gamma
z^{-k} \left\langle \Phi_{\mu}, V\left( \gamma,z \right) a_k
\Phi_{\nu}\right\rangle.
\end{eqnarray*}}
\hfill$\Box$\\[3mm]
Damit erhalten wir f"ur $V(\gamma ,z)$ folgende sch"one Form des Kommutators
mit $L_n$:
\begin{kor} Es gilt f"ur $\Phi \in {\cal F}$:
\bea\label{k005}
\lefteqn{L_n(\alpha+\gamma,\beta) V(\gamma,z)\Phi
-V(\gamma,z)L_n(\alpha,\beta)\Phi \equiv [L_n,V(\gamma,z)]
\Phi}\nonumber\\
&=& z^n \left\{ z \frac{\partial}{\partial z} + n(\frac{\gamma^2}{2} +
\beta \gamma) +\gamma\alpha + \frac{1}{2}\gamma^2 \right\} V(\gamma,z)
\Phi.
\eea
\end{kor}
{\bf Beweis.}\\
Aus Lemma \ref{kom01} und Lemma \ref{k002} folgt mit $\theta_{\pm k}=
\sqrt{k} \omega_{\pm k} =\gamma z^{\pm k}$
\begin{eqnarray*}
[L_n,V(\gamma,z)]&=& z^n \sum_{k=1}^{\infty} \gamma z^k a_{-k} V(\gamma,z)
\Phi + z^n \sum_{k=1}^{\infty} \gamma z^{-k} V(\gamma,z)a_k \Phi  \\
&&+
\left( \frac{n-1}{2}\gamma^2 z^n + \gamma^2 z^n + \gamma n \beta
z^n+\gamma\alpha z^n\right) V(\gamma,z) \Phi\\
&=& z^n\left\{  z \frac{\partial}{\partial z} + n(\frac{\gamma^2}{2} +
\beta \gamma) +\gamma\alpha + \frac{1}{2}\gamma^2 \right\} V(\gamma,z)
\Phi.
\end{eqnarray*}
\hspace*{\fill}$\Box$\\
Eine besonders einfache Form hat der Kommutator zwischen
$T_{\gamma}z^{\gamma(\alpha-\beta)} V(\gamma,z)$ und $L_n$, wir erhalten
dann:
\begin{kor}\label{primfeld} Es gilt stark auf $\cal F$
\ben\label{konglei}
[L_n,T_{\gamma} z^{\gamma(\alpha-\beta)} V(\gamma,z)]= z^n \left\{z
\frac{\partial}{\partial z} + (n+1)(\frac{\gamma^2}{2} + \beta \gamma)
\right\} T_{\gamma}z^{\gamma(\alpha-\beta)}V(\gamma,z).
\een
\end{kor}
{\bf Beweis.}\\
$z\frac{\partial}{\partial z} T_{\gamma}z^{\gamma(\alpha-\beta)}
V(\gamma,z)$ enth"alt nach der Produktregel den zus"atzlichen Term
$\gamma(\alpha-\beta) T_{\gamma}z^{\gamma(\alpha-\beta)} V(\gamma,z)$ der
(\ref{k005}) entsprechend modifiziert, es gilt
\[ [L_n,V]=z^n XV \Longleftrightarrow [L_n , z^{\lambda} V]=
z^n(X-\lambda)z^{\lambda} V.
\]
\hspace*{\fill}$\Box$
\begin{Def} Ein Quantenfeld $F(z)$ (eine operatorwertige Funktion), das
 f"ur ein $\kappa\in \CC$
\[ [L_n, F(z)] =z^n \left\{ z \frac{\partial}{\partial z} + (n+1)\kappa
\right\} F(z)
\]
f"ur alle  $n\in \ZZ$  erf"ullt, hei\3t konformes Feld vom Gewicht $\kappa$
\end{Def}
$T_{\gamma} z^{\gamma(\alpha-\beta)} V(\gamma,z)$ ist also ein konformes
Feld vom Gewicht $\kappa=\gamma^2/2+\beta \gamma$. Man bemerke, da\3 wir
im Fall $\kappa=1$  den Kommutator als eine Ableitung schreiben k"onnen,
es gilt dann $[L_n,F(z)]=\frac{\partial}{\partial z} \left(z^{n+1}
F(z)\right)$. Diese Gleichung wird f"ur die integrierten (``abgeschirmten'')
Vertex--Operatoren sehr wichtig werden.\\
Wir m"ochten nun Kommutatoren zwischen der Virasoro--Algebra und Produkten
von Vertex--Operatoren angeben. Wir verwenden dabei die folgende Form des
Vorfaktors von $V\left( \gamma_1, \ldots ,\gamma_r ; z_1, \ldots , z_r
\right)$: Sei
\[
F_{\lambda}\left(\gamma_1,\ldots ,\gamma_r;z_1, \ldots, z_r \right) =
\prod_{1\leq i<j\leq r}\left( z_i-z_j\right)^{\gamma_i \gamma_j}
\prod_{i=1}^r z_i^{\lambda \gamma_i}.
\]
Dabei w"ahlen wir den Zweig von $F_{\lambda}$ aus, der durch folgende
Bedingung festgelegt ist:\\
F"ur $0<z_r<z_{r-1}< \ldots <z_1$ sollen alle Logarithmen $\log (z_i-z_j)
\: (i<j)$ reell sein. Dieser Bereich ist genau dann der, auf dem
$V(\gamma_1,z_1) \cdots V(\gamma_r,z_r)$ existiert. F"ur andere Bereiche
ist $F_{\lambda}$ nat"urlich nur festgelegt, wenn wir den Pfad der
analytischen Fortsetzung angeben.\\
Was uns f"ur den Kommutator zwischen $L_n$ und $F_{\lambda}
V(\gamma_1,\ldots,\gamma_r;z_1,\ldots,\gamma_r)$ noch fehlt, ist die
passende Form von $z_i \frac{\partial}{\partial z_i} F_{\lambda} \left(
\gamma_1, \ldots , \gamma_r ; z_1, \ldots , z_r \right)$. Dazu verwenden
wir das folgende Lemma.
\begin{lem}\label{333}
F"ur $n \in \NN$ gilt
{\mathindent0mm\begin{eqnarray}\label{334}
\lefteqn{
\sum_{k=1}^r z^n_k \left( z_k \frac{\partial}{\partial z_k} +
\frac{\left(n+1\right)}{2} \gamma^2_k - \lambda \gamma_k \right)
F_{\lambda} \left( \gamma_1, \ldots , \gamma_r; z_1, \ldots , z_r \right)
 }\nonumber\\
&= &\frac{1}{2} \sum_{k=0}^n \left( \gamma_1 z^k_1 + \ldots + \gamma_r
z_r^k \right) \left( \gamma_1 z_1^{n-k}+ \ldots + \gamma_r z_r^{n-k}
\right) F_{\lambda} \left( \gamma_1, \ldots , \gamma_r; z_1, \ldots, z_r
\right).
\end{eqnarray}}
\end{lem}
{\bf Beweis.}\\
Nachrechnen.\hfill$\Box$\\[3mm]
Damit k"onnen wir den folgenden Satz beweisen.
\begin{satz} Sei kurz $F_{\lambda}:=F_{\lambda}(\gamma_1,\ldots, \gamma_r;
z_1,\ldots,z_r)$, $T=T_{\gamma_1+\ldots +\gamma_r}$ und $V=V(\gamma_1,
\ldots \gamma_r;z_1,\ldots,z_r)$. Dann gilt stark auf $\cal F$:
{\mathindent0mm\ben\label{335}
[L_n,F_{\lambda} T V]=\sum_{j=1}^r z_j^n\left\{ z_j
\frac{\partial}{\partial z_j} + (n+1)\frac{\gamma_j^2}{2} + n\beta
\gamma_j + \gamma_j(a_0-\Sigma_{j=1}^r\gamma_j-\lambda) \right\}
F_{\lambda} TV.
\een}
\end{satz}
{\bf Beweis.}\\
Aus Lemma \ref{kom01} folgt mit $\theta_k=\gamma_1 z_1^k + \ldots +
\gamma_r z_r^k$ stark auf $\cal F$
{\mathindent0mm\begin{eqnarray*}
[L_n,F_{\lambda}TV] &=& \sum_{k=1}^{\infty}(\gamma_1 z_1^{n+k} + \ldots +
\gamma_r z_r^{n+k} ) a_{-k} F_{\lambda} TV \\
&&+ \sum_{k=1}^{\infty} (\gamma_1 z_1^{n-k} + \ldots + \gamma_r z_r^{n-k}
)F_{\lambda} TV a_k \\
&&+  \frac{1}{2} \sum_{k=1}^{n-1} (\gamma_1 z_1^{k} + \ldots + \gamma_r
z_r^{k} )(\gamma_1 z_1^{n-k} + \ldots + \gamma_r z_r^{n-k} )F_{\lambda} TV
\\
&&+ (\gamma_1 z_1^{n} + \ldots + \gamma_r z_r^{n} )(\alpha+ \sum_{i=1}^r
\gamma_i + n \beta)  F_{\lambda} TV\\
&=& \sum_{j=1}^r z_j^n\left\{  \sum_{k=1}^{\infty} \gamma_j z_j^k a_{-k}
F_{\lambda} TV +  \sum_{k=1}^{\infty} \gamma_j z_j^{-k} F_{\lambda} TV
a_k\right\}\\
&&+ \frac{1}{2} \sum_{k=0}^n (\gamma_1 z_1^{k} + \ldots + \gamma_r z_r^{k}
)(\gamma_1 z_1^{n-k} + \ldots + \gamma_r z_r^{n-k} ) F_{\lambda} TV \\
&&+ (\gamma_1 z_1^{n} + \ldots + \gamma_r z_r^{n} )(\alpha + n \beta)
 F_{\lambda} TV.\\
\lefteqn{\mbox{ Nun k"onnen wir Lemma \ref{k002} und Lemma \ref{333}
anwenden und erhalten}}\\
&=& \sum_{j=1}^r z_j^n F_{\lambda}T (z_j\frac{\partial}{\partial z_j} V) +
\sum_{j=1}^r z_j^n (z_j \frac{\partial}{\partial z_j} F_{\lambda}) TV \\
&&+ \sum_{j=1}^r z_j^n \left\{ \frac{n+1}{2} \gamma_j^2 -\lambda \gamma_j
+ \gamma_j(\alpha+n\beta)\right\} F_{\lambda} TV\\
&=& \sum_{j=1}^r z_j^n \Bigg\{ z_j \frac{\partial}{\partial z_j} +
(n+1)\frac{\gamma_j^2}{2} + n \beta \gamma_j + \gamma_j(a_0-\Sigma
\gamma_i -\beta) \Bigg\} F_{\lambda} TV.
\end{eqnarray*}}
\hspace*{\fill}$\Box$\\[3mm]
Uns werden dabei insbesondere zwei Wahlen von $\lambda$ interessieren:
Setzen wir $\lambda=a_0-\Sigma \gamma_j -\beta=\alpha-\beta$, so erhalten
wir
\bea\label{336} \lefteqn{L_n(\alpha+\Sigma \gamma_i,\beta) F_{\alpha-\beta}
TV - F_{\alpha-\beta} TV L_n(\alpha,\beta) }\nonumber\\
&=& \sum_{j=1}^r z_j^n \left\{z_j\frac{\partial}{\partial z_j}+
(n+1)(\frac{\gamma_j^2}{2} + \beta \gamma_j) \right\} F_{\alpha-\beta} TV,
\eea
und f"ur den Fall $\gamma=\gamma_1=\ldots= \gamma_r$ und
$\lambda=-\frac{r-1}{2}\gamma$ erhalten wir
\bea\label{337}
\lefteqn{L_n(\alpha+r\gamma) F_{-\frac{r-1}{2}\gamma} T V -
F_{-\frac{r-1}{2}\gamma} T V L_n(\alpha,\beta)}\nonumber\\
&=& \sum_{j=1}^r z_j^n \left\{z_j\frac{\partial}{\partial z_j}+ n
(\frac{\gamma^2}{2} + \beta\gamma) +\gamma a_0 - r
\frac{\gamma^2}{2}\right\}F_{-\frac{r-1}{2}\gamma} T V.
\eea
Glei\-chung (\ref{336}) ist eine na\-t"ur\-liche Verallgemeinerung von
Korollar \ref{primfeld} auf $r$ Variable.
\section{Faktorisierung von Vertex--Operatoren}
Wir haben bis jetzt Vertex--Operatoren im Wesentlichen auf $\cal F$
definiert und nicht untersucht, wie gro\3 der Definitionsbereich der
Vertex--Operatoren wirklich ist. Nat"urlich k"onnen wir aus der Produktformel
f"ur Vertex--Operatoren  im Prinzip eine solche Aussage extrahieren,
 denn wir haben
gezeigt, da\3 unter bestimmten Voraussetzungen $V(\omega^1) V(\omega^2)$
auf $\cal F$ existiert, da\3 hei\3t $D(V(\omega^1)) \supset V(\omega)(
{\cal
F})$. Da wir aber die Menge $V(\omega)({\cal F}) $ nicht kennen, ist das
 wenig hilfreich. Wir werden nun unter gewissen zus"atzlichen Bedingungen
eine Faktorisierung von Vertex--Operatoren
$V(\omega) = Bc^{-N}$ beweisen, wobei $B$ ein Hilbert--Schmidt--Operator
ist, $N=\sum_{k>0} a_{-k}a_k$ der Teilchenzahloperator und
$c<1$. Damit sehen wir, da\3 der Definitionsbereich von $V(\omega)$
deutlich gr"o\3er als $\cal F$ ist.
\begin{satz}\label{340}
Sei $|\omega_i|\leq KR^i$ f"ur ein $0<R<1$ und $|\omega_{-i}| \leq KS^i$
f"ur ein $S>0$ und $K>1$. Weiter sei
$0<c<\left(1+K^2\left(R+S\right)^2\right)^{-\frac{1}{2}}$. Dann ist
\ben\label{G341}
B(\omega):=V(\omega) c^{N}
\een
ein Hilbert--Schmidt--Operator. F"ur dieses $c$ gilt
\ben\label{G342}
V(\omega) = V(\omega)c^N c^{-N} = B(\omega) c^{-N}
\een
auf $D(V(\omega)) \cap D(c^{-N})$. Wir k"onnen also $V(\omega)$ durch
die rechte Seite von (\ref{G342}) auf $D(c^{-N})$ definieren.
\end{satz}
{\bf Beweis.}\\
Wir m"ussen zeigen, da\3 f"ur geeignetes $c>0$
\[
\sum_{\eta ,\nu} |\nu_{\eta,\nu} (\omega) c^{\|\nu \|} |^2 < \infty
\]
gilt.
Wir setzen $M= \exp ( \sum^{\infty}_{i=1} |\omega_i|^2)$. Zun"achst
summieren wir "uber $\eta$. Aus dem Beweis von Satz \ref{dicht}
erhalten wir die Absch"atzung
\begin{eqnarray}\label{G343}
\sum_{\eta} |v_{\eta\nu} (\omega) c^{\|\nu\|}|^2
  &=& c^{2 \|\nu\|} \sum_{\eta}|\nu_{\eta\omega} (\omega)|^2 \nonumber\\
&\leq & \frac{M c^{2\|\nu\|}}{\nu!} \prod^{\infty}_{i=1} m_{\nu_i , \nu_j}
\left( |\omega_i|+|\omega_{-i}|, |\omega_i|+|\omega_{-i}|\right)
\nonumber\\
&\leq & M c^{2\|\nu\|} \prod_{i=1}^{\infty} \left\{ \frac{1}{\nu_i!}
\sum_{j=0}^{\nu_i} {\nu_i \choose j}^2 j! \left(|\omega_i|+
|\omega_{-i}|\right)^{2(\nu_{i-j})} \right\} \nonumber\\
&\leq & Mc^{2\|\nu\|} \prod_{i=1}^{\infty} \left\{ \sum_{j=0}^{\nu_i}
{\nu_i \choose j} \left(|\omega_i| + |\omega_{-i}|
\right)^{2(\nu_i-j)} \right\}  \nonumber\\
&=& M c^{2\|\nu\|} \prod^{\infty}_{i=1} \left(1+ \left(|\omega_i|
+|\omega_{-i}| \right)^2 \right)^{\nu_i}.
\end{eqnarray}
Bis jetzt war $c$ beliebig. Nach Voraussetzung k"onnen wir absch"atzen
\begin{eqnarray}\label{3.73}
c^{2i \nu_i} \left( 1+ \left(\|\omega_i\| +\|\omega_{-i}\right)^2
\right)^{\nu_i} &\leq & c^{2i \nu_i} \left( 1+K^2 \left(R^i+S^i \right)^2
\right)^{\nu_i}
\nonumber\\
&=& \left( c^{2i} \left(1+K^2 \left(R^i+S^i\right)^2
\right)\right)^{\nu_i}\nonumber\\
&=:&X_i^{\nu_i}.
\end{eqnarray}
F"ur die Summierbarkeit f"ur alle $i$ ben"otigen wir $|X_i|<1$ f"ur alle $i$.
Man "uberlegt sich leicht, da\3 falls
$c^2\left(1+K^2\left(R+S\right)^2\right)<1$ oder
\bea\label{G345}
c^2 < \frac{1}{1+K^2(R+S)^2}
\eea
gilt, aus den Voraussetzungen des Satzes  $1>X_1>X_2>
\ldots >X_i > \ldots $ folgt.\\
Wir w"ahlen ein $c$ das (\ref{G345}) erf"ullt und k"onnen so
weiter absch"atzen
\begin{eqnarray*}
\sum_{\eta,\nu} | v_{\eta,\nu}(\omega) c^{2\|\nu\|} |^2
&\leq & M \prod_{i=1}^{\infty} \sum^{\infty}_{\nu_i=0} X_i^{\nu_i}\\
&=& M \prod_{i=1}^{\infty} \frac{1}{1-X_i} < \infty,
\end{eqnarray*}
da $ \sum^{\infty}_{i=1} X_i < \infty $ ist. Damit ist $B(\omega) =
V(\omega) c^N$ ein Hilbert--Schmidt--Operator. Es ist klar, da\3
(\ref{G342}) auf ${\cal F}$ gilt, denn es ist
\begin{eqnarray*}
V(\omega) \Phi_{\nu} &=& \sum_{\eta} v_{\eta\nu}(\omega) \Phi_{\eta} \\
&=& \sum_{\eta} v_{\eta\nu} (\omega) c^{\|\nu\|}\; c^{-\|\nu\|}
\Phi_{\eta}\\
&=& B(\omega) c^{-N} \Phi_{\nu}.
\end{eqnarray*}
Weiter ist, da ${\cal F}\subset D(V(\omega))$ und $V(\omega)$ als
Matrixoperator definiert wurde, $V(\omega)$ ``abgeschlossen auf ${\cal
F}$'' in dem folgendem Sinne:\\
F"ur $\Phi = \sum c_{\nu} \Phi_{\nu} \in D(V(\omega))$ und $\Phi_L =
\sum_{\|\nu \|\leq L} c_{\nu} \Phi_{\nu}$ gilt $V(\omega) \Phi_L
\longrightarrow V(\omega) \Phi$. Damit folgt f"ur $\Phi \in D(V(\omega))
\cap D(c^{-N})$:
\begin{center}
\setsqparms[1`0`0`1;1000`400]
\square[V(\omega) \Phi_L`V(\omega) \Phi\qquad(L\to
\infty)`B(\omega)c^{-N}
\Phi_L`B(\omega)c^{-N} \Phi\qquad(L\to\infty),;`\|``]
\end{center}
da $c^{-N} $ abgeschlossen ist.
\hfill$\Box$\\[3mm]
Konkret f"ur $V(\gamma,z)$ bzw.
$V(\gamma_1,\ldots,\gamma_r;z_1,\ldots,z_r )$ mit $|z_i|<1$ erhalten wir
wegen $|\frac{\gamma}{\sqrt{i}} z^{\pm i} | \le \max\{|\gamma|,1\}
 |z^{\pm
1}|^i$ bzw. $|\frac{\gamma_1z_1^{\pm i}+ \ldots + \gamma_r z_r^{\pm
i}}{\sqrt{i}} |\le r\max\{|\gamma_j|,1\} (\max\{|z_j^{\pm 1}|\})^i$ die
folgende Bedingung f"ur $c$ aus (\ref{G345}):
\bea\label{G346}
c&<& \left( 1+\max\{|\gamma|,1\}^2(|z|+|z|^{-1})^2
\right)^{-\frac{1}{2}}.
\eea
Eine analoge Ungleichung erhalten wir f"ur
$V(\gamma_1,\ldots,\gamma_r;z_1,\ldots,z_r)$. Wichtig an (\ref{G346})
ist, da\3 $c$ unabh"angig von $\arg z$ ist, wir k"onnen eine Faktorisierung
von $V(\gamma,z)$ finden, so
da\3  $V(\gamma,z)=B(\gamma,z)c^{-N}$ f"ur alle $c_1<|z|<c_2$ f"ur ein
festes $c$ gilt. Eine analoge Aussage gilt f"ur
$V(\gamma_1,\ldots,\gamma_r;z_1,\ldots,z_r)$.

In Absch"atzung (\ref{3.73}) sieht man, da\3 f"ur jedes $c<S^{-1}$ ein $i_0$
existiert mit $|X_i|<1$ f"ur alle $i\ge i_0$. Anders gesagt, bedeutet das
(vgl. (\ref{06})):
\begin{kor}\label{kz1} Sei
\ben\label{3z1}
V_{i}(\omega)=\prod_{n\ge i} \exp\left(\frac{\omega_n}{\sqrt{n}}
a_{-n}\right) \exp\left(-\frac{\omega_{-n}}{\sqrt{n}} a_n\right)
\een
und $\omega$ erf"ulle die Voraussetzungen von Satz \ref{340}. Dann gibt es
f"ur jedes $c<S^{-1}$ ein $i_0$, so da\3 f"ur alle $i\ge i_0$ die
Matrixelemente von $V_i(\omega)c^{\sum_{j=i}^{\infty}-a_{-n}a_n}$ einen
Hilbert--Schmidt--Operator in $\cal H$ definieren.
\end{kor}
Wir vermuten, da\3 die Bedingung $c<S^{-1}$ die nat"urliche Voraussetzung
f"ur die G"ultigkeit von Satz \ref{340} ist. Das formulieren wir
folgenderma\3en:
\begin{ver}\label{verm1}
F"ur jedes $c<S^{-1}$ definieren die Matrixelemente von
\ben\label{3z2}
B(\omega)_n:=\exp\left(\frac{\omega_n}{\sqrt{n}} a_{-n}\right)
\exp\left(-\frac{\omega_{-n}}{\sqrt{n}} a_n\right) c^{-a_{-n} a_n}
\een
einen beschr"ankten Operator in $\cal H$.
\end{ver}
W"are diese  Vermutung wahr, so w"are f"ur jedes $c<S^{-1}$ der Operator
$V(\omega)c^{-N}$  Hilbert--Schmidt, denn $V(\omega)c^{-N}$ lie\3e sich dann
als ein Produkt eines Hilbert--Schmidt--Operators aus Korollar \ref{kz1} mit
endlich vielen beschr"ankten Operatoren darstellen. Im Falle von
$V(\gamma,z)$ w"urde folgen, da\3 $V(\gamma,z)c^{-N}$ f"ur jedes $c<|z|$ ein
Hilbert--Schmidt--Operator ist. In \cite{Fe2} wurde die (etwas schw"achere)
Behauptung aufgestellt, da\3 unter den genannten Voraussetzungen
$V(\omega)c^{-N}$ ein kompakter Operator ist.

Eine Bemerkung zum Beweis der Vermutung: Sei ${\cal H}_n=\overline{ \Lin
\{ \Phi_{\mu}\; : \; \mu_k=0 \;(k\ne n)\}}$ und $\hat{\cal H}_n =
\overline{\Lin \{ \Phi_{\mu} \; : \; \mu_n=0\}}$. Dann gilt ${\cal H} =
{\cal H}_n \otimes \hat{\cal H}_n$. Es reicht zu zeigen, da\3
\[
\check{B}(\omega)_n:=B(\omega)_n\restrict{{\cal H}_n} \; : {\cal H}_n \lra
{\cal H}_n
\]
ein beschr"ankter Operator ist, denn dann ist
$B(\omega)_n=\check{B}(\omega)_n \otimes 1$ ebenfalls beschr"ankt.
\section{Die Nichtabschlie\3barkeit von $V(\omega)$ und $\Phi(z)$}
Nach dem positiven Ergebnis des letzten Abschnittes wollen wir uns nun
einem negativen Resultat
zuwenden, der bereits erw"ahnten Nichtabschlie\3barkeit des freien Feldes
und der nicht verschmierten Vertex--Operatoren.\\
Man bemerke, da\3 auch bei der Faktorisierung $V(\omega) = B(\omega)
c^{-N}$ die rechte Seite trotz ihrer sch"onen Faktoren Hilbert--Schmidt
$\times$ Selbstadjungiert i.~allg. nicht abschlie\3bar sein mu\3 (und auch
nicht
sein kann).\\
Zum Beweis der Nichtabschlie\3barkeit ben"otigen wir eine neue Menge von
Elementen in $\cal H$, die sogenannten koh"arenten Zust"ande. Koh"arente
Zust"ande sind verwandt mit dem Vertex--Operatoren, denn wie wir gleich
sehen werden, ergibt ein Vertex--Operator, angewendet auf den
Grundzustand $\Phi_0$ in $\cal H$, einen koh"arenten Zustand. Eine gute
Einf"uhrung findet
man in dem Buch von J. Klauder und E. Sudarshan ``Quantum Optics''
\cite{KS} in Chapter 7.

Wir geben nun einen kurzen "Uberblick "uber die Eigenschaften koh"arenter
Zust"ande mit abz"ahlbar unendlich vielen Freiheitsgraden, soweit wir sie
ben"otigen.\\
F"ur $t=(t_1, \ldots ,t_l) \in \CC^l$ sei
\begin{eqnarray}
|t,l\rangle &:=&\label{G347} N_l \sum_{\mu_1=\ldots=\mu_l=0}^{\infty}
\prod^l_{i=1} \frac{t_i^{\mu_i}}{\sqrt{\mu_i! i^{\mu_i}}} \Phi_{\mu}\\
&=&\label{G348} N_l \exp \left( \sum^l_{i=1} \frac{t_i}{i} a_{-i} \right)
\Phi_0
\end{eqnarray}
mit $N_l =\exp \left(-\frac{1}{2} \sum^l_{i=1} \frac{|t_i|^2}{i} \right)$.
Das Exponential in (\ref{G348}) ist "uber die Potenzreihe definiert. Die
normierten Vektoren $|t,l\rangle$ hei\3en koh"arente Zust"ande, da sie
Eigenvektoren der Vernichteroperatoren sind. Das k"onnen wir entweder durch
eine explizite Rechnung sehen oder durch Anwendung von Lemma \ref{hdef}:\\
Wie (\ref{G348}) und (\ref{verallg}) zeigen, entsteht $|t,l\rangle$ durch
Anwendung des Vertex--Operators $V(\omega)$ mit
\[
\omega_i= \cases { \frac{t_i}{\sqrt{i}} & $i=1,\ldots, l,$\cr
                   0 & sonst }
\]
auf $\Phi_0$. Nach Lemma \ref{hdef} ist $V(\omega) \Phi_0 \in D(a_n)\: (n
\in \ZZ)$, und es gilt f"ur $l\geq n > 0$
\bea\label{G349}
a_n|t,l\rangle =N_l a_n V(\omega) \Phi_0= N_l V(\omega) (a_n+t_n)\Phi_0=
t_n N_lV(\omega)\Phi_0=t_n |t,l\rangle .
\eea
Es folgt also
\ben\label{G3410}
a_n|t,l\rangle = \cases { t_n |t,l\rangle & $0<n \leq l$ \cr
                          0 &               $n>l$.}
\een
Sei
$l_{2,h} = \{t=(t_i)_{i \in \NN} \subset \CC :
\|t\|_{2,h}:=\sum_{i=1}^{\infty} \frac{|t_i|^2}{i} < \infty \}$.
Nach Satz \ref{dicht} existiert f"ur $t \in l_{2,h}$ der Grenzwert
$\lim_{l\to \infty} |t,l\rangle =: |t\rangle$. Wiederum aus Lemma
\ref{hdef} folgt, da\3 $a_n|t\rangle = t_n|t \rangle $ f"ur alle $n>0$ gilt.
Eine wichtige Eigenschaft der koh"arenten Zust"ande ist:
\begin{lem}\label{L341}
Die Abbildung $l_{2,h} \lra S({\cal H}) \subset {\cal H}$ mit $ t
\longmapsto |t \rangle$ ist stetig, es gilt
\[
\|\:\: |t\rangle - |t'\rangle \|^2_{{\cal H}} \leq 2 \left( \|t\|_{2,h}+
\|t'\|_{2,h} \right) \|t-t'\|_{2,h}.
\]
\end{lem}
Einen Beweis findet man in \cite{KS}.\\
Au\3erdem werden wir die folgende Aussage ben"otigen:
\begin{lem}\label{L342}
F"ur jedes $t \in l_{2,h}$ ist die Menge der koh"arenten Zust"ande
\ben\label{G3411}
\{ |t+t_0 \rangle : t_0\in l_{2,0} \}
\een
total in $\cal H$.
\end{lem}
{\bf Beweis.}\\
Sei $t \in l_{2,h}$ beliebig und $\Phi$ fest gew"ahlt. Wir zeigen:\\
Falls $\langle \Phi |t+t_0\rangle =0$ f"ur alle $t_0 \in l_{2,0}$, folgt
$\Phi=0$. W"ahle $r \in \NN$ und $\mu_1 ,\ldots, \mu_r \in \NN$.
Weiter sei  $l>\mu_r$. F"ur beliebige $u_1, \ldots, u_r \in \CC$ gibt es
ein $t^{(l)}\in l_{2,0}$, so da\3
\[
t+t^{(l)}=( 0,
\ldots,u_1,0,\ldots,u_2,\ldots,0, u_r,0,\ldots,t_l,t_{l+1},\ldots ),
\]
wobei $u_i$ in der Position  $\mu_i$ stehen soll,
das hei\3t
\bea\label{G3412}
\left(t+t^{(l)}\right)_i = \cases{ u_j & $i=\mu_j$,  \cr
                                   t_i & $i \geq l$, \cr
                                   0 &    sonst. }
\eea
Da $t+t^{(l)} \stackrel{l \to \infty}{\lra} \left( 0, \ldots,u_1,
 \ldots,u_r,0,0,0, \ldots \right) =:u$ in $l_{2,h}$, folgt nach Lemma
\ref{L341} $|t+t^{(l)}\rangle \lra |u \rangle$ und folglich
\[
\langle \Phi|t+t^{(l)}\rangle \lra \langle \Phi|u\rangle.
\]
Dieses Skalarprodukt l"a\3t sich aber explizit angeben.\\
Zun"achst sei $\Phi=\Phi_{\eta}$ mit $\eta =(\eta_1, \ldots,
\eta_N,0,0,\ldots)$ ein Basisvektor. Dann gilt
\begin{eqnarray}\label{G3413}
\langle \Phi_{\eta} | u \rangle
&=& \left(\eta!I^{\eta} \right)^{-\frac{1}{2}} \left\langle
a^{\eta_N}_{-N} \ldots a_{-1}^{\eta_1} \Phi_0|u \right\rangle \nonumber\\
&=& \left( \eta! I^{\eta}\right)^{-\frac{1}{2}}
\left\langle\Phi_0|a_1^{\eta_1} \ldots a_N^{\eta_N}|u
\right\rangle\nonumber\\
&=& \prod^r_{i=1}
\frac{u_i^{\eta_{\mu_i}}}{\sqrt{\eta_{\mu_i}!\mu_i^{\eta_{\mu_i}}}}
\prod_{j\notin \{\mu_i\}} \delta_{0,\eta_j}\left\langle
\Phi_0|u\right\rangle \nonumber\\
&=& \prod^r_{i=1} \frac{u_i^{\eta_{\mu_i}}}{\sqrt{\eta_{\mu_i}!
\mu_i^{\eta_{\mu_i}}}}\prod_{j\notin \{\mu_i\}} \delta_{0,\eta_j}.
\end{eqnarray}
Die Kronecker--Deltas entstehen, da $a_n|t\rangle =0$ falls $t_n=0$ gilt
und das hier immer der Fall ist, falls $n\notin \{\mu_i\}$ ist. Es ist
also $\langle\Phi_{\eta}|u\rangle=0$, falls ein $\eta_j\ne 0$ mit
$j\notin \{\mu_i\}$ existiert. \\
Damit ergibt sich f"ur $\Phi=\sum_{\eta} c_{\eta} \Phi_{\eta}$
\begin{eqnarray}\label{G3414}
\left\langle \Phi|u\right\rangle &=& \sum_{\eta} \overline{c}_{\eta}
\left\langle \Phi_{\eta}|u\right\rangle \nonumber\\
&=& \sum^{\infty}_{\eta_{\mu_1}, \ldots, \eta_{\mu_r}=0}
\overline{c}_{\eta} \left\langle \Phi_{\eta} |u \right\rangle \nonumber\\
&=& \sum^{\infty}_{\eta_{\mu_1}, \ldots, \eta_{\mu_r}=0}
\overline{c}_{\eta} \prod^r_{i=1}
\frac{u_i^{\eta_{\mu_i}}}{\sqrt{\eta_{\mu_i}!\mu_i^{\eta_{\mu_i}}}}.
\end{eqnarray}
Dies ist aber eine ganze Funktion in den Variablen $u_1, \ldots, u_r$. Ist
nun $\langle\Phi |t+t^{(l)}\rangle=0$ f"ur alle $t^{(l)}\in l_{2,0}$, so
folgt $\langle \Phi |u \rangle=0$ f"ur alle $(u_1,\ldots, u_r) \in \CC^l$
und damit impliziert (\ref{G3414}) $c_{\eta}=0$ f"ur alle $\eta$ der Form
$\eta=(0,\ldots, \eta_{\mu_1},0,\ldots, \eta_{\mu_r},0,0)$. Da aber $r$ und
$\mu_1, \ldots,\mu_r$ beliebig waren, und alle $\eta$ von dieser Form f"ur
geeignete $r$ und $\mu_1,\ldots,\mu_r$ sind, folgt $c_{\eta}=0$ f"ur alle
$\eta$ und damit $\Phi=0$.
\hfill$\Box$\\[3mm]
Da wir Vertex--Operatoren auf koh"arente Zust"ande anwenden wollen,
ben"otigen wir noch das folgende Lemma:
\begin{lem}\label{L343}
Sei $\sum_{k=1}^{\infty} |\omega_k|^2 < \infty$. Dann gilt
\begin{itemize}
\item[i)] $|t,l\rangle \in D\left(V\left(\omega \right)\right)$ und
\item[ii)] $V(\omega)|t,l\rangle= K_l \exp \left( -\sum^l_{k=1}
\frac{\omega_k t_k}{\sqrt{k}} \right) | \tau,l \rangle$,
wobei $|\tau,l\rangle$ ein koh"arente Zustand ist, der durch
\[
a_n |\tau,l\rangle = \cases { \left( \sqrt{n} \omega_n +t_n\right)
|\tau,l\rangle & f"ur $0<n\leq l$ \cr
   \sqrt{n} \omega_n |\tau,l\rangle & f"ur $n>l$ }
\]
festgelegt ist. F"ur $K_l$ erhalten wir
\[
K_l^{(t)} =\exp \left( -\frac{1}{2} \sum^{\infty}_{k=1} \frac{ \left|
\sqrt{k} \omega_k +t_k \right|^2 - \left|t_k \right|^2}{k} \right),
\]
wobei wir $t_k=0$ f"ur $k>l$ gesetzt haben.
\end{itemize}
\end{lem}
{\bf Beweis.}\\
Eine direkte Anwendung von Satz \ref{produkt}, wenn wir $|t,l \rangle$
nach (\ref{G348}) durch einen Vertex--Operator erzeugt denken.
\hfill$\Box$\\[3mm]
Wir k"onnen nun ein Komplement zu Satz \ref{dicht} beweisen, hierzu
ben"otigen wir aber noch das folgende technische Lemma:
\begin{lem}\label{L344}
Sei $t=(t_n)_{n \in \NN} \subset \CC$. Ist $t \notin l_2$, so existiert
ein $\omega\in l_2$ mit $t\omega \notin l_1$ und $t_i \omega_i \geq 0$
f"ur alle $i$.
\end{lem}
{\bf Beweis.}\\
Eine einfache "Ubung.
\hfill$\Box$\\[3mm]
Ist $(\omega_i)_{i \in \NN} \in l_2$, so gilt ${\cal F} \subset
D\left(V(\omega)\right)$.
Wir wollen nun die Nichtabschlie\3barkeit dieses Operators, falls
$(\omega_{-i})_{i \in \NN} \notin l_2$ gilt, beweisen. Die Abschlie\3barkeit
eines Operators h"angt im allgemeinen vom Definitionsbereich des Operators
ab. Da wir $V(\omega)$ als den maximalen Operator definiert haben, k"onnte
eine Einschr"ankung von $V(\omega)$ abschlie\3bar sein, obwohl $V(\omega)$
nicht abschlie\3bar ist. Deshalb f"uhren wir auch einen ``minimalen
Operator'' ein,
\ben\label{G3415}
V(\omega)_0 := V(\omega)|_{{\cal F}}
\een
mit $D\left(V(\omega)_0\right)={\cal F}$. Es ist klar, da\3  jede
Fortsetzung von
$V(\omega)_0$ nicht abschlie\3bar ist, wenn
$V(\omega)_0$ nicht abschlie\3bar ist.
\begin{satz}\label{S345}
Sei $(\omega_i)_{i \in \NN} \in l_2$, aber $(\omega_{-i})_{i \in \NN}
\notin l_2$. Dann sind $V(\omega)_0$ und $V(\omega)$ nicht abschlie\3bare
(nicht abgeschlossene) Operatoren.
\end{satz}
Eine direkte Folgerung aus dem Beweis von Satz \ref{S345} ist:
\begin{kor}\label{K341}
$(\omega_i)_{i\in\NN}\notin l_2$, $(\omega_{-i})_{i\in\NN} \in l_2$. Dann
ist
\[
D\left( V(\omega)\right)= \{ 0\}
\]
und $V(\omega)$ trivialerweise abgeschlossen.
\end{kor}
{\bf Beweis.}\\
Es gilt $\left(V(\omega)_0\right)^* = V\left( \{ -\overline{\omega}_{-i}
\} \right)$ (\cite{Wd} Satz 6.20).
\hfill$\Box$\\
Zusammenfassend erhalten wir aus Satz \ref{dicht}, Satz \ref{S345} und
Korollar \ref{K341}:
\begin{kor}\label{S346}
\begin{itemize}
\item[i)] Sei $(\omega_{-i})_{i\in\NN} \in l_2$. Dann ist $V(\omega)$
 genau dann dicht definiert, wenn $(\omega_i)_{i
\in \NN} \in l_2$.
\item[ii)] Sei $(\omega_i)_{i \in \NN} \in l_2$. Dann ist $V(\omega)$
genau dann abgeschlossen, wenn
\[
(\omega_{-i})_{i \in \NN} \in l_2.
\]
\end{itemize}
\end{kor}
{\bf Beweis von Satz \ref{S345}.}\\
Wir beweisen zun"achst die Nichtabschlie\3barkeit von $V(\omega)$. Sei also
$\omega=(\omega_{\pm i})$ fest gew"ahlt mit
$(\omega_i)_{i \in \NN} \in l_2$,
$(\omega_{-i})_{i \in \NN} \notin l_2$.
Nach Lemma \ref{L344} existiert eine Folge $t=(t_n)_{n \in \NN} \in l_2$
mit
$\sum^N_{n=1} \frac{\omega_{-n} t_n}{\sqrt{n}} \lra +\infty$ und folglich
auch $ \exp \left( \sum_{n=1}^N \frac{\omega_{-n} t_n}{\sqrt{n}} \right)
\lra \infty (N \rightarrow \infty)$.\\
Sei nun $0 \ne \Phi \in {\cal H}$ beliebig. Wir wollen zeigen, da\3 $\Phi
\notin D\left(V(\omega)^*\right)$ gilt. Dazu zeigen wir, da\3 eine Folge
$(f_n) \subset D\left( V(\omega) \right)$, $\|f_n\| =1$ existiert mit
\[
\big| \langle\Phi,V(\omega) f_n\rangle \big| \lra \infty.
\]
Nach Lemma \ref{L343} gilt mit $F^{(t)}_l = \exp \left( \sum^l_{n=1}
\frac{\omega_{-n} t_n}{\sqrt{n}}\right)$
\[
V(\omega) |t,l\rangle = K_l^{(t)} F_l^{(t)} |\tau,l\rangle.
\]
Weiter existiert nach Lemma \ref{L342} ein $t'\in l_{2,0}$ mit
\[
\lim_{l\to \infty} \langle \Phi| \tau +t',l\rangle = \langle \Phi |\tau+t'
\rangle \ne 0.\]
Damit folgt
\[
\big| \left\langle \Phi \left| V(\omega)\right| t+t',l
\right\rangle\big| = \left| K_l^{(t+t')} F_l^{\left(t+t'\right)}
\left\langle \Phi|\tau,l \right\rangle\right| \to \infty \qquad (l \to
\infty),
\]
denn $K_l^{(t+t')}$ und $\langle\Phi|\tau,l\rangle$ konvergieren f"ur $l
\to \infty$ gegen endliche Zahlen $\not= 0$ und $F_l^{(t+ t')}$ ist
 divergent mit $l \to \infty$. Es folgt also $\Phi \notin
D\left(V(\omega)^*\right)$ und die Nichtabschlie\3barkeit von
$V(\omega)$.\\
Zu zeigen bleibt die Nichtabschlie\3barkeit von $V(\omega)_0$. Dazu
approximieren wir $|t,l\rangle$ durch Elemente aus ${\cal
F}=D\left(V(\omega)_0\right)$:\\
Sei f"ur $Q\in \NN$
\ben\label{G3417}
\left|t,l,Q\right\rangle := N_l \sum^Q_{\mu_1= \ldots =\mu_l=0}
\prod^l_{i=1} \frac{t_i^{\mu_i}}{\left(\mu_i!
i^{\mu_i}\right)^\frac{1}{2}} \Phi_{\mu}.
\een
Es gilt $|t,l,Q\rangle \lra |t,l\rangle \qquad (Q\to \infty)$ und wie wir
schon im Beweis von Satz \ref{340} bemerkt haben, folgt aus Satz
\ref{produkt}, da\3 sogar
\ben\label{G3418}
V(\omega)_0 |t,l,Q \rangle \lra V(\omega)|t,l\rangle \qquad (Q\to \infty)
\een
gilt.\\
Wir k"onnen also eine Folge $Q_l \to \infty \quad (l \to \infty )$ w"ahlen,
so da\3
\[
\big| \left\langle \Phi\left| V\left(\omega\right)_0 \right|
t,l,Q_l\right\rangle\big| \to \infty \qquad (l \to \infty)
\]
gilt, woran die Nichtabschlie\3barkeit von $V(\omega)_0$ folgt.
\hfill$\Box$\\[3mm]
Als letztes in diesem Abschnitt wollen wir den Beweis der
Nichtabschlie\3barkeit des
freien Feldes $\Phi (z)$ f"ur $|z|<1$ skizzieren.\\
Dazu sei $\Phi_-(z)= i \sum_{n>0} \frac{z^n}{n} a_{-n}$, $\Phi_+ (z) =-i
\sum_{n>0} \frac{z^{-n}}{n} a_n$ und $\Phi(z)= \Phi_-(z)+ \Phi_+(z)$.
Dabei haben wir im Vergleich zu (\ref{01}) die unwesentlichen Terme
$q-ia_0 \ln z$ weggelassen.\\
Die Nichtabschlie\3barkeit von $\Phi (z)$ k"onnen wir in folgenden
Schritten zeigen:
\begin{itemize}
\item[i)]
$\Phi_+(z)$ ist nicht abschlie\3bar:\\
Dazu berechnet man $\Phi_+(z)|t,l\rangle=\left(\sum_{n=1}^l \frac{t_n
z^n}{n}\right) |t,l\rangle$. Zu jedem $\Psi \in {\cal H}$ gibt es ein $t
\in l_{2,h}$ mit
\[
\left\langle \Psi| t\right\rangle \not= 0 \mbox{ und } \left| \sum^l_{n=1}
\frac{t_n z^{-n}}{n} \right| \to \infty \qquad (l\to\infty).
\]
Es folgt $\big| \left\langle \Psi| \Phi_+(z)|t,l\right\rangle\big| =
\left| \sum^l_{n=1} \frac{t_n z^{-n}}{n} \left\langle\Psi
|t,l\right\rangle\right| \to \infty \: (l \to \infty)$ und  deshalb $\Psi
\notin D\left(\Phi_+(z)^*\right)$.
\item[ii)] Es gilt $\left\| \Phi_-(z)|t,l \rangle \right\| \leq C(t)$
unabh"angig von $l$ und deshalb k"onnen wir absch"atzen
\begin{eqnarray*}
\left| \left\langle \Psi| \Phi(z)|t,l \right\rangle\right| &=&
\big| \left\langle \Psi|\Phi_-(z)|t,l\right\rangle + \left\langle \Psi
\left|\Phi_+(z)\right| t,l \right\rangle \big| \\
&\geq & \big|\langle \Psi | \Phi_+(z) | t,l \rangle \big| - \big\|
\Psi\big\|\; \big\| \Phi_-(z) |t,l \rangle \big\|\\
&\ge& \big| \langle \Psi|\Phi_+(z)|t,l\rangle \big| - C(t) \big\|\Psi\big\|
\to \infty\;(l\to \infty).
\end{eqnarray*}
Damit folgt auch die Nichtabschlie\3barkeit von $\Phi(z)$.
\end{itemize}
\chapter{Abgeschirmte Vertex--Operatoren}
\section{Der Ladungsoperator}
Als erstes Beispiel eines ``abgeschirmten'' Vertex--Operators definieren
wir den Ladungsoperator $Q$. Damit werden wir in der Lage sein, den Beweis
der Kac--Determinantenformel zu vervollst"andigen und die in Satz
\ref{Inter} behaupteten Intertwiner zu konstruieren.\\
Sei $C_z$ die  positiv orientierte Kurve aus Abbildung \ref{fig1}, die in
$z$ startet und wieder endet. \\
\begin{figure}[ht]\begin{center}\begin{picture}(500,500)
\put(000,400){\special{CS!m 0.2 fig1neu.gem}}
\put(600,500){\makebox(0,0){\scriptsize$z$}}
\put(400,250){\makebox(0,0){\scriptsize$0$}}
\end{picture}\end{center}\caption{Die Kontur $C_z$}\label{fig1}\end{figure}
Sei $ \Re \gamma^2>0$ und
\ben\label{G41}
1=|z_1|>|z_2|> \ldots >|z_r|>0.
\een
F"ur $n_1, \ldots,n_r \in \ZZ$ und $\Phi,\Psi\in {\cal F}$ sei
{\mathindent0mm\begin{eqnarray}\label{G42}
\lefteqn{\left\langle \Phi,Q \left( \gamma;n_1,\ldots,n_r \right) \Psi
\right\rangle :=}\nonumber\\
\lefteqn{\int_{C_1} \int_{C_{z_1}} \cdots \int_{C_{z_1}} \big\langle
\Phi,F_{-\frac{r-1}{2}\gamma} ( \gamma;z_1, \ldots ,z_r ) T_{r \gamma} V (
\gamma;z_1, \ldots ,z_r ) \Psi \big\rangle \prod_{j=1}^r z_j^{-n_j -1}
dz_r \cdots dz_1.}
\end{eqnarray}}
Das Integral ist dabei als ein $r$--faches Kurvenintegral zu verstehen,
die Integrale sind in der bezeichneten Reihenfolge auszuf"uhren. Nach dem
Satz von Fubini k"onnen wir aber die Integrationsreihenfolge von
$z_2,\ldots,z_r$ beliebig vertauschen. Die Integrationsmenge liegt nicht
in (\ref{G41}), ist aber (um die Phasen festzulegen) als Grenzwert aus
(\ref{G41}) zu verstehen.\\
Da wir $\Re \gamma^2 >0$ verlangen, existiert das Integral (\ref{G42}) f"ur
beliebige $\Phi,\Psi \in {\cal F}$, d.~h. (\ref{G42}) definiert eine Form
$Q(\gamma;n_1,\ldots,n_r)$ mit $D(Q(\gamma;n_1,\ldots,n_r))={\cal
F}\times{\cal F}$.
 Es gibt verschiedene M"oglichkeiten zu sehen, da\3
$Q(\gamma;n_1,\ldots,n_r)$ einen Operator zumindest auf $\cal F$
definiert. Wir werden zeigen, da\3 (\ref{G42}) ``fast immer'' Null ist,
weshalb die Form ohne weitere Probleme einen Operator auf $\cal F$
definiert. Wir k"onnten nat"urlich auch  die Ergebnisse aus Kapitel 3,
 Satz \ref{dicht} oder Satz \ref{340} anwenden, aber f"ur $Q$ w"are das
nicht angemessen. Auf diese Methode werden wir im n"achsten Abschnitt
zur"uckgreifen.
\begin{lem}\label{L41}
Seien $m_1, \ldots , m_r \in \ZZ$. Dann gilt
\begin{itemize}
\item[(i)]
\ben\label{G43}
\int_{c_1} \int_{c_{z_1}} \cdots \int_{c_{z_1}} F_{-\frac{r-1}{2} \gamma}
(\gamma; z_1, \ldots , z_r) \prod_{j=1}^r z^{m_j-1} dz_r \cdots dz_1=0
\een
falls $\sum^r_{j=1} m_j \not=0$.
\item[(ii)]
\ben\label{G44}
W(z):= \int_{c_{z_1}} \cdots \int_{c_{z_1}}  F_{-\frac{r-1}{2} \gamma}
(\gamma;z, z_2, \ldots , z_r) \prod_{j=1}^r z^{m_j-1} dz_r \cdots dz_2
\een
ist holomorph in $\CC \setminus \{0\}$.
\end{itemize}
\end{lem}
{\bf Beweis.}
\begin{itemize}
\item[(i)]
Sei $z_1= e^{i \varphi_1} = e^{i \tilde{\varphi}_1}$ mit $0 \leq
\tilde{\varphi}_1 \leq 2 \pi$, $dz_1=i e^{i \tilde{\varphi}_1 }
 d\tilde{\varphi}_1$ und $z_j=e^{i(\varphi_1 +\varphi_j)} =:e^{i
\tilde{\varphi}_j}$  mit $dz_j= i e^{i\tilde{\varphi}_j}
 d\tilde{\varphi}_j=e^{i\varphi_1} e^{i\varphi_j} d\varphi_j$ und $0 \leq
\varphi_j \leq 2 \pi \: (j \geq 2)$.
Wir haben zu berechnen $(\lambda = -\frac{r-1}{2} \gamma)$
{\mathindent4mm\begin{eqnarray*}
\lefteqn{\int^{2 \pi}_{0} \int^{\varphi_1 + 2 \pi}_{\varphi_1} \cdots
\int^{\varphi_1 +2\pi}_{\varphi_1} \prod_{1\le k<j\le r} \left( e^{i
\tilde{\varphi}_k} - e^{i\tilde{\varphi}_j} \right)^{\gamma^2}
\prod_{j=1}^r e^{i\tilde{\varphi}_j \left( \lambda \gamma+ m_j -1 \right)}
\prod_{j=1}^r ie^{i\tilde{\varphi}_j} d\tilde{\varphi}_r \cdots
d\tilde{\varphi}_1}\\
&=& \int_0^{2 \pi} \cdots \int_0^{2 \pi} \prod^r_{j=2} \left( e^{i
\varphi_1} \left( 1-e^{i\varphi_j} \right)\right)^{\gamma^2 } \prod_{2
\leq k<j \leq r} \left(e^{i \varphi_1} \left( e^{i\varphi_k} -
e^{i\varphi_j} \right)\right)^{\gamma^2} \times\\
& &  e^{i\varphi_1 \left( \lambda \gamma+m_1-1\right)} \prod^r_{j=2}
e^{i\left(\varphi_1+\varphi_j\right) \left( \lambda \gamma +m_j-1\right)}
e^{i\varphi_1\left(r-1\right)} \prod^r_{j=1} ie^{i\varphi_j} d\varphi_j.
\end{eqnarray*}
Sammeln der $e^{i\varphi_1}$--Terme ergibt:
\begin{eqnarray*}
\lefteqn{ = \int^{2 \pi}_0 \cdots\int_0^{2 \pi} e^{i\varphi_1 \left(
\sum^r_{j=1} m_j-1 \right) } \prod_{j \geq2} \left(1-e^{i\varphi_j}
\right)^{\gamma^2} \prod_{2 \leq k < j <r} \left( e^{i\varphi_k}
-e^{i\varphi_j} \right)^{\gamma^2} \times}\\
&&\prod_{j \geq 2} e^{i\varphi_j \left( \lambda \gamma+m_j-1\right)}
\prod^r_{j=1}ie^{i\varphi_j} d\varphi_j.
\end{eqnarray*}
Das $\varphi_1$--Integral faktorisiert und wir erhalten
\begin{eqnarray}\label{G44a}
 &=& \int^{2 \pi}_0 e^{i\varphi_1 \left( \sum^r_{j=1} m_j-1\right)}
ie^{i\varphi_1} d\varphi_1 \int_{[0,2 \pi]^{r-1}} \cdots \prod_{j\geq 2}
ie^{i\varphi_j} d\varphi_j
\nonumber\\
&=&\int_{S^1} z_1^{\sum m_i-1} dz_1 \int_{(S^1)^{r-1}} F_{-\frac{r-1}{2}
\gamma} \left( \gamma,1,z_2, \ldots ,z_r \right)  \prod_{j \geq 2}
z_j^{m_j-1} dz_r \ldots dz_2.
\end{eqnarray}}
Das Integral "uber $z_1$ ist Null, falls $\sum^r_{i=1}m_i \not=0$. Damit ist
(i) bewiesen.
\item[(ii)]
Der Fall $\gamma^2 \in \NN$ ist klar, wir setzen deshalb $\gamma^2 \notin
\NN$ voraus. $W(z)$ ist analytisch f"ur $z\not= 0$. Zu zeigen ist noch, da\3
$z=0$ kein Verzweigungspunkt von $W(z)$ ist, sondern h"ochstens eine
isolierte Singularit"at. F"ur jedes $\varepsilon >0$ definieren wir
$W_{\varepsilon}(z)$ als Integral "uber die  Konturen aus Abbildung
\ref{figwe}.\\
\begin{figure}[h]\begin{center}\begin{picture}(1600,1800)
\put(-0,800){\special{CS!m 0.4 fig34.gem}}
\put(1750,1550){\makebox(0,0){\scriptsize$z$}}
\put(1400,1650){\makebox(0,0)[l]{\scriptsize$z_2$}}
\put(1280,1380){\makebox(0,0)[l]{\scriptsize$z_{r-1}$}}
\put(1200,1250){\makebox(0,0)[l]{\scriptsize$z_r$}}
\put(1050,800){\makebox(0,0){\scriptsize$0$}}
\put(1370,1170){\makebox(0,0){\scriptsize$\varepsilon$}}
\put(1670,1500){\makebox(0,0){\scriptsize$\varepsilon$}}
\end{picture}\end{center}
\caption{Die Konturen von  $W_{\varepsilon}$}\label{figwe}
\end{figure}
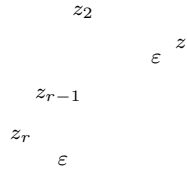
Wir haben also $|z_k-z_j| \geq |k-j| \varepsilon.$
Es gilt $\left|W(z) - W_{\varepsilon} (z) \right| \leq C\left(\left| z
\right| \right) \varepsilon^{r-1}$.  Wie aus einem
$\frac{\varepsilon}{3}$--Argument folgt, kann  $W(z)$ keinen
Verzweigungspunkt in Null haben, falls  $W_{\varepsilon}(z)$ keinen hat.\\
Den Integranden von $W_{\varepsilon}(z)$ k"onnen wir in binomischen Reihen
entwickeln, d.~h. jeder Faktor $( z_k- z_j)^{\gamma^2}= z_k^{\gamma^2}
\left(1- \frac{z_j}{z_k}\right)^{\gamma^2}$ wird in eine binomische Reihe
entwickelt. Da die Reihen auf der Integrationsmenge gleichm"a\3ig
konvergieren, k"onnen wir gliedweise integrieren.\\
Wir erhalten f"ur den Integranden
{\mathindent0mm\begin{eqnarray}\label{G45}
F_{-\frac{r-1}{2}\gamma} \left( \gamma;z_1, \ldots,z_r \right) \prod
z_j^{m_j-1}&=& \prod_{k<j}(z_k-z_j)^{\gamma^2} \prod_{j=1}^r
z_j^{-\frac{r-1}{2}\gamma^2 +m_j-1}\nonumber\\
&=& \prod^r_{j=1} z_j^{\frac{r-2j+1}{2} \gamma^2+m_j-1}
\prod_{k<j}(1-\frac{z_j}{z_k})^{\gamma^2}\nonumber\\
&=& \prod^r_{j=1} z_j^{\frac{r-2j+1}{2}\gamma^2 +m_j-1} \prod_{k<j} \Big\{
\sum^{\infty}_{n=0} C_n \left(\frac{z_j}{z_k} \right)^n \Big\},
\end{eqnarray}}
wobei $C_n= (-1)^n \frac{(\gamma^2-n+1)n}{n!}$ ist.\\
Das Produkt kann man (im Prinzip) ausmultiplizieren und erh"alt eine
Entwicklung der Form
\[
= \prod^r_{j=1} z_j^{\frac{r-2j+1}{2}\gamma} \times \mbox{
(Laurententwicklung in Null von } (z_1, \ldots,z_r)),
\]
g"ultig in $|z_1|-\varepsilon \geq |z_2| -\varepsilon \geq \cdots \geq
|z_{r-1}|-\varepsilon \geq |z_r|$. Nun gilt
\[
\int_{C_z} t^{\alpha} dt= \frac{e^{2\pi i(\alpha+1)}-1}{\alpha+1}
z^{\alpha+1},
\]
und folglich bleibt der nichtganzzahlige Anteil bei den Integrationen
erhalten. Wir erhalten also, da\3
\[
W_{\varepsilon}(z) = \int_{C_{z(1-\varepsilon)}} \cdots
\int_{C_{z(1-(r-1)\varepsilon)}} F_{-\frac{r-1}{2}\gamma}
(\gamma;z,z_2,\ldots , z_r ) \prod^r_{j=1} z_j^{m_j-1} dz_r \cdots dz_2
\]
bis auf eine Laurententwicklung in Null nur noch den Term
\[
\prod^r_{j=1} \left(z_1 \left( 1-\left(j-1 \right) \varepsilon \right)
\right)^{\frac{r-2j+1}{2}\gamma^2} = C(\varepsilon) z_1^{\sum_{j=1}^r
\frac{r-2j+1}{2} \gamma^2} =C(\varepsilon)
\]
enth"alt, also eine echte Laurententwicklung in Null besitzt und folglich
analytisch in $\CC \setminus \{0 \}$ ist.\hfill$\Box$
\end{itemize}
Mit Hilfe von Lemma \ref{L41} k"onnen wir nun zeigen, da\3 die
Matrixelemente (\ref{G42}) zu einem dicht definierten Operator geh"oren.
\begin{satz}\label{S42}
\begin{itemize}
\item[(i)]
Es ist $\left\langle \Phi_{\mu},Q \left( \gamma,n_1,\ldots, n_r \right)
\Phi_{\nu} \right\rangle =0 $, falls $\| \nu \| + \sum^r_{j=1} n_j \not= \|
\mu \|$.
\item[(ii)]
$Q\left( \gamma;n_1, \ldots,n_r\right) \Phi_{\nu} := \sum_{\mu}
\left\langle \Phi_{\mu} ,Q \left( \gamma,n_1,\ldots,n_r \right) \Phi_{\nu}
\right\rangle \Phi_{\mu}$ ist ein Operator vom Grad $\sum_j n_j$, d.~h. es
gilt $Q(\gamma;n_1, \ldots,n_r): {\cal F}(\alpha,\beta)_n \lra {\cal
F}(\alpha+ r\gamma,\beta)_{n+\sum n_j}$. $Q(\gamma;n_1, \ldots,n_r)$ ist
(als Matrixoperator) abgeschlossen.
\item[(iii)]
Es gilt stark auf ${\cal F}$
{\mathindent0mm\bea\label{G46}
\lefteqn{\left[L_n,Q(\gamma;n_1,\ldots,n_r)\right]}\nonumber\\
&=& \sum^r_{j=1} \left(n_j-n+\gamma a_0 - \frac{n}{2} \gamma^2 +n(\beta
\gamma+ \frac{\gamma^2}{2} )\right) Q (\gamma;n_1, \ldots,n_j-n_i, \ldots
,n_r ).
\eea}
\end{itemize}
\end{satz}
\begin{bem}\label{bem41} \begin{itemize}\item[(i)] Im Falle $n_1= \ldots
=n_r=s$ schreiben wir $Q(\gamma;r,s)$ f"ur
$Q(\gamma;\underbrace{s,\ldots,s}_{r-\mbox{\scriptsize viele}})$.
\item[(ii)] W"ahlt man $\beta\gamma+ \frac{\gamma^2}{2}=1$ und
$s=\frac{r}{2} \gamma^2-\gamma a_0$, so ist $[L_n,Q(\gamma;r,s)]=0$, d.~h.
$Q(\gamma;r,s)$ ist ein Intertwiner zwischen den Fock--Moduln ${\cal
H}(\alpha-r\gamma,\beta)$ und ${\cal H}(\alpha,\beta)$.
\end{itemize}\end{bem}
{\bf Beweis von Satz \ref{S42}} (i).
{\mathindent0mm\bea\label{G47}
\lefteqn{ \left\langle \Phi_{\mu},Q\left( \gamma;n_1, \ldots,n_r\right)
\Phi_{\nu} \right\rangle }\nonumber\\
&=& \int_{C_1} \cdots \int_{C_{z_1}} \langle
\Phi_{\mu},F_{-\frac{r-1}{2}\gamma} ( \gamma,z_1, \ldots,z_r) T_{r \gamma}
V(\gamma,z_1, \ldots ,z_r) \Phi_{\nu} \rangle \prod z_i^{-n_i-1} dz_r
\cdots dz_1 \nonumber\\
&=& \frac{1}{\sqrt{\mu! \nu!}} \int_{C_1}\cdots \int_{C_{z_1}}
F_{-\frac{r-1}{2}\gamma} (\gamma; z_1,
\ldots, z_r)\times \nonumber\\
&&\prod^{\infty}_{i=1} m_{\mu_i ,\nu_i} ( \frac{\gamma}{\sqrt{i}} (z_1^i+
\ldots+ z_r^i), -\frac{\gamma}{\sqrt{i}} (z_1^{-i}+ \ldots + z_r^{-i}))
\prod z_i^{-n_i-1} dz_r \cdots dz_1.
\eea}
Sei $\deg (\prod z_i^{k_i}):= \sum k_i$, dann gilt
\[
\deg \left(\left( z_1^i+ \ldots + z^i_r \right)^{n-j} \left( z_1^{-i}+
\ldots+ z^{-i}_r \right)^{m-j}\right)=i \left(n-m\right)
\]
und deshalb
\[
\deg m_{\mu_i ,\nu_i} \left( \frac{\gamma}{\sqrt{i}} \left( z^i_1+
\ldots+z^i_r\right),-\frac{\gamma}{\sqrt{i}} \left( z_1^{-i}+ \ldots +
z^{-i}_r \right)\right)= i \left( \mu_i-\nu_i \right).
\]
Das Produkt in $(\ref{G47})$ hat also den Grad $\sum_i
i(\mu_i-\nu_i)=\|\mu\|- \|\nu\|$. Aus Lemma \ref{L41}, (i) folgt darum
$\langle \Phi_{\mu},Q( \gamma;n_1, \ldots,n_r) \Phi_{\nu} \rangle =0$,
 falls  $0\not=  \| \mu \|- \| \nu \|- \sum n_i$
oder "aquivalent $\| \mu \| \not= \| \nu \| + \sum n_i$. \\
(ii) folgt unmittelbar aus (i), die Abgeschlossenheit folgt aus Lemma
\ref{seppel}.\\
(iii): Da wir $Q$ als ein Integral "uber die  Ma\-trix\-ele\-men\-te von\\
$F_{-\frac{r-1}{2}\gamma} (\gamma;z_1, \ldots,z_r) T_{r \gamma}
V(\gamma;z_1, \ldots,z_r)$ definiert haben, ist sind die Matrixelemente
von $[L_n,Q]$ als das schwache Integral von $[L_n,FTV]$ gegeben. Wir
erhalten mit $\kappa = \gamma a_0 - \frac{r}{2} \gamma^2 +n (\beta\gamma+
\frac{\gamma^2}{2})$
{\mathindent0mm\begin{eqnarray*}
\lefteqn{\left\langle \Phi_{\mu}, \left[ L_n, Q \left(\gamma;n_1,
\ldots,n_r\right) \right] \Phi_{\nu} \right\rangle}\\
&=& \int_{C_1} \int_{C_{z_1}} \cdots \int_{C_{z_1}} \sum^r_{j=1} z_j^n
\Big\{ \big(z_j \frac{\partial}{\partial z_j} + \kappa \big)
\big\langle \Phi_{\mu},FTV \Phi_{\nu} \big\rangle \Big\} \prod^r_{i=1}
z_i^{-n_i-1} dz_r \cdots dz_1.
\end{eqnarray*}
Wir k"onnen in $z_r, \ldots ,z_2$ partiell integrieren und da
$F_{-\frac{r-1}{2} \gamma} (\gamma;z_1, \ldots ,z_r)$ in den Endpunkten
$z_i=z_1$ eine Nullstelle hat, treten keine Randterme auf. In $z_1$
k"onnen
wir wegen Lemma \ref{L41} (ii) ohne Randterme partiell integrieren.
Insgesamt ergibt sich:
\begin{eqnarray*}
\lefteqn{\left\langle \Phi_{\mu}, \left[ L_n, Q \left(\gamma;n_1,
\ldots,n_r\right) \right] \Phi_{\nu} \right\rangle}\\
&=& \sum^r_{j=1} \int_{C_1} \int_{C_{z_1}} \cdots \int_{C_{z_1}} \big( (
-\frac{\partial}{\partial z_j}  z_j^{n-n_j}) + \kappa z_j^{n-n_j-1}\big)
\big\langle \Phi_{\mu},FTV \Phi_{\nu} \big\rangle \prod_{i\ne j}
z_i^{-n_i-1} dz_r \cdots dz_1 \\
&=& \sum^r_{j=1} \int_{C_1} \int_{C_{z_1}} \cdots \int_{C_{z_1}}\left(
n_j-n +\kappa \right) z_j^{n-n_j-1} \big\langle \Phi_{\mu},FTV \Phi_{\nu}
\big\rangle \prod_{i\ne j} z_i^{-n_i-1} dz_r \cdots dz_1\\
&=&\sum_{j=1}^r\left(n_j-n+\kappa\right)\left\langle
\Phi_{\mu},Q\left(\gamma;n_1,\ldots,n_j-n,\ldots,n_r\right)
\Phi_{\nu}\right\rangle.
\end{eqnarray*}}
Diese Gleichung gilt aber sogar stark auf ${\cal F}$, da f"ur jedes
$\Phi_{\nu}$ nur f"ur endlich viele $\mu$ $\left\langle\Phi_{\mu},
Q\Phi_{\nu}\right\rangle \ne 0$ ist.
\hfill$\Box$\\[3mm]
 Zum Beweis von Satz \ref{Inter} fehlt nur
noch die folgende Aussage "uber die Nichttrivialit"at von $Q(\gamma;r,s)$.
\begin{lem}\label{L401}
Sei $s \in \ZZ,\: r\in \NN$. Ist $\gamma^2 \notin \QQ$, so ist
$Q(\gamma;r,s)$ nicht die Nullabbildung,
insbesondere ist f"ur $s\ge 0$ $v_{\alpha-r\gamma,\beta} \notin \ker
Q(\gamma;r,s)$. Weiter ist f"ur $s<0$ $v_{\alpha,\beta}\notin \coker
Q(\gamma;r,s)$.
\end{lem}
Damit erhalten wir:
\begin{kor}\label{K402}
Sei $\beta\gamma + \frac{\gamma^2}{2} =1$ und $\alpha=-\frac{s}{\gamma} +
\frac{r}{2} \gamma$ und $\gamma^2 \notin \QQ$.
\begin{itemize}
\item[(i)] Sei weiter $s\ge 0$. Dann ist
$Q(\gamma;r,s)v_{\alpha-r\gamma,\beta}$
ein singul"arer Vektor vom Grad $r  s$ in ${\cal H} (\alpha,\beta)$.
\item[(ii)] Ist $s<0$, so gibt es einen Vektor vom Grad $-r s$ in
${\cal H}(\alpha-r\gamma,\beta)$ mit $Q(\gamma; r,s)w=v_{\alpha,\beta}$.
\end{itemize}
\end{kor}
Satz \ref{Inter} ist bewiesen.\\
{\bf Beweis von Lemma \ref{L401}.}\\
Sei $s\ge 0$.\\
Wir werden ein bestimmtes Skalarprodukt
$\langle\Phi,Q(\gamma;r,s)v_{\alpha-r\gamma,\beta}\rangle$ explizit
berechnen k"onnen. Dazu w"ahlen wir ein $\Phi =\sum c_{\mu} \Phi_{\mu} \in
{\cal F}(\alpha,\beta)_{r \cdot s}$ so, da\3 der Integrand m"oglichst
einfach wird. Zun"achst ist
\begin{eqnarray}\label{G401}
\langle \Phi,V(\gamma;z_1,\ldots,z_r)v_{\alpha-r\gamma,\beta}\rangle &=&
\sum_{\mu: \|\mu\|=rs} \overline{c}_{\mu} \langle
\Phi_{\mu},V(\gamma;z_1,\ldots,z_r)v_{\alpha-r\gamma,\beta}
\rangle\nonumber\\
&=& \sum_{\mu:\|\mu\|=rs} \overline{c}_{\mu} \prod_{i=1}^{rs}
\frac{1}{\sqrt{\mu_i !}} \left( \frac{\gamma}{\sqrt{i}} \left(z_1^i+
\ldots+z_r^i\right)\right)^{\mu_i}\nonumber\\
&=:& \sum_{\mu:\|\mu\|=rs} \overline{c}_{\mu} P_{\mu}\left(z_1,
\ldots,z_r\right).
\end{eqnarray}
$P_{\mu}(z_1, \ldots,z_r)$ ist ein Polynom in $z_1, \ldots,z_r$ vom Grad
$rs$, falls wir wieder $\deg\left(\prod z_i^{n_i}\right) = \sum n_i$
setzen. Die Polynome vom Grad $rs$ bilden einen $rs$--dimensionalen Raum.
Deshalb gibt es ein $\Phi= \sum c_{\mu} \Phi_{\mu}$, so da\3
\ben\label{G402}
\langle\Phi,V(\gamma;z_1, \ldots, z_r) v_{\alpha-r\gamma,\beta} \rangle
=z_1^s \cdots z_r^s
\een
gilt. Mit diesem $\Phi$ gilt dann
{\mathindent0mm\begin{eqnarray}
\lefteqn{\langle\Phi,Q(\gamma;r,s)v_{\alpha-r\gamma,\beta} \rangle
}\nonumber \\
&=& \int_{C_1} \int_{C_{z_1}} \cdots \int_{C_{z_1}}
    \langle \Phi,V(\gamma;z_1, \ldots,z_r) v_{\alpha-r\gamma,\beta}\rangle
    F_{-\frac{r-1}{2} \gamma} (\gamma;z_1, \ldots ,z_r) \prod^r_{i=1}
    z_i^{-s-1} dz_r \cdots dz_1 \nonumber\\
&=& \int_{C_1} \int_{C_{z_1}} \cdots \int_{C_{z_1}}F_{-\frac{r-1}{2}
    \gamma} (\gamma;z_1, \ldots ,z_r) \prod^r_{i=1} z_i^{-1} dz_r \cdots
    dz_1.\nonumber
\end{eqnarray}}
Wir k"onnen nun Lemma  \ref{L41} anwenden und das Integral "uber $z_1$
faktorisieren und erhalten aus (\ref{G44a})
\bea
\lefteqn{= 2 \pi i \int_{(C_1)^{r-1}} F_{-\frac{r-1}{2} \gamma}
(\gamma;1,z_2,
\ldots ,z_r) \prod^r_{i=1} z_i^{-1} dz_r \cdots dz_1 .}\nonumber
\eea
Wir ersetzen $z_i \to z_{i-1}$, aus der Definition von $F_{\lambda}$
folgt weiter
\bea
\label{G403}
\lefteqn{= 2 \pi i \int_{(C_1)^{r-1}} \prod_{j=1}^{r-1} (1-z_j)^{\gamma^2}
\prod_{1\leq k<j\leq r-1} (z_k-z_j)^{\gamma^2} \prod_{j=1}^{r-1}
z_j^{-\frac{r-1}{2} \gamma^2 -1} dz_{r-1} \cdots dz_1.}
\eea
Dieses Integral ist zu berechnen. Wir betrachten statt (\ref{G403})
allgemeiner f"ur $\alpha,\beta, \kappa \in \CC$, $z \in \CC^n$
\ben\label{G404}
I\: (\alpha,\beta,\kappa) := 2 \pi i \int_{(C_1)^{n}}
F(\alpha,\beta,\kappa;z) dz_n \cdots dz_1
\een
mit $F(\alpha,\beta,\kappa;z)= \prod^n_{j=1} z^{\alpha}_j \prod_{k<j}
(z_k-z_j)^{\beta} \prod^n_{j=1}(1-z_j)^{\kappa}$, wobei f"ur
$0<z_n<\ldots<z_1<1$ alle auftretenden Logarithmen ihren Hauptwert
annehmen sollen. Sei $\Re\alpha >0$, $\Re \beta>0$, $\Re\kappa>0$. Dann
existiert (\ref{G404}).\\
Zum Gl"uck k"onnen wir (\ref{G404}) auf ein bekanntes Integral zur"uckf"uhren.
A. Selberg \cite{Se} berechnete das folgende Integral.
\begin{satz}\label{S403}
Seien $\alpha,\beta,\kappa \in\CC$ mit $\Re \alpha >-1$, $\Re\kappa >-1$
und $\Re\beta > \min \{\frac{1}{n}$, $\Re \frac{\alpha+1}{n-1}$, $\Re
\frac{\kappa+1}{n-1} \}$. Dann konvergiert das folgende uneigentliche
(reelle) Integral und ist
\begin{eqnarray}\label{G405}
\lefteqn{ \int^1_0 \int^{t_1}_0 \cdots \int^{t_{n-1}}_0
F(\alpha,\beta,\kappa) dt_n \cdots dt_1 }\nonumber\\
&=& \frac{1}{n!} \prod^n_{j=1} \frac{\Gamma (j\beta+1)
     \Gamma((j-1)\beta+\alpha+1)
\Gamma((j-1)\beta+\kappa+1)}{\Gamma(\beta+1) \Gamma
((n+j-2)\beta+\alpha+\kappa+2)}.
\end{eqnarray}
\end{satz}
F"ur $n=1$ ist (\ref{G405}) nichts anderes als die Integraldarstellung der
Beta--Funktion
\[
B(\alpha+1, \beta+1) = \int^1_0 t^{\alpha} (1-t)^{\beta} dt =
\frac{\Gamma(\alpha+1) \Gamma(\beta+1)}{ \Gamma (\alpha+\beta+2)}.
\]
Wir m"ochten nun (\ref{G404}) auf (\ref{G405}) zur"uckf"uhren. Das ist eine
Verallgemeinerung der bekannten Tatsache, da\3 das Integral einer
symmetrischen Funktion von $n$ Variablen "uber den Einheitsw"urfel das
$n!$--fache des Integrals "uber dem Einheitssimplex ist. Wir haben hier
keine symmetrische Funktion als Integrand, aber eine bis auf Phasen
symmetrische Funktion.\\
Zun"achst k"onnen wir (\ref{G404}) zu einem reellen Integrand machen, indem
wir die Integrationswege von unten und oben auf die reelle Achse, genauer
gegen das Intervall $[0,1]$ dr"ucken. Dabei verlassen wir den Sektor
 $1>|z_n| > \cdots >|z_1|>0$.
Jeder Konfiguration der reellen Variablen $t_1, \ldots ,t_n$ im
Intervall $[0,1]$ $0<t_{\pi (1) }< \cdots <t_{\pi (n)}<1$ k"onnen wir eine
Phase  zuordnen, so da\3 der Integrand entlang der auf die Achse gedr"uckten
Integrationsweg analytisch ist. Das Integral "uber eine vorgegebene
Konfiguration ist bis auf diese  Phase gleich dem Integral "uber dem
Einheitssimplex. Diese Phasen k"onnen wir aufsummieren und erhalten
folgenden Zusammenhang zwischen (\ref{G404}) und (\ref{G405}):
\ben\label{G406}
I\: (\alpha,\beta,\kappa) = \varphi (\alpha,\beta) \int^1_0 \int^{t_1}_0
\cdots \int_0^{t_{n-1}} F(\alpha,\beta, \kappa,t) dt_n \cdots dt_1
\een
mit
\ben\label{G407}
\varphi(\alpha,\beta) =(-2i)^n e^{i \pi n \alpha} e^{i\pi n(n-1)\beta}
\prod^n_{l=1} \frac{\sin \pi l \beta}{\sin \pi \beta} \prod_{l=0}^{n-1}
\sin (\pi (\alpha+  l\beta)).
\een
Ist nun $\gamma^2 \notin \QQ$, so folgt wegen $\alpha=\beta=\gamma^2$,
$\kappa= -\frac{r-1}{2} \gamma^2-1$, da\3 (\ref{G406}) nicht verschwinden
kann, denn weder (\ref{G405}) noch (\ref{G407}) k"onnen dann Null sein.\\
Im Falle $s<0$ betrachten wir das Skalarprodukt $\langle
v_{\alpha,\beta}, Q(\gamma;r,s) \Phi \rangle $ f"ur $\Phi \in {\cal
F}(\alpha-r\gamma,\beta)_{-rs}$. Die Argumentation verl"auft dann v"ollig
analog zu dem Fall $s\ge 0$.
\hspace*{\fill}$\Box$
\section{Abgeschirmte Vertex--Operatoren}
Wir wollen nun etwas allgemeiner als im letzten Abschnitt
Vertex--Operatoren integrieren. Diese Operatoren werden in bestimmten
F"allen weitere prim"are Felder definieren. Im Gegensatz zum letzten
Abschnitt, wo wir von Lemma \ref{L41} ohne weitere Funktionalanalysis
geschenkt bekommen haben, da\3 $Q(\gamma,n_1,\ldots,n_r)$ vern"unftige
Operatoren sind, m"ussen wir hier etwas mehr arbeiten, um die integrierten
Vertex--Operatoren zu definieren. Sei $\bgamma:=(\gamma_0,
\ldots,\gamma_r)$ und zun"achst $\Re \gamma_i\gamma_j >0$ f"ur alle $i,j$.
Weiter sei ${\bf z}:=(w,z_1,\ldots,z_r)$. Wir wollen
\ben\label{G421}
\int_{(C_w)^r} F_{\alpha-\beta}(\bgamma;{\bf  z}) T_{\Sigma \gamma_j}
V(\bgamma;{\bf  z})\:dz_r \cdots dz_1
\een
als Operator  ${\cal H}(\alpha,\beta) \lra {\cal H}(\alpha+\sum
\gamma_i,\beta)$ einen Sinn geben. Wir k"onnen auf drei Arten versuchen,
(\ref{G421}) einen Sinn zu geben. Das Integral kann schwach, stark oder
im Operatorsinne interpretiert werden, wir werden es meist stark
interpretieren.\\
Sei $\Phi \in {\cal F}$. Dann ist nach Satz \ref{dicht}
\ben\label{G422}
\| V(\bgamma,{\bf z})\Phi\| \le C_{\Phi}(|w|,|z_1|,\ldots,|z_r|),
\een
die Norm des Integranden in (\ref{G421}), angewendet auf $\Phi\in {\cal
F}$, ist also gleichm"a\3ig beschr"ankt auf dem Integrationsgebiet. Au\3erdem
ist
$V(\bgamma,{\bf z})\Phi$ nach Lemma \ref{k002} eine holomorphe Funktion.
Folglich definiert
\ben\label{G423}
{\bf V}({\bgamma},w)\Phi:= \int_{(C_w)^r} F_{\alpha-\beta}({\bgamma},{\bf
z}) T_{\Sigma \gamma_j} V({\bgamma},{\bf z}) \Phi \:dz_r \cdots dz_1
\een
einen linearen Operator auf ${\cal F}$. Das Integral kann dabei als
Riemann-- oder als Bochner--Integral aufgefa\3t werden.

Wir k"onnen aber auch die Ergebnisse "uber die Faktorisierung von
Vertex--Operatoren anwenden. Wie wir in Abschnitt 3.5 gezeigt haben, gilt
\ben\label{G424}
V({\bgamma},{\bf z}) = B_c({\bgamma},{\bf z}) c^{-N}
\een
f"ur ein hinreichend kleines $c>0$ f"ur alle $(z_1,\ldots,z_r)\in (C_w)^r$.
Da die Hilbert--Schmidt--Norm von $B_c({\bgamma},{\bf z})$ entlang der
Integrationswege gleichm"a\3ig beschr"ankt ist (und $B_2({\cal H})$ separabel
ist), sind die Abbildungen $z_i \mapsto F_{\alpha-\beta}({\bgamma},{\bf
z}) B_c({\bgamma},{\bf z})$ holomorph und Bochner--integrierbar
(\cite{DU}) und wir
k"onnen alternativ mittels
\ben\label{G425}
{\bf V}^B({\bgamma},w):= \left(\int_{(C_w)^r)}
F_{\alpha-\beta}({\bgamma},{\bf z}) B_c({\bgamma},{\bf z})\:dz_r\cdots
dz_1\right) T_{\Sigma \gamma_j} c^{-N}
\een
die integrierten Vertex--Operatoren  mit $D(V^B(\bgamma,\omega))=D(c^{-N})$
definieren. Wegen (\ref{G342}) gilt
\ben\label{G426}
{\bf V}^B({\bgamma},w)\Phi={\bf V}({\bgamma},w)\Phi
\een
f"ur alle $\Phi\in {\cal F}$.

Leider ist die Bedingung $\Re \gamma_j \gamma_k >0$ eine Einschr"ankung,
die wir bei der Anwendung auf die konforme Quantenfeldtheorie nicht
aufrecht erhalten k"onnen, dort treten notwendigerweise auch negative
Exponenten in $F_{\alpha-\beta}({\bgamma},{\bf z})$ auf. Deshalb wollen
wir (\ref{G423}) analytisch in den Exponenten fortsetzen. Wir verwenden
die folgende einfache Methode, bekannt von der Beta--Funktion, indem wir
die Konturen $C_w$ durch Pochhammer--Konturen ersetzen. Dazu sei $P_w$ die
Pochhammer--Kontur um Null und $w$. Weiter seien $\gamma_j\gamma_0 \notin
\ZZ$ und  $(\alpha-\beta)\gamma_j \notin \ZZ$. Dann gilt f"ur $\Phi\in
{\cal F}$
{\mathindent5mm\bea\label{G427}
{\bf V}({\bgamma},w)\Phi &=& \prod_{j=1}^r\frac{1-e^{2 \pi i
(\alpha-\beta)\gamma_j)}}{1-e^{2\pi i \gamma_j
\gamma_0}}
\int_{(P_{w})^r} F_{\alpha-\beta}({\bgamma},{\bf z}) T_{\Sigma \gamma_j}
V({\bgamma},{\bf z}) \Phi \:dz_r \cdots dz_1.
\eea}
Dabei sollen die noch nicht integrierten Variablen $z_{i-1},\ldots,z_1$
bei der Integration von $z_i$ au\3erhalb der Pochhammer--Kontur liegen.
Gl. (\ref{G427}) liefert eine analytische Fortsetzung von (\ref{G423})
und existiert unter den genannten Voraussetzungen an die Exponenten. Wir
k"onnen nat"urlich auch die Konturen $P_w$ und $C_w$ mischen, falls die
Exponenten geeignete Bedingungen erf"ullen.\\
Genauso k"onnen wir in (\ref{G425}) die Konturen $C_{w}$ durch
Pochhammer--Konturen  $P_w$ in der Form von Abbildung \ref{figur4}
ersetzen. Wir erhalten
den gleichen Zusammenhang wie in (\ref{G427}).
\begin{figure}[ht]\begin{center}\begin{picture}(1000,1000)
\put(-200,500){\special{CS!m 0.3 fig41.gem}}
\put(1100,550){\makebox(0,0){\scriptsize$w$}}
\put(600,550){\makebox(0,0){\scriptsize$0$}}
\end{picture}\end{center}\caption{Die Pochhammer--Kontur f"ur ${\bf
V}^B$}\label{figur4}\end{figure}
 Insgesamt haben wir gezeigt:
\begin{satz}\label{S421}
Sei $\Re \gamma_k\gamma_j >0$ f"ur alle $k,j$ bzw. $\gamma_j\gamma_0\notin
\ZZ$ und $(\alpha-\beta)\gamma_j\notin \ZZ$ f"ur alle $j$. Weiter sei
$|w|<1$. Dann definiert (\ref{G423}) bzw. (\ref{G427}) einen  Operator
${\bf V}({\bgamma},w)$ mit $D({\bf V}({\bgamma},w))={\cal F}$.
Unter denselben Voraussetzungen an die Exponenten k"onnen wir den Operator
${\bf V}^B(\bgamma,w)$ mit $D({\bf V}^B(\bgamma,w))=D(c^{-N})$ f"ur ein
hinreichend kleines $c>0$ definieren. ${\bf V}^B(\bgamma,w)$ ist eine
Fortsetzung von ${\bf V}(\bgamma,w)$.
\end{satz}
Als n"achstes m"ochten wir den Kommutator zwischen ${\bf V}({\bgamma},w)$ und
$L_n$ angeben. Wir erhalten:
\begin{satz}\label{S422} Unter den Voraussetzungen von Satz \ref{S421} an
${\bf V}({\bgamma},w)$ gilt f"ur $\Phi\in {\cal F}$
{\mathindent0mm\bea\label{G428}
\lefteqn{\left[L_n,{\bf V}({\bgamma},w)\right]\Phi = w^n \left( w
\frac{\partial}{\partial w} + (n+1)\left(\frac{\gamma_0^2}{2} +
\beta\gamma_0\right) \right) {\bf V}({\bgamma},w)\Phi}\nonumber \\
&&+ \sum_{j=1}^r\left((n+1)(\frac{\gamma_j^2}{2} +
\beta\gamma_0)-n-1\right) \int z_j^n F_{\alpha-\beta}({\bgamma},{\bf z})
T_{\Sigma \gamma_j}  V({\bgamma},{\bf z}) \Phi\: dz_r \cdots dz_1.
\eea}
Ist insbesondere $\frac{\gamma_j^2}{2} + \beta\gamma_j =1$ f"ur
$j=1,\ldots,r$ erhalten wir
\ben\label{G429}
\left[L_n,{\bf V}({\bgamma},w)\right]\Phi = w^n \left( w
\frac{\partial}{\partial w} + (n+1)\left(\frac{\gamma_0^2}{2} +
\beta\gamma_0\right) \right) {\bf V}({\bgamma},w)\Phi.
\een
${\bf V}({\bgamma},w)$ ist in diesem Fall ein konformes Feld vom Gewicht
$\frac{\gamma_0^2}{2} + \beta\gamma_0$.
\end{satz}
Man kann sich nun fragen, was wir in (\ref{G429}) im Vergleich zu
(\ref{konglei}) gewonnen haben, dort haben wir ebenfalls konforme Felder
mit denselben Gewichten erhalten, ohne integrieren zu m"ussen. Der
entscheidende Unterschied ist, da\3 (\ref{G429}) im Vergleich zu
(\ref{konglei}) zwischen anderen Fock--Moduln abbilden, es gilt
\[
{\bf V}({\bgamma},w) \,:\,{\cal H}(\alpha,\beta) \lra {\cal H}(\alpha+
\sum_{j=1}^r\gamma_j,\beta),
\]
wogegen $T_{\gamma}z^{\gamma(\alpha-\beta)}V(\gamma,z) \,:\,{\cal
H}(\alpha,\beta) \lra {\cal H}(\alpha+ \gamma,\beta)$ gilt. ${\bf
V}({\bgamma},w)$ ist ein weiterer Baustein zur Konstruktion der  konformen
Felder im physikalischen Hilbertraum. Zu diesen Fragen kommen wir im
n"achsten Kapitel.

\noindent{\bf Beweis.}\\
Formal ist (\ref{G428}) klar, die Gleichung entsteht durch Integration von
(\ref{336}) und $r$ partiellen Integrationen analog zum Beweis von
(\ref{G46}). Dabei entstehen keine Randterme, da wir entweder zwischen
Nullstellen des Integranden oder auf einer geschlossenen Kontur
integrieren. Es bleibt zu zeigen, da\3 f"ur $\Phi\in {\cal F}$ (in
Kurzschreibweise) $\int [L_n,V]\Phi= [L_n,\int V]\Phi$ gilt. Es ist
zun"achst wegen $L_n\,:\, {\cal F}\lra {\cal F}$ klar, da\3 $\int VL_n \Phi$
existiert. Da die $L_n$ abgeschlossene Operatoren sind, folgt aus dem Satz
von Hille, Thm.~6, Chapter II \cite{DU} (die Me\3barkeit aller zu
integrierenden Abbildungen ist dabei trivial), da\3 $L_n\int V\Phi=\int L_n
V$ gilt. Wir erhalten also
\[
[L_n,\int V] \Phi=L_n\int V\Phi - \int V L_n \Phi = \int (L_n V- VL_n)
\Phi.\]
\hspace*{\fill}$\Box$
\section{Produkte abgeschirmter Vertex--Operatoren}
Wir wollen nun untersuchen, unter welche Vorausetzungen wir Produkte
abgeschirmter Vertex--Operatoren bilden k"onnen. Da wir ${\bf
V}(\bgamma,w)$ nur auf $\cal F$ definiert haben, ist klar, da\3 das
Produkt
dieser Operatoren i.~allg. nicht existiert, denn man kann nicht erwarten,
da\3
${\bf V}(\bgamma,w):{\cal F} \lra {\cal F}$ gilt. Deshalb kann nur
\ben\label{G431}
{\bf V}^B(\bgamma_1,w_1) {\bf V}(\bgamma_2,w_2)\Phi
\een
Sinn machen. Wir machen zun"achst eine einfache Beobachtung.
\begin{lem}\label{L431} Sei $\sum_{i=1}^{\infty} |
\omega_i c^{-i}|^2 <\infty$. Dann gilt
$V(\omega)({\cal F}) \subset D(c^{-N})$.\end{lem}
{\bf Beweis.} Eine einfache Variante vom Beweis von Satz \ref{dicht}.
\hfill$\Box$\\[3mm]
Falls nun $c$ gleichzeitig die Voraussetzungen von Lemma \ref{L431} und
von Satz \ref{340} erf"ullt, ist (\ref{G431}) gut definiert, denn es ist
leicht zu sehen, da\3 auch ${\bf V}(\bgamma,w_2)\Phi \in D(c^{-N})$ gilt.
Dazu m"ussen allerdings $w_1$ und $w_2$ ``weit genug'' auseinander sein,
wir erhalten so also eine st"arkere Bedingung an $w_1$ und $w_2$ als die
radiale Ordnung $|w_1|>|w_2|$, wie wir sie von den Produkten der freien
Vertex--Operatoren kennen.
 Seien die Integranden von ${\bf V}^B(\bgamma_1,w_1)$ und $ {\bf
V}(\bgamma_2,w_2)$ kurz mit $V(\bgamma_1,{\bf z}_1)$ und
$V(\bgamma_2,{\bf
z}_2)$ bezeichnet. Weiter sei $V(\bgamma_1,{\bf z}_1)= B(\bgamma_1,{\bf
z}_1) c^{-N}$. Es folgt dann weiter
\bea\label{G432}
(\int V(\bgamma_1,{\bf z}_1 ) \int V(\bgamma_2,{\bf z}_2) \Phi &=&\int\int
B(\bgamma_1,{\bf z}_1) c^{-N} V(\bgamma_2,{\bf
z}_2)\\\nonumber
&=&\int \int
V(\bgamma_1,{\bf z}_1) V(\bgamma_2,{\bf z}_2) \Phi,
\eea
denn alle Integrale existieren nach Voraussetzung, und eine zweimalige
Anwendung des Satzes von Hille liefert (\ref{G432}). Andererseits
existiert die rechte Seite von (\ref{G432}) f"ur beliebige $|w_1|<|w_2|$,
nur k"onnen wir ohne die Faktorisierung von ${\bf V}^B(\bgamma_1,w_1)$
(\ref{G432}) nicht beweisen, denn der
Satz von Hille ist wegen der Nichtabschlie\3barkeit des Integranden nicht
anwendbar. W"urde aber Vermutung \ref{verm1} gelten, so w"urde
 f"ur alle $|w_2|<c<|w_1|$ die Faktorisierung
\ben\label{G433}
V(\bgamma_1,{\bf z}_1 )  V(\bgamma_2,{\bf z}_2) \Phi = B(\gamma_1,{\bf
z}_1) c^{-N} V(\bgamma_2,{\bf z}_2) \Phi
\een
gelten und  k"onnten wir wieder (\ref{G432}) beweisen.
Zusammenfassend haben wir:
\begin{satz} \label{abprod}
Sei $0<|w_2|<c<|w_1|<1$ so, da\3 $V(\bgamma_1,{\bf z}_1)$ die
Voraussetzungen von Satz \ref{340} erf"ullt. Dann existiert
\ben\label{G434}
{\bf V}^B(\bgamma_1,w_1 )  {\bf V}(\bgamma_2,w_2) \Phi
\een
f"ur alle $\Phi \in {\cal F}$. Gilt Vermutung \ref{verm1}, so existiert
(\ref{G434}) f"ur alle $0<|w_2|<|w_1|<1$.
\end{satz}
Au\3erhalb der G"ultigkeit von Satz \ref{abprod} {\em definieren} wir, falls
das Produkt der Integranden existiert, das
Produkt der abgeschirmten Vertex--Operatoren   "uber die rechte Seite von
(\ref{G432}).
\newpage\thispagestyle{plain}\hbox{}\newpage
\chapter{Konforme Quantenfeldtheorie}
\section{Einleitung}
In diesem nichtmathematischen Abschnitt wollen wir kurz die von uns
ben"otigten Begriffe und Sprechweisen aus der konformen Quantenfeldtheorie
in zwei Dimensionen einf"uhren, um danach zu zeigen, wie die bisher
eingef"uhrten Operatoren bei der Konstruktion von konkreten Modellen
n"utzlich sind. Eine ausf"uhrliche Einleitung findet man z.~B. in
\cite{gin,st.a}. Ganz allgemein ist zu sagen, da\3 die im folgenden
aufgestellten Behauptungen aus der axiomatischen Quantenfeldtheorie
(d.~h. entweder den Wightman--Axiomen und konformer Invarianz oder den
Osterwalder--Schrader--Axiomen und konformer Invarianz) folgen,  falls
man Unitarit"at der Theorie verlangt, vgl. dazu \cite{FFK} bzw.
\cite{Mack}. Diese Forderung wird aber meist nicht gestellt.
F"ur masselose Theorien (wie es die konformen Theorien sind) ist es
n"utzlich, Lichtkegelkoordinaten $t+x$, $t-x$ zu verwenden. Im Falle
euklidischer Theorien k"onnen wir die Koordinaten $w=t+ix$ und
$\bar{w}=t-ix$ verwenden. Es ist ein Charakteristikum der konformen
Theorien und bei konkreten Rechnungen "au\3erst n"utzlich,  die Variablen $w$
und $\bar{w}$ als unabh"angige komplexe Variable zu interpretieren. Zur
physikalischen Interpretation k"onnen $w$ und $\bar{w}$ wieder auf den
entsprechenden Unterraum in $\CC^2$ eingeschr"ankt werden.

Da masselose Theorien typischerweise infrarot divergent sind, wird der
Raum kompaktifiziert, was wir durch die Periodizit"atsbedingung $w+2\pi=w$,
$\bar{w}+2\pi=\bar{w}$ ausdr"ucken. $w$ ist also eine Koordinate von einem
Zylinder, wobei die Kurven $t$=const. einem Gro\3kreis auf dem Zylinder
entsprechen. Mit Hilfe der konformen Abbildung $z=e^w$,
$\bar{z}=e^{\bar{w}}$ f"uhren wir den Zylinder in die komplexe Ebene "uber,
die Kurven $t=$const. sind nun konzentrische Kreise mit Radius $e^t$. Die
Operation der Zeitumkehr $t \to -t$ entspricht nun $z \to 1/\bar{z}$ und
der Generator der Dilatationen $z\to e^a z$ (d.~h. $w\to a+w$) entspricht
dem Hamiltonoperator des Systems. Diese Beschreibung der Theorie wird
radiale Quantisierung genannt.

Die wichtigste Observable ist der Noethersche Strom, der zur Invarianz
unter der Poincar\'{e}-Gruppe assoziert ist, der Energie--Impuls Tensor
$T$ mit vier Komponenten. Aus der konformen Invarianz und der
Kontinuit"atsgleichung f"ur $T$ folgt, da\3 $T$ Spur Null und nur zwei
unabh"angige Komponenten $T(z)$ und $\overline{T}(\bar{z})$ hat, wobei
$\frac{\partial}{\partial\bar{z}}T(z)=0$ und $\frac{\partial}{\partial z}
\overline{T}(\bar{z})=0$ gilt. $T(z)$ ist also analytisch und
$\overline{T}(\bar{z})$ antianalytisch. Das Theorem von L"uscher und Mack
\cite{Mack} sagt nun, da\3 die folgende ``Operatorproduktentwicklung'' f"ur
$T(z)$  gilt. ($\langle\cdot\rangle$ ist der Vakuumerwartungswert.)
\bea\label{E1}
\langle T(z_0) T(z_1)\rangle&=&\frac{c/2}{(z_0-z_1)^4} + \left(
\frac{2}{(z_0-z_1)^2} +\frac{\partial}{\partial z_1} \right)\langle
T(z_1)\rangle,\\ \label{E1a}
\langle \overline{T}(\bar{z}_0)
\overline{T}(\bar{z}_1)\rangle&=&\frac{\bar{c}/2}{(\bar{z}_0-\bar{z}_1)^4} +
\left( \frac{2}{(\bar{z}_0-\bar{z}_1)^2} +\frac{\partial}{\partial
\bar{z}_1} \right)\langle \overline{T}(\bar{z}_1)\rangle,
\eea
wobei $c$ ein modellabh"angiger Parameter ist. Wie man sieht, unterscheiden
sich (\ref{E1}) und (\ref{E1a}) nur dadurch, da\3 ein Querstrich "uber
(fast) alles gezogen wird, oft werden wir deshalb nur eine der Gleichungen
aufschreiben.
Man f"uhrt nun eine operatorwertige Laurententwicklung
\ben\label{E2}
T(z)=\sum_{n\in\ZZ} L_n z^{-n-2},\qquad \overline{T}(\bar{z})=\sum_{n\in\ZZ}
\overline{L}_n \bar{z}^{-n-2}
\een ein, dann sind (\ref{E1}) und (\ref{E1a}) "aquivalent dazu, da\3 $L_n$
und $\overline{L}_n$ Darstellungen der Virasoro--Algebra sind, wobei der
zentrale Term $z$ aus (\ref{virrel}) durch $c$ bzw. $\bar{c}$ dargestellt
wird. In einer unit"aren Theorie sollen die euklidischen Felder zur Zeit
$t=0$ symmetrisch sein, hier w"urde diese Forderung bedeuten $T(z)^*=T(z)$
f"ur $|z|=1$ oder "aquivalent $L_n^*=L_{-n}$. Die Operatoren $L_n$ m"ussen
also eine unit"aren Darstellung von Vir sein. Diese Forderung werden wir
 nicht stellen. Die Klasse der minimalen Modelle, die wir untersuchen
werden, enth"alt aber die unit"aren Modelle. $L_0+\overline{L}_0$ ist der
 konforme Hamiltonoperator.

Wenn wir annehmen, da\3 der Energie--Impuls--Tensor die einzige Observable
ist, folgt, da\3 $T$ die Observablenalgebra erzeugt und diese als die
universelle einh"ullende Algebra von Vir gew"ahlt werden kann. Der Raum der
physikalischen Zust"ande zerf"allt dann in ``Superauswahlsektoren'', die
eine irreduzible Darstellung von $\Vir\plus \overline{\Vir}$ tragen. F"ur
die bereits erw"ahnten minimalen Modelle verlangt man, da\3 es nur endlich
viele Superauswahlsektoren gibt. Der physikalische Hilbertraum ist dann
\ben\label{E2a}
\HH_{\mbox{\scriptsize phys}}=\plus_{j=1}^N \HH_j \otimes \overline{\HH}_j,
\een
wobei $\HH_j$ und $\overline{\HH}_j$ irreduzible Vir--Moduln sind. Wir
 werden der Einfachheit halber nur Modelle mit Spin Null, d.~h.
$c=\bar{c}$ und damit $\HH_j\simeq\overline{\HH}_j$ betrachten. Da
$H=L_0+\overline{L}_0$ nach unten halbbeschr"ankt sein soll und $\HH_j$
einen zyklischen Vakkuumvektor besitzen soll, sind $\HH_j$
H"ochstgewichtsmoduln, d.~h. irreduzible Vir--Moduln von einem Typ $(h,c)$.
$c$ ist ja bereits durch $T(z)$ festgelegt und $h$ ist ein weiterer
Parameter, der durch $j$ parametrisiert wird.

In \cite{BPZ} wurde eine weitere Forderung aufgestellt: Es sollen Felder
$\Phi_{\alpha}$ existieren, die sich tensoriell unter konformen
Transformationen $z\to w(z)$ und $\bar{z}\to \bar{w}(\bar{z})$ nahe der
Identit"at transformieren, d.~h. es soll
\ben\label{E3}
U(w,\bar{w})\Phi_{\alpha}(z,\bar{z}) U(w,\bar{w})^{-1} = \left(
\frac{dw}{dz}\right)^{h_{\alpha}}(z) \left(
\frac{d\bar{w}}{d\bar{z}}\right)^{\bar{h}_{\alpha}}(\bar{z})
\Phi_{\alpha}(w,\bar{w})
\een
f"ur unit"ares $U(w,\bar{w})$ gelten. Ein solches Feld wird prim"ares Feld vom
Gewicht $(h_{\alpha},\bar{h}_{\alpha})$ genannt. F"ur Felder von Spin 0 gilt
$h_{\alpha}=\bar{h}_{\alpha}$. Die infinitesimale Variante von (\ref{E3})
ist
\ben\label{E4}
\left[L_k\plus 1,\Phi_{\alpha}(z,\bar{z})\right]=
\left(z^{k+1}\frac{\partial}{\partial z}
+ (k+1)z^n h_{\alpha} \right) \Phi_{\alpha}(z,\bar{z})
\een
und eine analoge Gleichung f"ur den Kommutator mit $\overline{L}_k$. Aus
(\ref{E4}) folgt, da\3 $\Phi_{\alpha}(0,0)\Omega:=\lim_{z,\bar{z}\to 0}
\Phi_{\alpha}(z,\bar{z})\Omega$ ($\Omega$ sei der Vakuumvektor) ein
H"ochstgewichtsvektor f"ur $\Vir\oplus\Vir$ mit $(L_0 \plus
1)\Phi_{\alpha}(0,0)\Omega=h_{\alpha}\Phi_{\alpha}(0,0)\Omega$ und
$(1\plus\overline{L}_0)\Phi_{\alpha}(0,0)\Omega=\bar{h}_{\alpha}
\Phi_{\alpha}(0,0)\Omega$ ist. Der von diesem Vektor erzeugte irreduzible
H"ochstgewichtsmodul bezeichnen wir mit $\HH_{\alpha}$. Eine Theorie hei\3t
nun minimal, wenn
\begin{itemize}
\item[(i)] (\ref{E2a}) erf"ullt ist, und
\item[(ii)] zu jedem Summanden in (\ref{E2a}) genau ein prim"ares Feld
$\Phi_{\alpha}$  mit $\HH_j\otimes \overline{\HH}_j \simeq \HH_{\alpha}$
existiert.
\end{itemize}
Die "ubrigen Felder werden  aus den prim"aren Feldern auf die folgende
Weise erzeugt: Sei
\ben\label{E5}
L_{-k}(z)=\int_{|\xi-z|=\varepsilon} \frac{T(z)}{(\xi-z)^{k+1}}\; d\xi
\een
und
\ben\label{E6}
\Phi_{\alpha}^{(k_1,\ldots,k_N)}(z)=L_{-k_1}(z) \cdots L_{-k_N}(z)
\Phi_{\alpha}(z).
\een
Die Felder in (\ref{E6}) werden sekund"are Felder genannt. Es reicht, die
Korrelationsfunktionen der prim"aren Felder zu kennen, die der sekund"aren
Felder ergeben sich dann durch die Anwendung eines zu $L_{-k_1}(z) \cdots
L_{-k_N}(z)$ assozierten Differentialoperators auf die Korrelation der
zugeh"origen prim"aren Felder.

Die Bestimmung der Korrelationen der prim"aren Felder f"ur die minimalen
Modelle "uber die sogenannte Feigin--Fuks--Integraldarstellung  geht auf
Dotsenko und Fateev (\cite{DF1,DF2}) und G. Felder \cite{Fe1} zur"uck. Dazu
ist es wesentlich, die Bausteine des physikalischen Hilbertraumes durch
einen (Subquotienten eines) Fock--Raum(es) zu ersetzen. Die sogenannte
Coulomb--Gas--Konstruktion erlaubt es dann die prim"aren Felder  zu
konstruieren.
\section{Die minimalen Modelle}
Um die Notationen zu fixieren wiederholen wir kurz die f"ur den n"achsten
Abschnitt relevanten Daten. \\
Die minimalen Modelle leben in Vir--Moduln vom Typ $III_-$, deshalb sind
sie
fixiert durch die Wahl von $p, p' \in \NN$ mit $p,p'$ relativ prim. Dadurch
ist $c$  durch
\ben\label{G51}
c=c(p',p)=1-\frac{6(p'-p)^2}{p'p}
\een
festgelegt. Weiter ist
\ben\label{G52}
h_{n'n}=\frac{(np'-n'p)^2-(p'-p)^2}{4pp'}
\een
und
\bea\label{G53}
\HH_{{\mbox{\scriptsize phys}}}= \plus_{h, \overline{h}} \HH_h \otimes
\HH_{\overline{h}} = \plus_{n'n} \HH_{h_{n'n}} \otimes
\overline{\HH}_{h_{n'n}}=: \plus_{n'n}\HH_{n'n} \otimes
\overline{\HH}_{n'n}
\eea
mit $\HH_{n'n}= V(h_{n'n},c) / M(h_{n'n},c)$ dem eindeutigen
irreduziblen Vir--Modul vom Typ $(h_{n'n},c)$.
$M(h_{n'n},c)$ ist der maximale nichttriviale Untermodul von
$V(h_{n'n},c)$ und nach Satz \ref{klasse} von zwei singul"aren Vektoren
erzeugt.
Die direkte Summe in (\ref{G53})
geht "uber $1\leq n' \leq p'-1$, $1\leq n \leq p'-1$ und $n'p \le np'$.\\
$\HH_{n'n}$ sind nach Satz \ref{unirep} nur im Fall
$|p-p'| =1$ unit"are Vir--Moduln, in den anderen F"allen ist die
Shapovalov--Form auf $\HH_{n'n}$ zwar nicht ausgeartet, aber nicht
positiv
definit. Leider ist $\HH_{n'n}$ versehen mit der Shapovalov--Form
i.~allg. kein (Pr"a--)Krein--Raum, denn $\HH_{n'n}$ zerf"allt nicht in eine
direkte Summe von Vektoren positiver bzw. negativer L"ange. Siehe dazu
Beispiel \ref{notkrein}.  Obwohl wir also vom ``physikalischen
Hilbertraum'' gesprochen haben, ist zun"achst nicht klar, wie wir
$\HH_{n'n}$ in einen Hilbertraum einbetten k"onnen. Da wir aber sehen
werden, da\3 $\HH_{n'n}$ als Vir--Modul zu einem Unterraum gewisser
Fock--Moduln isomorph ist, k"onnen wir auf diese Weise eine
Vervollst"andigung von $\HH_{n'n}$ angeben.\\
%
Zu jedem Summand in (\ref{G53}) gibt es ein prim"ares Feld
$\Phi_{n'n}(z,\overline{z})$ vom Gewicht $(h_{n'n},\bar{h}_{n'n})$, das
die Gleichung (\ref{E4}) erf"ullt.
Wir haben die folgende Zerlegung von $\Phi_{n'n}(z,\overline{z})$
gem"a\3 (\ref{G53}). \\
Sei $P_{r'r}: \HH_{\mbox{\scriptsize phys}} \lra \HH_{r'r}$ die
kanonische Projektion. Wir k"onnen schreiben
\ben\label{G55}
\Phi_{n'n} (z,\overline{z})= \sum_{r',r,s',s} C^{r'r}_{n'ns's}
\varphi^{r'r}_{n'ns's} (z) \otimes \overline{\varphi}^{r'r}_{n'ns's}
(\overline{z})
\een
mit $\varphi^{r'r}_{n'ns's} (z)=: P_{r'r}\Phi_{n'n} (z,\overline{z})
P_{s's}$. Wir k"onnen $\varphi^{r'r}_{n'ns's}(z)$ auch als Abbildung
zwischen $\HH_{s's}$ und $\HH_{r'r}$ auf\/fassen. Die Strukturkonstanten
$C^{r'r}_{n'ns's}$ sind eindeutig durch die Normierung $\langle
v_{r'r},\varphi^{r'r}_{n'ns's} (1) v_{s's} \rangle =1$ festgelegt.
$\varphi^{n'r}_{n'n s's} (z)$ wird als konformes Feld bezeichnet und ist
eine i.~allg. vieldeutige operatorwertige Funktion.\\
$\varphi_{n'ns's}^{r'r}(z)$ erf"ullt entsprechend zu (\ref{E4})
\ben\label{G57}
\left[L_k,\varphi_{n'ns's}^{r'r}(z)\right] = \left( z^{k+1} \frac{d}{dz} +
(k+1) h_{n'n}z^k\right) \varphi_{n'ns's}^{r'r}(z).
\een
Mit Hilfe von (\ref{G55}) k"onnen wir nat"urlich auch Korrelationen von
prim"aren Feldern auf Korrelationen von konformen Feldern zur"uckf"uhren.
 Dazu sei $\mu=(m_1^{'},m_1^{\phantom{'}},m_2^{'},\ldots,m_{k-1}^{'},
m_{k-1}^{\phantom{'}})$. Es gilt
\ben\label{G56}
\langle \Phi_{n_1^{'}n_1^{\phantom{'}}}(z_1,\overline{z}_1)\cdots
\Phi_{n^{'}_k
n_k^{\phantom{'}}} (z_k,\overline{z}_k)\rangle = \sum_{\mu} \lambda_{\mu}
F_{\mu}(\{z_j\}
) F_{\mu}(\{ \overline{z}_j\})
\een
mit
\ben\label{G56a}
F_{\mu}(\{z_j\})=\langle v_{1,1},\varphi_{n_1^{'} n_1^{\phantom{'}}
m_1^{'} m_1^{\phantom{'}}}^{11} (z_1) \varphi_{n_2^{'} n_2^{\phantom{'}}
m_2^{'} m_2^{\phantom{'}}}^{m_1^{'} m_1^{\phantom{'}}} (z_2)\cdots
\varphi_{n_k^{'} n_k^{\phantom{'}}11}^{m_{k-1}^{'}
m_{k-1}^{\phantom{'}}}(z_k)v_{11}\rangle.
\een
$F_{\mu}$ wird als konformer Block bezeichnet, $\lambda_{\mu}$ entstehen
als Produkte der Strukturkonstanten.
\section{Die Konstruktion der prim"aren Felder im Fock--Raum}
Wir wollen nun die konformen Bl"ocke
$\varphi_{n'ns's}^{r'r}(z)$ mit Hilfe der Ergebnisse aus Kapitel 3 und 4
explizit konstruieren, d.~h. wir definieren Operatoren, die (\ref{G57})
erf"ullen. In einem zweiten Schritt zeigen wir dann, da\3 diese Operatoren
auch auf den irreduziblen H"ochstgewichtskomponenten der beteiligten
Fock--R"aume wohldefiniert sind.

Wir setzen $\gamma_+=\sqrt{\frac{2p'}{p}}$ und
$\gamma_-=-\sqrt{\frac{2p}{p'}}$. Es gilt $\gamma_+\gamma_-=-2$.
Weiter setzen wir  $\alpha_{n'n}=\frac{n'p-np'}{\sqrt{2p'p}}$,
$\beta=\frac{p-p'}{\sqrt{2p'p}}$ und
$\gamma_{n'n}=\alpha_{n'n}-\beta=\frac{1}{2} (1-n')\gamma_- +
\frac{1}{2}(1-n) \gamma_+$ (vgl. dazu (\ref{G28})). Dann sind ${\cal
H}_{n'n}:={\cal H}(\alpha_{n'n},\beta)$ Fock--Moduln vom Typ $(h_{n'n},c)$
und vom Typ $III_-$ der Feigin--Fuks Klassifikation Satz \ref{fockklasse}
und es gilt $h_{n'n}=\frac{\gamma_{n'n}^2}{2}+ \beta \gamma_{n'n}$ und
$\frac{\gamma_{\pm}^2}{2} + \beta \gamma_{\pm}=1$.\\
Sei
\bea\label{G59a}
0<n<p,\quad 0< n^{'}< p^{'},\quad 0\le r < n,\quad 0\le r^{'} <
n^{'}\mbox{ und}
\\
\label{G59}
\bgamma_{n'n}^{r'r}=(\gamma_{n'n},\underbrace{\gamma_-,\ldots,\gamma_-}_{r'
-\mbox{
\scriptsize viele}},\underbrace{\gamma_+,\ldots,\gamma_+}_{r-\mbox{
\scriptsize viele}}).
\eea
Die Exponenten $\bgamma_{n'n}^{r'r}$ erf"ullen die Voraussetzungen von Satz
\ref{S421}, folglich k"onnen wir f"ur $\Phi \in {\cal F}$ und $|z|<1$
\ben\label{G510}
D(V_{n'n}^{r'r}(z)):={\cal F}, \qquad V_{n'n}^{r'r}(z)\Phi := {\bf
V}(\bgamma_{n'n}^{r'r},z) \Phi
\een
definieren. Die Bedingungen an die Anzahl der Integrale $r$ bzw. $r'$ in
(\ref{G59a}) entstehen aus den sogenannten Fusionsregeln f"ur die minimalen
Modelle, siehe dazu \cite{Fe1,Mat}. Nach Satz \ref{S422} erf"ullen
$V_{n'n}^{r'r}$
\ben\label{G511}
\left[L_k,V_{n'n}^{r'r}(z)\right]=\left( z^{k+1} \frac{d}{dz} + (k+1)
h_{n'n} z^k\right) V_{n'n}^{r'r}(z),
\een
und es gilt $V_{n'n}^{r'r}(z): {\cal H}_{m'm}\lra {\cal
H}_{m'+n'-2r'-1,m+n-2r-1}$. Damit erf"ullt $V_{n'n}^{r'r}(z)$ die richtigen
Vertauschungsrelationen mit Vir zwischen Vir--Moduln vom richtigen Typ. Es
bleibt zu zeigen, da\3 wir von den Fock--R"aumen zu den irreduziblen
H"ochstgewichtsmoduln "ubergehen k"onnen. Dazu wird uns wieder der in
Abschnitt 4.1 eingef"uhrte Ladungsoperator n"utzlich sein. Sei dazu f"ur $\Phi
\in {\cal F}$
\ben\label{G512}
Q_{m'm}\Phi := \frac{e^{\pi i m\gamma_+^2 }-1}{ m (e^{\pi i \gamma_+^2}
-1)}Q(\gamma_+; m, -m')\Phi
\een
(zur Schreibweise siehe Bemerkung \ref{bem41}). Nach Konstruktion ist
$Q_{m'm}$ ein Intertwiner vom Grad $-m'm$ zwischen Fock--Moduln; es gilt
\[
Q_{m'm} : {\cal H}_{m'm}\lra {\cal H}_{m',-m}.
\]
$Q_{m'm}$ ist ein abschlie\3barer Operator, den Abschlu\3  bezeichnen wir
ebenfalls mit $Q_{m'm}$.

Wir betrachten nun die folgende Sequenz von Fock--Moduln
\ben\label{G513}
{\cal H}_{m'-p',p-m} \stackrel{Q_{m'-p',p-m}}{\strich\strich\lra} {\cal
H}_{m'-p',m-p} \simeq {\cal H}_{m'm}\stackrel{Q_{m'm}}{\strich\strich\lra}
{\cal H}_{m',-m}.
\een
Diese Sequenz kann in eine beidseitig unendliche Sequenz eigebettet werden,
wir ben"otigen hier aber nur diesen Ausschnitt.\\
Wir beweisen:
\begin{lem}\label{L51}
\begin{itemize}\item[(i)] $Q_{m'm} Q_{m'-p',p-m}=0$
\item[(ii)]\ben\label{G513a} B_{m'm} := \quot{\ker Q_{m'm}\cap {\cal
F}}{\im Q_{m'-p',p-m}\cap {\cal F}} \simeq \HH_{m'm}
\een
als Vir--Moduln. Es gilt $(B_{m'm}, \langle \cdot,\cdot\rangle_J) \simeq
(\HH_{m'm}, \langle \cdot,\cdot \rangle)$.
\end{itemize}
\end{lem}
{\bf Beweis.}\\
Aus Satz \ref{fockklasse} kennen wir die Struktur der Fock--Moduln. Da die
Kerne und Bilder von $Q$ Untermoduln sind, m"ussen sie von einer Teilmenge
der Vektoren der Diagramme aus Satz \ref{fockklasse} erzeugt werden. Wir
werden die folgenden Abbildungseigenschaften von $Q_{m'-p',p-m}$ und
$Q_{m'm}$ beweisen.
\ben\label{G514}
\parbox{28em}{
\setlength{\unitlength}{.008em}
\begin{picture}(3500,3100)
\multiput(450,450)(0,500){4}{\vector(-1,-1){400}}
\multiput(450,50)(0,500){3}{\vector(-1,1){400}}
\multiput(0,60)(0,1000){2}{\vector(0,1){390}}
\put(0,950){\vector(0,-1){390}}
\multiput(500,450)(0,1000){2}{\vector(0,-1){390}}
\multiput(500,560)(0,1000){2}{\vector(0,1){390}}
\put(0,0){\makebox(0,0){\small$\bar{w}_7$}}
\put(0,500){\makebox(0,0){\small$\bar{w}_5$}}
\put(0,1000){\makebox(0,0){\small$\bar{w}_3$}}
\put(0,1500){\makebox(0,0){\small$\bar{w}_1$}}
\put(500,00){\makebox(0,0){\small$\bar{w}_8$}}
\put(500,500){\makebox(0,0){\small$\bar{w}_6$}}
\put(500,1000){\makebox(0,0){\small$\bar{w}_4$}}
\put(500,1500){\makebox(0,0){\small$\bar{w}_2$}}
\put(500,2000){\makebox(0,0){\small$\bar{w}_0$}}
\multiput(1500,450)(0,1000){3}{\vector(0,-1){400}}
\multiput(1500,550)(0,1000){2}{\vector(0,1){400}}
\multiput(2000,50)(0,1000){2}{\vector(0,1){400}}
\multiput(2000,950)(0,1000){2}{\vector(0,-1){400}}
\multiput(1950,50)(0,500){5}{\vector(-1,1){400}}
\multiput(1950,450)(0,500){4}{\vector(-1,-1){400}}
\put(1500,0){\makebox(0,0){\small$w_9$}}
\put(1500,500){\makebox(0,0){\small$w_7$}}
\put(1500,1000){\makebox(0,0){\small$w_5$}}
\put(1500,1500){\makebox(0,0){\small$w_3$}}
\put(1500,2000){\makebox(0,0){\small$w_1$}}
\put(1500,2500){\makebox(0,0){\small$w_0$}}
\put(2000,0){\makebox(0,0){\small$w_{10}$}}
\put(2000,500){\makebox(0,0){\small$w_8$}}
\put(2000,1000){\makebox(0,0){\small$w_6$}}
\put(2000,1500){\makebox(0,0){\small$w_4$}}
\put(2000,2000){\makebox(0,0){\small$w_2$}}
\multiput(3000,950)(0,1000){2}{\vector(0,-1){390}}
\multiput(3000,60)(0,1000){2}{\vector(0,1){390}}
\multiput(3500,560)(0,1000){1}{\vector(0,1){390}}
\multiput(3500,450)(0,1000){2}{\vector(0,-1){390}}
\multiput(3450,50)(0,500){4}{\vector(-1,1){400}}
\multiput(3450,450)(0,500){3}{\vector(-1,-1){400}}
\put(3000,0){\makebox(0,0){\small$\tilde{w}_7$}}
\put(3000,500){\makebox(0,0){\small$\tilde{w}_5$}}
\put(3000,1000){\makebox(0,0){\small$\tilde{w}_3$}}
\put(3000,1500){\makebox(0,0){\small$\tilde{w}_1$}}
\put(3000,2000){\makebox(0,0){\small$\tilde{w}_0$}}
\put(3500,0){\makebox(0,0){\small$\tilde{w}_8$}}
\put(3500,500){\makebox(0,0){\small$\tilde{w}_6$}}
\put(3500,1000){\makebox(0,0){\small$\tilde{w}_4$}}
\put(3500,1500){\makebox(0,0){\small$\tilde{w}_2$}}
\multiput(600,0)(0,500){5}{\vector(1,0){800}}
\multiput(600,-10)(0,500){5}{\line(0,1){20}}
\multiput(2100,0)(0,500){5}{\vector(1,0){800}}
\multiput(2100,-10)(0,500){5}{\line(0,1){20}}
\put(250,2750){\makebox(0,0){${\cal H}_{m'-p',p-m}$}}
\put(1750,2750){\makebox(0,0){${\cal H}_{m'm}$}}
\put(3250,2750){\makebox(0,0){${\cal H}_{m',-m}$}}
\multiput(600,2750)(1500,0){2}{\vector(1,0){800}}
\put(1000,2850){\makebox(0,0){$Q_{m'-p',p-m}$}}
\put(2500,2850){\makebox(0,0){$Q_{m'm}$}}
\end{picture}
\setlength{\unitlength}{0.01em}}
\een
Wir bemerken zun"achst, da\3 die Gewichte der Vektoren nach (\ref{gewicht})
zusammenpassen, es gilt ${\rm wt}(\bar{w}_{2k})+ (p'-m')(p-m) ={\rm
wt}(w_{2k+1})$ u.s.w. Die Behauptung aus Diagramm (\ref{G514}) ist mit
anderen Worten:
\begin{itemize}
\item[(i)] $\ker Q_{m'-p',p-m}$ wird von $\bar{w}_{2k+1}\; (k\ge 0)$
erzeugt.
\item[(ii)] $\im Q_{m'-p',p-m}$ wird von $w_{2k+1}\; (k\ge1)$ erzeugt.
\item[(iii)] $\ker Q_{m'm}$ wird von $w_0$ und $w_{2k+1}\; (k\ge 0)$
erzeugt.
\item[(iv)] $\im Q_{m'm}$ wird von $\tilde{w}_0$ und $\tilde{w}_{2k+1} \;
(k\ge 0)$ erzeugt.
\end{itemize}
Damit folgt unmittelbar Behauptung (i)  und weiter
\ben
\quot{\ker Q_{m'm}}{\im Q_{m'-p',p-m}} \simeq \quot{[w_0]}{[w_1]} \simeq
\HH_{m'm}.
\een
Die Aussage "uber die Formen $\langle\cdot,\cdot\rangle_J$ und
$\langle\cdot,\cdot\rangle$ folgt unmittelbar aus der Vir--Invarianz
beider Formen. Zu zeigen bleibt nur (\ref{G514}).\\
Wir untersuchen zun"achst $Q_{m'-p',p-m} \bar{w}_0$ und $Q_{m'm}w_2$.
Analog zu Korollar \ref{K402} ist zu zeigen, da\3 diese Vektoren nicht Null
sind.  Korollar \ref{K402} l"a\3t sich aber nicht direkt anwenden, denn es
ist $\gamma_+^2=\frac{2p'}{p}\in \QQ$. Wir m"ussen also in (\ref{G405}),
(\ref{G406}) und (\ref{G407}) noch einmal etwas genauer hinschauen.
\begin{kor}[zu Lemma \ref{L401}]\label{k51}
Sei $\Phi$ wie im Beweis von Lemma \ref{L401} und $m'<0$. Dann gilt
\bea\label{G515}
\langle \Phi,Q_{-m',m} v_{m'm}\rangle &=& \langle v_{m',-m},Q_{m'm}
\Phi\rangle\nonumber\\
 &=& \frac{(2\pi i)^{m-1}}{m!} \frac{\Gamma(1+m\frac{\gamma_+^2}{2})}{
\Gamma(1+\frac{\gamma_+^2}{2})^m} \prod_{j=1}^{m} \frac{\sin(\pi j
\frac{\gamma_+^2}{2})}{\sin(\pi \frac{\gamma_+^2}{2}}.
\eea
Dieser Ausdruck ist insbesondere nicht Null, falls $1\le m\le p-1$ gilt.
\end{kor}
Damit folgt unmittelbar, da\3 $Q_{m'-p',p-m}\bar{w}_0$ ein singul"arer Vektor
in ${\cal H}_{m'm}$ ist und damit proportional zu $w_1$ sein mu\3. Weiter
mu\3 ein Vektor $w_2\in {\cal H}_{m'm}$ existieren mit
$Q_{m'm}w_2=\tilde{w}_0$.

Nun k"onnen wir uns in (\ref{G514})  auf folgende einfache Weise
weiterhangeln. Wir beschr"anken uns nun auf die Aussagen "uber
$Q_{m'-p',p-m} =: Q$,
die Aussagen "uber $Q_{m'm}$ folgen analog.

Da die Vektoren $\bar{w}_{4k+1}$ f"ur $k\ge0$ singul"ar sind, ist klar, da\3
$Q\bar{w}_{4k+1}=0\;(k\ge0)$ gelten mu\3, denn im Bild gibt es keine
singul"aren Vektoren mit den entsprechenden Graden.\\
Wir untersuchen nun $Q\bar{w}_2$. Aus $Q\bar{w}_2=0$ w"urde wegen $\bar{w}_0
\in [\bar{w}_2]$ $Q\bar{w}_0=0$ folgen, ein Widerspruch. Wir setzen $w_3=Q
\bar{w}_2$. (Erinnern wir uns dabei daran,  da\3 die Vektoren in diesen
Diagrammen nur bis auf "Aquivalenzklassen eindeutig sind; $Q\bar{w}_2$ ist
in der selben "Aquivalenzklasse wie $w_3$, da beide Vektoren singul"ar
werden, wenn wir den Quotienten mit $[w_1]$ bilden.)

Weiter gilt $w_5\in [w_3]$, es gibt demnach ein $L\in \frU(\Vir)$ mit
$w_5=Lw_3$. Da $Q$ ein Intertwiner ist, folgt weiter
\[
w_5=Lw_3= LQ\bar{w}_2 = QL\bar{w}_2.
\]
Wir setzen $\bar{w}_4:=L\bar{w}_2$. Au\3erdem folgt $Q\bar{w}_3=0$, denn im
Bild gibt es keinen Vektor mit entsprechenden Eigenschaften.\\
Diese Schritte induktiv fortgesetzt beweisen (\ref{G514}) und damit Lemma
\ref{L51}.\hfill$\Box$\\[5mm]
Die Sequenz (\ref{G513}) liefert damit eine Auf\/l"osung der irreduziblen
H"ochstgewichtsmoduln durch bestimmte Fock--Moduln. Als n"achsten Schritt
m"ochten wir zeigen, da\3 die von uns konstruierten konformen Felder
(\ref{G510}) auch auf $B_{m'm}$ wohldefiniert sind. Zun"achst
vervollst"andigen wir $B_{m'm}$, diese Vervollst"andigung k"onnen wir
identifizieren mit $\overline{B}_{m'm}:= \ker Q_{m'm} \ominus \overline{\im
Q_{m'-p',p-m}}$. Da $B_{m'm}$ aus dem Vakuumvektor in ${\cal H}_{m'm}\cap
{\cal F}$ durch die Virasoro--Algebra erzeugt wird und $L_n: {\cal F} \lra
{\cal F}$ gilt, ist ${\cal F}\cap B_{m'm}=B_{m'm}$ dicht in
$\overline{B}_{m'm}$. Wir werden zeigen, da\3 $V_{n'n}^{r'r}(z)$
Kettenabbildungen sind, d.~h. da\3 sie mit $Q$ vertauschen.
\begin{lem}\label{L53}
Es gilt  das folgende bis auf Phasen kommutative Diagramm. Sei
$l=m+n-2r-1$ und $l^{'}=m^{'}+ n^{'}-2r^{'}-1$.
\ben\label{G516}\parbox{32em}{
\xext=3200\yext=1100
\begin{picture}(\xext,\yext)
\setsqparms[1`1`1`1;900`900]
\putsquare(700,100)[{\cal H}_{m',-m+2p}`{\cal H}_{m'm}`{\cal
H}_{l',-l+2p}`{\cal H}_{l'l};Q_{m'-p',p-m}``V_{n'n}^{r'r}(z)`
Q_{l'-p',p-l}]
\putsquare(1600,100)[\phantom{{\cal H}_{m'm}}`{\cal H}_{m',-m}`\phantom{
{\cal H}_{l'l}}`{\cal H}_{l'l};Q_{m'm}``V_{n'n}^{r',n-r-1}(z)`Q_{l'l}]
\setsqparms[1`0`0`1;700`900]
\putsquare(2500,100)[\phantom{{\cal H}_{m',-m}}`\cdots`\phantom{{\cal
H}_{l'l}}`\cdots; Q_{m'm}```Q_{l',-l}]
\putsquare(0,100)[\cdots`\phantom{{\cal H}_{m',-m+2p}}`
\cdots`\phantom{{\cal
H}_{l',-l+2p}};Q_{m',2p-m}``V_{n'n}^{r'n-r}(z)`Q_{l',2p-l}]
\end{picture}
}\een
Genauer gilt f"ur $\Phi \in {\cal F}$
\ben\label{G517}
Q_{m',m+n-2r-1} V_{n'n}^{r'r}(z)\Phi =e^{\pi i \gamma_{n'n}\gamma_+
(m+n-2r-1)} V_{n'n}^{r',n-r-1}(z) Q_{m'm}\Phi.
\een
\end{lem}
Die  Phase in (\ref{G517}) k"onnen wir eliminieren, indem wir die Operatoren
$V_{n'n}^{r'r}(z)$ mit der Phase $\exp(\pi i \gamma_{n'n}\gamma_+
(r-m/2))$ multiplizieren.\\[3mm]
{\bf Beweis.}\\
Das Diagramm (\ref{G516}) beweisen wir in zwei Schritten, wir behaupten f"ur
$\Phi \in {\cal F}$
\ben\label{G5b1}
V_{n'n}^{r'r}(z) Q_{m'm} \Phi = V_{n'n}^{r',r+m}(z)\Phi
\een
und
\ben\label{G5b2}
Q_{m'm} V_{n'n}^{r'r}(z)\Phi = e^{\pi i m \gamma_{n'n}\gamma_+}
V_{n'n}^{r',r+m}(z) \Phi.
\een
Diese beiden Gleichungen implizieren (\ref{G517}) und damit (\ref{G516}).
Wir nehmen an, da\3 die beteiligten Operatoren durch (\ref{G423}) bzw. durch
(\ref{G425}) definiert sind, d.~h. da\3 $\Re \gamma_i\gamma_j >0$ f"ur alle
Exponenten gilt. Aus der G"ultigkeit von (\ref{G5b1}) und (\ref{G5b2}) in
diesem Fall folgt, da\3 auch die analytisch in den Exponenten fortgesetzten
Operatoren diese Gleichungen erf"ullen.

Das Produkt der Operatoren ist im Sinne von Abschnitt 4.3 zu verstehen.
Nach Lemma \ref{L41} k"onnen wir in (\ref{G5b1}) den Punkt $1$ als Start--
und Endpunkt der letzten Integration von $Q_{m'm}$ durch einen beliebigen
Punkt in der komplexen Ebene ersetzen. Wir w"ahlen $z$ als diesen Punkt; die
Integrationskonturen von $Q_{m'm}$ sollen innerhalb denen von
$V_{n'n}^{r'r}(z)$ liegen, um die radiale Ordnung zu gew"ahrleisten.
Genauso lassen wir die Konturen von $Q_{m'm}$ in (\ref{G5b2}) von au\3en
gegen $C_z$ schrumpfen.

Es reicht zu zeigen, da\3 (\ref{G5b1}) und (\ref{G5b2}) schwach auf $\cal F
\times F$ gelten. Nach Definition der Produkte der abgeschirmten
Vertex--Operatoren ist klar, da\3 die Integranden von (\ref{G5b1}) und
(\ref{G5b2}) bis auf die Normierungskonstante von $Q_{m'm}$ aus
(\ref{G512}) "ubereinstimmen; diese Konstante wurde eingef"uhrt, damit
(\ref{G5b1}) und (\ref{G5b2}) in der behaupteten Form stimmen. Wir
konzentrieren uns nun auf den Fall (\ref{G5b1}).

Sei  ${\bf w}=(z,u_1,\ldots,u_{r'},\ldots,u_{r'+r+m})$. Dann sind die
Matrixelemente von $V_{n'n}^{r'r}(z) Q_{m'm}$ und $V_{n'n}^{r',r+m}(z)$
endliche Summen von Integralen "uber
\ben\label{G5b3}
F_{\gamma_{n'n}}(\bgamma_{n'n}^{r',r+m};{\bf w})\prod_{i=1}^{r'+r+m}
u_i^{k_i} z^{k_0}
\een
gegeben. Der einzige Unterschied besteht in den Integrationskonturen, die
von $V_{n'n}^{r'r}(z) Q_{m'm}$ sind in Abbildung \ref{konto1} und die von
$V_{n'n}^{r',r+m}(z)$  sind in Abbildung \ref{konto2} angedeutet.
\begin{figure}[p]
\begin{center}\begin{picture}(1500,1900)
\put(-700,-700){\special{CS!m 0.7 kontur1.gem}}
\put(1100,1100){\makebox(0,0){\small$0$}}
\put(2150,1100){\makebox(0,0){\small$z$}}
\put(1000,2050){\makebox(0,0){\small$\vdots$}}
\put(1050,2000){\makebox(0,0)[l]{$\bigg\}$\small $l$--viele}}
\end{picture}\end{center}\caption{Die Konturen
$I_{l}(z)$}\label{konto1}\end{figure}
\begin{figure}[p]\begin{center}\begin{picture}(1500,2000)
\put(-700,-700){\special{CS!m 0.7 kontur2.gem}}
\put(1100,1100){\makebox(0,0){\small$0$}}
\put(2150,1100){\makebox(0,0){\small$z$}}
\put(1000,2080){\makebox(0,0){$\scriptscriptstyle\vdots$}}
\put(1050,2020){\makebox(0,0)[l]{$\bigg\}$\small $l$--viele}}
\put(1050,1640){\makebox(0,0)[l]{$\Big\}$\small $m$--viele}}
\put(1000,1680){\makebox(0,0){$\scriptscriptstyle\vdots$}}
\multiput(1680,1010)(50,0){8}{\line(25,0){25}}
\end{picture}\end{center}\caption{Die Konturen $I_{l,m}(z)$ (Die durch die
gestrichelte Linie verbundenen Punkte sind zu
identifizieren.)}\label{konto2}
\end{figure}
Wir haben damit zu zeigen
\bea\label{G5b4}
\lefteqn{\frac{e^{\pi i \gamma_+^2 m} -1}{m(e^{\pi i \gamma_+^2}-1)} \int
\cdots \int_{I_{r'+r,m}(z)} F_{\gamma_{n'n}}(\bgamma_{n'n}^{r',r+m},{\bf
w}) \prod_{i=1}^{r'+r+m} u_i^{k_i} du_i}\nonumber\\
 &=& \int \cdots\int_{I_{r'+r+m}(z)}
 F_{\gamma_{n'n}}(\bgamma_{n'n}^{r',r+m},{\bf w}) \prod_{i=1}^{r'+r+m}
u_i^{k_i} du_i.
\eea
Zum Beweis von (\ref{G5b4}) verwendet man dieselbe Technik, die wir schon
im Beweis von Lemma \ref{L401} nur angedeutet haben. Auf beiden Seiten
werden die Integrationskonturen auf die Gerade durch $0$ und $z$ gedr"uckt
und dann die Integrale zerlegt in Beitr"age "uber geordnete Konfigurationen
der Variablen $u_i$. Diese Beitr"age k"onnen bis auf eine Phase auf das
Integral "uber das ``Einheitssimplex'' zur"uckgef"uhrt werden. Die Phasen
k"onnen aufsummiert werden und liefern so den Zusammenhang zwischen den
Integralen, wie er in (\ref{G5b4}) behauptet wird.

Beim Beweis von (\ref{G5b2}) ist zus"atzlich noch zu beachten, da\3 auf der
linken Seite von (\ref{G5b2}) die Integrale nicht in der gleichen Ordnung
wie auf der rechten Seite stehen. Die Umordnung der Integrale auf der
linken Seite liefert genau den Phasenfaktor in
(\ref{G5b2}).\hfill$\Box$\\[5mm]
Mit Lemma \ref{L53} folgt insgesamt:
\begin{satz}\label{S54}
\begin{itemize}
\item[(i)] $\overline{B}_{m'm} = \ker Q_{m'm} \ominus \overline{\im
Q_{m'-p',p-m}}$ ist ein irreduzibler H"ochstgewichtsmodul, es gilt
$\overline{B}_{m'm}\cap {\cal F} \stackrel{\simeq}{\lra}
\quot{V(h_{m'm},c)}{M(h_{m'm},c)} = \HH_{m'm}$.
\item[(ii)] Die Operatoren $V_{n'n}^{r'r}(z)$ induzieren dicht definierte
Operatoren zwischen $\overline{B}_{m'm}$ und $\overline{B}_{l'l}$ mit
$l'=m'+n'-2r'-1,l=m+n+2r-1$, die wieder konforme Felder vom Gewicht
$h_{n'n}$ sind.
\end{itemize}
\end{satz}
{\bf Beweis.}\\
Aus Lemma \ref{L53} folgt, da\3
\[
V_{n'n}^{r'r}(z): \ker Q_{m'm} \cap {\cal F} \lra \ker Q_{l',l}
\]
gilt. Weiter folgt aus $\Phi\in \im Q_{m'm}\cap {\cal F}$, da\3 ein $\Psi
\in \cal F$ mit $\Phi=Q_{m'm}\Psi$ existiert. Es folgt demnach
\[
V_{n'n}^{r',n-r-1}(z)\Phi = Q_{l'l} V_{n'n}^{r'r}(z)\Psi \in \im Q_{l'l}.
\]
Damit induziert $V_{n'n}^{r'r}$ einen Operator
\[
\tilde{V}_{n'n}^{r'r}(z):\overline{B}_{m'm}\lra \overline{B}_{l'l}
\]
mit $D(\tilde{V}_{n'n}^{r'r}(z))=\overline{B}_{m'm} \cap {\cal F} = B_{m'm}$,
$\tilde{V}_{n'n}^{r'r}(z)$ sind also dicht definiert. Da $\overline{B}_{m'm}$
ein Untermodul ist, erf"ullen die Operatoren (\ref{G511}) stark auf
$B_{m'm}$, sie sind damit konforme Felder vom Gewicht
$h_{n'n}$.\hfill$\Box$\\[5mm]
Also sind
\bea
\varphi_{n'nm'm}^{l'l}(z) &:& \HH_{m'm} \lra \HH_{l'l} \mbox{ und
}\nonumber\\
V_{n'n}^{r'r}(z)&:& \overline{B}_{m'm} \lra \overline{B}_{l'l}\nonumber
\eea
f"ur $m^{(')}+n^{(')}-2r^{(')}-1=l^{(')}$ proportional. Die
Proportionalit"atskonstanten wurden in \cite{Fe1} angegeben. In \cite{DF1}
wurden die Strukturkonstanten berechnet, wir haben also unter Verwendung
dieser Ergebnisse die prim"aren Felder
\[
\Phi_{n'n}(z,\bar{z})=\sum_{m'nl'l}C_{n'nm'm}^{l'l} \varphi_{n'nm'm}^{l'l}
(z) \otimes \varphi_{n'nm'm}{l'l}(\bar{z})
\]
in
\[ \plus_{n'n} \HH_{n'n}\otimes \overline{\HH}_{n'n}
\]
explizit konstruiert. Bei der Bildung der konformen Bl"ocke (\ref{G56a})
ist zu beachten, da\3 wir  zur Berechnung der Korrelationen konformer Felder
die Form $\langle\cdot,\cdot\rangle_J$ verwenden m"ussen.
\newpage\thispagestyle{plain}\hbox{}\newpage
\newpage
\section*{Ausblick}
\addcontentsline{toc}{chapter}{Ausblick}
\setcounter{equation}{0}\stepcounter{chapter}
\mymark{Ausblick}
Ausgangspunkt dieser Arbeit waren die Arbeiten von V.~Dotsenko und
V.~Fateev \cite{DF1,DF2} und G.~Felder \cite{Fe1}, in denen die
sogenannten Feigin--Fuks--Integraldarstellungen f"ur die  Korrelationen
prim"arer Felder f"ur Vir--minimale Modelle gefunden wurden.

In diesen Arbeiten wird die ``Coulomb--Gas--Darstellung'' der prim"aren
Felder, d.~h. die Fock--Raum--Realisierung der prim"aren Felder "uber
abgeschirmte Vertex--Operatoren, auf einer formalen Ebene verwendet. In
\cite{DF1}
wird des weiteren der Unterschied zwischen irreduziblen
H"ochstgewichtsmoduln und den Fockmoduln der Virasoro--Algebra ignoriert.
Felder korrigierte dies durch seine Einf"uhrung des BRST--Operators $Q$,
die es erlaubt einen invarianten Unterraum in den Fock--Moduln zu finden,
den wir mit den irreduziblen Vir--Moduln identifizieren k"onnen.

Der Beweis dieser Eigenschaften von $Q$ verwendet tief\/liegende
Eigenschaften der Fock--Moduln, die wir wegen der Unzug"anglichkeit der
Originalliteratur \cite{FF} in Kapitel 2 noch einmal aufbereitet haben.

Ein Ziel dieser Arbeit war es, zu zeigen, da\3 sich die Behandlung der
beteiligten Operatoren auf mathematisch rigorose Weise durchf"uhren l"a\3t.
Dazu war es notwendig, Vertex--Operatoren im Fock--Raum zu untersuchen.
Mit Hilfe der Ergebnisse aus Kapitel 3 und 4 ist es  uns gelungen, die
prim"aren Felder der minimalen Modelle der konformen Quantenfeldtheorie im
Hilbertraum zu konstruieren.

Ein wesentliches Hilfsmittel, die Konstruktion der BRST--Symmetrie, also
von Fock--Raum Auf\/l"osungen der irreduziblen Vir--Moduln, ist inzwischen
in verschieden Arbeiten verallgemeinert worden. Hier sei insbesondere die
Arbeit von P.~Bouwknegt, J.~McCarthy und J.~Pilch \cite{BMP1} erw"ahnt, in
der Fock--Raum--Auf\/l"osungen f"ur alle Moduln mit $c\le1$ bewiesen wurden.
Der Unterschied zu dem von uns betrachteten Fall $III_-$ besteht im
wesentlichen darin, da\3 man die Integrationswege geeignet variieren mu\3,
um die Nichttrivialit"at der Intertwiner zu garantieren.\\
Die Autoren untersuchen allerdings nicht die Konstruktion der prim"aren
Felder, hier wird man i.~allg.  Satz \ref{S421} nicht anwenden k"onnen, da
die Exponenten die Vor\-aussetzung\-en nicht immer erf"ullen werden.

Eine andere Verallgemeinerung besch"aftigt sich mit den WZNW--Modellen, die
 konforme Quantenfeldtheorien sind, wobei dann die Symmetriealgebra ein
semidirektes Produkt einer Kac--Moody--Algebra und der Virasoro--Algebra
 ist. Die Observablen--Algebra ist dann von $T(z)$ und einem weiteren
Strom $J_a(z)$ erzeugt.

In \cite{BMP2} und \cite{Kur} werden Fock--Raum--Darstellungen von
Kac--Moody--Algebren, die Wakimoto--Moduln \cite{Wak} eingef"uhrt und
Intertwiner zwischen diesen Moduln mit Hilfe von Vertex--Operatoren
eingef"uhrt.  In den genannten Arbeiten werden die Vertex--Operatoren
allerdings nur "uber formale Potenzreihen definiert. Die dort verwendeten
Vertex--Operatoren haben gro\3e "Ahnlichkeit mit den von uns untersuchten,
insbesondere in Hinsicht auf die funktionalanalytischen Eigenschaften
(siehe \cite{Boe}), es sollte m"oglich sein die von uns bewiesenen
Eigenschaften auf die dort verwendeten Vertex--Operatoren zu
verallgemeinern.

Ein Punkt, der in dieser Arbeit "uberhaupt nicht auftaucht, ist die
St"orungstheorie von konformen Quantenfeldtheorien. Dies ist ein recht
neues und noch wenig homogenes Forschungsgebiet. In einer Arbeit von V.
Yurov und Al. Zamolodchikov \cite{YZ} wurde ein
Wechselwirkungs--Hamilton--Operator der Form
\ben\label{D1}
H=H_0 + \int_{|z|=\mbox{\scriptsize const.}} \Phi_{n'n}(z,\bar{z}) dz
\een
f"ur das konforme Modell $c=2/5$ untersucht, wobei $H_0=L_0+\bar{L}_0$ ist
und $\Phi$ ein prim"ares Feld ist. M. L"assig e.a. \cite{LMC} haben diese
Arbeit verallgemeinert und das Modell $c=7/10$ untersucht. Die Vermutung
der Physiker ist, da\3  die konforme Invarianz zwar zerst"ort wird, aber da\3
unter diesen St"orungen das gest"orte System integrabel ist. Beide
Autorengruppen legen die Vermutung nahe, (\ref{D1}) sei mit analytischer
St"orungstheorie zu behandeln. Wir haben zwar nicht die
Nichtabschlie\3barkeit der St"orung in (\ref{D1}) zeigen k"onnen, aber
immerhin die Nichtabschlie\3barkeit des Integranden der St"orung von
(\ref{D1}). Dieses und auch die Aussagen "uber den Definitionsbereich von
Vertex--Operatoren (wir k"onnen im wesentlichen  die St"orung auf
$D(e^{H_0})$ definieren) lassen zum jetzigen Zeitpunkt eine mathematische
Behandlung von (\ref{D1}) im Hilbertraum nicht zu. Als positives Ergebnis
sei hier die Arbeit von F. Constantinescu und R. Flume \cite{CF} erw"ahnt,
in der ein positiver Konvergenzradius der Gell--Mann--Low--Reihe f"ur das
Modell $c=7/10$ gest"ort durch $\Phi_{2,1}$ bewiesen wird. Was diese
Aussage allerdings f"ur die entsprechende St"orungstheorie f"ur (\ref{D1})
bedeutet, ist nicht klar.
\addcontentsline{toc}{chapter}{Literaturverzeichnis}
\mymark{Literaturverzeichnis}

\end{document}